\documentclass[12pt]{article}
\usepackage{xr-hyper}

\usepackage{changepage}

\usepackage[dvipsnames]{xcolor}
\usepackage[T1]{fontenc}
\usepackage[utf8]{inputenc}
\usepackage[english]{babel}
\usepackage{indentfirst}
\usepackage{arydshln}

\usepackage[normalem]{ulem}
\usepackage{amsfonts}
\usepackage{amssymb}
\usepackage{enumitem}
\usepackage{amsmath}
\usepackage{adjustbox}
\usepackage{adjustbox}
\usepackage{bbm}
\usepackage{mathtools}
\usepackage{caption}
\usepackage{subcaption}
\usepackage{relsize,exscale}
\usepackage{multirow}
\usepackage{booktabs}
\usepackage[nodisplayskipstretch]{setspace}
\usepackage{graphicx}
\usepackage{fullpage}

\usepackage{soul}
\usepackage{xspace}
\usepackage{csquotes}
\usepackage[margin = .9in]{geometry}
\usepackage{placeins}
\usepackage[footfont=normalsize,font=normalsize]{floatrow}
\usepackage{lscape}
\usepackage{mathabx}
\usepackage{accents}
\usepackage{float}
\usepackage{pgffor}
\usepackage{tikz}
\usetikzlibrary{calc,intersections}
\usepackage{xfrac}
\usepackage{bigints}

\usepackage[longnamesfirst]{natbib}
\definecolor{darkblue}{rgb}{0.0, 0.0, 0.55}
\definecolor{darkgreen}{rgb}{0.0, 0.2, 0.13}

\usepackage{hyperref}

\hypersetup{
	colorlinks=true,
	linkcolor=darkgreen,
	filecolor=magenta,
	urlcolor=darkgreen,
	citecolor = darkgreen
}
\usepackage{cleveref}

\usepackage{amsthm}

\newtheorem{theorem}{Theorem}[section]
\newtheorem{assumption}{Assumption}

\newtheorem{proposition}{Proposition}[section]
\newtheorem{lemma}{Lemma}[section]
\newtheorem{corollary}{Corollary}[section]
\newtheorem{algorithm}{Algorithm}[section]

\theoremstyle{definition}
\newtheorem{remark}{Remark}

\renewenvironment{proof}[1][Proof]{\noindent \textbf{#1.} }{\  \rule{0.5em}{0.5em}}

\usepackage{ragged2e}
\newlength\ubwidth

\usepackage[bottom]{footmisc}
\usepackage{xfrac}
\usepackage{bigints}
\allowdisplaybreaks
\expandafter\def\expandafter\normalsize\expandafter{%
	\normalsize%
	\setlength\abovedisplayskip{4pt}%
	\setlength\belowdisplayskip{4pt}%
	\setlength\abovedisplayshortskip{4pt}%
	\setlength\belowdisplayshortskip{4pt}%
}

\newcommand{\bracks}[1]{\left[#1\right]}
\newcommand{\expe}[1]{\mathbb{E}\bracks{#1}}

\newcommand\independent{\protect\mathpalette{\protect\independenT}{\perp}}
\def\independenT#1#2{\mathrel{\rlap{$#1#2$}\mkern2mu{#1#2}}}

\usepackage[utf8]{inputenc}
\usepackage[T1]{fontenc}
\usepackage{titlesec}
\newsavebox{\tablebox} \newlength{\tableboxwidth}
\numberwithin{equation}{section}

\titleclass{\subsubsubsection}{straight}[\subsection]

\newcounter{subsubsubsection}[subsubsection]
\renewcommand\thesubsubsubsection{\thesubsubsection.\arabic{subsubsubsection}}

\titleformat{\subsubsubsection}
  {\normalfont\normalsize\bfseries}{\thesubsubsubsection}{1em}{}
\titlespacing*{\subsubsubsection}
{0pt}{3.25ex plus 1ex minus .2ex}{1.5ex plus .2ex}
\makeatletter
\def\toclevel@subsubsubsection{4}
\def\l@subsubsubsection{\@dottedtocline{4}{7em}{4em}}
\makeatother
\title{ Was Javert right to be suspicious? Marginal Treatment Effects with Duration Outcomes\thanks{We thank the editor, Peter Hull, and three anonymous referees for excellent comments and suggestions. We also thank Xiaohong Chen, Jesse Bruhn, Paul Goldsmith-Pinkham, Peter Hull, Toru Kitagawa, Helena Laneuville, Elizabeth Luh, Andriy Norets, Jonathan Roth, Susanne Schennach, Henrik Sigstad, Alexander Torgovitsky, and Edward Vytlacil, seminar participants at Brandeis University, Brown University, Georgia Tech University, Southern Methodist University, Syracuse University, University of Chicago, University of Georgia at Athens and Yale University, and the institutional support of Associação Brasileira de Jurimetria. We thank Joana Getlinger for providing excellent research assistance.}}

\author{ Santiago Acerenza\thanks{Universidad ORT Uruguay. Email: \href{mailto:acerenza@ort.edu.uy}{acerenza@ort.edu.uy}} \and Vitor Possebom\thanks{Sao Paulo School of Economics - FGV. Email: \href{mailto:vitor.possebom@fgv.br}{vitor.possebom@fgv.br} } \and {Pedro H. C. Sant'Anna}\thanks{Emory University. Email: \href{mailto:pedro.santanna@emory.edu}{pedro.santanna@emory.edu}}
}

\begin{document}
\maketitle
\thispagestyle{empty}
\begin{abstract}
    We identify the distributional and quantile marginal treatment effect functions when the outcome is right-censored. Our method requires a conditionally exogenous instrument and random censoring. We propose asymptotically consistent semi-parametric estimators and valid inferential procedures for the target functions. To illustrate, we evaluate the effect of alternative sentences (fines and community service vs. no punishment) on recidivism in Brazil. Our results highlight substantial treatment effect heterogeneity: we find that people whom most judges would punish take longer to recidivate, while people who would be punished only by strict judges recidivate at an earlier date than if they were not punished.


\end{abstract}

\newpage

\setcounter{page}{1}
\begin{quote}
    \small \textit{To owe his life to a malefactor [...]; to be, in spite of himself, on a level with a fugitive from justice [...]; to allow it to be said to him, ``Go,'' and to say to the latter in his turn: ``Be free''; [...]---this was what overwhelmed him.}

    \begin{flushright}
    \emph{Les Misérables} by Victor Hugo
    \end{flushright}
\end{quote}

\section{Introduction}\label{Sintro}

Many relevant applications in the causal inference literature face two simultaneous identification challenges: endogenous selection into treatment and right-censored data. For example, in crime economics, when analyzing the effect of a punishment on defendants' time-to-recidivism, a researcher has to consider that judges observe more information than the econometrician when making their decisions and that time-to-recidivism is a right-censored variable \citep{Huttunen2020,Giles2021,Possebom2022,Lieberman2023}. A similar problem arises when analyzing the effect of rehabilitation programs on recidivism \citep{Alsan2024}. In labor economics, the same identification challenges appear when analyzing the effect of receiving unemployment benefits on unemployment spells \citep{Delgado2022}. In the health sciences, when studying the effect of a drug on survival time, a researcher has to address both identification problems too \citep{Sullivan1993}.

In this article, we are interested in addressing these two problems and, simultaneously, unpacking treatment effect heterogeneity with respect to unobserved, individual-specific resistance to treatment. Towards this end, we study the identification, estimation, and inference for the distributional marginal treatment effect ($DMTE$) and the quantile marginal treatment effect ($QMTE$) functions when the outcome variable is right-censored. In the crime economics example, these functions capture the distributional effects of receiving a punishment on time-to-recidivism for the defendant who is at the margin of being fined, conditional on the amount of evidence in her case.

By analyzing these distributional treatment effect parameters at different values of judge leniency, a researcher uncovers a detailed picture of how punishments heterogeneously affect recidivism. This picture can be used to design better sentencing criteria and/or train judges to follow a specific protocol. For example, suppose that one finds negatively sloped $QMTE$ functions with some positive and negative effects. This finding would suggest that defendants who would be punished even by very lenient judges---i.e., defendants with low unobserved punishment resistance---would take more time to recidivate as a result of the punishment (punishment is working as intended). On the other hand, defendants who would be punished only by very strict judges---i.e., defendants with high unobserved punishment resistance---would recidivate sooner than if they were not punished (punishment is not effective, perhaps because of scaring effects of a criminal record). Such degree of heterogeneity is usually washed out when using more standard summaries of treatment effects such as local average treatment effect (LATE) \citep{Imbens1994} or the IV estimand.\footnote{To be fair, we stress that LATE was not meant to highlight this degree of heterogeneity and that it has the advantage of only requiring binary instruments. However, when the empirical setting presents a continuous instrument, the definition of a ``complier'' is less clear than in the binary instrument case, potentially making the LATE results more challenging to interpret formally.} Thus, $DMTE$ and $QMTE$ parameters can be used to unpack treatment effect heterogeneity and provide additional economic insights.

Our methodological results highlight that extending the marginal treatment effect ($MTE$) framework of \citet{Heckman2006} to deal with a duration variable subject to right-censoring introduces some interesting challenges depending on the censoring mechanism. For instance, if censoring is independent of potential outcomes, we can point-identify the distributional marginal treatment effect and quantile marginal treatment effect functions for some, but not necessarily all, support points and quantiles. Nonparametrically identifying the entire $DMTE$ and $QMTE$ functions is only possible if the support of the censoring variable is at least as large as the support of the duration outcome. Thus, in practice, in the presence of censoring, the conditions for point-identification of (average) $MTE$ parameters are arguably too restrictive in many applications.\footnote{When this support restriction is not satisfied, one can nonetheless nonparametrically point-identify truncated $MTE$ functions, which are still well-defined causal parameters. See Appendix \ref{AppAverage} for more details.} As we discuss in detail,  working with $QMTE$ and $DMTE$ side-step these issues. We propose semiparametric estimators and inference procedures for the $DMTE$ and $QMTE$ functions and establish their large sample properties.

We also discuss two potential avenues to handle cases where censoring possibly depends on the potential outcomes. First, we leverage a negative regression dependence between potential outcomes and censoring variables, which can be justified when defendants commit fewer crimes over time. Second, we discuss how one can continuously relax the independent censoring assumption. Both strategies, which do not impose exogenous censoring, lead to partial identification of the causal parameters and are discussed in Appendix \ref{AppPartialID}.

As highlighted above, explicitly handling right-censored outcomes introduces new econometric challenges. One question that naturally arises is whether empirical researchers can bypass (or ignore) these challenges and use standard causal inference tools. In the crime economics example, one common way to avoid such challenges is to restrict sample construction and focus on recidivism within a given time frame, say two years. This essentially changes the outcome of interest from time-to-recidivism to whether one recidivates within 2 years. The censoring issue is avoided if one follows \emph{all} defendants for at least two years. Although this is convenient and generically valid, this procedure has drawbacks: (a) it changes the question of interest by changing the outcome variable, and (b) the choice of the cutoff (two years in this example) is arbitrary. For instance, it may be that punishment has no effect on recidivism within two years but then has an effect within three years (or one year).

These concerns are minor if one is interested only in recidivism within a known time frame. One potential way to assess whether this is the case is to ask ourselves: if censoring were not a concern, would we use time-to-recidivism or recidivism within two years as the outcome? If the answer is only the latter, then standard practice is justified. If not, we caution researchers that the conclusions of their analysis may be sensitive to the cutoff used for ``binarization'' of the time-to-recidivism outcome, and that directly using the time-to-recidivism outcome may yield different conclusions. See Appendix \ref{AppIllustration} for a simple example of this.

If the conclusions of a study might be sensitive to the cutoff used for binarization, a natural next step would be for empirical researchers to consider a handful of cutoffs and demonstrate that the results are ``robust'' to the cutoff. We note that this happens often in practice. However, when doing this, researchers often ensure that the sample used for the entire analysis is not contaminated by compositional changes, which can lead to a loss of power at shorter horizons. To be more concrete, suppose that researchers considered three thresholds for the ``binarization'': two, three, and five years. To ensure that the same defendants are analyzed across all transformed outcomes, researchers commonly restrict the sample to those they have observed in the data for at least 5 years. As a result, they drop available observations for shorter horizons, leading to a loss of statistical power. In contrast, our proposed method enables researchers to maximize the use of their data by leveraging longer observation periods for earlier-observed individuals and shorter observation periods for later-observed individuals. This advantage increases as we include more later-observed individuals in our sample, since it allows us to use more observations.

This discussion leads to the next natural question: what if we considered a large set of cutoffs and did not restrict the sample across cutoffs? Heuristically, one can interpret our $DMTE$ results as doing precisely that: avoiding choosing arbitrary cutoffs and considering recidivism within $y$ periods for a continuum of $y\in \mathbb{R}_+$. Our $QMTE$ results ``transform'' our $DMTE$ results so the underlying treatment effects are expressed in the same units as the time-to-recidivism outcome, leading to additional insights. Here, though, it is worth stressing that we explicitly tackle the censoring problem when considering the continuum of cutoffs by adapting the \citeauthor{Frandsen2015}'s (\citeyear{Frandsen2015}) reweighing approach to our context. Erroneously ignoring the censoring problem can indeed lead to misleading conclusions.

We show the appeal of our causal inference tools by evaluating the effect of fines and community service sentences (possibly combined with a ``conviction marker'') as a form of punishment on time-to-recidivism in the State of São Paulo, Brazil, between 2010 and 2019.\footnote{São Paulo is the largest state in Brazil, with a population above 44 million people according to the Brazilian Census in 2022. Moreover, analyzing the impact of judicial policies on criminal behavior in this state is relevant due to its relatively high criminality. For example, according to São Paulo Public Safety Secretary, there were 12.56 murders, 1261.95 thefts, and 498.82 robberies per 100,000 inhabitants in 2023. Importantly, theft is one of the most common crimes in our sample.} Our treated group (punished group) contains the defendants who were fined or sentenced to community services due to either a conviction or a non-prosecution agreement, and our untreated group (unpunished group) contains defendants who were acquitted or whose cases were dismissed.\footnote{Our sample only contains cases whose punishment must be a fine or community service sentence. In 1998, a new law established that criminal charges whose maximum prison sentence is less than four years in the 1940 Criminal Law Code must be punished with a fine or a community service sentence from that point onwards. As we are particularly interested in the effect of fines and community service sentences, we focus on these specific criminal cases and define them as misdemeanor offenses.} To measure recidivism, we check whether the defendant's name appears in any criminal case within the sample period after the final sentence's date. More precisely, our outcome variable is the time between the final sentence and a subsequent criminal case. Since the sampling period is finite, the outcome variable is right-censored.

To deploy our proposed methodology, we need a continuous instrumental variable since we face endogenous selection into punishment.\footnote{{ If the researcher is comfortable with parametric assumptions, it is possible to use a discrete instrument as suggested by \citet{Brinch2017}.}} We use the trial judge's leave-one-out rate of punishment (or ``leniency rate'') as an instrument for the trial judge's decision \citep{Bhuller2019,Agan2021}. Importantly, this instrumental variable is continuous with large support and is independent of the defendant's counterfactual criminal behavior because judges are randomly assigned to cases conditional on court districts according to state law in São Paulo. Our outcome data --- time-to-recidivism --- is right-censored by construction, requiring a methodology that accounts for this identification challenge.

Empirically, we find that the cross-district average QMTE functions for $.10, .15, .25, .40, .50$, and $.75$ quantiles are heterogeneous with respect to unobserved punishment resistance. The treatment effects are estimated to be sometimes positive and sometimes negative. More precisely, we find that people who would be punished by most judges (those with low punishment resistance) take longer to recidivate as a consequence of the punishment, while people who would be punished only by strict judges (high punishment resistance) recidivate at an earlier date than if they were not punished. This result suggests that designing sentencing guidelines that encourage strict judges to become more lenient could increase time-to-recidivism.

Lastly, we compare our results with methods that ignore that time-to-recidivism is right-censored. If one used the typical judge-fixed-effect regression, one would find that treatment increases the likelihood of short-term recidivism. If one used IV quantile regressions (ignoring censoring), one would find that treatment effects are slightly negative. In either case, the researcher would not be able to highlight heterogeneity as in the $DMTE$ and $QMTE$ functions, which show that treatment benefits some defendants while harming others. These differences highlight that our tools can indeed bring new insights to policy discussions. 

\textbf{Related literature:} This article contributes to different branches of literature. Concerning its theoretical contribution, our work contributes to the literature on MTE by extending the MTE framework of \citet{Heckman2006} and \citet{carneirolee2009} to a setting with right-censored data.
We also contribute to the literature on duration outcomes; see, e.g., \citet{Frandsen2015}, \citet{tchetgen2015instrumental}, \citet{Santanna2021}, \citet{Delgado2022}. None of these papers consider MTE-type parameters as we do. Among these, the closest work to ours is \citet{Frandsen2015}, which considers the case where the censoring variable is observed and shows how one can identify distributional and quantile local treatment effects, assuming that censoring is exogenous. Our results can be interpreted as an extension of \citet{Frandsen2015} to the MTE framework, possibly allowing for endogenous censoring. 

Concerning its empirical contribution, our work is inserted in the literature about the effect of fines and community service sentences on future criminal behavior; see, e.g., \citet{Huttunen2020}, \citet{Giles2021}, \citet{Possebom2022}, and \citet{Lieberman2023}.\footnote{This literature focuses on non-incarceration punishments for individuals who are already being prosecuted. The reader who is interested in the effect of alternative solutions to prosecution may check the recent work developed by \citet{Agan2021} and \citet{ShemTov2024}. Readers interested in incarceration's effect may check the recent work of \citet{Rose2021}, \citet{Humphries2023} and \citet{Kamat2023}.} They all focus on binary variables indicating recidivism within a pre-specified period. Within these, as we build on his dataset, \citet{Possebom2022} is the closest to ours. However, his focus differs greatly from ours, and he does not handle duration outcomes as we do.

\textbf{Organization of the paper:} Section \ref{Squestion} defines the causal parameters of interest. Section \ref{Sframework} presents our model, discusses our identifying assumptions, and provides our identification results with a right-censored outcome variable. Section \ref{Sestimation} explains how to semiparametrically estimate the objects necessary to implement the identification strategy described in the previous section. Furthermore, Section \ref{Sempirical} discusses the empirical context, the data, and our results. Lastly, Section \ref{Sconclusion} concludes. This paper also contains an online supporting appendix.

\section{Causal Questions of Interest}\label{Squestion}

In this section, we define our causal questions of interest in terms of treatment effect parameters. To make them concrete and intuitive, we use our empirical application as a running example. In this applied exercise, we study the effect of alternative sentences in the form of fines and community service on time-to-recidivism in the state of São Paulo, Brazil. 

For each observation $i$, let $Y_i^*(1)$ be the potential outcome if treated and let $Y_i^*(0)$ be the potential outcome if untreated. In our empirical example, these unobservable variables capture, respectively, the potential time-to-recidivism if defendant $i$ were punished with a fine or community service either due to a conviction or a non-prosecution agreement, and the potential time-to-recidivism if defendant $i$ were not punished with a fine or community service either due to acquitall or dismissal.\footnote{Alternatively, a researcher might be interested in analyzing the effect of alternative sentences on total crime counts or severity-weighted offenses. We focus on time-to-recidivism because this variable facilitates the understanding of the censoring issues that are the focus of our proposed identification method. Importantly, total crime counts or severity-weighted offenses must be measured within a sample period (e.g., total crime counts within 2 or 3 years) and, consequently, also suffer from censoring. Adapting our censoring-focused methods to encompass these more complex outcome variables is an interesting future line of research. Additionally, a researcher might be interested in ``true criminal behavior (even if not observed by the police)'' as an outcome variable instead of ``criminal behavior observed by the court system''. We discuss this case in Appendix \ref{AppLeeBounds}.} Defendant $i$'s treatment effect is therefore $\theta_i = Y_i^*(1) - Y^*_i(0)$. Ideally, we would like to learn $\theta_i$ for all defendants. However, that is very challenging (if not impossible) when we allow for (a) heterogeneous treatment effects across defendants and (b) whether a defendant is punished or not being related to $\theta_i$ (``essential heterogeneity'' as defined by \citet{Heckman2006}). 

Due to these challenges, it is common for researchers to focus on aggregated summary measures of $\theta_i$, such as the average treatment effect among ``compliers'' \citep{Imbens1994}, defined as $LATE = \expe{Y^*(1) - Y^*(0)|\text{Compliers}}$ (see, e.g., \citealp{Agan2021,Bhuller2019,Huttunen2020}).\footnote{As discussed before, these papers use a different outcome of interest $Y^*$ that bypass the censoring issues we face. However, we can ignore these censoring issues while discussing our causal questions of interest (as this does not play a prominent role in it).} Although interesting and policy-oriented, such aggregated measures of causal effects are unsuitable for highlighting important types of treatment effect heterogeneity. In particular, these parameters cannot answer whether defendants with high punishment resistance (i.e., defendants who would only be punished by very strict judges) would, on average, take longer to recidivate if they were punished. The same goes for defendants with lower punishment resistance. These are the exact types of causal questions that interest us in this paper. We want to go beyond LATE-type parameters and provide a more detailed picture of how alternative punishments heterogeneously affect recidivism with respect to the defendant's (unobserved) punishment resistance, which we denote by $V$.\footnote{In general, our punishment resistance variable, $V$, is commonly called unobserved treatment resistance or latent cost of treatment.} Here, punishment resistance may capture the evidence gathered against the defendant and additional defendant-specific characteristics.

{
	
	One can measure the causal effect of fines and community services on time-to-recidivism for defendants with a given punishment resistance using the notion of distribution and quantile treatment effects.\footnote{In Appendix \ref{AppAverage}, we also explore how to define and identify types of average treatment effects. Importantly, censoring causes technical issues when defining parameters based on average effects, and we propose two alternative solutions that overcome those challenges.}  All these causal parameters build on the Marginal Treatment Effects framework of \citet{Heckman2006} and can be used to answer complementary policy-relevant questions (\citealp{Heckman2006}, \citealp{carneirolee2009}).  We now carefully define and interpret them.
	
	For treatment status $d \in \{0, 1 \}$, let the distributional and quantile marginal treatment response functions be defined as
	\begin{eqnarray}
		DMTR_{d}\left(y, v\right) &\coloneqq& \mathbb{P}\left[\left.Y^*(d) \leq y \right\vert V = v\right],\label{EqDMTR} \\
		QMTR_{d}\left(\tau, v\right) &\coloneqq& \inf\{y\in \mathbb{R}_+ \colon \mathbb{P}\left[\left.Y^*(d) \leq y \right\vert V=v\right]\geq \tau \}, \label{EqQMTR}
	\end{eqnarray}
	where $y \in \mathbb{R}_+$, $\tau \in (0,1)$ and $v \in \left[0, 1\right]$. All these counterfactual parameters have a clear interpretation. For instance, $DMTR_{d}\left(y, v\right)$ gives the proportion of defendants with punishment resistance $v$ who would have already recidivated after $y$ periods since the court's final ruling if they were punished ($d=1$) or not ($d=0$). Analogously, $QMTR_{d}\left(\tau, v\right)$ provides the $\tau$-th quantile of the time-to-recidivism under treatment $d$, among defendants with punishment resistance $v$.
	
	Based on these counterfactual objects, it is straightforward to define the Distributional and Quantile Marginal Treatment Effect functions:
	\begin{eqnarray}
		DMTE\left(y,v\right) &\coloneqq& DMTR_{1}\left(y,v\right) - DMTR_{0}\left(y,v\right), \label{EqDMTE}\\
		QMTE\left(\tau, v\right) &\coloneqq& QMTR_{1}\left(\tau, v\right) - QMTR_{0}\left(\tau, v\right).\label{EqQMTE}
	\end{eqnarray}
	
	Positive values of the QMTE function indicate that punishment by fines and community services increases the defendant's time-to-recidivism compared to no punishment (so treatment is working as intended). On the other hand, positive values of the DMTE function indicate that punishment by fines and community services increases the proportion of defendants who recidivate by time $y$ compared to no punishment (so treatment is not working as intended). For policy effectiveness in our context, positive values of QMTE are ``good'', while negative values of DMTE are ``good''.
	
	\begin{remark}
		When analyzing the impact of judicial decisions on recidivism, many papers focus on distributional marginal treatment effects for a small set of  values of $y$.\footnote{See, e.g., \citet{Agan2021}, \citet{Bhuller2019}, \citet{Giles2021}, \citet{Huttunen2020}, and \citet{Possebom2022}.} Here, we entertain the possibility of moving beyond small set of horizons by considering a continuous set of cutoffs $y$'s or by focusing on different target parameters, such as the quantile marginal treatment effects on time-to-recidivism. See Appendix \ref{AppIllustration} for a simple illustration of the appeal of our approach compared to the traditional ``small-set-of-horizons'' approach.
	\end{remark}
	
}

\section{Econometric framework and identification results} \label{Sframework}
We face some challenges in identifying the causal parameters of interest described in the previous section. As usual, potential outcomes are only (potentially) observed under one treatment status, i.e., $Y^{*}_i  = Y^{*}_i(1) \cdot D_i + Y^{*}_i(0) \cdot \left(1 - D_i\right)$, where $D_i$ is the treatment indicator variable. Furthermore, our outcome of interest is subject to right-censoring, implying that we do not always observe $Y^{*}$ but rather observe $Y_i = \min \{Y^*_i, C_i\}$, where $C$ is the censoring variable. In the case of draws, we assume that $Y^*_i$ happens before $C_i$, as is customary in survival analysis. Finally, we also expect that treatment statuses are related to the potential outcomes and possibly related to the censoring variable.

To tackle all these issues, we build on the MTE framework of \citet{Heckman2006} and extend it to handle duration outcomes. 
Toward this end, we consider a threshold-crossing treatment selection model
\begin{align}
	\label{EqTreatment} D & = \mathbf{1}\left\lbrace P\left(Z, C\right) \geq V \right\rbrace,
\end{align}
where $Z$ is an observed instrumental variable (with support $\mathcal{Z} \subset \mathbb{R}$), $C$ is an observed censoring variable (with support $\mathcal{C}\subset \mathbb{R}_+$), and $V$ is a latent heterogeneity variable that captures the unobserved treatment resistance. The function $P: \mathcal{Z} \times \mathcal{C} \rightarrow \mathcal{P}$ is unknown and captures the willingness to take the treatment for each value of $Z$ and $C$. Importantly, our treatment selection model \eqref{EqTreatment} imposes monotonicity \citep{Vytlacil2002}.

To better understand Equation \eqref{EqTreatment}, let us go back to our empirical context and explain each component of it. Our instrumental variable $Z$ is a measure of the trial judge's leniency, which arguably does not affect time-to-recidivism other than through the judge's decision to punish or not. Our censoring variable $C$ captures the time between the defendant's sentence date and the end of our sampling period, and it is observed for all defendants. $C$ can also capture seasonality patterns, as it is fully determined depending on the sentence's date.  The function $P$ captures the trial judge's punishment criteria, and it allows trial judges to update their punishment criteria over time \citep{Bhuller2022}, as it includes $C$ as an argument.\footnote{Since the end of the sample period is the same for all defendants, we can equivalently express the decision rule for $D$ in terms of the sentencing date $S_{i}$, where the sentencing date equals the fixed calendar date when the sample period ends $\left(\overline{T}\right)$ minus the censoring variable $C_{i}$, i.e., $S_{i} = \overline{T} - C_{i}$.  This equality implies that the propensity score can be equivalently expressed as a function of the sentencing date since $P(Z_i,C_i)=P(Z_i,T - S_i) \equiv \tilde{P}(Z_i,S_i)$. This way of writing the propensity score function highlights that trial judges may update their punishment criteria over time in our setting.}\textsuperscript{,}\footnote{We emphasize that we allow function $P$ to depend on the censoring variable, but we do not impose that the censoring variable has an impact on this function. This type of connection is testable through the first-stage regression.} Finally, the variable $V$ can be interpreted as unobserved punishment resistance, and it captures, among other things, the amount of criminal evidence in the defendant's favor. The higher the $V$, the less likely the defendant will be punished, all else equal. As already discussed, $Y^{*}$ captures the length of time between the defendant's sentence date and her next criminal case's starting date, and $Y$ is the minimum of $Y^*$ and time from the sentence's date to the end of our sampling period, $C$.

\subsection{Assumptions}
In our setup, the available data for the researcher are $\left\{Y_i,C_i,D_i,Z_i\right\}_{i=1}^{n}$, while $Y^{*}_i\left(0\right)$, $Y^{*}_i\left(1\right)$, $Y^{*}_i$ and $V_i$ are latent variables. Henceforth, we assume that $\left\{Y_i,C_i,D_i,Z_i\right\}_{i=1}^{n}$ are independently and identically distributed as $\left(Y,C,D,Z\right)$. For simplicity, we drop exogenous covariates from the model and focus on the case with a single instrument. All results derived in the paper hold conditionally on covariates and can be extended to the case with multiple instruments.\footnote{In our empirical application, the conditioning covariates are a full set of court district indicators.} Since we deal with a time-to-event outcome, $Y$ is non-negative by construction.

{
	In what follows, we present a set of five assumptions (Assumptions \ref{AsIndependence}-\ref{AsCensoring}) that will allow us to point-identify the DMTE and the QMTE functions across a range of thresholds and quantile points. These assumptions are related to those imposed by \citet{Heckman2006} and \citet{Frandsen2015} and involve assuming that censoring is not related to the potential outcomes $Y^*(d)$. 
	
	We now state our five assumptions and contextualize each of them to our empirical setup.
	
}
\begin{assumption}[Random Assignment]\label{AsIndependence}
	Conditional on C, the potential outcomes $Y^{*}\left(0\right)$, $Y^{*}\left(1\right)$ and $V$ are independent of the instrument $Z$, i.e., $\left. Z \independent \left(Y^{*}\left(0\right), Y^{*}\left(1\right), V\right) \right\vert C.$
\end{assumption}

Assumption \ref{AsIndependence} is an exogeneity assumption and is common in the literature about instrumental variables with censored outcomes \citep{Frandsen2015}. In our empirical application, this assumption holds conditional on the court district because trial judges are randomly assigned to cases within each court district. { Importantly, judges are assigned to criminal cases based on a computer algorithm that creates a lottery of judges and there is no suspicion in the press that this software is manipulated.}

Note also that Assumption \ref{AsIndependence} allows the instrument to depend on the censoring variable. In our empirical application, this flexibility is useful because the trial judge's punishment rate may depend on the case's sentence date if judges who entered the Judiciary more recently are more lenient than judges who retired at the beginning of our sampling period, for example.

\begin{assumption}[Propensity Score is Continuous]\label{AsRank}
	Conditional on $C$, $P(z,c)$ is a nontrivial function of $z$ and the random variable $\left. P\left(Z,c\right) \right\vert C = c$ is absolutely continuous in $Z$, with support given by an interval $\mathcal{P} \coloneqq \left[\underline{p}, \overline{p}\right]$ for any $c \in \mathcal{C}$.\footnote{The assumption that $\mathcal{P}$ is an interval is made for notational simplicity. All the proofs can be easily extended to the case where $\mathcal{P}$ is a set with a non-empty interior.}
\end{assumption}

Assumption \ref{AsRank} is a rank condition, intuitively imposing that the instrument is locally relevant. In addition, we implicitly assume that the support of the propensity score does not vary with the value of $C$. In our application, this implicit assumption is plausible because the judges are mostly the same over time. Furthermore, the judge's lenience rate has a good amount of variation { since we observe 525 judges who use many different punishment criteria (Figure \ref{FigPS})}.\footnote{{ Since our main identification result is fully nonparametric, Assumption \ref{AsRank} requires sufficient (continuous) variation in the instrument, which induces continuous variation in the propensity score, which generates the variation we need on the outcomes to identify the effects for every value of $V$. Nevertheless, our proposed estimation procedure is semi-parametric and relies on series approximations for the propensity score. If one treats the series as fixed (meaning that the first stage is parametric), our approach allows for discrete instruments \citep{Brinch2017}. The extrapolation and interpolation induced by the parametric second stage permit $Z$ (and $P$) to vary discretely while still being used to identify the effects of interest.}}

\begin{assumption}[$V$ is continuous]\label{AsContinuous}
	The distribution of the latent heterogeneity variable $V$ conditional on $C$ is absolutely continuous with respect to the Lebesgue measure.
\end{assumption}

Assumption \ref{AsContinuous} is a regularity condition that allows us to normalize the marginal distribution of $\left. V \right\vert C$ to be the standard uniform. Consequently, we can write $P\left(z, c\right) = \mathbb{P}\left[\left. D = 1 \right\vert Z = z, C = c\right]$ for any $z \in \mathcal{Z}$ and $c \in \mathcal{C}$. 

\begin{assumption}[Overlap]\label{AsPositive}
	Conditional on $C$, all treatment groups exist, i.e., $\mathbb{P}\left[\left. D = d \right\vert C = c \right] \in \left(0, 1\right)$ for any $d \in \left\lbrace 0, 1 \right\rbrace$ and any $c \in \mathcal{C}$.
\end{assumption}

Assumption \ref{AsPositive} is a regularity condition about overlap. It imposes that there is strictly positive mass in both treatment groups for every value of the censoring variable.

{
	
	\begin{assumption}[Random Censoring]\label{AsCensoring}
		The censoring variable is independent of the uncensored potential outcomes given the latent heterogeneity $V$, i.e., $\left. C \independent \left(Y^{*}\left(0\right), Y^{*}\left(1\right)\right) \right\vert V.$
	\end{assumption}
	
	Assumption \ref{AsCensoring} is an exogeneity assumption and is common in the literature about duration outcomes \citep{Frandsen2015,Delgado2022}. When combined with Assumption \ref{AsContinuous}, Assumption \ref{AsCensoring} implies that $C$ is unconditionally independent of the uncensored potential outcomes, i.e., $C \independent \left(Y^{*}\left(0\right), Y^{*}\left(1\right)\right)$. In our empirical application, this restriction imposes that the case's sentence date is independent of the defendant's decision to commit another crime in the future, { i.e., potential recidivism is stationary.}
	
	Importantly, Assumption \ref{AsCensoring} requires that controlling for $V$ accounts for all sources of endogeneity coming through the censoring variable. This assumption can be restrictive, as endogeneity may persist even after controlling for latent heterogeneity. For example, in our empirical setting, potential recidivism may not be stationary if legislation or inputs to the production of Justice change over time. These inputs may include the number of police officers, judges, or public defenders. Appendix \ref{AppRobustness} has a detailed discussion about these inputs in our empirical application.
	
	If the researcher believes that Assumption \ref{AsCensoring} is too strong in a particular application, she can use alternative assumptions that are sufficient to partially identify the distributional marginal treatment effect and some quantile marginal treatment effects when the outcome variable is right-censored. We discuss two alternative partial identification strategies in Appendix \ref{AppPartialID}.\footnote{Moreover, Appendix \ref{AppAltAs} discusses the costs of not imposing restrictions on the censoring mechanism, and Appendix \ref{AppModelJustification2} cautions about simply treating the censoring variable as an additional covariate.}

}

\subsection{Identification}\label{Sidentification}

We present our point-identification results that rely on Assumptions \ref{AsIndependence}-\ref{AsCensoring}. 

First, define $\gamma_{d}(y,v,c) = \dfrac{d}{d v}\mathbb{P}\left[\left. Y\leq y, D = d \right\vert P\left(Z,C\right) = v, C = c\right]. $
We now state our main identification result: point-identification of the $DMTR$ functions. 
\begin{proposition}\label{PropDMTR}
	Suppose that Assumptions \ref{AsIndependence}-\ref{AsCensoring} hold. Then, for any $d \in \left\lbrace 0, 1 \right\rbrace$, $y < \gamma_{C}$ and $v \in \mathcal{P}$, $DMTR_{d}\left(y, v\right) = \left(2 d - 1\right) \cdot \expe{\gamma_{d}(y,p,C) | P(Z,C) = v, C > y},$
	{ where $\gamma_{C} \coloneqq \inf\left\lbrace c \in \overline{\mathbb{R}} \colon \mathbb{P}\left[C \leq c\right] = 1 \right\rbrace$ is the upper bound of the support of the censoring variable $C$.}
\end{proposition}
\begin{proof}
	See Appendix \ref{AppProofPropDMTR}.
\end{proof}

The above proposition shows how we can point-identify the distributional marginal treatment response for a given unobserved treatment resistance $v$. It involves first taking the derivative of the conditional joint distribution of the realized outcome $Y$ and treatment status $D$ given the propensity score $P=v$ and the censoring variable being above $y$ ($C = y + \delta$ for $\delta > 0$) with respect to $v$, and then integrating over all values $\delta > 0$ such that the $y+\delta$ remains in the support of the censoring variable $C$. Differently from the results in \citet{carneirolee2009}, we need to tackle the right-censoring problem, which manifests in our results by having to condition on $C=y+\delta$, so that $C > y$, and then integrating over $\delta$.

Furthermore, our results are specific to the $DMTR_d(y,v)$ function, and not for a generic transformation of $Y^*(d)$, say $G(Y^*(d))$ as in \citet{carneirolee2009}. This follows from the fact that we may not be able to identify the $DMTR_d(y,v)$ over all values of $y$ in the support of $Y^*(d)$, as a consequence of the censoring problem. Having said that, there are several functions that we can actually nonparametrically point-identify without additional restrictions and under standard regularity conditions, including the QMTE functions for a range of quantiles. We state these results{, which use only Assumptions \ref{AsIndependence}-\ref{AsCensoring} and a regularity condition, as a corollary for convenience. The proof is a direct consequence of the previous proposition and the definition of quantiles.
	
	\begin{corollary}\label{CorQMTE}
		Suppose that Assumptions \ref{AsIndependence}-\ref{AsCensoring} and Assumption \ref{AsRC4} listed in the Appendix \ref{Ssemiparaconsistency} hold. Then, the $QMTE\left(\tau,v\right)$ function (Equation \eqref{EqQMTE}) is point-identified for any $v \in \mathcal{P}$ and $\tau \in \left(0, \overline{\tau}\left(v\right)\right)$, where $\overline{\tau}\left(v\right) \coloneqq \min\left\lbrace \overline{\tau}_{0}\left(v\right), \overline{\tau}_{1}\left(v\right) \right\rbrace$ and $\overline{\tau}_{d}\left(v\right) \coloneqq DMTR_{d}\left(\gamma_{C}, v\right)$ for any $d \in \left\lbrace 0, 1 \right\rbrace$.
	\end{corollary}
	
}
Notice that the right-tail of the $DMTR_d(\cdot, v)$ may be differentially affected by the censoring problem, implying that $\overline{\tau}_{1}\left(v\right)$ may be different from $\overline{\tau}_{0}\left(v\right)$. As a consequence, we can only identify the $QMTE\left(\tau,v\right)$ over the common range of identified quantiles among treated and untreated units.

\section{Estimation and inference}\label{Sestimation}

In this section, we provide an algorithm on how to semiparametrically estimate the DMTE and QMTE functions based on the identification results described in Proposition \ref{PropDMTR} and Corollary \ref{CorQMTE}. Section \ref{SecCourtDistricts} first discusses that, in our empirical setting, judges are randomly assigned within court districts and then explains the consequences of this mechanism for our identification and estimation procedures. Afterwards, we provide practical and formally justified estimation and inference procedures based on a semi-parametric class of estimators for the nuisance functions (Section \ref{SestSemiPara}).\footnote{We refer the reader to Appendix  \ref{SecGenericEst} for a generic procedure to estimate the marginal treatment effect functionals while remaining agnostic about the type of estimators used to estimate the nuisance functions.} Lastly, Section \ref{AppCovariateAggregation} discusses how to aggregate our target parameters across covariate values.

\subsection{Conditioning on Covariates: Judges are randomly assigned within court districts}\label{SecCourtDistricts}

All the results derived in Section \ref{Sframework} could be interpreted as conditional on covariates. This possibility is particularly relevant in our application since random judge assignment is legally guaranteed within the court district where the crime occurred. Thus, conditioning on a full set of dummy variables indicating each one of the court districts in São Paulo is fundamental in our empirical setting.

Consequently, we want to estimate court-district-specific $DMTE$ and $QMTE$ functions capturing the effect of alternative sentences on time-to-recidivism. Since non-parametric estimation in the presence of covariates can be demanding due to the curse of dimensionality, we propose an easy-to-implement semiparametric estimator in Section \ref{SestSemiPara}.

However, in our empirical setting, there are 193 court districts in the State of São Paulo. Consequently, our semi-parametric procedure estimates 193 $DMTE$ functions and 193 $QMTE$ functions. To facilitate interpretation of these functions, we propose summary functions that aggregate across covariate values. We do so using the proportion of cases per court district as weights and explain the details of this procedure as well as its asymptotic validity in Section \ref{AppCovariateAggregation}.

\subsection{Semiparametric estimation and inference procedures}\label{SestSemiPara}

This section provides a specific procedure for semiparametrically estimating marginal treatment effect functionals. We have data on $\left\lbrace Y_{i}, C_{i}, D_{i}, Z_{i}, X_{i}' \right\rbrace_{i = 1}^{n}$.\footnote{Appendix \ref{AppSemiparametricdetails} explains the technical details behind our estimation procedure.} In the context of our application, $X$ is a set of 193 court district indicators.

Similar to \citet{carneirolee2009}, we model $P(Z,C,X) \coloneqq \expe{D|Z,C,X}$ using an additive partially linear series regression
\begin{equation}\label{EqPartiallinear}
	P(Z,C,X) = {\alpha}_{0} + {X'\alpha}_{X} + C {\alpha}_{C} + \varphi(Z),
\end{equation}
where $(\alpha_0, \alpha_X', \alpha_C)'$ are unknown finite-dimensional parameters, and $\varphi$ is an unknown (infinite-dimension) function. In our context, the partially linear additive specification  (Equation \eqref{EqPartiallinear}) allows one to pool information from different court districts and run a single propensity score model for all courts.\footnote{Alternatively, one could use a semiparametric logit model. In Appendix \ref{AppMainRegularity}, we show that our regularity conditions hold for this model as well. The same is true for a fully nonparametric series estimator.} 

We pick a polynomial basis function to approximate $\varphi(\cdot)$, $\psi^L(z) = \left(z, z^2, \dots, z^L\right)'$, though other options, such as B-splines, are also possible. Note that all the series coefficients can be estimated via ordinary least squares to obtain an estimate $(\widehat{\alpha}_0, \widehat{\alpha}_X', \widehat{\alpha}_C, \widehat{\alpha}_Z))'$  and compute:\footnote{See Appendix \ref{AppPS} for details.}
\begin{equation}\label{EqLogit_OLS}
	\widetilde{P}_i  = \widetilde{P}(Z_i,C_i,X_i) = \widehat{\alpha}_{0} + X_{i}'\widehat{\alpha}_{X} +  C_{i}\widehat{\alpha}_{C} + \psi^L(Z_i)' \widehat{\alpha}_Z.
\end{equation}
Since, in finite samples, $\widetilde{P}_i$ might be negative or larger than one, we follow \citet{carneirolee2009} and use the trimmed version of $\widetilde{P}_i$ as our estimator, 
\begin{equation}\label{EqLogit_trimmed}
	\widehat{P}_i  = \widetilde{P}_i + (1 - \epsilon - \widetilde{P}_i) \cdot \mathbf{1}\{ \widetilde{P}_i>1\} + (\epsilon - \widetilde{P}_i) \cdot \mathbf{1}\{ \widetilde{P}_i<0\} ,
\end{equation}
for a sufficiently small $\epsilon$.\footnote{ Our application uses $\epsilon = 0.01$, though this is not material as only one observation is trimmed.}

Next, we move into the estimation of the conditional distribution function of $Y \cdot \mathbf{1}\left\lbrace D = d \right\rbrace$ given $P, C, X$ for $d \in \{0,1\}$. Here, we impose the following model:\footnote{See Appendix \ref{AppCDF} for additional details.} 
\begin{eqnarray}
	{\Gamma}({P},C,X;y,d) &\coloneqq& \mathbb{E}\left[ \mathbf{1}\left\lbrace Y \leq y, D = d \right\rbrace\vert P, C,X\right] \nonumber \\
	&=&\Lambda \left(	\beta_0 \left(y,d\right) +X^{\prime }\beta_{X}(y,d) + C \beta_C(y,d) + P \beta_P (y,d)\right) 
	\text{ a.s.}  \label{EqDistReg}
\end{eqnarray}
where $\theta_0(\cdot, \cdot) = (\beta_0 \left(\cdot,\cdot\right), \beta_X \left(\cdot,\cdot\right)', \beta_C \left(\cdot,\cdot\right), \beta_P \left(\cdot,\cdot\right))' \mapsto \Theta
\subseteq \mathbb{R}^{3+k_X}$ is a vector of nonparametric functions, $k_X$ is the dimension of $X$, and $\Lambda$ is a known link function.\footnote{This class of distribution regression models nests and extends many traditional duration models such as the proportional hazard model and the accelerated time model. See \citet{Delgado2022} for a discussion.} For concreteness,  we focus on a logistic link function, $\Lambda(\cdot) = \exp(\cdot)\big{/}(1 + \exp(\cdot))$.

To estimate these unknown functions, we first need to acknowledge that the propensity score $P_i$ is not observed. However, we can use the ``generated regressor'' $\widehat{P}_i$ from Equation \eqref{EqLogit_trimmed}. Once we replace $P_i$ with $\widehat{P}_i$, we can then leverage the insights of \citet{Foresi1995} and \citet{Chernozhukov2013a} by noticing that, for a fixed $y$ and $d$, Equation \eqref{EqDistReg} is a binary regression problem. Consequently, we can pointwise estimate these parameters by maximizing the (feasible) conditional likelihood function and obtain a distribution regression estimate of $\theta_{0}(y,d)$ denoted by $\hat{\theta}\left( y,d\right)$.\footnote{ 
	Computing the distribution regression estimators for many $(y,d)$ points only requires running a sequence of binary regressions.}

Next, note that the derivative of $\Gamma$ (Equation \eqref{EqDistReg}) with respect to $P$ can computed in closed-form for each $(y,d)$, 
\begin{eqnarray}
	\gamma_{d}(y,v,c,x) &\coloneqq& \dfrac{d}{d v}{\Gamma}(v,c,x;y,d) = \beta_{P}(y,d) \cdot {\Gamma}(v,c,x;y,d)(1 - {\Gamma}(v,c,x;y,d)),\label{Eq_gamma}
\end{eqnarray}
where we explored that $\Lambda$ is the logistic link function. Denote the estimated fitted values of $\gamma_{d}(y,v,c,x)$ by 
\begin{eqnarray}
	\widehat{\gamma}_{d}(y,v,c,x) = \widehat{\beta}_{P}(y,d) \cdot \widehat{\Gamma}(v,c,x;y,d)(1 - \widehat{\Gamma}(v,c,x;y,d)),\label{EqEstimated-Derivatives}
\end{eqnarray}
where the distribution regression coefficients are the components of $\hat{\theta}\left( y,d\right)$ , and 
\begin{eqnarray*}
	\widehat{\Gamma}(v,c,x;y,d) = \dfrac{\exp\left(\widehat{\beta}_0 \left(y,d\right) +x^{\prime }\widehat{\beta}_{X}(y,d) + c \widehat{\beta}_C(y,d) + v \widehat{\beta}_P(y,d)\right)}{1 + \exp\left(\widehat{\beta}_0 \left(y,d\right) +x^{\prime }\widehat{\beta}_{X}(y,d) + c \widehat{\beta}_C(y,d) + v \widehat{\beta}_P(y,d)\right)}. 
\end{eqnarray*}

Using Equation \eqref{EqEstimated-Derivatives}, we can estimate $DMTR_d (y,v,x) \coloneqq  \mathbb{P}\left[\left.Y^*(d) \leq y \right\vert V = v, X=x\right].$ To do so, let $n_{d,x,y} = \sum_{i=1}^n \mathbf{1}\{D_i=d, X_i=x, C_{i} > y \}$ denote the sample size with treatment status $d$, covariate value $x$, and censoring variable above $y$. Our proposed estimator for $DMTR_d (y,v,x)$ is given by\footnote{Since $\widehat{DMTR}_d (y,v,x)$ is an estimator for a conditional distribution, it needs to be non-decreasing in $y$ for all $(d,v,x) \in \{0,1\} \times \mathcal{P} \times \mathcal{X}$. However, this may not be the case in finite samples. We recommend using the rearrangement procedure of \citet{Chernozhukov2009b}.} 
\begin{eqnarray}
	\widehat{DMTR}_d (y,v,x) = (2d - 1) \dfrac{\sum_{i=1}^n \mathbf{1}\{D_i=d, X_i=x, C_{i} > y \}~ \widehat{\gamma}_{d}(y,v,C_i,x)}{n_{d,x,y}}. \label{EqDMTR_d_estimator}
\end{eqnarray}

Based on Equation \eqref{EqDMTR_d_estimator}, we can then estimate $DMTE(y,v,x) \coloneqq DMTR_1(y,v,x) - DMTR_0(y,v,x)$ using \begin{eqnarray}
	\widehat{DMTE}(y,v,x) \coloneqq \widehat{DMTR}_1(y,v,x) - \widehat{DMTR}_0(y,v,x). \label{DMTE_semi}
\end{eqnarray}
Analogously, one can estimate $QMTE(\tau,v,x)$ functionals using 
\begin{eqnarray}
	\widehat{QMTE}(\tau,v,x) &=& \widehat{QMTR}_1(\tau,v,x) - \widehat{QMTR}_0(\tau,v,x), \label{QMTE_semi}
\end{eqnarray}
where $\widehat{QMTR}_d(\tau,v,x) = \inf\{y\in \mathbb{R}_+ \colon \widehat{DMTR}_d(y,v,x)\geq \tau \}$.

We summarize all these estimation steps in Algorithm \ref{algo:semi} in Appendix \ref{AppAlgo}.

The next theorem establishes the large-sample properties of our proposed estimators. We defer all the regularity conditions to the appendix to streamline the presentation. Let $\overline{\tau}\left(v,x\right) \coloneqq \min\left\lbrace \overline{\tau}_{0}\left(v,x\right), \overline{\tau}_{1}\left(v,x\right) \right\rbrace$ and $\overline{\tau}_{d}\left(v,x\right) \coloneqq DMTR_{d}\left(\gamma_{C}, v,x\right)$ for any $d \in \left\lbrace 0, 1 \right\rbrace$.

\begin{theorem}\label{Thm:semi}
	Suppose that Assumptions \ref{AsIndependence}-\ref{AsCensoring} and Assumptions \ref{AsSP1}-\ref{AsRC4} listed in Appendices \ref{AppMainRegularity} and \ref{AppAdditionalRegularity} hold. Then, as $n \rightarrow \infty$,
	
	\begin{enumerate}
		\item[(a)] for each fixed $y < \gamma_{C}$, $v \in \mathcal{P}$, and $x \in \mathcal{X}$, $\sqrt{n}\left( \widehat{DMTE}(y,v,x) - {DMTE}(y,v,x) \right)$ ${\overset{d}{\rightarrow}} N(0,V_{y,v,x}^{dmte}),$
		with $\widehat{DMTE}(y,v,x)$ in Equation \eqref{DMTE_semi} and $V_{y,v,x}^{dmte}$ in Appendix \ref{AppProofVariance}.
		\item[(b)] for each fixed $ \tau \in (0,\overline{\tau}\left(v,x\right))$, $v \in \mathcal{P}$, and $x \in \mathcal{X}$, $\sqrt{n}\left( \widehat{QMTE}(\tau,v,x) - {QMTE}(\tau,v,x) \right)$ ${\overset{d}{\rightarrow}} N(0,V_{\tau,v,x}^{qmte}),$
		with $\widehat{QMTE}(\tau,v,x)$ in Equation \eqref{QMTE_semi} and $V_{\tau,v,x}^{qmte}$ in Appendix \ref{AppProofVariance}.
	\end{enumerate}
\end{theorem}

Theorem \ref{Thm:semi} follows from first deriving the influence function of the $DMTR$ functions (Equation \eqref{EqDMTR_d_estimator}), paying particular attention to quantifying the estimation effect arising from replacing the true propensity score with the estimated one. After this step, all the results follow from the functional delta method and the continuous mapping theorem. The proof strategy is similar to \citet{Rothe2009}.

Although Theorem \ref{Thm:semi} indicates that one can potentially conduct inference using plug-in estimates of the variance, this procedure would involve estimating additional nuisance functions and could be cumbersome in practice. To avoid this issue, we propose using a weighted bootstrap procedure as in \citet{Ma2005} and \citet{Chen2009}. This bootstrap procedure is very straightforward to implement and the details can be found in Algorithm \ref{algo:bootstrap} in Appendix \ref{AppAlgo}.

\subsection{Covariate Aggregation}\label{AppCovariateAggregation}
In our empirical setting, there are 193 court districts in the State of São Paulo. Consequently, our semi-parametric procedure estimates 193 $DMTE$ functions and 193 $QMTE$ functions. To facilitate interpretation of these functions, we propose summary functions that aggregate across covariate values.

To do so, we aggregate the court-district-specific DMTE functionals across court districts using the proportion of cases per court district as weights.\footnote{Another possible covariate aggregation and its limitations are discussed in Appendix \ref{AppCovariateAggregationextra}.} Let $w_{x} = \mathbb{P}(X=x)$ denote the probability of a covariate $X$ taking the value $x$, which, in our case, denotes the true proportion of cases assigned to a court district $x$. Let $\widehat{w}_{x} = n^{-1}\sum_{i=1}^n \mathbf{1}\{X_i=x\}$ be the plug-in estimator of $w_{x}$.

For each $d \in \{0,1\}$, $y \in \mathcal{Y}$ and $v\in \mathcal{P}$, let $$DMTR_d^{avg}(y,v) = \expe{DMTR_d (y,v,X)} = \sum_{x\in \mathcal{X}}w_x~DMTR_d (y,v,x).$$

Analogously, let $DMTE^{avg}(y,v) = DMTR_1^{avg}(y,v) - DMTR_0^{avg}(y,v)$ and $QMTE^{avg}(\tau,v) = QMTR_1^{avg}(\tau,v) - QMTR_0^{avg}(\tau,v)$, where $    QMTR_{d}^{avg}\left(\tau, v\right) \coloneqq \inf\{y\in \mathbb{R}_+ \colon DMTR_d^{avg}(y,v)\geq \tau \}. $ 
All these functionals can be straightforwardly estimated using functionals of $ \widehat{DMTR}_d^{avg}(y,v) = \sum_{x\in \mathcal{X}}\widehat{w}_x~\widehat{DMTR}_d (y,v,x),$ with  $\widehat{DMTR}_d (y,v,x)$ as in Equation \eqref{EqDMTR_d_estimator}, just like in Equations \eqref{DMTE_semi}-\eqref{QMTE_semi}. Their large-sample properties follow from the delta method and are summarized in the following corollary.

\begin{corollary}\label{cor:semi}
	Suppose that Assumptions \ref{AsIndependence}-\ref{AsCensoring} and Assumptions \ref{AsSP1}-\ref{AsRC4} listed in Appendices \ref{AppMainRegularity} and \ref{AppAdditionalRegularity} hold. Then,  as $n \rightarrow \infty$,
	
	\begin{enumerate}
		\item[(a)] for each fixed $y < \gamma_{C}$, and  $v \in \mathcal{P}$, $\sqrt{n}\left( \widehat{DMTE}^{avg}(y,v) - {DMTE}^{avg}(y,v) \right) {\overset{d}{\rightarrow}} N(0,V_{y,v}^{dmte, {avg}}).$
		\item[(b)] for each fixed $ \tau \in (0,\overline{\tau}\left(v,x\right))$, and $v \in \mathcal{P}$, $\sqrt{n}\left( \widehat{QMTE}^{avg}(\tau,v) - {QMTE}^{avg}(\tau,v) \right)$ ${\overset{d}{\rightarrow}} N(0,V_{\tau,v}^{qmte, avg}).$
	\end{enumerate}
\end{corollary}

It is also straightforward to construct a weighted-bootstrap confidence interval for these functionals by using $\widehat{w}^*_{x} = n^{-1}\sum_{i=1}^n \omega_i~\mathbf{1}\{X_i=x\}$ as weights for the MTE functionals, where $\omega_{i}$ is defined in Algorithm \ref{algo:bootstrap}. We omit a detailed description to avoid repetition.

\section{Effect of alternative sentences on time-to-recidivism}\label{Sempirical}

Our empirical application answers the question: ``Do alternative sentences such as fines and community service impact time-to-recidivism in São Paulo, Brazil?''. To do so, we start by explaining our empirical context and data in Subsection \ref{Scontext}. Then, Subsection \ref{SVarInterest} defines the variables of interest and argues that long-run recidivism is a relevant problem in Brazil. Lastly, Subsection \ref{Sresults} presents the results of our empirical analysis using our proposed tools. Subsection \ref{Sresults} also compares our methods against more traditional methods, highlighting the ability of our proposed tools to unpack treatment effect heterogeneity. For the interested reader, we assess the plausibility of our identifying assumptions in Appendix \ref{Sdescriptive}, and provide robustness checks in Appendices \ref{AppRobustness} and \ref{AppCaseProcessingTimeMainResults}.

\subsection{Empirical Context and Data}\label{Scontext}

To study the effect of alternative sentences in the form of fines and community service on time-to-recidivism, we collect data from all criminal cases brought to the Justice Court System in the State of São Paulo, Brazil, between January 4\textsuperscript{th}, 2010, and December 3\textsuperscript{rd}, 2019.\footnote{See Appendix \ref{AppData} for an overview of the data-construction.} According to a Brazilian law from 1998, criminal charges whose maximum prison sentence is less than four years in the 1940 Criminal Law Code must, from that year onwards, be punished with a fine or a community service sentence if the defendant is found guilty. As we are particularly interested in the effect of these alternative sentences, we focus on these specific criminal cases { and define them as misdemeanor offenses}. We also restrict our sample to cases that started between 2010 and 2017. Based on these restrictions, the most common types of crime in our sample are theft and domestic violence.

There are 332 court districts in the state of São Paulo. Criminal complaints are analyzed by a trial judge working at the court with geographic jurisdiction over the location of the alleged offense.\footnote{{ In Brazil, instead of being elected, judges are appointed for life based on their performance in a civil service exam and frequently serve as judges until retirement \citep{Laneuville2024breaking}. Furthermore, they make decisions about conviction, sentence type and sentence intensity in all cases but ``crimes against life'' (murder, attempted murder, manslaughter, incentivizing or assisting suicide, and abortion). Importantly, our sample does not contain ``crime against life'' cases. Therefore, all our cases are entirely decided by appointed judges.}} Moreover, there are 862 trial judges during our sample period. We keep 642 judges who analyzed more than 20 cases out of these. In court districts that have more than one judge, the case is randomly allocated to one of the judges { according to a computer algorithm that creates a lottery of judges}. Of the 332 court districts in our sample, 193 have more than one judge who analyzed more than 20 cases. Given that our econometric procedure explores the random allocation of judges to criminal cases and their different leniency levels, we restrict our attention to court districts with more than two judges who analyzed 20 cases or more. After imposing these two restrictions, our sample has 525 trial judges from 193 different court districts, handling 43,468 cases in total.\footnote{{ We treat each case-defendant pair as a separate observation. Consequently, defendants involved with more than one case will appear more than once in our dataset. Appendix \ref{AppRepeatedOffenders} analyzes the question of repeated offenders in detail.}} Appendix \ref{AppDescriptiveCourtJudge} discusses the distribution of judges per court district and cases per judge.

\subsection{Defining the Variables of Interest}\label{SVarInterest}

In our dataset, we observe the defendant's full name, the defendant's court district, the case's starting date, the assigned trial judge's full name, the case's final ruling, and the case's final ruling's date. All our variables of interest will be constructed from these pieces of information. Henceforth, let $X$ denote the full set of court district dummies, which will play the role of covariates in our analysis.

Let us start with our treatment variable, $D$, which denotes the final ruling in the case. Defendants who were fined or sentenced to community services because they were either convicted or signed a non-prosecution agreement according to the final ruling in their case belong to our treatment group, $D=1$. Defendants who were acquitted or their cases were dismissed according to the final ruling in their case belong to our comparison group, $D=0$.

Our outcome of interest, $Y^*$, is the ``time-to-recidivism'', i.e., the number of days it takes for a defendant to appear in court once again after the case's final ruling's date. Here, note that our outcome of interest is a duration variable and that some defendants may not recidivate by the end of our sampling period, though they may recidivate later. Putting it simply, we do not always observe $Y^*$, but rather observe a right-censored version of $Y^*$: $Y=min(Y^*, C)$, where $C$ is a right-censoring variable.\footnote{Appendix \ref{AppDescriptiveOutcome} presents summary statistics about our outcome variable.} In our context, $C$ is the follow-up period for each defendant, i.e., the number of days from their case's final ruling date to December 3\textsuperscript{rd}, 2019. 

Besides the censoring problem, it is important to be explicit about how we define recidivism. In this paper, a defendant $i$ in a case $j$ recidivated by the end of our sample if and only if defendant $i$'s full name appears in a case $\bar{j}$ whose starting date is after case $j$'s final sentence's date.\footnote{To match defendants' names across cases, we follow the same procedure as in \citet{Possebom2022} and define a fuzzy match if the similarity between full names in two different cases is greater than or equal to 0.95 using the Jaro–Winkler similarity metric. Appendix \ref{AppNames} discusses the number of words per defendant's name.}\textsuperscript{,}\footnote{In this manuscript, our definition of recidivism is being prosecuted for another crime. In Appendix \ref{AppLeeBounds}, we discuss how to combine the methods proposed here with the methods proposed by \cite{Bartalotti2021} to identify the effect of judicial decisions on committing a crime even if the police are not able to observe all crimes. A similar issue arises when defendants move to other states and commit crimes in different jurisdictions. Even though these out-of-state recidivism events are not captured by our outcome variable, we do not believe they represent a relevant concern because out-migration in São Paulo is low, accounting for less than 2\% of the state's population according to data from the 2022 Census.} Then, we measure our outcome variable as the number of days between case $j$'s final ruling's date and case $\bar{j}$'s starting date.\footnote{Case $\bar{j}$ can be about any type of crime, including more severe crimes whose maximum sentence is over four years, while case $j$ has to be about a crime whose maximum sentence is at most four years.} If defendant $i$ did not recidivate by the end of the sampling period, then $Y = C$.

At this stage, it is important to stress that we are not adopting a more restrictive notion of ``short-run'' recidivism based on a fixed period, say two years, which could potentially allow us to ``ignore'' the censoring problem. Instead, we focus on time-to-recidivism directly, which, in our view, entails some important advantages. For instance, we do not need to pick a threshold to define (short-run) recidivism arbitrarily. Doing so may lead to potentially sensitive conclusions, as illustrated in an example in Appendix \ref{AppIllustration}. 

\begin{figure}[p]
	\begin{center}
		\includegraphics[width = .55\textwidth, keepaspectratio]{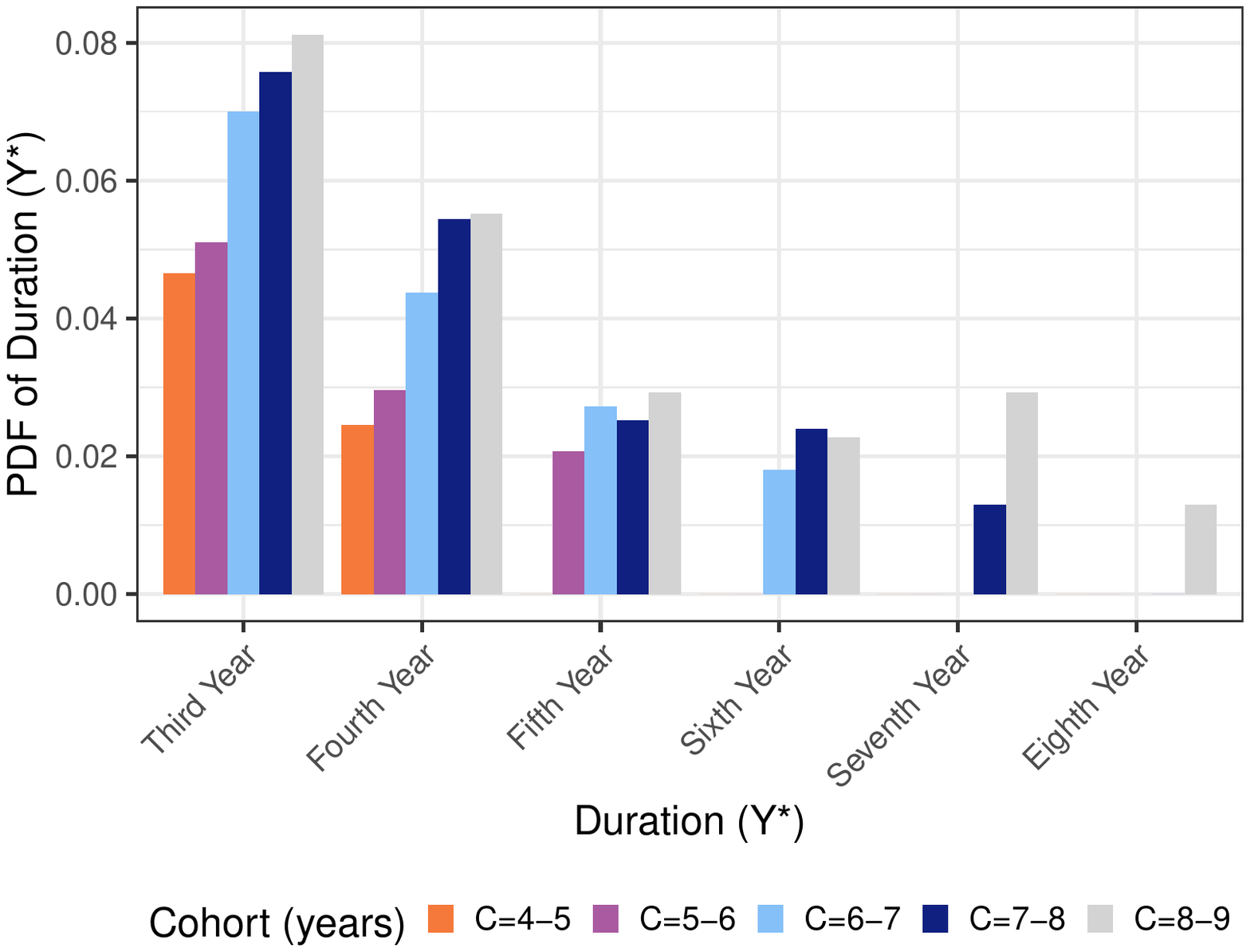}
		\caption{PDF of the Uncensored Outcome given the Defendant's Cohort}
		\label{FigPDF}
	\end{center}
	\justifying
	{Notes: This figure shows the right tail of the probability mass function of the uncensored potential outcome ($Y^{*}$) given cohorts based on the censoring variable, $\mathbb{P}\left[\left. y_{1} \leq Y^{*} \leq y_{2} \right\vert C\right]$ where $y_{1}$ and $y_{2}$ define the bins indicated in the x-axis. In particular, these conditional PDFs are evaluated at six bins of the uncensored potential outcome. For example, ``Third Year'' denotes that the first recidivism event occurred between 730 days ($ = y_{1}$) and 1095 days ($ = y_{2}$), while ``Fourth year'' denotes that the first recidivism event occurred between 1095 days ($ = y_{1}$) and 1460 days ($ = y_{2}$). Each color denotes a different cohort: orange denotes defendants who are observed for at least four years and at most five years during our sampling period, purple denotes defendants who are observed for at least five years and at most six years, light blue denotes defendants who are observed for at least six years and at most seven years, dark blue denotes defendants who are observed for at least seven years and at most eight years, and gray denotes defendants who are observed for at least eight years and at most nine years.  The y-axis denotes the value of the PDF.}
\end{figure}

However, if almost all defendants who recidivate do it in the short run, then focusing on short-run measures would be sufficient. But this is an empirical matter and should be handled as such. To assess if this is the case in our data, Figure \ref{FigPDF} displays estimates of the right tail of the probability distribution function (PDF) of the uncensored potential outcome ($Y^{*}$) among cohorts defined based on the censoring variable. These descriptive results reveal that, in the case of the state of São Paulo, a non-negligible share of defendants have their first recidivism event in their fifth, sixth, or seventh year after their sentence's date, implying that analyzing long-term recidivism is practically relevant. Consequently, we must tackle the censoring problem directly. See Appendix \ref{AppModelJustification} for additional motivations for leveraging time-to-recidivism as a key outcome of interest from a welfare maximizer decision-maker perspective.

As it will be clear in the next sections, our causal inference procedures leverage the availability of an instrumental variable $Z$ with large support. In our context, the instrument $Z$ is the trial judge's leniency rate. This variable equals the leave-one-out rate of punishment for each trial judge, where the defendant's own decision is excluded from this average.\footnote{{ Similarly to \citet{DiTella2013} and \citet{Bhuller2019}, we use the simple leave-one-out rate of punishment for each trial judge as our instrumental variable. Alternatively, we could have used the residualized leave-one-out rate of punishment as done by \citet{Agan2021}, who remove court-district averages before computing each decision maker's rate. We choose to use the simple leave-one-out rate because we already include court-district fixed effects in our regression specifications, and each judge analyzes many cases as shown in Figure \ref{FigJudges}.}} We ensure that the minimum and maximum values of the $Z$ are the same across both treatment arms to enforce better overlap properties. 

\subsection{Empirical results}\label{Sresults}

In this section, we present our empirical results.\footnote{Appendix \ref{Sfirststage} contains information about the first stage of our estimation procedure (Equation \eqref{EqPartiallinear}), which relates how censoring, court district dummies, and the judge's leniency rate affect the final ruling of the case.} To estimate the DMTE and QMTE functions in our empirical application, we flexibly account for court district fixed effects. More precisely, we estimate 193 district-specific functions for each of our treatment effect parameters (Theorem \ref{Thm:semi}). Although very flexible, this strategy makes it challenging to concisely report summary results. The way we proceeded was to average these district-specific functions over court districts using the proportion of cases per court district as weights, as in Corollary \ref{cor:semi}. We report the average DMTE function in Section \ref{SresultsDMTE} and the average QMTE function in Section \ref{SresultsQMTE}.\footnote{{ Moreover, Appendix \ref{AppCaseProcessingTime} discusses whether our main results are robust to including case-processing time as an additional covariate.}} Moreover, we compare our proposed methods against standard methods in the literature in Section \ref{Scomparisons}.

\subsubsection{Estimated DMTE function}\label{SresultsDMTE}

\begin{figure}[!p]
	\begin{center}
		\begin{subfigure}[t]{0.47\textwidth}
			\centering
			\includegraphics[width = \textwidth]{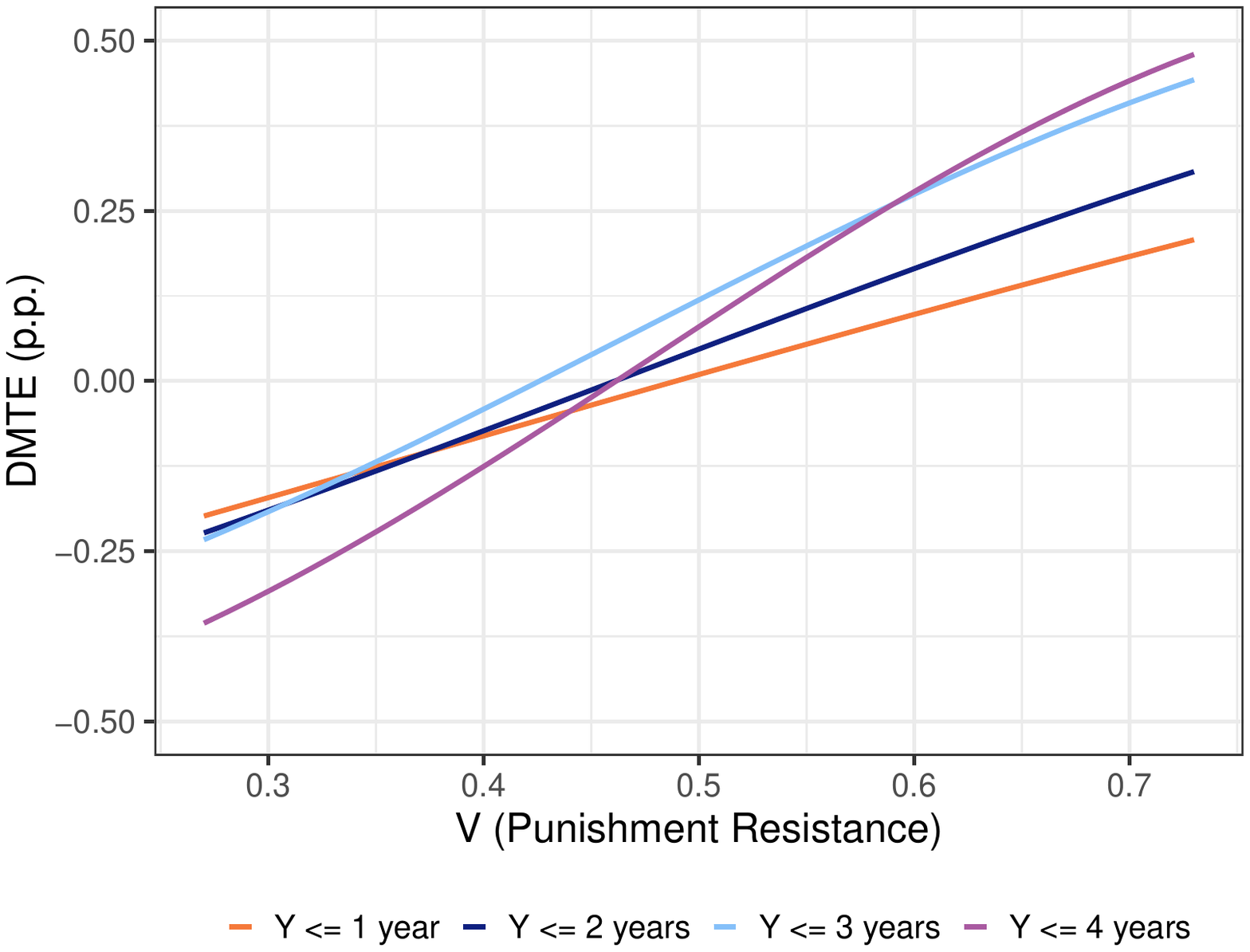}
			\caption{$DMTE\left(y,\cdot\right)$ for $y \in \left\lbrace 1, 2, 3, 4 \right\rbrace$}
			\label{DMTE1-4}
		\end{subfigure}
		\hfill
		\begin{subfigure}[t]{0.47\textwidth}
			\begin{center}
				\includegraphics[width = \textwidth]{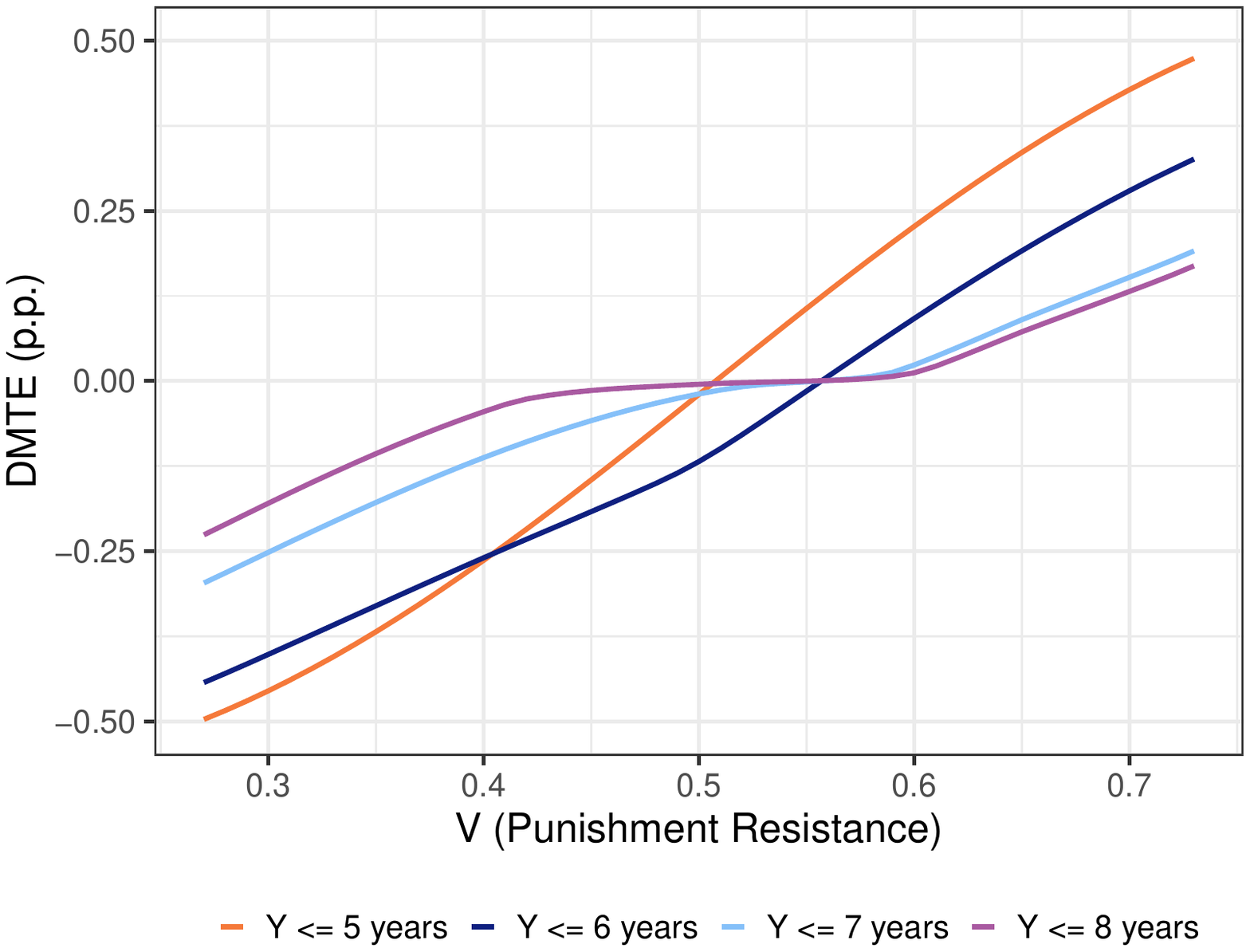}
				\caption{$DMTE\left(y,\cdot\right)$ for $y \in \left\lbrace 5, 6, 7, 8 \right\rbrace$}
				\label{DMTE5-8}
			\end{center}
		\end{subfigure}
		\begin{subfigure}[t]{0.47\textwidth}
			\centering
			\includegraphics[width = \textwidth]{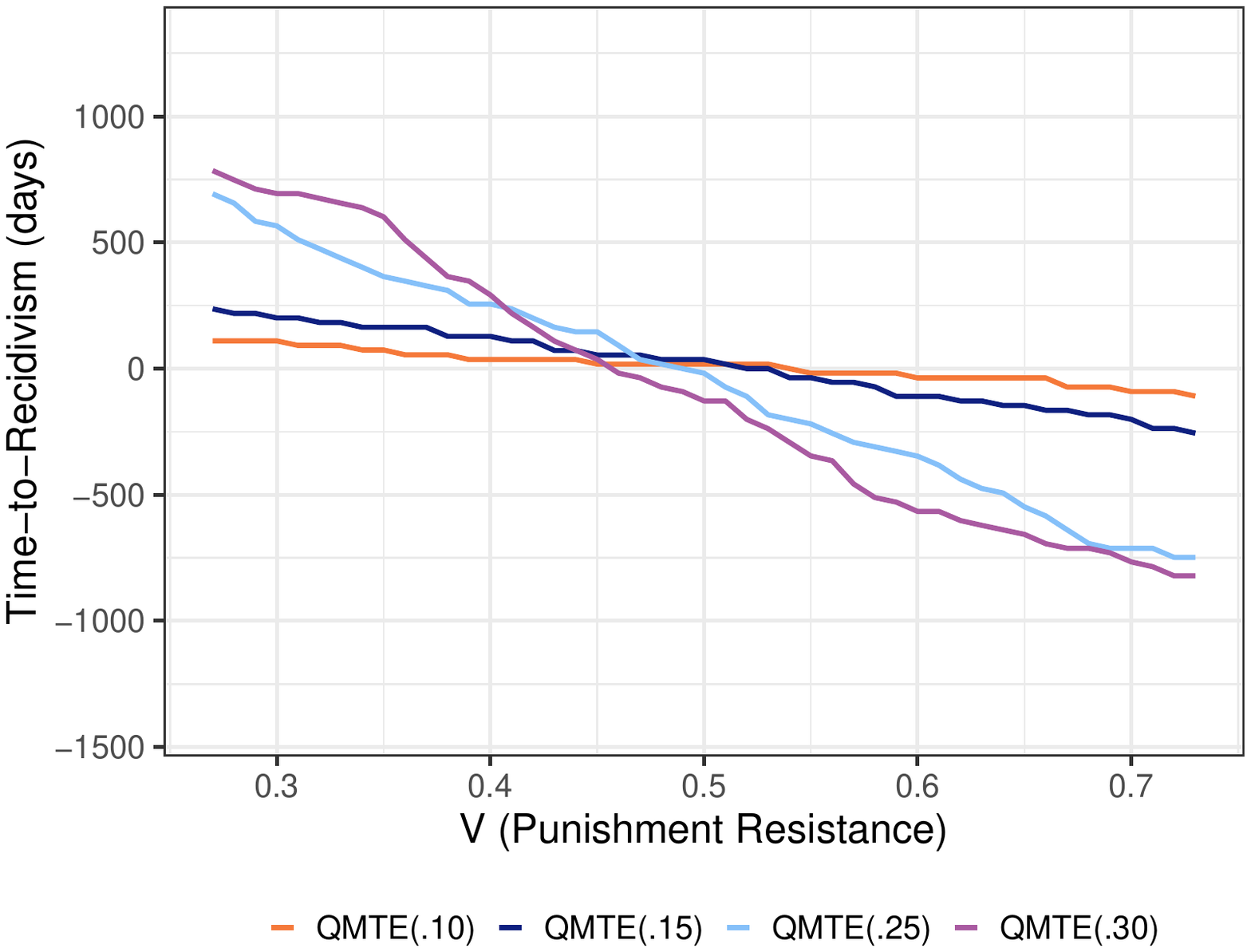}
			\caption{$QMTE\left(\tau,\cdot\right)$ for $\tau \in \left\lbrace .10, .15, .25, .30 \right\rbrace$}
			\label{QMTE10-30}
		\end{subfigure}
		\hfill
		\begin{subfigure}[t]{0.47\textwidth}
			\begin{center}
				\includegraphics[width = \textwidth]{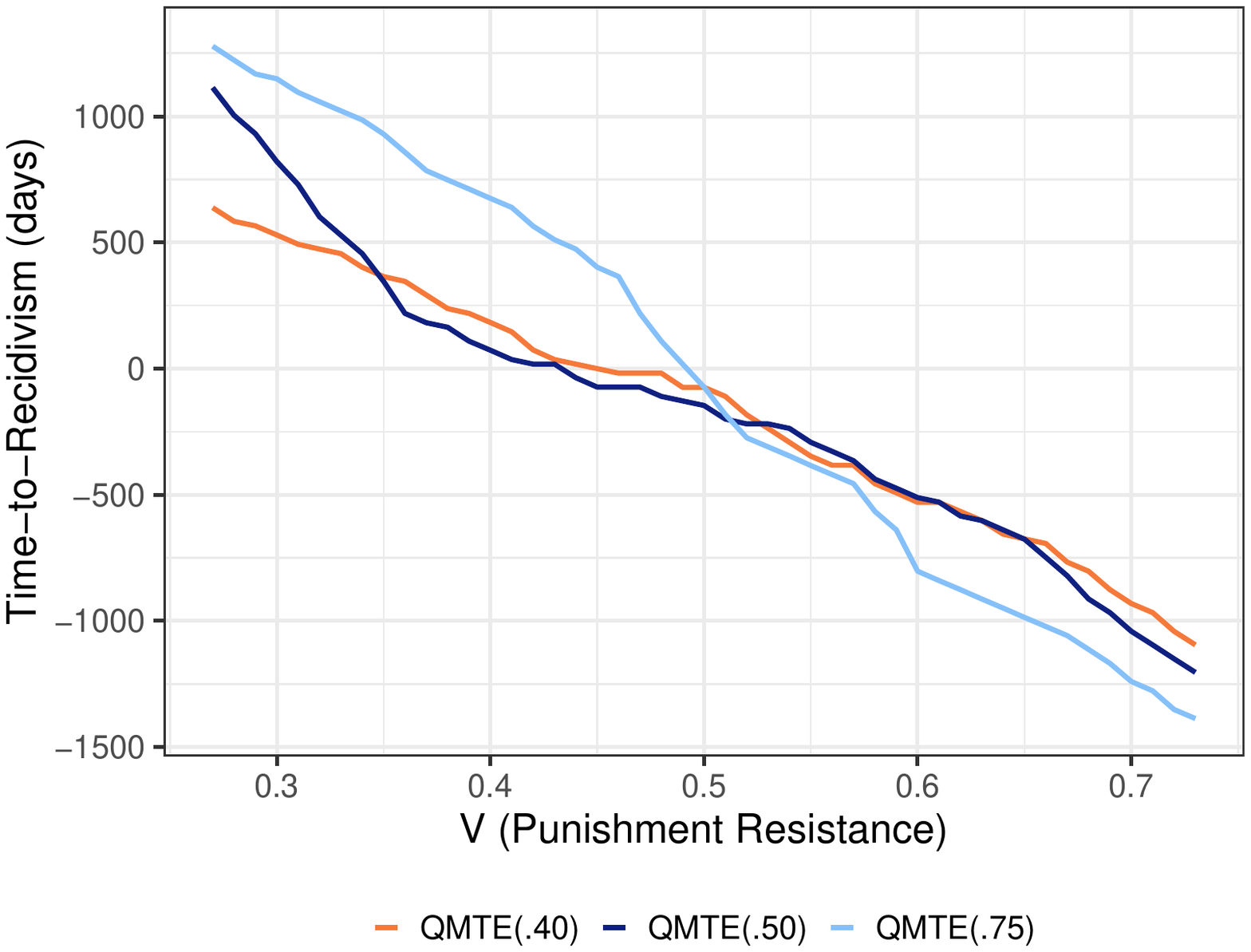}
				\caption{$QMTE\left(\tau,\cdot\right)$ for $\tau \in \left\lbrace .40, .50, .75, \right\rbrace$}
				\label{QMTE40-75-RMTE}
			\end{center}
		\end{subfigure}
		\caption{$DMTE\left(y,\cdot\right)$ for $y \in \left\lbrace 1, 2, \ldots, 8 \right\rbrace$ and $QMTE\left(\tau,\cdot\right)$ for $\tau \in \left\lbrace .10, .15, .25, .30, .40, .50, .75 \right\rbrace$}
		\label{FigDMTE}
	\end{center}
	\justifying
	{Notes: In Figures \ref{DMTE1-4} and \ref{DMTE5-8}, solid lines are the point estimates for average $DMTE\left(y,\cdot\right)$ functions. In Figures \ref{QMTE10-30} and \ref{QMTE40-75-RMTE}, solid lines are the point estimates for average $QMTE\left(\tau,\cdot\right)$ functions. All results are based on Corollary \ref{cor:semi}. Moreover, point-wise 90\%-confidence intervals are reported in Appendix \ref{AppCIs}. These confidence intervals were computed using the weighted bootstrap, clustered at the court district level (Algorithm \ref{algo:bootstrap}). 
	}
\end{figure}

Figures \ref{DMTE1-4} and \ref{DMTE5-8} shows the estimated average $DMTE\left(y,\cdot\right)$ functions for $y \in \left\lbrace 1, 2, \ldots, 8 \right\rbrace$, where instead of measuring time-to-recidivism in days we measured it in years (to enhance readability).\footnote{In our data, we observe time-to-recidivism in days. To illustrate the readability improvements of writing the $DMTE$ function in years instead of days, we focus on one value of the time-to-recidivism variable. The $DMTE\left(y,\cdot\right)$ function when $y = 2$ shows the distributional marginal treatment effect given by $\mathbb{P}\left[\left.Y^*(1) \leq 2 \cdot 365 \text{ days } \right\vert V = v\right] - \mathbb{P}\left[\left.Y^*(0) \leq  2 \cdot 365 \text{ days } \right\vert V = v\right]$. } These point estimates show relevant heterogeneity with respect to the treatment resistance (horizontal axis denotes values of $V$) and with respect to the recidivism horizon (different colors denote different values of $y$).

First, the $DMTE\left(y,\cdot\right)$ functions are increasing. This functional behavior indicates that defendants whom almost all judges would punish are less likely to recidivate, while defendants who would be punished only by tough judges are more likely to recidivate compared to situations in which they would not be punished.\footnote{Appendix \ref{AppPartialContinuous} shows that these results are robust to violations of Assumption \ref{AsCensoring}.} This conclusion is supported by our 90\%-confidence intervals (Figures \ref{DMTE1-CI}-\ref{DMTE8-CI}). { Importantly, these point-wise confidence intervals suggest that constant treatment effects are implausible in our empirical context. Hence, they highlight the importance of taking idiosyncratic latent heterogeneity seriously through an analysis of ``MTE-like'' parameters.}

Second, the $DMTE\left(y,\cdot\right)$ functions are steeper for $y \in \left\lbrace 3, 4, 5, 6\right\rbrace$ than for $y \in \left\lbrace 1, 2, 7, 8\right\rbrace$. This functional behavior indicates that the effect of alternative sentences on recidivism is more intense for extreme cases (small or large punishment resistance levels) in the mid-run than it is in the short or long-run horizon.

This rich heterogeneity underscores the importance of accounting for different levels of treatment resistance. Our point estimates suggest that designing sentencing guidelines that encourage strict judges to become more lenient could increase time to recidivism. However, the $DMTE$ functions do not allow us to quantify this impact directly.

For this reason, $DMTE$ functions may not be the ideal way to convey the main takeaway of the application, even though they answer well-posed and policy-relevant questions. In what follows, we show that this limitation can be minimized by focusing on other functionals of interest, such as the QMTE, which are measured in days instead of percentage points.

\subsubsection{Estimated QMTE function}\label{SresultsQMTE}

To better understand the time trade-offs associated with the effect of punishment on time to recidivism, we now focus on the average quantile marginal treatment effect functions. These functionals are easier to interpret than the DMTE functions because they express the underlying treatment effects in the same units as the time-to-recidivism outcomes, i.e., days before the first recidivism event.

Figures \ref{QMTE10-30} and \ref{QMTE40-75-RMTE} show the estimated average QMTE$\left(\tau,\cdot\right)$ functions for $\tau \in \left\lbrace .10, .15, .25, .30, .40, .50, .75 \right\rbrace$. Once more, these point estimates show relevant heterogeneity with respect to the punishment resistance (horizontal axis denotes values of $V$).

Although the level of the estimated $QMTE\left(\tau,\cdot\right)$ functions depends on the quantile, all functions are decreasing in the unobserved resistance to punishment. These point estimates suggest that defendants whom almost all judges would punish would take longer to recidivate when punished. In contrast, defendants who would be punished only by tough judges would recidivate faster compared to situations in which they would not be punished. This result is statistically significant for $\tau \in \left\lbrace .10, .15, .25, .30, .40, .50, .75 \right\rbrace$ at the 10\% significance level, according to Figures \ref{FigQMTE-CI-10-30} and \ref{QMTE40-CI}-\ref{QMTE75-CI} in the Appendix. { Interestingly, these point-wise confidence intervals suggest that constant treatment effects are likely invalid in our empirical setting. Hence, they emphasize the importance of taking essential heterogeneity seriously through an analysis of ``MTE-like'' parameters.}

We reach a similar conclusion when we analyze the $QMTE\left(\cdot, v\right)$ as a function of the quantiles for specific values of unobserved resistance to treatment. Figure \ref{FigQMTEfunctionTau} in the Appendix shows the average $QMTE\left(\cdot, v\right)$ for $v \in \left\lbrace .3, .4, .5, .6, .7 \right\rbrace$. Our results suggest that this function is always positive for small values of the unobserved resistance to punishment, while it is always negative for large values of $v$. 

Overall, our QMTE point estimates suggest that designing sentencing guidelines that encourage strict judges to become more lenient could increase time-to-recidivism.

\subsubsection{Comparison with other available methods}\label{Scomparisons}
Here, we compare our proposed methods against other available methods in the literature. Differently from our approach, these estimates ignore that the outcome variable is right-censored and provide different conclusions when compared against our proposed estimator. First, we compare our DMTE methods against alternative methods, including the typical judge-fixed-effect regressions that drop many observations to avoid directly handling censored outcomes. Second, we compare our QMTE methods against alternative methods.

\begin{figure}[!p]
	\begin{center}
		\begin{subfigure}[t]{0.47\textwidth}
			\centering
			\includegraphics[width = \textwidth]{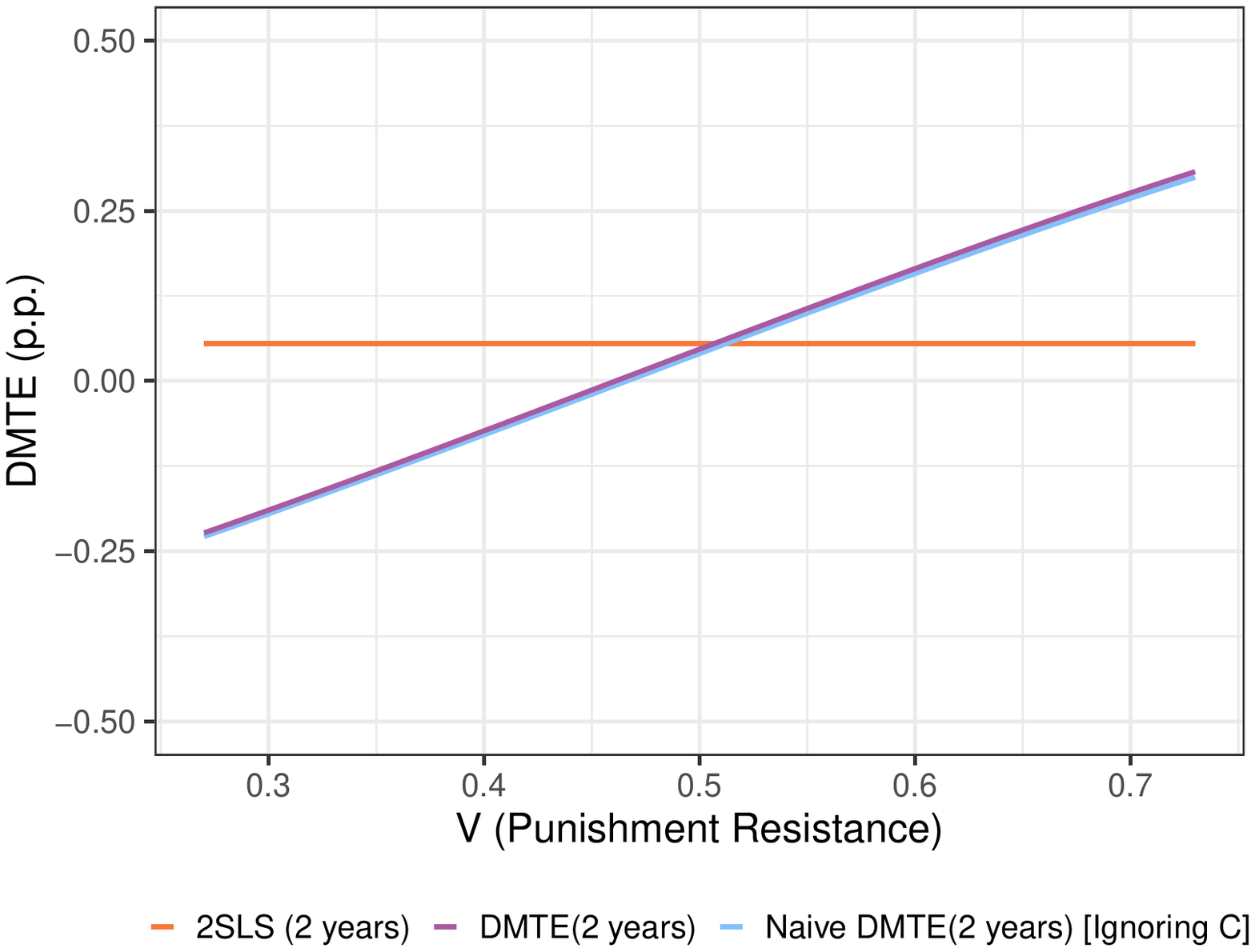}
			\caption{$DMTE\left(2, \cdot \right)$ v. Other Methods}
			\label{DMTE2comparison}
		\end{subfigure}
		\hfill
		\begin{subfigure}[t]{0.47\textwidth}
			\begin{center}
				\includegraphics[width = \textwidth]{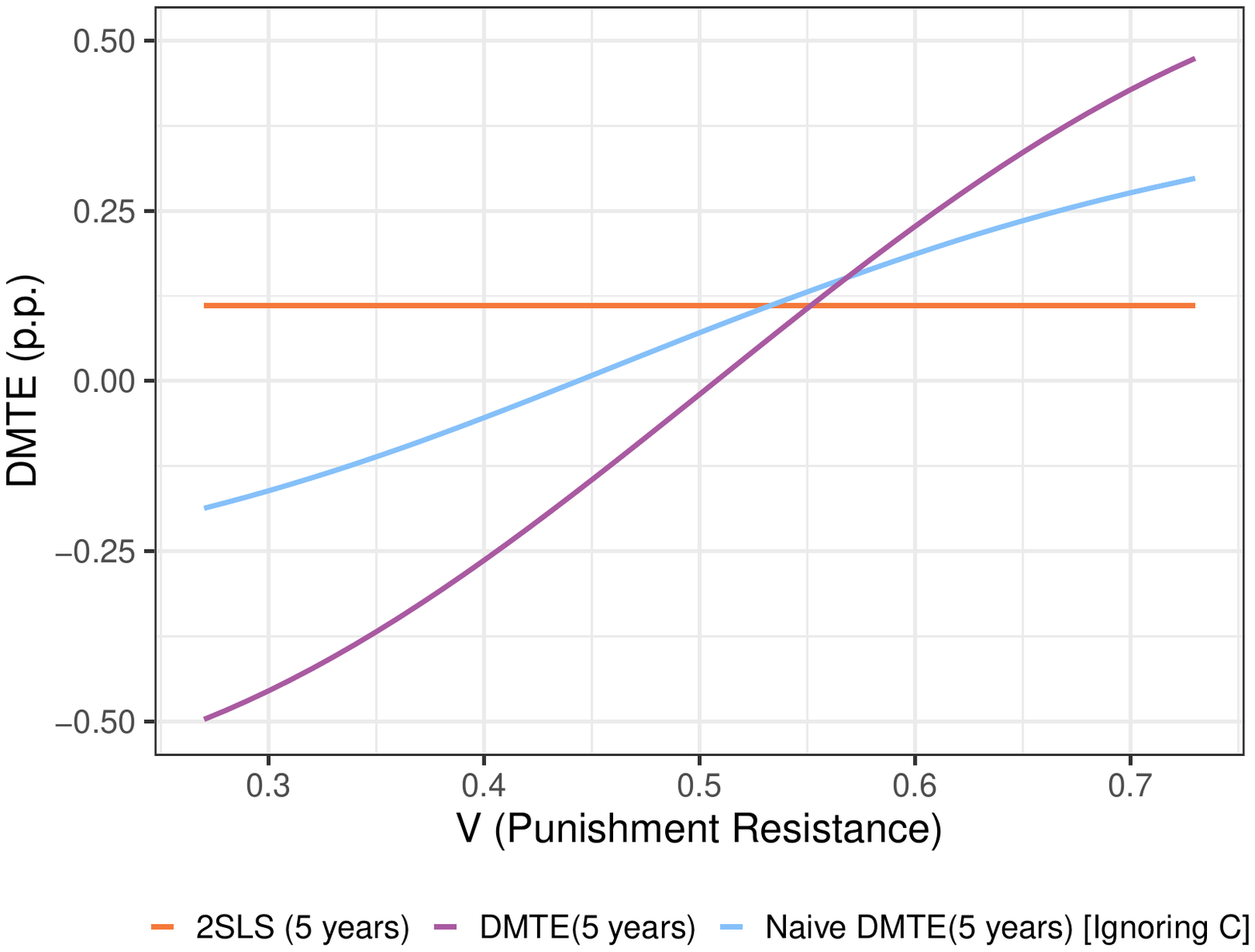}
				\caption{$DMTE\left(5, \cdot \right)$ v. Other Methods}
				\label{DMTE3comparison}
			\end{center}
		\end{subfigure}
		\begin{subfigure}[t]{0.47\textwidth}
			\centering
			\includegraphics[width = \textwidth]{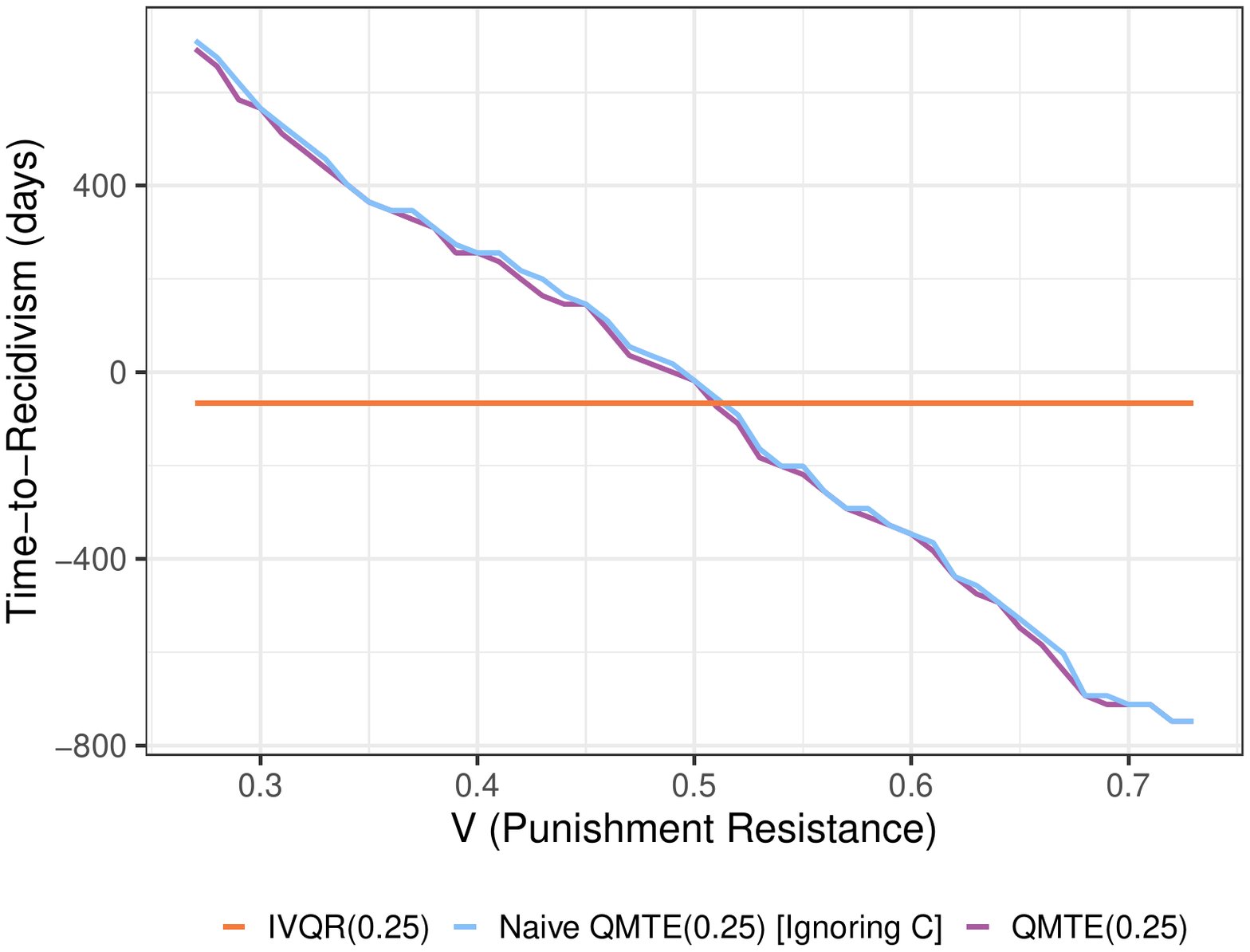}
			\caption{$QMTE\left(.25, \cdot \right)$ v. Other Methods}
			\label{QMTEcomparison}
		\end{subfigure}
		\hfill
		\begin{subfigure}[t]{0.47\textwidth}
			\begin{center}
				\includegraphics[width = \textwidth]{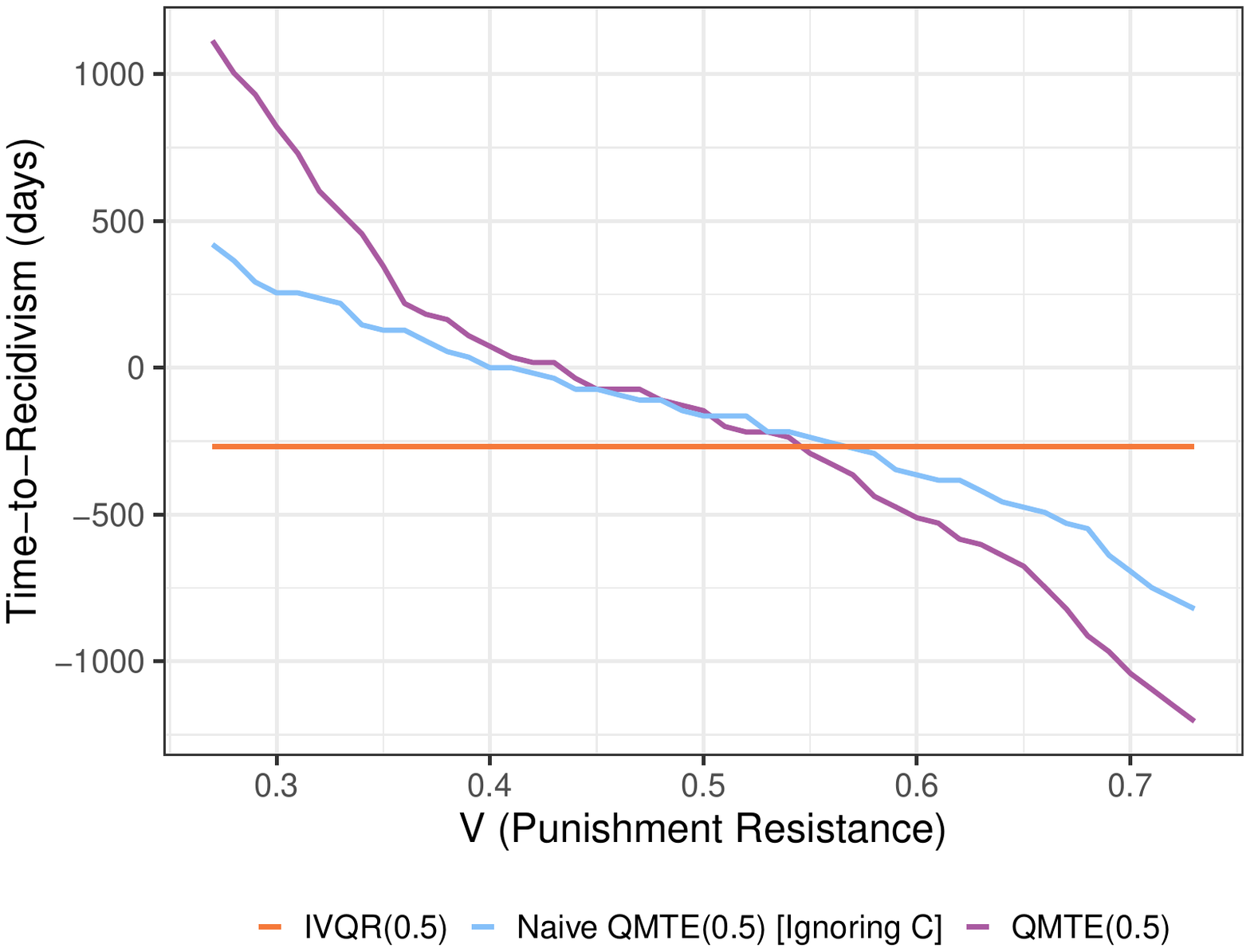}
				\caption{$QMTE\left(.50, \cdot \right)$ v. Other Methods}
				\label{QMTE50comparison}
			\end{center}
		\end{subfigure}
		\caption{Comparing our $DMTE$ and $QMTE$ Methods against Other Available Methods}
		\label{FigComparisonsDMTE}
	\end{center}
	\justifying
	{Notes: Our methods are the purple lines. The light blue lines denote a version of our estimators that ignores censoring. The orange lines in Figures \ref{DMTE2comparison} and \ref{DMTE3comparison} are the 2SLS coefficients that use each judge's punishment rate to instrument for final punishment and include court district fixed effects. The outcome variable indicates recidivism within 2 years in Figure \ref{DMTE2comparison} and recidivism within 5 years in Figure \ref{DMTE3comparison}. The orange lines in Figures \ref{QMTEcomparison}-\ref{QMTE50comparison} illustrate IV quantile regressions \citep{Kaplan2017}.
	}
\end{figure}

When analyzing distributional effects, we focus on recidivism within 2 and 5 years in Figures \ref{DMTE2comparison} and \ref{DMTE3comparison}. Our proposed methods are illustrated by the purple lines. We present the cross-district average $DMTE\left(2,\cdot \right)$ function in Figure \ref{DMTE2comparison} and the cross-district average $DMTE\left(5,\cdot \right)$ function in Figure \ref{DMTE3comparison} (Corollary \ref{cor:semi}). The light blue lines denote a ``naive'' version of our estimators that follows the same steps as described in Section \ref{SestSemiPara}, but does not condition on the censoring variable, i.e., it removes the terms associated with $C$ from Equations \eqref{EqPartiallinear}-\eqref{EqDMTR_d_estimator}. The orange lines are the treatment coefficients of two-stage least squares regressions that use each judge's punishment rate to instrument for the defendant's final punishment and include a full set of court district fixed effects. In Figure \ref{DMTE2comparison}, the outcome variable is an indicator equal to 1 if the defendant recidivated within 2 years, and the regression includes 43,468 observations. In Figure \ref{DMTE3comparison}, the outcome variable is an indicator equal to 1 if the defendant recidivated within 5 years, and the regression includes only 14,608 observations because it has to remove 28,860 defendants who are observed for less than 5 years.

The comparison between the orange and purple lines in Figures \ref{DMTE2comparison} and \ref{DMTE3comparison} illustrates the drawbacks of using typical methods rather than embracing essential heterogeneity and addressing censoring directly, as we advocate. First, the regression in the orange line in Figure \ref{DMTE3comparison} has 28,860 fewer observations than the regression in the orange line in Figure \ref{DMTE2comparison}. This reduction in sample size shows that the typical approach either risks compositional changes (i.e., different samples for each time horizon) or must lose statistical power for shorter time horizons. Second, the typical judge-fixed effect regression does not capture the rich heterogeneity behind the treatment effects of fines and community service. In particular, the typical regression estimates suggest a positive effect, ignoring that the treatment decreases the probability of recidivism for some defendant types. Importantly, these regression estimates do not lie entirely within the 90\%-confidence intervals of the correctly estimated $DMTE\left(2,\cdot\right)$ and $DMTE\left(5,\cdot\right)$ functions (Figures \ref{DMTE2-CI} and \ref{DMTE5-CI}).

Moreover, in Figure \ref{DMTE2comparison}, we observe that our proposed estimator (purple line) and its naive version (light blue line) reach very similar point estimates. This finding is unsurprising because all defendants are observed for at least 2 years, implying that the censoring problem is not binding for short horizons.

However, the censoring problem is binding when we analyze longer horizons. In Figure \ref{DMTE3comparison}, we focus on the $DMTE\left(5,\cdot\right)$ function and find that our proposed estimator (purple line) and its naive version (light blue line) differ in relevant ways. For example, the naive estimator finds a $DMTE$ function that is less steep, implying smaller treatment effects for extreme values of punishment resistance. Importantly, the naive estimates do not lie entirely within the 90\%-confidence intervals of the correctly estimated $DMTE\left(5,\cdot\right)$ function (Figure \ref{DMTE5-CI}).

Now, we compare our QMTE methods against other available methods in the literature. Differently from our approach, these estimates ignore that the outcome variable is right-censored and provide different conclusions when compared against our proposed estimator. For brevity, we focus our attention on the effects on the 25th and 50th percentiles ($QMTE\left(.25, \cdot \right)$ and $QMTE\left(.50, \cdot \right)$ functions) in Figures \ref{QMTEcomparison} and \ref{QMTE50comparison}. Our proposed methods are illustrated by the purple lines. We have the cross-district average $QMTE\left(.25,\cdot \right)$ function in Figure \ref{QMTEcomparison} and the cross-district average $QMTE\left(.50,\cdot \right)$ function in Figure \ref{QMTE50comparison} (Corollary \ref{cor:semi}). The light blue lines denote a ``naive'' version of our estimators that follows the same steps as described in Section \ref{SestSemiPara}, but does not condition on the censoring variable, i.e., it removes the terms associated with $C$ from Equations \eqref{EqPartiallinear}-\eqref{EqDMTR_d_estimator}. The orange lines denote the standard method in the IV literature that accounts for endogenous selection into treatment but ignores (or aggregates) treatment effect heterogeneity with respect to unobserved resistance to treatment. The orange line in Figure \ref{QMTEcomparison} is the treatment coefficient of an IV quantile regression \citep{Kaplan2017} for the 25th percentile, while the orange line in Figure \ref{QMTE50comparison} is the treatment coefficient of an IV quantile regression \citep{Kaplan2017} for the 50th percentile. Both IV quantile regressions use the censored outcome variable as the left-hand side variable, control for court district fixed effects, and use the judge's punishment rate as the instrument for the defendant being punished.

Analyzing Figures \ref{QMTEcomparison} and \ref{QMTE50comparison}, we find that the IV quantile regression does not capture the rich heterogeneity behind the treatment effects of fines and community service. In particular, the IV quantile regression estimates suggest a negative effect, ignoring that the treatment increases time-to-recidivism for some defendant types. Importantly, the IV quantile regression estimates do not lie entirely within the 90\%-confidence intervals of the correctly estimated $QMTE\left(.25,\cdot\right)$ and $QMTE\left(.50,\cdot\right)$ functions (Figures \ref{QMTE25-CI} and \ref{QMTE50-CI}).

Moreover, in Figure \ref{QMTEcomparison}, we observe that our proposed estimator (purple line) and its naive version (light blue line) reach similar point estimates. This finding is unsurprising because the estimated $QMTR_{d}\left(.25,\cdot\right)$ functions are always smaller than 2.5 years, and all defendants are observed for at least 2 years. Consequently, the censoring problem is not binding for low percentiles.

However, the censoring problem is binding for higher percentiles. In  Figure \ref{QMTE50comparison}, we focus on the $QMTR_{d}\left(.50,\cdot\right)$ function and find that our proposed estimator (purple line) and its naive version (light blue line) differ in relevant ways. For example, the naive estimator finds a $QMTE$ function that is less steep, implying smaller treatment effects for extreme values of punishment resistance. Importantly, the naive estimates do not lie entirely within the 90\%-confidence intervals of the correctly estimated $QMTE\left(.50,\cdot\right)$ function (Figure \ref{QMTE50-CI}).

All in all, our results indicate that our proposed tools can provide detailed measures of treatment effect heterogeneity of punishing misdemeanor offenses on time-to-recidivism that other methods are not meant to capture. 

\section{Conclusion}\label{Sconclusion}
In this paper, we identify the distributional marginal treatment effect ($DMTE$) and the quantile marginal treatment effect ($QMTE$) functions when the outcome variable is right-censored. To do so, we extend the MTE framework \citep{Heckman2006,carneirolee2009} to scenarios with duration outcomes. In this section, we deepen our empirical discussion.

Concerning its empirical contribution, our work is inserted in the literature about the effect of fines and community service sentences on future criminal behavior. Four recent papers in this field were written by \citet{Huttunen2020}, \citet{Giles2021}, \citet{Possebom2022}, and \citet{Lieberman2023}. They all focus on binary variables indicating recidivism within a pre-specified period. \citet{Huttunen2020} and \citet{Giles2021} find that this type of punishment increases the probability of recidivism in Finland and Milwaukee (a city in the State of Wisconsin in the U.S.), respectively. \citet{Possebom2022} finds that this type of punishment has a small and statistically insignificant effect on the probability of recidivism in São Paulo, Brazil. Finally, \citet{Lieberman2023} analyzes five American states and finds that court fees do not impact recidivism.

Unlike these four papers, our outcome variable is time-to-recidivism. Using a continuous outcome instead of binary indicators allows for a finer analysis of the heterogeneous effects of fines and community service sentences on future criminal behavior, and may reconcile the conflicting results in the previous literature. For example, we find that this type of punishment increases time-to-recidivism for some individuals while decreasing it for other individuals. If the first type of individual is more common in the states analyzed by \cite{Lieberman2023} than in Milwaukee and Finland, our focus on essential heterogeneity may explain these results.

As future work, we believe that analyzing the effects of conviction decisions on felonies (rather than misdemeanors) is interesting. Answering this type of question requires the use of censoring-conscious tools (such as the ones developed here) and the collection of longer and more detailed datasets. For example, to disentangle incapacitation effects from rehabilitation, one would need data on the length of prison sentences. It may also require more complex heterogeneity analyses based on defendant's age and incarceration length.

\singlespace

\bibliography{references_all}


\newpage
\pagenumbering{arabic}
\renewcommand*{\thepage}{A-\arabic{page}}

\setcounter{table}{0}
\renewcommand\thetable{A.\arabic{table}}

\setcounter{figure}{0}
\renewcommand\thefigure{A.\arabic{figure}}

\setcounter{equation}{0}
\renewcommand\theequation{A.\arabic{equation}}

\appendix

\begin{center}
{\LARGE {Was Javert right to be suspicious? Marginal Treatment Effects with Duration Outcomes
}} \medskip \\ \Large Online Appendix \bigskip \\
\large Santiago Acerenza  \hspace{0.3cm} Vitor Possebom  \hspace{0.3cm} Pedro H. C. Sant'Anna \medskip\\
\today
\bigskip
\end{center}

This online appendix contains proofs and additional results for the paper ``Was Javert right to be suspicious? Marginal Treatment Effects with Duration Outcomes'' by Santiago Acerenza, Vitor Possebom, and Pedro H. C. Sant'Anna. Section \ref{AppProofs} contains proofs and auxilliary lemmas for results stated in the main text. Section \ref{Ssemiparaconsistency} contains additional details about our estimation and inference methods, including standard regularity conditions for these methods to work. Section \ref{AppData} explains the construction of our dataset in detail. Section \ref{AppExtraEmpirical} presents additional results for our empirical application. Section \ref{AppTM} provides additional motivation and theoretical arguments for focusing on duration (time-to-event) outcomes instead of short-term outcomes. Section \ref{AppAltAs} discusses the costs of not imposing restrictions on the censoring mechanism. Section \ref{AppPartialID} extends our results to some types of dependent censoring and displays partial identification of the causal parameters of interest. Section \ref{AppAverage} discusses the average marginal treatment effect function. Section \ref{AppModelJustification2} analyzes the drawback of simply treating the censoring variable as an additional covariate and should be seen as a cautionary tale. Section \ref{AppCaseProcessingTime} shows that our empirical results are robust to adding case processing time as an additional covariate. Section \ref{AppLeeBounds} discusses how to adapt trimming bounds to our context so that we can analyze the effect of alternative sentences on true criminal behavior even if crimes are not always observed by the police.

\clearpage

\section{Proofs of the main results} \label{AppProofs}

\setcounter{table}{0}
\renewcommand\thetable{A.\arabic{table}}

\setcounter{figure}{0}
\renewcommand\thefigure{A.\arabic{figure}}

\setcounter{equation}{0}
\renewcommand\theequation{A.\arabic{equation}}

\setcounter{theorem}{0}
\renewcommand\thetheorem{A.\arabic{theorem}}

We start by stating an auxiliary lemma that will be used to derive our main identification results.
\begin{lemma}\label{LePYD}
	If Assumptions \ref{AsIndependence}-\ref{AsPositive} hold, then, for any $y  < \gamma_{C}$, $v \in \mathcal{P}$ and $\delta \in \mathbb{R}_{++}$ such that $y + \delta \in \mathcal{C}$,
	\begin{equation}\label{EqPYD1}
		\mathbb{P}\left[\left. Y\leq y, D = 1 \right\vert P\left(Z,C\right) = v, C = y+\delta\right] = \int_0^{ v} \mathbb{P}\left[\left. Y^*(1) \leq y  \right\vert C = y+\delta, V=v\right]dv
	\end{equation}
	and
	\begin{equation}\label{EqPYD0}
		\mathbb{P}\left[\left. Y\leq y, D = 0 \right\vert P\left(Z,C\right) = v, C = y+\delta\right] = \int_{v}^{1} \mathbb{P}\left[\left.  Y^*(0) \leq y  \right\vert C = y+\delta, V=v\right]dv,
	\end{equation}

	If Assumption \ref{AsCensoring} holds too, then, for any $y < \gamma_{C}$, $v \in \mathcal{P}$ and $\delta \in \mathbb{R}_{++}$ such that $y + \delta \in \mathcal{C}$,
	\begin{equation}\label{EqPYD1A6}
		\mathbb{P}\left[\left. Y\leq y, D = 1 \right\vert P\left(Z,C\right) = v, C = y+\delta\right] = \int_0^{v} \mathbb{P}\left[\left. Y^*(1) \leq y  \right\vert V=v\right]dv
	\end{equation}
	and
	\begin{equation}\label{EqPYD0A6}
		\mathbb{P}\left[\left. Y\leq y, D = 0 \right\vert P\left(Z,C\right) = v, C = y+\delta\right] = \int_{v}^{1} \mathbb{P}\left[\left.  Y^*(0) \leq y  \right\vert V=v\right]dv.
	\end{equation}
\end{lemma}

\subsection{Proof of Lemma \ref{LePYD}}\label{AppProofLePYD}

Fix $y  < \gamma_{C}$, $v \in \mathcal{P}$ and $\delta \in \mathbb{R}_{++}$ such that $y + \delta \in \mathcal{C}$. To prove \eqref{EqPYD1}, note that
\begin{align*}
	& \mathbb{P}\left[\left. Y\leq y, D = 1 \right\vert P\left(Z,C\right) = v, C = y+\delta\right] \\
	& \hspace{20pt} = \mathbb{E}\left[\left. \mathbf{1}\left\lbrace Y \leq y \right\rbrace\mathbf{1}\left\lbrace P\left(Z, C\right) \geq V \right\rbrace \right\vert P\left(Z,C\right) = v, C = y+\delta\right] \\
	& \hspace{40pt} \text{by \eqref{EqTreatment}} \\
	& \hspace{20pt} = \mathbb{E}\left[\left. \mathbf{1}\left\lbrace Y^*(1) \leq y \right\rbrace\mathbf{1}\left\lbrace v \geq V \right\rbrace \right\vert P\left(Z,y + \delta\right) = v, C = y+\delta\right] \\
	& \hspace{40pt} \text{because $\mathbf{1}\left\lbrace Y(1) \leq y \right\rbrace = \mathbf{1}\left\lbrace Y^*(1) \leq y \right\rbrace$ when $C > y$} \\
	& \hspace{20pt} = \int_0^1\mathbb{E}\left[\left. \mathbf{1}\left\lbrace Y^*(1) \leq y \right\rbrace\mathbf{1}\left\lbrace v \geq \tilde{v} \right\rbrace \right\vert P\left(Z,y + \delta\right) = \tilde{v}, C = y+\delta, V=\tilde{v}\right]d\tilde{v} \\
	& \hspace{40pt} \text{by the Law of Iterated Expectations and Assumption \ref{AsContinuous}} \\
	& \hspace{20pt} = \int_0^1 \mathbf{1}\left\lbrace v \geq \tilde{v} \right\rbrace \mathbb{E}\left[\left. \mathbf{1}\left\lbrace Y^*(1) \leq y \right\rbrace \right\vert P\left(Z,y + \delta\right) = v, C = y+\delta, V=\tilde{v}\right]d\tilde{v} \\
	& \hspace{20pt} = \int_0^{v} \mathbb{E}\left[\left. \mathbf{1}\left\lbrace Y^*(1) \leq y \right\rbrace \right\vert P\left(Z,y + \delta\right) = v, C = y+\delta, V=\tilde{v}\right]d\tilde{v} \\
	& \hspace{20pt} = \int_0^{ v} \mathbb{P}\left[\left. Y^*(1) \leq y  \right\vert C = y+\delta, V=\tilde{v}\right]d\tilde{v} \\
	& \hspace{40pt} \text{by Assumption \ref{AsIndependence}}.
\end{align*}

We can prove \eqref{EqPYD0} analogously.

To prove \eqref{EqPYD1A6}, observe that
\begin{align*}
	& \mathbb{P}\left[\left. Y\leq y, D = 1 \right\vert P\left(Z,C\right) = v, C = y+\delta\right] \\
	& \hspace{20pt} =  \int_0^{v} \mathbb{P}\left[\left. Y^*(1) \leq y  \right\vert C = y+\delta, V=\tilde{v}\right]d\tilde{v} \\
	& \hspace{20pt} = \int_0^{v} \mathbb{P}\left[\left.Y^*(1) \leq y \right\vert V=\tilde{v}\right]d\tilde{v} \\
	& \hspace{40pt} \text{by Assumption \ref{AsCensoring}}.
\end{align*}

We can prove \eqref{EqPYD0A6} analogously.

\subsection{Proof of Proposition \ref{PropDMTR}}\label{AppProofPropDMTR}
Fix $y  < \gamma_{C}$, $v \in \mathcal{P}$ and $\delta \in \mathbb{R}_{++}$ such that $y + \delta \in \mathcal{C}$.

First, note that Equations \eqref{EqPYD1A6} and \eqref{EqPYD0A6} imply that
\begin{equation}\label{EqPYD1A6derivative}
	\dfrac{\partial \mathbb{P}\left[\left. Y\leq y, D = 1 \right\vert P\left(Z,C\right) = v, C = y+\delta\right]}{\partial v} = \mathbb{P}\left[\left. Y^*(1) \leq y  \right\vert V = v\right]
\end{equation}
and
\begin{equation}\label{EqPYD0A6derivative}
	\dfrac{\partial \mathbb{P}\left[\left. Y\leq y, D = 0 \right\vert P\left(Z,C\right) = v, C = y+\delta\right]}{\partial z} = - \mathbb{P}\left[\left.  Y^*(0) \leq y  \right\vert V = v\right]
\end{equation}
according to the Leibniz Integral Rule.

Combining Equations \eqref{EqDMTR} and \eqref{EqPYD1A6derivative}-\eqref{EqPYD0A6derivative}, we prove that
\begin{equation}
	\label{EqTesting}
	DMTR_{d}\left(y, v\right) = \left(2 d - 1\right) \cdot \dfrac{\partial \mathbb{P}\left[\left. Y\leq y, D = d \right\vert P\left(Z,C\right) = v, C = y+\delta\right]}{\partial v}
\end{equation}
for any $d \in \left\lbrace 0, 1 \right\rbrace$.

Since the last equation holds for any $\delta \in \mathbb{R}_{++}$ such that $y + \delta \in \mathcal{C}$, we have that
\begin{equation*}
	DMTR_{d}\left(y, v\right) = \left(2 d - 1\right) \cdot \int_{y}^{+\infty} \dfrac{\partial \mathbb{P}\left[\left. Y\leq y, D = d \right\vert P\left(Z,C\right) = v, C = c\right]}{\partial v} \cdot f_{\left. C \right\vert P\left(Z,C\right) = v, C > y} \left(c\right) \, dc
\end{equation*}
for any $d \in \left\lbrace 0, 1 \right\rbrace$.

\newpage
\section{Regularity Conditions and Semiparametric Estimation}\label{Ssemiparaconsistency}

\setcounter{table}{0}
\renewcommand\thetable{B.\arabic{table}}

\setcounter{figure}{0}
\renewcommand\thefigure{B.\arabic{figure}}

\setcounter{equation}{0}
\renewcommand\theequation{B.\arabic{equation}}

\setcounter{theorem}{0}
\renewcommand\thetheorem{B.\arabic{theorem}}

\setcounter{assumption}{0}
\renewcommand\theassumption{B.\arabic{assumption}}

\setcounter{algorithm}{0}
\renewcommand\thealgorithm{B.\arabic{algorithm}}

In this appendix, we discuss algorithms on how to semiparametrically estimate the DMTE and QMTE functions based on the identification results described in Proposition \ref{PropDMTR} and Corollary \ref{CorQMTE}. We discuss two sets of results in this appendix. First, we present generic algorithms to estimate the marginal treatment effect functionals that remain agnostic about the type of estimators used to estimate the nuisance functions (Appendix \ref{SecGenericEst}). These results are useful for pinpointing intuition and providing templates for flexible estimation procedures. However, pinning down the asymptotic properties of such estimators at this level of generality is rather challenging, especially regarding inference. To ameliorate this, we provide practical and formally justified estimation and inference procedures based on a more restricted class of estimators for the nuisance functions (Section \ref{SestSemiPara}). Importantly, Appendix \ref{AppSemiparametricdetails} elaborates on standard regularity conditions for our proposed identification and estimation results to work as well as collects the theoretical results and implementation details of the algorithm proposed in Section \ref{SestSemiPara}. Moreover, Appendix \ref{AppCovariateAggregationextra} discusses alternative procedures to aggregate the $DMTE$ and $QMTE$ functions across covariates.

\subsection{Generic estimation procedure}\label{SecGenericEst}
In this appendix, we present a generic algorithm to estimate the marginal treatment effect functionals that remain agnostic about the type of estimators used to estimate the nuisance functions. These results are useful for pinpointing intuition and providing templates for flexible estimation procedures. 

We first present a generic algorithm one can use to estimate DMTE functions across a grid of threshold points $\left\lbrace y_{k} \right\rbrace_{k = 0}^{K}$. The algorithm will make use of an estimator for the propensity score, $P(Z,C) = \expe{D|Z,C}$, and the conditional distribution of $Y \cdot \mathbf{1}\left\lbrace D = d \right\rbrace$ given $P\left(Z,C\right)$ and $C$ for any $d \in \{0,1\}$. This algorithm builds on Proposition \ref{PropDMTR}. Recall that our data consist of $i.i.d.$ observations $\left\lbrace Y_{i}, C_{i}, D_{i}, Z_{i} \right\rbrace_{i = 1}^{n}$, where $n$ is the sample size.

\begin{algorithm}[Generic Estimation of DMTE function]\label{algo:generic}\phantom{a}
	\begin{enumerate}
		\item Semiparametrically (or nonparametrically) estimate the propensity score $P\colon\mathcal{Z}\times\mathcal{C}\rightarrow\left[0,1\right]$. Denote the fitted propensity score values by $\widehat{P}_i$.
		
		\item Define a grid of values for the duration outcome $Y$, $\left\lbrace y_{k} \right\rbrace_{k = 0}^{K}$, such that $y_{k} > y_{k-1}$ for any $k \in \left\lbrace 1, \ldots, K \right\rbrace$ and $K \in \mathbb{N}$.
		
		\item For each $k \in \left\lbrace 0, \ldots, K \right\rbrace$ and each $d \in \left\lbrace 0, 1 \right\rbrace$, estimate the conditional distribution function of $Y \cdot \mathbf{1}\left\lbrace D = d \right\rbrace$ given $P\left(Z,C\right)$, and $C$, that is, $${\Gamma}({P},C;y_k,d) = \mathbb{E}\left[ \mathbf{1}\left\lbrace Y \leq y_{k}, D = d \right\rbrace\vert P, C\right].$$ Since the propensity score for unit $i$, $P_i$, is unknown, use the estimated fitted values from Step 1. Denote the estimated fitted values by $\widehat{\Gamma}(\widehat{P}_i,C_i;y_k,d) = \widehat{\Gamma}_{d, k,i}$. 
		
		
		\item For each $k \in \left\lbrace 0, \ldots, K \right\rbrace$ and $d \in \{0, 1 \}$, estimate the derivative of ${\Gamma}({P},C;y_k,d)$ with respect to $P$.  Since ${\Gamma}({P},C;y_k,d)$ is unknown, use the estimated $\widehat{\Gamma}_{d, k,i}$. Denote the estimated derivative evaluated at $P=v,C=c$ by $\widehat{\gamma}_{d}(y_k,v,c)$, where $v \in \mathcal{P}$, and $c \in \mathcal{C}$.
		
		\item For each $k \in \left\lbrace 0, \ldots, K \right\rbrace$ and each $d \in \left\lbrace 0, 1 \right\rbrace$, estimate $DMTR_{d}\left(y_{k}, v\right)$ by averaging $(2d - 1) \widehat{\gamma}_{d}\left(y_k, v,c\right)$ over values of $c$ such that $c > y_{k}$, 
		$$\widehat{DMTR}_{d}\left(y_{k},v\right) = (2d - 1)\widehat{\mathbb{E}}\left[\left. \widehat{\gamma}_{d}\left(y_k,v,C\right)\right|C > y_k,\widehat{P} = v\right], $$
		where $\widehat{\mathbb{E}}[\cdot | \cdot]$ is a generic estimator for a conditional expectation.
		
		\item\label{StepMonotonic} For each value $v\in \mathcal{P}$ and $d\in \{0,1\}$, ensure that $\widehat{DMTR}_{d}\left(y_{k}, v\right)$ is non-decreasing in $y_{k}$, and bounded between zero and one.

		\item For each $k \in \left\lbrace 0, \ldots, K \right\rbrace$, estimate $DMTE\left(y_{k}, v\right)$ using $$\widehat{DMTE}\left(y_{k},v\right) \coloneqq \widehat{DMTR}_{1}\left(y_{k},v\right) - \widehat{DMTR}_{0}\left(y_{k},v\right).$$
		
	\end{enumerate}
	
\end{algorithm}

Building on Algorithm \ref{algo:generic}, it is straightforward to compute the QMTE functions. More specifically, from Step \ref{StepMonotonic}, we have that for both $d=1$ and $d=0$, $DMTR_{d}\left(y, v\right)$ is monotone in $y$ for a given $v$, as any cumulative distribution function should be. Thus, one can invert these to compute the quantile marginal treatment response functions, and then take their differences to compute the QMTE function. More precisely, for each $d \in \left\lbrace 0, 1 \right\rbrace$ and a quantile $\tau$, a generic estimator of the $QMTR_{d}\left(\tau, v\right)$ is given by $$\widehat{QMTR}_{d}\left(\tau, v\right) \coloneqq \min_{k \in \left\lbrace 0, \ldots, K \right\rbrace}\left\lbrace y_{k} \colon \widehat{DMTR}_{d}\left(y_{k}, v\right) \geq \tau \right\rbrace.$$  The QMTE estimator for any $\tau \in [0, \overline{\tau}\left(v\right))$ is given by $$\widehat{QMTR}\left(\tau, v\right) =  \widehat{QMTR}_{1}\left(\tau, v\right) - \widehat{QMTR}_{0}\left(\tau, v\right).$$

\subsection{Details of the Semi-parametric estimation procedure}\label{AppSemiparametricdetails}
\subsubsection{Propensity Score}\label{AppPS}

As noted in the main text, we estimate the propensity score imposing the semiparametric structure: 
\begin{equation}\label{EqPartiallineara}
	P(Z,C,X) = {\alpha}_{0} + {X'\alpha}_{X} + C {\alpha}_{C} + \varphi(Z),
\end{equation}
where $(\alpha_0, \alpha_X', \alpha_C)'$ are unknown finite-dimensional parameters, and $\varphi$ is an unknown (infinite-dimension) function. 
Note that all the series coefficients can be estimated via ordinary least squares, i.e., 
\begin{equation}
	\hat{\theta}^{fs} =\underset{\theta^{fs} \in \Theta^{fs} }{\arg \min }~ n^{-1}\sum_{i=1}^n \left( D_i -  {\alpha}_{0} - X_{i}'{\alpha}_{X} -  C_{i}{\alpha}_{C} - \psi^L(Z_i)' {\alpha}_Z,\right)^2
\end{equation}%
where $\hat{\theta}^{fs} = (\widehat{\alpha}_0, \widehat{\alpha}_X', \widehat{\alpha}_C, \widehat{\alpha}_Z))'$.\footnote{In practice, we approximate $\varphi(\cdot)$ using a linear combination of the vector of basis functions 
	$\psi^L(z) = \left(\psi_{1}(z), \psi_{2}(z), \dots, \psi_{L}(z)\right)'$ for $L \in \mathbb{N}$ as explained by \citet{Chen2007}. In other words, $\varphi(z) \approx \psi^L(z)' \alpha_Z $, such that, the approximation error shrinks to zero as $L \rightarrow \infty$.} Thus, we can compute the fitted propensity score values  from Equation \eqref{EqLogit_OLS}. 

\subsubsection{Conditional Distribution Function}\label{AppCDF}
For the conditional distribution function of $Y \cdot \mathbf{1}\left\lbrace D = d \right\rbrace$ given $P, C, X$ for $d \in \{0,1\}$,  we follow the distribution regression approach introduced by \citet{Foresi1995} and further formalized by \citet{Chernozhukov2013a}. The idea is to pose a model with ``varying coefficients'' for the conditional distribution of $Y \cdot \mathbf{1}\left\lbrace D = d \right\rbrace$, $d \in \{0,1\}$, 
\begin{eqnarray}
	{\Gamma}({P},C,X;y,d) &\equiv& \mathbb{E}\left[ \mathbf{1}\left\lbrace Y \leq y, D = d \right\rbrace\vert P, C,X\right] \nonumber \\
	&=&\Lambda \left(	\beta_0 \left(y,d\right) +X^{\prime }\beta_{X}(y,d) + C \beta_C(y,d) + P \beta_P (y,d)\right) 
	\text{ a.s.}  \label{EqDistRega}
\end{eqnarray}
where $\theta_0(\cdot, \cdot) = (\beta_0 \left(\cdot,\cdot\right), \beta_X \left(\cdot,\cdot\right)', \beta_C \left(\cdot,\cdot\right), \beta_P \left(\cdot,\cdot\right))' \mapsto \Theta
\subseteq \mathbb{R}^{3+k_X}$ is a vector of nonparametric functions, $k_X$ is the dimension of $X$, and $\Lambda$ is a known link function.\footnote{This class of distribution regression models nests and extends many traditional duration models such as the proportional hazard model \citep{Cox1972} and the accelerated time model \citep{Kalbfleisch1980}. See \citet{Delgado2022} for a discussion.} For concreteness,  we focus on a logistic link function, $\Lambda(\cdot) = \exp(\cdot)\big{/}(1 + \exp(\cdot))$.

The form of the feasible likelihood function in this context is: 
\begin{equation}
	\hat{Q}(\theta; y,d)=\frac{1}{n}\sum_{i=1}^{n}\ln
	\ell_\theta(\mathbf{1}\{Y_i \leq y, D_i=d \}, X_i, C_i, \widehat{P}_i; y,d)
\end{equation}%
with 
\begin{equation*}
	\ell_{\theta}(b,x,c,p;y,d)=\Lambda \left( {w}'\theta \right)
	^{b}\left( 1-\Lambda \left( {w}'\theta \right) \right)
	^{1-b},
\end{equation*}%
and $w= (1, x', c, p)'$. Thus, the distribution regression (DR) estimator of $\theta_{0}(y,d)$ is given by
\begin{equation}
	\hat{\theta}\left( y,d\right) =\underset{\theta \in \Theta }{\arg \max }%
	\text{ }\hat{Q}_{n}\left( \theta; y,d \right) . \label{EqDR_coeffs}
\end{equation}%
Notice that computing the distribution regression estimators for several $(y,d)$ points only requires running a sequence of binary regressions. This can be performed in any statistical software.

\subsubsection{Algorithms}\label{AppAlgo}

We summarize all the algorithms for  estimation and inference in this section.

\begin{algorithm}[Semiparametric Estimation of DMTE and QMTE functionals]\label{algo:semi}\phantom{a}
	\begin{enumerate}
		\item Semiparametrically estimate the propensity score using the series partially linear model in Equation \eqref{EqPartiallinear}. Denote its trimmed fitted propensity score values by $\widehat{P}_i$ as defined in Equation \eqref{EqLogit_trimmed}.
		
		\item Define a grid of values for the duration outcome $Y$: $\left\lbrace y_{k} \right\rbrace_{k = 0}^{K}$ such that $y_{k} > y_{k-1}$ for any $k \in \left\lbrace 1, \ldots, K \right\rbrace$ and $K \in \mathbb{N}$.
		
		\item For each $k \in \left\lbrace 0, \ldots, K \right\rbrace$ and each $d \in \left\lbrace 0, 1 \right\rbrace$, estimate the conditional distribution function of $Y \cdot \mathbf{1}\left\lbrace D = d \right\rbrace$ given $P\left(Z,C\right)$, $C$, and $X$ using the distribution regression model in Equation \eqref{EqDistReg}, with estimated coefficients in Equation \eqref{EqDR_coeffs}.
		
		\item For each $k \in \left\lbrace 0, \ldots, K \right\rbrace$ and $d \in \{0, 1 \}$, estimate the derivative of the distribution regression model with respect to $P$ as in Equation \eqref{Eq_gamma}. Denote its estimated fitted value for a given $x$ by $\widehat{\gamma}_{d}(y_k,v,c,x)$ as in Equation \eqref{EqEstimated-Derivatives}.
		
		\item\label{StepMean} For each $k \in \left\lbrace 0, \ldots, K \right\rbrace$ and each $d \in \left\lbrace 0, 1 \right\rbrace$, compute $\widehat{DMTR}_d (y_k,v,x)$ as in Equation \eqref{EqDMTR_d_estimator}.
		
		\item For each value $v\in \mathcal{P}$, $d\in \{0,1\}$, and for a given $x$, ensure that $\widehat{DMTR}_d (y_k,v,x)$ is non-decreasing in $y_{k}$, and bounded between zero and one.
		
		\item For each $k \in \left\lbrace 0, \ldots, K \right\rbrace$, estimate the $DMTE\left(y_{k}, v, x\right)$ using Equation \eqref{DMTE_semi}.
		
		\item Estimate the $QMTE\left(\tau, v, x\right)$ using Equation \eqref{QMTE_semi}, with $$\widehat{QMTR}_d(\tau,v,x) = \min\{y\in \{y_k\}_{k=0}^K \colon \widehat{DMTR}_d(y,v,x)\geq \tau \},~ d\in \{0,1\}.$$
	\end{enumerate}
\end{algorithm}

\begin{algorithm}[Weighted-Bootstrap Implementation]\label{algo:bootstrap}\phantom{a}
	\begin{enumerate}
		\item Estimate DMTE and QMTE according to Algorithm \ref{algo:semi}.
		\item Generate $\{\omega_{i}, i=1,\dots,n\}$ as a sequence of independent and identically distributed non-negative random variables with mean one, variance one, and finite third moment (e.g., $\omega_i \sim Exp\left(1\right)$).
		\item Compute the propensity score coefficients associated with Equation \eqref{EqPartiallinear} by minimizing the weighted least squares function, i.e, \begin{equation}
			\hat{\theta}^{fs, *} =\underset{\theta^{fs} \in \Theta^{fs} }{\arg \min }~ n^{-1}\sum_{i=1}^n \omega_i \left( D_i -  {\alpha}_{0} - X_{i}'{\alpha}_{X} -  C_{i}{\alpha}_{C} - \psi^L(Z_i)' {\alpha}_Z,\right)^2
		\end{equation}%
		where $\hat{\theta}^{fs,*} = (\widehat{\alpha}_0^*, \widehat{\alpha}^{*,\prime}_X, \widehat{\alpha}^*_C, \widehat{\alpha}^{*}_Z))'$.
		Denote its trimmed fitted propensity score values by $\widehat{P}_i^*$ as defined in Equation \eqref{EqLogit_trimmed}, but with $\hat{\theta}^{fs, *}$ in place of $\hat{\theta}^{fs}$.
		
		\item Consider the same grid of values for the duration outcome $Y$ as defined in Step 2 of Algorithm \ref{algo:semi}.
		
		\item For each $k \in \left\lbrace 0, \ldots, K \right\rbrace$ and each $d \in \left\lbrace 0, 1 \right\rbrace$, estimate the conditional distribution function of $Y \cdot \mathbf{1}\left\lbrace D = d \right\rbrace$ given $P\left(Z,C\right)$, $C$, and $X$ using the distribution regression model (Equation \eqref{EqDistReg}) with estimated coefficients 
		\begin{equation}
			\hat{\theta}^*\left( y,d\right) =\underset{\theta \in \Theta }{\arg \max }%
			\text{ }\frac{1}{n}\sum_{i=1}^{n} \omega_i~\ln
			\ell_\theta(\mathbf{1}\{Y_i \leq y, D_i=d \}, X_i, C_i, \widehat{P}_i^*; y,d) . \label{EqDR_coeffs_boot}
		\end{equation}
		\item Follow Steps 4-9 of Algorithm \ref{algo:semi} using $\hat{\theta}^*\left( y,d\right)$ instead of $\hat{\theta}\left( y,d\right)$. Denote by $\widehat{DMTE}^*(y_k,v,x)$ and $\widehat{QMTE}^*(\tau,v,x)$ the distributional and quantile marginal treatment effects estimates. 
		\item Repeat Steps 2-6 $B$ times, e.g., $B=399$, and collect $\left\{  \left(\widehat{DMTE}^*(y_k,v,x)\right)  _{b},b=1\dots,B\right\}$. Do the same for the $\widehat{QMTE}^*(\tau,v,x)$.
		
		\item Obtain the $\left(  1-\alpha\right)  $ quantile of $\left\{ \left| \left(\widehat{DMTE}^*(y_k,v,x) - \widehat{DMTE}(y_k,v,x)\right)_{b}\right|,b=1\dots,B\right\}$, $c^{dmte,\ast}(y_k,v,x;{\alpha})$. Compute the analogous critical values based on $\widehat{QMTE}^*(\tau,v,x)$.
		
		\item Construct the $1-\alpha$ (pointwise) confidence interval for ${DMTE}(y_k,v,x)$ as $\widehat{C}^{dmte}(y_k,v,x) = [\widehat{DMTE}(y_k,v,x) \pm c^{dmte,\ast}(y_k,v,x;{\alpha})]$. Define $\widehat{C}^{qmte}(\tau,v,x;\alpha)$ analogously.\footnote{{If the researcher wants to compute uniform confidence bands instead of pointwise confidence intervals, it is straightforward to adapt the sup-t procedure proposed by \citet{Olea2018} to our setting.}}
	\end{enumerate}
\end{algorithm}

Validity of the weighted bootstrap procedure for asymptotically correct coverage is collected in the following theorem.
\begin{theorem}\label{Thm:semi_boot}
	Under the assumptions of Theorem \ref{Thm:semi}, for any $0<\alpha<1$, and for each $v\in \mathcal{P}$, $x \in \mathcal{X}$, $y < \gamma_C$, and $\tau \in \left( 0,\overline{\tau}(v,x) \right)$, for $n \rightarrow \infty$,
	\begin{enumerate}
		\item[(a)]$\mathbb{P} \left(DMTE(y_k,v,x) \in \widehat{C}^{dmte}(y_k,v,x;\alpha) \right) \rightarrow 1 - \alpha$,
		\item[(b)] $\mathbb{P} \left(QMTE(\tau,v,x) \in \widehat{C}^{qmte}(\tau,v,x;\alpha) \right) \rightarrow 1 - \alpha$.
	\end{enumerate}
\end{theorem}

\textbf{Proof:} See Appendix \ref{AppProofBootstrap}.

\subsection{Other possible covariate aggregation}\label{AppCovariateAggregationextra}
We note that other aggregations of the covariate-specific marginal treatment effect functionals do exist, but may be more challenging to estimate. For instance, one may be interested in functionals of the ``unconditional'' $DMTR_d (y,v)$, defined as
\begin{eqnarray*}
	DMTR_d (y,v)&=& \expe{DMTR_d (y,P,X)|D=d, P=v} \\
	&=& \int DMTR_d (y,v,\bar{x})f_{X|D,P}(\bar{x}|D=d,P=v)d\bar{x}.
\end{eqnarray*}
It should be clear that $DMTR_d (y,v)$ is different from $DMTR_d^{avg} (y,v)$, though the latter can be more easily estimated as it does not require estimation of conditional densities as the former does.   If one were to focus on $DMTR_d (y,v)$, all the uncertainty in estimating it would come from the estimation of the conditional density of $X$. This follows from Theorem $\ref{Thm:semi}$ establishing that $DMTR_d (y,v,{x})$ is $\sqrt{n}$-consistent with discrete $X$'s, while $f_{X|D,P}$ converges at slower rates. Developing a higher-order asymptotic analysis for estimators for $DMTR_d (y,v)$ would be interesting. We leave a detailed analysis of it for future research.

\subsection{Main Regularity Conditions}\label{AppMainRegularity}
Although identification does not rely on any parametric assumption, some of them aid the estimation procedure. Covariates are easily incorporated when semiparametric assumptions are made and  the curse of dimensionality is avoided. Additionally, semiparametric assumptions demand less data. In this appendix, we follow \citet{Rothe2009} closely, but adapt his setting for the case where the link function is known instead of unknown. For the rest of the section, we assume an $i.i.d$ sample. In this context, we introduce the following assumption:
\begin{assumption}[Semiparametric CDF]\label{AsSP1}
	Let $\mathbb{P}\left[\left. Y\leq y, D=d \right\vert P, C, X\right]=\Lambda(\beta_{0}\left(d,y\right)+\beta_{C}\left(d,y\right)C+\beta_{P}\left(d,y\right)P+X^{\prime} \beta_{X}\left(d,y\right))$, where $\Lambda()$ is a known link function up to a finite dimensional vector (such as the logistic link), which is continuously differentiable in the index.  Let $\Lambda^{\prime}(.)$ be the derivative of $\Lambda(.)$, which is continuous.
\end{assumption}

For the sake of exposition, let $W_{y,d}=1\{Y\leq y, D=d \}$, $H=\{1,C,P,X\}, \hat{H}=\{1,C,\hat{P},X\}$, {$H_{v}=\{1,C,v,X\}$, $\beta_{d,y} \coloneqq \left(\beta_{0}\left(d,y\right),\beta_{C}\left(d,y\right),\beta_{P}\left(d,y\right), \beta_{X}\left(d,y\right)\right)$ for any $y$ and $d \in \left\lbrace 0,1 \right\rbrace$}.
Taking the derivative with respect to $P$ for $\Lambda(\cdot)$ for both $W_{y,1}$ and $W_{y,0}$, we get the $DMTE(y,v)$ as $$DMTR_1(y,v)-DMTR_0(y,v)=\Lambda^{\prime}(\beta_{1,y}H_{v})\beta_{P}\left(1,y\right)-\Lambda^{\prime}(\beta_{0,y}H_{v})\beta_{P}\left(0,y\right)$$ If $P$ was known, it would be easy to estimate the DMTE as in the parametric part.

Since $P$ is not known, we can estimate $P$ in a semiparametric first stage, and obtain estimates for $\beta_{d,y}$ from the following maximum-likelihood procedure.\footnote{In the semiparametric first stage, we can estimate $P$ using a standard series estimator.} We focus on $d=1$ for the sake of exposition and denote the semiparametric first-stage estimates by $\hat{P}$. Define
\begin{eqnarray} \label{Log}
	Ln(\beta_{1,y}, \hat{P})=\max_{\beta_{1,y}} \frac{1}{N} \sum_i W_{y,1,i} log[\Lambda(\beta_{1,y} \hat{H}_i)]+(1-W_{y,1,i})log[1-\Lambda(\beta_{1,y} \hat{H}_i)]
\end{eqnarray}
with solution $\hat{\beta}_{1,y}(\hat{P})$. If $P$ was known, we could use the following unfeasible standard maximum likelihood procedure:
\begin{eqnarray} \label{Logun}
	Ln(\beta_{1,y}, P)=\max_{\beta_{1,y}} \frac{1}{N} \sum_i W_{y,1,i} log[\Lambda(\beta_{1,y} H_i)]+(1-W_{y,1,i})log[1-\Lambda(\beta_{1,y} H_i)]
\end{eqnarray}
with solution $\hat{\beta}_{1,y}(P)$.

To analyze our semiparametric estimator \eqref{Log}, we need to ensure that the unfeasible estimator in \eqref{Logun} is well-behaved. To do so, we impose the following assumption:
\begin{assumption}[Unfeasible Likelihood]\label{AsSP2}
	The maximum likelihood estimator of \eqref{Logun}, follows standard regularity conditions from \citet{NeweyMcFadden1994} for consistency and asymptotic normality.
\end{assumption}
Assumption \ref{AsSP2} ensures that standard parametric inference could be performed if $P$ was observed, implying that $\hat{\beta}_{1,y}(P) \xrightarrow{p} \beta_{1,y}$. Since $\Lambda$ is the logistic link, the result is standard.

To ensure that our semiparametric estimator is consistent and derive its asymptotic distribution, we need to ensure that our propensity score estimator converges sufficiently fast and satisfies some regularity conditions. To do so, we follow \citet{Rothe2009} and impose the following assumption.
\begin{assumption}[First stage assumptions]\label{AsSP3}
	Let $\hat{P}$ satisfy:
	\begin{enumerate}
		\item $\hat{P}_i-P_i=\frac{1}{N}\sum_jw_n(Z_i,C_i,X_i,Z_j,C_j,X_j)\phi_j+r_{in}$ with $\max_i||r_{in}||=o_p(N^{-\frac{1}{2}})$ and $\max_i|\hat{P}_i-P_i|=o_p(N^{-\frac{1}{4}})$ where $\phi_j=\phi(D_j,Z_j,C_j,X_j)$ is an influence function with $\mathbb{E}\left[\phi_j|Z_j,C_j,X_j\right]=0$ and $\mathbb{E}\left[\phi_j^2|Z_j,C_j,X_j\right]< \infty$ and weights $w_n(Z_i,C_i,X_i,Z_j,C_j,X_j)=o(N)$.
		
		\item There exists a space $\mathcal{P}$ such that $\mathbb{P}(\hat{P} \in \mathcal{P}) \rightarrow 1$ and $\int_0^\infty \sqrt{\log N(\lambda, \mathcal{P}, ||\cdot||_\infty)d\lambda}<\infty$ where $N(\lambda, \mathcal{P}, ||\cdot||_\infty)$ is the covering number with respect to the $L_{\infty}$ norm of the class of functions $\mathcal{P}$.
	\end{enumerate}
\end{assumption}
{ Assumption \ref{AsSP3} is analogous to Assumption 8 in \citet{Rothe2009} and can be interpreted as a high-level condition on the propensity score estimator.} Assumption \ref{AsSP3}(i) states that the estimator admits a certain asymptotic expansion. Assumption \ref{AsSP3}(ii) requires the propensity score estimator to take values in some well-behaved function space with probability approaching 1. These are relatively mild conditions. Indeed, it is easy to show that series-based and kernel-based estimators satisfy Assumption \ref{AsSP3}; see, e.g., \citet[page 55]{Rothe2009}, for a discussion.

\subsection{Additional Regularity Conditions}\label{AppAdditionalRegularity}
Besides the previously mentioned conditions, which are the key components of the semi-parametric procedure, we need to add additional regularity conditions to ensure that our procedures work.

\begin{assumption}\label{AsRC1}
	Assume that $\mathbb{P}\left[\left. Y\leq y, D=d \right\vert P, C, X\right], \mathbb{P}\left[\left. D=d \right\vert P, C, X\right]$ are twice continuously differentiable in $P$. 
\end{assumption}

\begin{assumption}\label{AsRC2}
	Assume that the support of $Z$ is known and is a Cartesian
	product of compact connected intervals on which $Z$ has a
	probability density function that is bounded away from zero. 
\end{assumption}

\begin{assumption}\label{AsRC3}
	Assume that $\psi_{l}(z)$, for $l \in L$ are $r_{\psi}$-times continuously differentiable on the support of $Z$ for $r_{\psi}\geq 2$, where $\psi_l(z)$ is used to approximate   $\varphi(z)$, an unknown function.
\end{assumption}

\begin{assumption}\label{AsRC4}
	Assume that $Y^*(d)$ is continuous with respect to the Lebesgue measure. 
\end{assumption}

\begin{assumption}\label{AsRC5}
	Assume that  $C$ and $Z$ have a bounded support
\end{assumption}

\begin{assumption}\label{AsRC6}
	Assume that  the distribution of $C,Z,D$ belongs to an exponential family.
\end{assumption}

Assumption \ref{AsRC1} assures that we can apply the Leibniz Integral Rule to identify the $DMTR$ functions. 
Assumptions \ref{AsRC2} and \ref{AsRC3} are standard in the series estimation literature. In particular, Assumption \ref{AsRC3} implies that the asymptotic bias
(due to the series approximation by regression splines) converges to zero at a rate of $L^{-r_{\psi}}$
as the number of approximation functions,
$L$, diverges to infinity. Assumption \ref{AsRC4} is a regularity condition that ensures point identification for our quantile results.
Assumption \ref{AsRC5} assures that the conditions from Assumption 2 of \cite{IchimuraNewey2022} hold.  Assumption \ref{AsRC6} ensures identification via a moment equation of the propensity score in the spirit of \cite{newey2003instrumental}, exploiting bounded completeness. 

\subsection{Proof of Theorems \ref{Thm:semi} and \ref{ThmAsymptoticRMTE}: Consistency and Asymptotic Normality}\label{AppProofVariance}
To ensure our estimators of the $DMTE, QMTE, MTE$ are consistent and asymptotically normal, we first need to ensure consistency of the feasible estimator of all the components of the DMTE. Then, we use functional  approximation results to show asymptotic results for $DMTE, QMTE, MTE$.
\par 
\subsubsection{Consistency of the $\beta_{d,y}$ estimators}
We need to prove asymptotic equivalence between the solution of Equations \eqref{Log} and \eqref{Logun}. Then, by Assumption \ref{AsSP2}, we get the consistency of the feasible semiparametric estimator.
Note that
\footnotesize
\begin{align*}
	& \sup_{\beta_{1,y}}|Ln(\beta_{1,y}, \hat{P})-Ln(\beta_{1,y}, P)| \\
	& \hspace{20pt} \leq \Big[\inf_{\beta_{1,y}} \min_i \{\Lambda(\beta_{1,y} \hat{H}_i), \Lambda(\beta_{1,y} H_i), 1-\Lambda(\beta_{1,y} \hat{H}_i), 1-\Lambda(\beta_{1,y} H_i) \} \big( \sup_{\beta_{1,y}} \max_i |\Lambda(\beta_{1,y} \hat{H}_i)-\Lambda(\beta_{1,y} H_i)|\big)  \Big] \\
	& \hspace{20pt} \leq \Big[ O(1)\big( \sup_{\beta_{1,y}} \max_i |\Lambda(\beta_{1,y} \hat{H}_i)-\Lambda(\beta_{1,y} H_i)|\big)  \Big]  \\
	& \hspace{20pt} = o_p(1),
\end{align*}
\normalsize
where the first inequality can be derived using standard algebraic manipulations. Moreover, the second inequality holds because $\Lambda(\cdot) \in (0,1)$. Furthermore, note that $\Lambda(\cdot)$ is continuous and $\max_i |\hat{H}_i-H_i|$ converges due to Assumption \ref{AsSP3}, implying that $\max_i |\Lambda(\beta_{1,y} \hat{H}_i)-\Lambda(\beta_{1,y} H_i)|$ converges due to the continuous mapping theorem. Finally, since the supremum over $\beta_{1,y}$ in the third line is also continuous, we can apply the continuous mapping theorem again to prove the last equality.

Furthermore, $Ln(\beta_{1,y}, P)$ is a standard parametric likelihood, implying that it converges uniformly in $\beta_{1,y}$ to its expectation \citep[Lemma 2.4]{NeweyMcFadden1994}. Formally, we have that
\begin{eqnarray*}
	\sup_{\beta_{1,y}}|Ln(\beta_{1,y}, P)-L(\beta_{1,y})|=o_p(1)
\end{eqnarray*}
where $L(\beta_{1,y})=\mathbb{E}\left[Ln(\beta_{1,y}\right]=\mathbb{E}\left[W_{y,1,i}log(\Lambda(\beta_{1,y}H))
+ (1-W_{y,1,i})log(1-\Lambda(\beta_{1,y}H))\right]$
is a non-random function that is continuous in $\beta_{1,y}$. Taken together,
it follows from the triangle inequality that
\begin{eqnarray*}
	\sup_{\beta_{1,y}}|Ln(\beta_{1,y}, \hat{P})-L(\beta_{1,y})|=o_p(1)
\end{eqnarray*}
implying that $\hat{\beta}_{1,y}(P)$ is consistent whenever $L(\beta_{1,y})$ attains a unique maximum at the true value of the parameter, which is the case by our identification results and Assumption \ref{AsSP1}.

As a consequence, the consistency of our feasible semiparametric estimator follows from Theorem 2.1 by \citet{NeweyMcFadden1994} via Assumption \ref{AsSP2}.

\subsubsection{Asymptotic distribution of the $\beta_{d,y}$ estimators}
Now, we derive the asymptotic distribution of our semiparametric estimator in \eqref{Log}. Let $Ln(\beta_{1,y}, \hat{P}_i)_{\beta}, Ln(\beta_{1,y}, P_i)_{\beta}, L(\beta_{1,y},P_i)_{\beta} $ be the derivative with respect to $\beta$ of the individual's feasible log-likelihood, unfeasible log-likelihood and true log-likelihood respectively (the score). Define similarly the second-order derivative.
\par
From a standard second-order Taylor expansion of the semiparametric log-likelihood around $\beta_{1,y}$, we have that
\begin{eqnarray}\label{Asymdis1}
	\sqrt{N}(\hat{\beta}_{1,y}(\hat{P})-\beta_{1,y})=\left[\frac{1}{N} \sum_i Ln(\Bar{\beta}_{1,y}, \hat{P}_i)_{\beta,\beta}\right]^{-1}\sqrt{N}\frac{1}{N}\sum_{i} Ln(\beta_{1,y}, \hat{P}_i)_{\beta},
\end{eqnarray}
where $\Bar{\beta}_{1,y}$ is between the estimated and true values. By the first part of Assumption \ref{AsSP3} and the consistency of $\hat{\beta}_{1,y}(\hat{P})$, we know that, $$\left[\frac{1}{N} \sum_i Ln(\Bar{\beta}_{1,y}, \hat{P}_i)_{\beta,\beta}\right]^{-1} \xrightarrow{p} \mathbb{E}\left[L(\beta_{1,y},P_i)_{\beta,\beta}\right]^{-1} \eqqcolon \Sigma.$$

Now, we focus on the last term in \eqref{Asymdis1}:
\begin{eqnarray*}
	\sum_{i} Ln(\beta_{1,y}, \hat{P}_i)_{\beta}&=&\sum_{i}W_{y,1,i} \frac{\partial log[\Lambda(\beta_{1,y} \hat{H}_i)]}{\partial \beta}+(1-W_{y,1,i})\frac{\partial log[1-\Lambda(\beta_{1,y} \hat{H}_i)]}{\partial \beta} \\
	&=& \sum_{i}W_{y,1,i} \begin{bmatrix}
		\frac{\Lambda^{\prime}(\beta_{0}\left(1,y\right)+\beta_{C}\left(1,y\right)C_i+\beta_{P}\left(1,y\right)\hat{P}_i+\beta_{X}\left(1,y\right)X_i)}{\Lambda(\beta_{0}\left(1,y\right)+\beta_{C}\left(1,y\right)C_i+\beta_{P}\left(1,y\right)\hat{P}_i+\beta_{X}\left(1,y\right)X_i)}\\
		\frac{\Lambda^{\prime}(\beta_{0}\left(1,y\right)+\beta_{C}\left(1,y\right)C_i+\beta_{P}\left(1,y\right)\hat{P}_i+\beta_{X}\left(1,y\right)X_i}{\Lambda(\beta_{0}\left(1,y\right)+\beta_{C}\left(1,y\right)C_i+\beta_{P}\left(1,y\right)\hat{P}_i+\beta_{X}\left(1,y\right)X_i)}C_i\\
		\frac{\Lambda^{\prime}(\beta_{0}\left(1,y\right)+\beta_{C}\left(1,y\right)C_i+\beta_{P}\left(1,y\right)\hat{P}_i+\beta_{X}\left(1,y\right)X_i)}{\Lambda(\beta_{0}\left(1,y\right)+\beta_{C}\left(1,y\right)C_i+\beta_{P}\left(1,y\right)\hat{P}_i+\beta_{X}\left(1,y\right)X_i)}\hat{P}_i \\
		\frac{\Lambda^{\prime}(\beta_{0}\left(1,y\right)+\beta_{C}\left(1,y\right)C_i+\beta_{P}\left(1,y\right)\hat{P}_i+\beta_{X}\left(1,y\right)X_i)}{\Lambda(\beta_{0}\left(1,y\right)+\beta_{C}\left(1,y\right)C_i+\beta_{P}\left(1,y\right)\hat{P}_i+\beta_{X}\left(1,y\right)X_i)} X_i\\
	\end{bmatrix}\\
	&+&(1-W_{y,1,i})\begin{bmatrix}
		\frac{-\Lambda^{\prime}(\beta_{0}\left(1,y\right)+\beta_{C}\left(1,y\right)C_i+\beta_{P}\left(1,y\right)\hat{P}_i+\beta_{X}\left(1,y\right)X_i)}{1-\Lambda(\beta_{0}\left(1,y\right)+\beta_{C}\left(1,y\right)C_i+\beta_{P}\left(1,y\right)\hat{P}_i+\beta_{X}\left(1,y\right)X_i)}\\
		\frac{-\Lambda^{\prime}(\beta_{0}\left(1,y\right)+\beta_{C}\left(1,y\right)C_i+\beta_{P}\left(1,y\right)\hat{P}_i+\beta_{X}\left(1,y\right)X_i)}{1-\Lambda(\beta_{0}\left(1,y\right)+\beta_{C}\left(1,y\right)C_i+\beta_{P}\left(1,y\right)\hat{P}_i+\beta_{X}\left(1,y\right)X_i)}C_i\\
		\frac{-\Lambda^{\prime}(\beta_{0}\left(1,y\right)+\beta_{C}\left(1,y\right)C_i+\beta_{P}\left(1,y\right)\hat{P}_i+\beta_{X}\left(1,y\right)X_i)}{1-\Lambda(\beta_{0}\left(1,y\right)+\beta_{C}\left(1,y\right)C_i+\beta_{P}\left(1,y\right)\hat{P}_i+\beta_{X}\left(1,y\right)X_i)}\hat{P}_i \\
		\frac{-\Lambda^{\prime}(\beta_{0}\left(1,y\right)+\beta_{C}\left(1,y\right)C_i+\beta_{P}\left(1,y\right)\hat{P}_i+ \beta_{X}\left(1,y\right)X_i)}{1-\Lambda(\beta_{0}\left(1,y\right)+\beta_{C}\left(1,y\right)C_i+\beta_{P}\left(1,y\right)\hat{P}_i+\beta_{X}\left(1,y\right)X_i)} X_i\\
	\end{bmatrix}
\end{eqnarray*}

Considering the path $P_e=(1-e)P+e[\hat{P}-P]$, we take the path-wise derivative of  $\sum_{i} Ln(\beta_{1,y}, P_i)_{\beta}$ at direction $\hat{P}-P$ (the derivative of the submodel $P_e$ evaluated at $e=0$). This object is denoted by $\sum_{i} Ln(\beta_{1,y}, P_i)_{\beta,P_i}$ and is equal to
\footnotesize
\begin{eqnarray}\label{U-stat}
	\sum_{i} Ln(\beta_{1,y}, P_i)_{\beta,P_i}&=&\sum_{i}W_{y,1,i} \begin{bmatrix}
		\frac{\Lambda^{\prime\prime}(\beta_{1,y}H_i)\Lambda(\beta_{1,y}H_i)-\Lambda^{\prime}(\beta_{1,y}H_i)^2}{\Lambda(\beta_{1,y}H_i)^2}\beta_{P}\left(1,y\right)[\hat{P}_i-P_i]\\
		\frac{\Lambda^{\prime\prime}(\beta_{1,y}H_i)\Lambda(\beta_{1,y}H_i)-\Lambda^{\prime}(\beta_{1,y}H_i)^2}{\Lambda(\beta_{1,y}H_i)^2}C_i\beta_{P}\left(1,y\right)[\hat{P}_i-P_i] \\
		\left[\frac{\Lambda^{\prime\prime}(\beta_{1,y}H_i)\Lambda(\beta_{1,y}H_i)-\Lambda^{\prime}(\beta_{1,y}H_i)^2}{\Lambda(\beta_{1,y}H_i)^2}P_i\beta_{P}\left(1,y\right)+\frac{\Lambda^{\prime}(\beta_{1,y}H_i)}{\Lambda(\beta_{1,y}H_i)}\right][\hat{P}_i-P_i]\\
		\frac{\Lambda^{\prime\prime}(\beta_{1,y}H_i)\Lambda(\beta_{1,y}H_i)-\Lambda^{\prime}(\beta_{1,y}H_i)^2}{\Lambda(\beta_{1,y}H_i)^2}X_i\beta_{P}\left(1,y\right)[\hat{P}_i-P_i]
	\end{bmatrix}\\
	&+& \sum_{i}(1-W_{y,1,i}) \begin{bmatrix}
		\frac{-\Lambda^{\prime\prime}(\beta_{1,y}H_i)[1-\Lambda(\beta_{1,y}H_i)]-\Lambda^{\prime}(\beta_{1,y}H_i)^2}{[1-\Lambda(\beta_{1,y}H_i)]^2}\beta_{P}\left(1,y\right)[\hat{P}_i-P_i] \\
		\frac{-\Lambda^{\prime\prime}(\beta_{1,y}H_i)[1-\Lambda(\beta_{1,y}H_i)]-\Lambda^{\prime}(\beta_{1,y}H_i)^2}{[1-\Lambda(\beta_{1,y}H_i)]^2}C_i\beta_{P}\left(1,y\right)[\hat{P}_i-P_i] \\
		\left[\frac{-\Lambda^{\prime\prime}(\beta_{1,y}H_i)[1-\Lambda(\beta_{1,y}H_i)]-\Lambda^{\prime}(\beta_{1,y}H_i)^2}{[1-\Lambda(\beta_{1,y}H_i)]^2}P_i\beta_{P}\left(1,y\right)+\frac{-\Lambda^{\prime}(\beta_{1,y}H_i)}{1-\Lambda(\beta_{1,y}H_i)}\right][\hat{P}_i-P_i] \\
		\frac{-\Lambda^{\prime\prime}(\beta_{1,y}H_i)[1-\Lambda(\beta_{1,y}H_i)]-\Lambda^{\prime}(\beta_{1,y}H_i)^2}{[1-\Lambda(\beta_{1,y}H_i)]^2}X_i\beta_{P}\left(1,y\right)[\hat{P}_i-P_i]
	\end{bmatrix} \nonumber
\end{eqnarray}
\normalsize
We also define $\mathbb{E}\left[Ln(\beta_{1,y}, P_i)_{\beta,P_i}\right]$ analogously.

With these results in hand, we go back to \eqref{Asymdis1} and expand around the deviations of the true first stage:
\begin{eqnarray}\label{Asymdis2}
	\sqrt{N}(\hat{\beta}_{1,y}(\hat{P})-\beta_{1,y})=\Sigma \cdot \sqrt{N}\left(\frac{1}{N}\sum_{i} Ln(\beta_{1,y},P_i)_{\beta}+\frac{1}{N}\sum_{i} Ln(\beta_{1,y}, P_i)_{\beta,P_i}\right)+o_p(1)
\end{eqnarray}
where $\sum_{i} \frac{1}{N}Ln(\beta_{1,y},P_i)_{\beta}$ is the usual estimate of the score, which has mean $0$. Thus, if we can show that the second term also has mean $0$, the asymptotic normality of our semiparametric estimator follows by a standard multivariate $CLT$ for the vector $\left[\frac{1}{N}\sum_{i} Ln(\beta_{1,y},P_i)_{\beta}, \frac{1}{N}\sum_{i} Ln(\beta_{1,y}, P_i)_{\beta,P_i}\right]$.

Since all the components of \eqref{U-stat} have a similar structure, we can focus on one of them and the results are symmetric for the rest. 
\par 
Consider
$\frac{1}{N}\sum_{i}W_{y,1,i}
\frac{\Lambda^{\prime\prime}(\beta_{1,y}H_i)\Lambda(\beta_{1,y}H_i)-\Lambda^{\prime}(\beta_{1,y}H_i)^2}{\Lambda(\beta_{1,y}H_i)^2}\beta_{P}\left(1,y\right)[\hat{P}_i-P_i]$. 
\par 
For notation simplicity, let $\frac{\Lambda^{\prime\prime}(\beta_{1,y}H_i)\Lambda(\beta_{1,y}H_i)-\Lambda^{\prime}(\beta_{1,y}H_i)^2}{\Lambda(\beta_{1,y}H_i)^2}\beta_{P}\left(1,y\right) \eqqcolon A(\beta_{1,y}H_i)$. 
\par 
Note that
\begin{align*}
	& \frac{1}{N}\sum_{i} W_{y,1,i}
	A(\beta_{1,y}H_i)[\hat{P}_i-P_i]  \\
	& \hspace{20pt} = \frac{1}{N^2}\sum_{i}\sum_j w_n(Z_i,C_i,Z_j,C_j)W_{y,1,i}A(\beta_{1,y}H_i)\phi_j+o_p(N^{-\frac{1}{2}}) \\
	& \hspace{20pt} = \frac{1}{N}\sum_{i}\mathbb{E}\left[ w_n(Z_i,C_i,Z,C)\mathbb{E}\left[W_{y,1,i}A(\beta_{1,y}H_i)|H,Z,C\right]|Z_i,C_i\right]\phi_i+o_p(N^{-\frac{1}{2}})
\end{align*}
where the first equality is due to Assumption \ref{AsSP3} and the second equality is due to the $U$-statistics Hajek projection.

Now, by a standard law of large numbers, we have that
\begin{align*}
	& \frac{1}{N}\sum_{i}\mathbb{E}\left[ w_n(Z_i,C_i,Z,C)\mathbb{E}\left[W_{y,1,i}A(\beta_{1,y}H_i)|H,Z,C\right]|Z_i,C_i\right]\phi_i+o_p(N^{-\frac{1}{2}}) \\
	& \hspace{20pt} \xrightarrow{p} \mathbb{E}[\mathbb{E}\left[ w_n(Z_i,C_i,Z,C)\mathbb{E}\left[W_{y,1,i}A(\beta_{1,y}H_i)|H,Z,C\right]|Z_i,C_i\right]\phi_i] \\
	& \hspace{20pt} = \mathbb{E}[\mathbb{E}[\mathbb{E}\left[ w_n(Z_i,C_i,Z,C)\mathbb{E}\left[W_{y,1,i}A(\beta_{1,y}H_i)|H,Z,C\right]|Z_i,C_i\right]\phi_i|Z_i,C_i]] \\
	& \hspace{20pt} = \mathbb{E}[\mathbb{E}\left[ w_n(Z_i,C_i,Z,C)\mathbb{E}\left[W_{y,1,i}A(\beta_{1,y}H_i)|H,Z,C\right]|Z_i,C_i\right]\mathbb{E}[\phi_i|Z_i,C_i]] \\
	& \hspace{20pt} = 0
\end{align*}
where the last equality is due to Assumption \ref{AsSP3}. Thus, a standard $CLT$ assures the asymptotic normality of the estimator for the parametric part.
\subsubsection{Asymptotic behaviour of the DMTE estimator}

We derive the influence function of our estimator to be able to express the variance and to apply functional limit theory results. Since DMTE is the difference of two $DMTR$ functions, we can focus on the influence function of one of the $DMTR$ and then apply linearity to get the influence function of the estimator for the DMTE. We will express the $DMTR$ as an unconditional moment that depends on two parameters ($\beta_{d,y}, P$) that are themselves expressed as unconditional moments.
\par 
In particular, $P(Z,C,X)$ is such that $\mathbb{E}(D|Z,C,X)=P(Z,C,X)$. Omitting $X$ for simplicity, we can note that:
\begin{eqnarray*}
	\mathbb{E}[CZ(D-P(Z,C))]=\mathbb{E}[CZ(\mathbb{E}(D-P(Z,C)|Z,C))]   
\end{eqnarray*}
Note that under Assumption \ref{AsRC6} and since $CZ\neq 0$
$$\mathbb{E}[CZ(D-P(Z,C))]=0 \iff \mathbb{E}(D-P(Z,C)|Z,C)=0. $$

Then we can express the moment that determines $P$  as: 
\begin{eqnarray}\label{MOM1}
	\mathbb{E}[CZ(D-P(Z,C))]=0  
\end{eqnarray}
where  $CZ$ is normalized to be negative to avoid negative weights issues later. 

Similarly, recall that $L(\beta_{1,y})=\mathbb{E}\left[Ln(\beta_{1,y})\right]=\mathbb{E}\left[W_{y,1,i}log(\Lambda(\beta_{1,y}H))
+ (1-W_{y,1,i})log(1-\Lambda(\beta_{1,y}H))\right]$
is a non-random function that is continuous and differentiable in $\beta_{1,y}$. Note $L(\beta_{1,y})$ is itself a function of $P$ since $P$ is inside $H$.  We can rewrite this as a moment equality using the score as:  
\begin{eqnarray}\label{MOM2}
	\mathbb{E}[S(\beta_{1,y}[P])]=\mathbb{E}\left[ W_{y,1,i}\frac{\partial log(\Lambda(\beta_{1,y}H))}{\partial \beta_{1,y}}
	+ (1-W_{y,1,i})\frac{\partial log(1-\Lambda(\beta_{1,y}H))}{\partial \beta_{1,y} } \right] =0  
\end{eqnarray}
Let $N = \left(P, \mathbf{1} \left\lbrace C > y \right\rbrace \right)$. Note that $DMTR_{1}(y,v)$ is a function of $v$. To properly establish the asymptotic distribution for the
pointwise estimator, one must account for the localization at $P = v$.
Recall $DMTR_{1}(y,v)= \Lambda^{\prime}(\beta_{0}\left(1,y\right)+\beta_{C}\left(1,y\right)C+\beta_{P}\left(1,y\right)v)\beta_{P}\left(1,y\right)$. Thus, to obtain the IF, we can apply the chain rule of influence functions.  The key issue is that there are several estimated parameters for which their respective IF need to account for. In particular, these depend on a non-parametric component.  

To be more specific, the relevant moments that define the parameters of interest are
\begin{align}
	0 & = \mathbb{E}[CZ(D-P)]  \label{Moment1} \\ ~ \nonumber \\
	0 & = \mathbb{E}\left[ W_{y,1}\dfrac{\Lambda^{\prime}((\beta_{0}\left(1,y\right)+\beta_{C}\left(1,y\right)C+\beta_{P}\left(1,y\right)P)}{\Lambda((\beta_{0}\left(1,y\right)+\beta_{C}\left(1,y\right)C+\beta_{P}\left(1,y\right)P)}  \right. \nonumber \\
	& \hspace{40pt} \left. - (1-W_{y,1})\frac{\Lambda^{\prime}((\beta_{0}\left(1,y\right)+\beta_{C}\left(1,y\right)C+\beta_{P}\left(1,y\right)P)}{1-\Lambda((\beta_{0}\left(1,y\right)+\beta_{C}\left(1,y\right)C+\beta_{P}\left(1,y\right)P)}  \right] \label{Moment2a} \\ ~ \nonumber \\
	0 & = \mathbb{E}\left[ W_{y,1}\dfrac{\Lambda^{\prime}((\beta_{0}\left(1,y\right)+\beta_{C}\left(1,y\right)C+\beta_{P}\left(1,y\right)P)}{\Lambda((\beta_{0}\left(1,y\right)+\beta_{C}\left(1,y\right)C+\beta_{P}\left(1,y\right)P)} C  \right. \nonumber \\
	& \hspace{40pt} \left. - (1-W_{y,1})\frac{\Lambda^{\prime}((\beta_{0}\left(1,y\right)+\beta_{C}\left(1,y\right)C+\beta_{P}\left(1,y\right)P)}{1-\Lambda((\beta_{0}\left(1,y\right)+\beta_{C}\left(1,y\right)C+\beta_{P}\left(1,y\right)P)} C  \right] \label{Moment2b} \\ ~ \nonumber \\
	0 & = \mathbb{E}\left[ W_{y,1}\dfrac{\Lambda^{\prime}((\beta_{0}\left(1,y\right)+\beta_{C}\left(1,y\right)C+\beta_{P}\left(1,y\right)P)}{\Lambda((\beta_{0}\left(1,y\right)+\beta_{C}\left(1,y\right)C+\beta_{P}\left(1,y\right)P)} P  \right. \nonumber \\
	& \hspace{40pt} \left. - (1-W_{y,1})\frac{\Lambda^{\prime}((\beta_{0}\left(1,y\right)+\beta_{C}\left(1,y\right)C+\beta_{P}\left(1,y\right)P)}{1-\Lambda((\beta_{0}\left(1,y\right)+\beta_{C}\left(1,y\right)C+\beta_{P}\left(1,y\right)P)} P \right] \label{Moment2c}
\end{align}
To aid ourselves with the mechanics behind the derivation in a more manageable scenario, suppose we also care about the  parameter defined by: 
\begin{eqnarray}\label{MOM3}
	\mathbb{E}[(\Lambda^{\prime}(\beta_{0}\left(1,y\right)+\beta_{C}\left(1,y\right)C+\beta_{P}\left(1,y\right)P)\beta_{P}\left(1,y\right)-ADMTR_1) \cdot P \cdot \mathbf{1} \left\lbrace C > y \right\rbrace]=0  
\end{eqnarray} 
which identifies a weighted average of $DMTR$  over the distribution of $P$. This would add to the previous moments: 
\begin{align}
	0 & = \mathbb{E}\left[ (\Lambda^{\prime}(\beta_{0}\left(1,y\right)+\beta_{C}\left(1,y\right)C+\beta_{P}\left(1,y\right)P)\beta_{P}\left(1,y\right)-ADMTR_1) \cdot P \cdot \mathbf{1}\left\lbrace C > y \right\rbrace \right] \label{Moment3}
\end{align}

We now follow \citet{Newey1994}, \citet{IchimuraNewey2022}, and \citet{ACKERBERGetal2014}. We will assume that standard conditions for the interchange of integration and differentiation hold (such as dominated convergence theorem conditions). 
\par 
Note that:
\begin{eqnarray}\label{Scoreauxiliar}
	\mathbb{E}[CZ(D-P)]=\int_{C\times Z \times D} cz(d-P)f(c,z,d)d \mu(c,z,d)=\int_{\mathcal{D}}cz(d-P)f(\delta)d \mu(\delta)
\end{eqnarray}
This also holds for the rest of the moments, where $\delta$ is the full data vector. This is useful to be able to replicate the form of Equation (3.10) in \citet{Newey1994}, which is instrumental in deriving the influence functions.
\par 
We want to derive the influence function of $ADMTR_1$. Thus, we can start with \eqref{Moment3} and consider a parametric submodel $t$ for the nuisance parameters and differentiate. Let $s(\delta)$ be the score of the data. 
\begin{align*}
	& \frac{\partial \mathbb{E}_t[(\Lambda^{\prime}(\beta_{0}\left(1,y\right)(t)+\beta_{C}\left(1,y\right)(t)C+\beta_{P}\left(1,y\right)P(t))\beta_{P}\left(1,y\right)(t)-ADMTR_1)P(t)\mathbf{1} \left\lbrace C > y \right\rbrace ]}{\partial t}\bigg \vert_{t=0}\\
	& \hspace{20pt} = \mathbb{E}[(\Lambda^{\prime}(\beta_{0}\left(1,y\right)+\beta_{C}\left(1,y\right)C+\beta_{P}\left(1,y\right)P)\beta_{P}\left(1,y\right)-ADMTR_1)P\mathbf{1} \left\lbrace C > y \right\rbrace s(\delta)]\\
	& \hspace{40pt} + \frac{\partial \mathbb{E}[(\Lambda^{\prime}(\beta_{0}\left(1,y\right)(t)+\beta_{C}\left(1,y\right)(t)C+\beta_{P}\left(1,y\right)(t)P)\beta_{P}\left(1,y\right)(t)-ADMTR_1)P\mathbf{1} \left\lbrace C > y \right\rbrace ]}{\partial t}\bigg \vert_{t=0} \\
	& \hspace{40pt} + \frac{\partial \mathbb{E}[(\Lambda^{\prime}(\beta_{0}\left(1,y\right)+\beta_{C}\left(1,y\right)C+\beta_{P}\left(1,y\right)P(t))\beta_{P}\left(1,y\right)-ADMTR_1)P(t)\mathbf{1} \left\lbrace C > y \right\rbrace ]}{\partial t}\bigg \vert_{t=0} 
\end{align*}
By the implicit function theorem, we have that
\begin{align*}
	& \frac{\partial ADMTR_1 }{\partial t}\bigg \vert_{t=0}  \\
	& \hspace{10pt} = [-\mathbb{E}[P\mathbf{1} \left\lbrace C > y \right\rbrace ]]^{-1}\Big[\mathbb{E}[(\Lambda^{\prime}(\beta_{0}\left(1,y\right)+\beta_{C}\left(1,y\right)C+\beta_{P}\left(1,y\right)P)\beta_{P}\left(1,y\right)-ADMTR_1)P\mathbf{1} \left\lbrace C > y \right\rbrace s(\delta)] \\
	& \hspace{40pt} + \frac{\partial \mathbb{E}[(\Lambda^{\prime}(\beta_{0}\left(1,y\right)(t)+\beta_{C}\left(1,y\right)(t)C+\beta_{P}\left(1,y\right)(t)P)\beta_{P}\left(1,y\right)(t)-ADMTR_1)P\mathbf{1} \left\lbrace C > y \right\rbrace ]}{\partial t}\bigg \vert_{t=0} \\
	& \hspace{40pt} + \left. \frac{\partial \mathbb{E}[(\Lambda^{\prime}(\beta_{0}\left(1,y\right)+\beta_{C}\left(1,y\right)C+\beta_{P}\left(1,y\right)P(t))\beta_{P}\left(1,y\right)-ADMTR_1)P(t)\mathbf{1} \left\lbrace C > y \right\rbrace ]}{\partial t}\bigg \vert_{t=0}  \right]. 
\end{align*}
Now, we need to express the second two components as products with the score of the data to apply Equation (3.10) of \citet{Newey1994}.  We start with  $$\frac{\partial \mathbb{E}[(\Lambda^{\prime}(\beta_{0}\left(1,y\right)(t)+\beta_{C}\left(1,y\right)(t)C+\beta_{P}\left(1,y\right)(t)P)\beta_{P}\left(1,y\right)(t)-ADMTR_1)P\mathbf{1} \left\lbrace C > y \right\rbrace ]}{\partial t},$$ which is the parametric part of the model. We have that
\begin{align*}
	& \left. \frac{\partial \mathbb{E}[(\Lambda^{\prime}(\beta_{0}\left(1,y\right)(t)+\beta_{C}\left(1,y\right)(t)C+\beta_{P}\left(1,y\right)(t)P)\beta_{P}\left(1,y\right)(t)-ADMTR_1)P\mathbf{1} \left\lbrace C > y \right\rbrace ]}{\partial t}\right\vert_{t=0} \\
	& \hspace{20pt} = \mathbb{E} \Bigg[ \Lambda^{\prime \prime }(\beta_{0}\left(1,y\right)+\beta_{C}\left(1,y\right)C+\beta_{P}\left(1,y\right)P) \cdot\beta_{P}\left(1,y\right) \cdot P \cdot \mathbf{1}\left\lbrace C > y \right\rbrace \\
	& \hspace{40pt} \cdot \left( \left.\frac{\partial \beta_{0}\left(1,y\right)(t)}{\partial t}\right\vert_{t=0}+ \left. \frac{\partial \beta_{C}\left(1,y\right)(t)}{\partial t}\right\vert_{t=0} \cdot C+\left. \frac{\partial \beta_{P}\left(1,y\right)(t)}{\partial t}\right\vert_{t=0} \cdot P \right)  \\
	& \hspace{20pt} + [(\Lambda^{\prime}(\beta_{0}\left(1,y\right)+\beta_{C}\left(1,y\right)C+\beta_{P}\left(1,y\right)P)P\mathbf{1} \left\lbrace C > y \right\rbrace \left. \frac{\partial \beta_{P}\left(1,y\right)(t)}{\partial t}\right\vert_{t=0} ]\Bigg]
\end{align*}

We claim, and will become evident further on that is indeed true, that, using Equations \eqref{Moment1}-\eqref{Moment2c}, we can express $\frac{\partial \beta_{A}\left(1,y\right)(t)}{\partial t}=\mathbb{E}[IF_{ \beta_{A}\left(1,y\right)}s(\delta)]$, where $IF_{ \beta_{A}\left(1,y\right)}$ is the influence function of $\beta_{A}\left(1,y\right)$ and $A$ is a random variable.
\par 
Then, we have that
\begin{align*}
	& \frac{\partial ADMTR_1 }{\partial t}\bigg \vert_{t=0} \\
	& \hspace{20pt} = [-\mathbb{E}[P\mathbf{1} \left\lbrace C > y \right\rbrace ]]^{-1} \\
	& \hspace{40pt} \cdot \Big[E \big [(\Lambda^{\prime}(\beta_{0}\left(1,y\right)+\beta_{C}\left(1,y\right)C+\beta_{P}\left(1,y\right)P)\beta_{P}\left(1,y\right)-ADMTR_1)P\mathbf{1} \left\lbrace C > y \right\rbrace s(\delta)]\\
	& \hspace{60pt} + E\big [\Lambda^{\prime \prime }(\beta_{0}\left(1,y\right)+\beta_{C}\left(1,y\right)C+\beta_{P}\left(1,y\right)P)\beta_{P}\left(1,y\right) P \mathbf{1} \left\lbrace C > y \right\rbrace  \\
	& \hspace{100pt} \cdot (\mathbb{E}[IF_{ \beta_{0}\left(1,y\right)}s(\delta)] + \mathbb{E}[IF_{ \beta_{C}\left(1,y\right)}s(\delta)]C+\mathbb{E}[IF_{ \beta_{P}\left(1,y\right)}s(\delta)]P) \big] \\
	& \hspace{60pt} E[(\Lambda^{\prime}(\beta_{0}\left(1,y\right)+\beta_{C}\left(1,y\right)C+\beta_{P}\left(1,y\right)P)P\mathbf{1} \left\lbrace C > y \right\rbrace \mathbb{E}[IF_{ \beta_{P}\left(1,y\right)}s(\delta)] ]
	\\
	& \hspace{60pt} \left. + \frac{\partial \mathbb{E}[(\Lambda^{\prime}(\beta_{0}\left(1,y\right)+\beta_{C}\left(1,y\right)C+\beta_{P}\left(1,y\right)P(t))\beta_{P}\left(1,y\right)-ADMTR_1)P(t)\mathbf{1} \left\lbrace C > y \right\rbrace ]}{\partial t}\bigg \vert_{t=0}  \right],
\end{align*}

or, equivalently,

\begin{align*}
	& \frac{\partial ADMTR_1 }{\partial t}\bigg \vert_{t=0} \\
	& \hspace{20pt} = [-\mathbb{E}[P\mathbf{1} \left\lbrace C > y \right\rbrace ]]^{-1}\Big[E\big[\{(\Lambda^{\prime}(\beta_{0}\left(1,y\right)+\beta_{C}\left(1,y\right)C+\beta_{P}\left(1,y\right)P)\beta_{P}\left(1,y\right)-ADMTR_1)P\mathbf{1} \left\lbrace C > y \right\rbrace  \\
	& \hspace{80pt} + \mathbb{E}[\Lambda^{\prime \prime }(\beta_{0}\left(1,y\right)+\beta_{C}\left(1,y\right)C+\beta_{P}\left(1,y\right)P)\beta_{P}\left(1,y\right))P\mathbf{1} \left\lbrace C > y \right\rbrace )]IF_{ \beta_{0}\left(1,y\right)} \\
	& \hspace{80pt} + \mathbb{E}[\Lambda^{\prime \prime }(\beta_{0}\left(1,y\right)+\beta_{C}\left(1,y\right)C+\beta_{P}\left(1,y\right)P)\beta_{P}\left(1,y\right))P\mathbf{1} \left\lbrace C > y \right\rbrace )C]IF_{ \beta_{C}\left(1,y\right)} \\
	& \hspace{80pt} + \mathbb{E}[\Lambda^{\prime \prime }(\beta_{0}\left(1,y\right)+\beta_{C}\left(1,y\right)C+\beta_{P}\left(1,y\right)P)\beta_{P}\left(1,y\right))P\mathbf{1} \left\lbrace C > y \right\rbrace )P]IF_{ \beta_{P}\left(1,y\right)} \\
	& \hspace{40pt}+ \mathbb{E}[(\Lambda^{\prime}(\beta_{0}\left(1,y\right)+\beta_{C}\left(1,y\right)C+\beta_{P}\left(1,y\right)P)P\mathbf{1} \left\lbrace C > y \right\rbrace   ] IF_{ \beta_{P}\left(1,y\right)} \}s(\delta) \big]
	\\
	& \hspace{40pt} + \left. \frac{\partial \mathbb{E}[(\Lambda^{\prime}(\beta_{0}\left(1,y\right)+\beta_{C}\left(1,y\right)C+\beta_{P}\left(1,y\right)P(t))\beta_{P}\left(1,y\right)-ADMTR_1)P(t)\mathbf{1} \left\lbrace C > y \right\rbrace ]}{\partial t}\bigg \vert_{t=0}  \right].
\end{align*}

Now, we need to derive the non-parametric component $$\frac{\partial \mathbb{E}[(\Lambda^{\prime}(\beta_{0}\left(1,y\right)+\beta_{C}\left(1,y\right)C+\beta_{P}\left(1,y\right)P(t))\beta_{P}\left(1,y\right)-ADMTR_1)P(t)\mathbf{1} \left\lbrace C > y \right\rbrace ]}{\partial t}\bigg \vert_{t=0}. $$
\par 
We assume that the conditions for the Riesz Representation Theorem hold for $$\frac{\partial \mathbb{E}[(\Lambda^{\prime}(\beta_{0}\left(1,y\right)+\beta_{C}\left(1,y\right)C+\beta_{P}\left(1,y\right)P(t))\beta_{P}\left(1,y\right)-ADMTR_1)P(t)\mathbf{1} \left\lbrace C > y \right\rbrace ]}{\partial t}. $$
See, for example, \citet{ACKERBERGetal2014} for such conditions as being a linear bounded functional. Then, there is a unique $b$ in a properly defined space with the inner product $<b_1,b_2 >=\mathbb{E}[b_1b_2]$ such that:
\begin{align}
	& \frac{\partial \mathbb{E}[(\Lambda^{\prime}(\beta_{0}\left(1,y\right)+\beta_{C}\left(1,y\right)C+\beta_{P}\left(1,y\right)P(t))\beta_{P}\left(1,y\right)-ADMTR_1)P(t)\mathbf{1} \left\lbrace C > y \right\rbrace ]}{\partial t} \nonumber \\
	& \hspace{20pt} = \mathbb{E}\left[b\frac{ \partial \left\lbrace CZ(D-P(t)) \right\rbrace }{\partial t}\right] \nonumber \\
	& \hspace{20pt} = \frac{\partial \mathbb{E}[bCZ(D-P(t))]}{ \partial t} \nonumber \\
	& \hspace{20pt} = \mathbb{E}\left[\tilde{b}(\delta)\frac{ \partial P(t)}{\partial t}\right] \nonumber \\
	& \hspace{20pt} = \frac{\partial \mathbb{E}[\tilde{b}(\delta)P(t)]}{\partial t} \label{Rieszrepresenter}
\end{align}
where $\tilde{b}(\delta)=-bCZ$.
\par 
To find the $IF$ following \citet{IchimuraNewey2022},  we need to find a $\phi(\delta, P,\alpha)$ such that: 
\begin{align}
	& \frac{\partial \mathbb{E}[(\Lambda^{\prime}(\beta_{0}\left(1,y\right)+\beta_{C}\left(1,y\right)C+\beta_{P}\left(1,y\right)P(t))\beta_{P}\left(1,y\right)-ADMTR_1)P(t)\mathbf{1} \left\lbrace C > y \right\rbrace ]}{\partial t} \nonumber \\
	&\hspace{30pt} = \int \phi(\delta, P,\alpha) G(d \delta) \label{Newey}
\end{align}
where $G$ is a perturbation from the true $CDF$. 
\par 
Furthermore, from \eqref{Moment1}, we can see that 
\begin{eqnarray*}
	0&=&\frac{\partial \mathbb{E}_t[CZ(D-P(t))]}{\partial t}= \int CZ(D-P)G(d \delta)+\frac{\partial \mathbb{E}[CZ(D-P(t))]}{\partial t}
\end{eqnarray*}
or, equivalently, 
\begin{eqnarray*}
	\int CZ(D-P)G(d \delta)=-\frac{\partial \mathbb{E}[CZ(D-P(t))]}{\partial t}.
\end{eqnarray*}
\par 
Thus, following \citet{IchimuraNewey2022}, if we find an $\alpha(\delta)$ such that: 
\begin{align}
	& \frac{\partial \mathbb{E}[(\Lambda^{\prime}(\beta_{0}\left(1,y\right)+\beta_{C}\left(1,y\right)C+\beta_{P}\left(1,y\right)P(t))\beta_{P}\left(1,y\right)-ADMTR_1)P(t)\mathbf{1} \left\lbrace C > y \right\rbrace ]}{\partial t} \nonumber \\
	& \hspace{20pt} = -\frac{\partial \mathbb{E}[\alpha(\delta)CZ(D-P(t))]},{\partial t} \label{Newey3}
\end{align}
then we will get \eqref{Newey}. 
\par 
Furthermore, from \eqref{Moment1}, we can see that 
\begin{eqnarray}\label{Newey2}
	\frac{\partial \mathbb{E}[CZ(D-P(t))]}{\partial t}=\mathbb{E}\left[\frac{ \partial CZ(D-P(t))}{\partial t}\right]= \mathbb{E}\left[a(\delta) \frac{ \partial P(t))}{\partial t}\right]= \frac{\partial \mathbb{E}[a(\delta)P(t)] }{\partial t}
\end{eqnarray}
where $a(\delta)=CZ$. 
\par 
From \eqref{Rieszrepresenter} and \eqref{Newey2} combined with \eqref{Newey3}: 
\begin{eqnarray}\label{Newey4}
	\frac{\partial \mathbb{E}[\tilde{b}(\delta)P(t)]}{\partial t}=  -\frac{\partial \mathbb{E}[\alpha(\delta)a(\delta)P(t)]}{\partial t}   
\end{eqnarray}
The equality in Equation  \eqref{Newey4} will be satisfied if it holds for every $t$. Since $P(t)$ is in the space of possible propensity scores, the condition will be satisfied if it holds for any $P$ in the space of possible propensity scores. Thus: 
\begin{eqnarray}\label{Newey5}
	\mathbb{E}[\tilde{b}(\delta)P]=  -\mathbb{E}[\alpha(\delta)a(\delta)P]
\end{eqnarray}
or, equivalently, 
\begin{eqnarray}\label{Newey6}
	0=\mathbb{E}[\{ \tilde{b}(\delta)+\alpha(\delta)a(\delta) \}P]= \mathbb{E}\left[(-a(\delta))\left\lbrace \frac{-\tilde{b}(\delta)}{a(\delta)}-\alpha(\delta) \right\rbrace  P \right] 
\end{eqnarray}
Thus, as in \citet{IchimuraNewey2022}, the $\alpha(\delta)$ that minimizes $\mathbb{E}\left[(-a(\delta))\left\lbrace \frac{-\tilde{b}(\delta)}{a(\delta)}-\alpha(\delta) \right\rbrace^2\right]$ satisfies 
\par 
$$\phi(\delta, P,\alpha)=\alpha(\delta)CZ(D-P).$$
\par 
Combining this result with the previous display, the Influence Function of $ADMTR_1$ is given by 
\begin{align}
	& IF_{ADMTR_1} \nonumber \\
	& \hspace{20pt} = [-\mathbb{E}[P\mathbf{1} \left\lbrace C > y \right\rbrace ]]^{-1}\Big[(\Lambda^{\prime}(\beta_{0}\left(1,y\right)+\beta_{C}\left(1,y\right)C+\beta_{P}\left(1,y\right)P)\beta_{P}\left(1,y\right)-ADMTR_1)P\mathbf{1} \left\lbrace C > y \right\rbrace  \nonumber \\
	& \hspace{40pt} + \mathbb{E}[\Lambda^{\prime \prime }(\beta_{0}\left(1,y\right)+\beta_{C}\left(1,y\right)C+\beta_{P}\left(1,y\right)P)\beta_{P}\left(1,y\right))P\mathbf{1} \left\lbrace C > y \right\rbrace )]IF_{ \beta_{0}\left(1,y\right)} \nonumber\\
	& \hspace{40pt} + \mathbb{E}[\Lambda^{\prime \prime }(\beta_{0}\left(1,y\right)+\beta_{C}\left(1,y\right)C+\beta_{P}\left(1,y\right)P)\beta_{P}\left(1,y\right))P\mathbf{1} \left\lbrace C > y \right\rbrace )C]IF_{ \beta_{C}\left(1,y\right)} \nonumber \\
	& \hspace{40pt} + \mathbb{E}[\Lambda^{\prime \prime }(\beta_{0}\left(1,y\right)+\beta_{C}\left(1,y\right)C+\beta_{P}\left(1,y\right)P)\beta_{P}\left(1,y\right))P\mathbf{1} \left\lbrace C > y \right\rbrace )P]IF_{ \beta_{P}\left(1,y\right)}  \nonumber  \\
	& \hspace{40pt}+ \mathbb{E}[(\Lambda^{\prime}(\beta_{0}\left(1,y\right)+\beta_{C}\left(1,y\right)C+\beta_{P}\left(1,y\right)P)P\mathbf{1} \left\lbrace C > y \right\rbrace   ] IF_{ \beta_{P}\left(1,y\right)} \\
	& \hspace{40pt} + \phi(\delta, P,\alpha) \Big] 
\end{align}

The influence function will depend on the influence functions of the estimated coefficients. Their influence function could be derived similarly to the nonparametric effect of $P$  on $ADMTR_1$, but using Equations \eqref{Moment1}-\eqref{Moment2c}. We could then use a similar logic as we did following \citet{IchimuraNewey2022} and a form of Riezs representation.  That would fully complete the representation, implying that our previous claim holds.
\par 
If we compute all the $IF$, we can then compute the influence function of $DMTR_1(y,v)$ by the chain rule as: 

\begin{align}
	& IF_{DMTR_1} \nonumber \\
	& \hspace{20pt} =[\Lambda^{\prime \prime }(\beta_{0}\left(1,y\right)+\beta_{C}\left(1,y\right)C+\beta_{P}\left(1,y\right)v)\beta_{P}\left(1,y\right))]IF_{ \beta_{0}\left(1,y\right)} 
	\nonumber \\
	& \hspace{20pt} +[\Lambda^{\prime \prime }(\beta_{0}\left(1,y\right)+\beta_{C}\left(1,y\right)C+\beta_{P}\left(1,y\right)v)\beta_{P}\left(1,y\right))]CIF_{ \beta_{C}\left(1,y\right)} 
	\nonumber \\
	& \hspace{20pt} +\Bigg[[\Lambda^{\prime \prime }(\beta_{0}\left(1,y\right)+\beta_{C}\left(1,y\right)C+\beta_{P}\left(1,y\right)v)\beta_{P}\left(1,y\right))]v \nonumber \\
	& \hspace{20pt} +[\Lambda^{ \prime }(\beta_{0}\left(1,y\right)+\beta_{C}\left(1,y\right)C+\beta_{P}\left(1,y\right)v)\left(1,y\right))]\Bigg]IF_{ \beta_{P}\left(1,y\right)} 
\end{align}

Since $DMTE=DMTR_1-DMTR_0$, we can then say that: 
\begin{eqnarray}\label{IF}
	IF_{DMTE}=IF_{DMTR_1}-IF_{DMTR_0} 
\end{eqnarray}
where $IF_{DMTR_0}$ can be derived analogously to  $IF_{DMTR_1}$. 
\par 
Then, we know that 
$$\sqrt{N}(\widehat{DMTE}-DMTE)=\frac{1}{\sqrt{n}}\sum_i^n IF_{DMTE}(\delta_i)+o_p(1) \xrightarrow{d} N(0,Var(DMTE))$$ 
and 
$$Var(DMTE)=\mathbb{E}[IF_{DMTE}^2],$$
where convergence is due to standard results on influence functions \citep{vanwellner1996,vaart_1998}. 
\par 
Following \citet{Frandsen2015} and our influence function calculations, we can recover the asymptotic distribution of the $QMT\mathbb{E}\left[\tau,p\right]$. The $QMTR_d$ as quantiles have the following standard and known influence functions:
$$IF_{QMTR_d(\tau,p)}=-\frac{1\{DMTR_d^{-1}(\tau,p)>y \}-\tau}{\frac{\partial DMTR_d(DMTR_d^{-1}(\tau,p),p)}{\partial y}},$$
implying that $IF_{QMTE(\tau,p)}=IF_{QMTR_1(\tau,p)}-IF_{QMTR_0(\tau,p)}$, 
$$\sqrt{N}(\widehat{QMTE}-QMTE)=\frac{1}{\sqrt{n}}\sum_i^n IF_{QMTE}(\delta_i)+o_p(1) \xrightarrow{d} N(0,Var(QMTE))$$ 
and 
$$Var(QMTE)=\mathbb{E}[IF_{QMTE}^2],$$
where convergence is due to standard results on influence functions \citep{vanwellner1996,vaart_1998}. 
\par 
By recalling that $MTE(p)=\int_0^1 QMTE(\tau,p) d \tau$ and the fact that we just provided asymptotic normality for $QMTE(\tau,p)$, we can recover the distribution of the $MTE(p)$ by an application of the chain rule of influence functions again. 

Consequently, 
$$IF_{MTE(p)}=\int_0^1 IF_{QMTE(\tau,p)} d\tau $$
\par 
Although relevant to show the results for the $MTE$, in our text, we also focus on the $RMTE(v)$ to avoid additional support assumptions. Thus, we can similarly derive the asymptotic results for the RMTE, since  $RMTE(v)= -\int_{0}^{\gamma_C} DMTE\left(y,v\right)dy.$
Then, by the chain rule of influence functions, 
$$IF_{RMTE(p)}=-\int_0^{\gamma_C} IF_{DMTE(y,p)} dy. $$
\subsection{Proof of Theorems \ref{Thm:semi_boot} and \ref{ThmBootRMTE}: Validity of the Weighted Bootstrap}\label{AppProofBootstrap}

We generate $\{V_{i}, i=1,\dots,n\}$ as a sequence of independent and identically distributed non-negative random variables with mean one, variance one, and finite third moment (e.g., $V_i \sim Exp\left(1\right)$).
\par 
Based on this we estimate  the propensity score by minimizing the weighted version of the standard series minimization criteria,
\begin{equation}
	\hat{\theta}^{fs, *} =\underset{\theta^{fs} \in \Theta^{fs} }{\arg \min }~ n^{-1}\sum_{i=1}^n V_i \left( D_i -  {\alpha}_{0} - X_{i}'{\alpha}_{X} -  C_{i}{\alpha}_{C} - \psi^L(Z_i)' {\alpha}_Z,\right)^2
\end{equation}
where $\hat{\theta}^{fs,*} = (\widehat{\alpha}_0^*, \widehat{\alpha}^{*,\prime}_X, \widehat{\alpha}^*_C, \widehat{\alpha}^{*}_Z))'$ and thus $\hat{P}_i^*$ is a function of $\hat{\theta}^{fs,*}$. Note then that by Corollary 3 in \citet{MAKosorok} or Corollary 3.2.3 and Theorem 3.2.5 of \citet{vanwellner1996}, $\hat{P}_i^*$  converges to  $P_i$. 
\par 
For the estimation of the $\beta$s,  let: 
\begin{eqnarray} 
	Ln^*(\beta_{1,y}, \hat{P})=\max_{\beta_{1,y}} \frac{1}{N} \sum_iV_i W_{y,1,i} log[\Lambda(\beta_{1,y} \hat{H}_i)]+V_i(1-W_{y,1,i})log[1-\Lambda(\beta_{1,y} \hat{H}_i)]
\end{eqnarray}
And, 
\begin{eqnarray} 
	Ln^*(\beta_{1,y},P)=\max_{\beta_{1,y}} \frac{1}{N} \sum_iV_i W_{y,1,i} log[\Lambda(\beta_{1,y} H_i)]+V_i(1-W_{y,1,i})log[1-\Lambda(\beta_{1,y}H_i)]
\end{eqnarray}
By a similar display as in the unweighted case, we know that: 
\begin{eqnarray*}
	\sup_{\beta_{1,y}}|Ln^*(\beta_{1,y}, \hat{P})-Ln^*(\beta_{1,y},P)|=o_p(1)
\end{eqnarray*}
Where $Ln^*(\beta_{1,y},P)$ is a standard, but weighted parametric likelihood and thus by a second application of  Corollary 3 in \citet{MAKosorok} or Corollary 3.2.3 and Theorem 3.2.5 of \citet{vanwellner1996}, $\hat{\beta}_{1,y}^*$  converges to  $\beta_{1,y}$.
\par 
It then remains to show asymptotic normality of $\hat{\beta}_{1,y}^*$ and then of $DMTR^*$. Since the $V_i$ are independent of everything and $Var\left(V_{i}\right) = 1$, we have $\sqrt{N}(\hat{\beta}_{1,y}^*-\beta_{1,y})$ is asymptotically normal as long as a version  of Assumption \ref{AsSP3} holds with $\hat{P}_i^*$. 
\par 
Furthermore, since $V_i$ is independent of everything, similar calculations of the influence functions for $DMTE,QMTE, \text{ and } RMTE$ can be provided to obtain asymptotic normality of the bootstrapped versions of these. Alternatively, one can note that $DMTE,QMTE, \text{ and } RMTE$ are all continuous and Hadamard differentiable functions of  $\beta_{1,y}, P_i$, with $\hat{P}^*$  slower than root-$n$ via a bootstrap version of Assumption \ref{AsSP3}  and thus functional delta methods and continuous mapping theorems hold.

		\newpage
		\section{Constructing the Dataset}\label{AppData}
		\setcounter{table}{0}
		\renewcommand\thetable{C.\arabic{table}}
		
		\setcounter{figure}{0}
		\renewcommand\thefigure{C.\arabic{figure}}
		
		\setcounter{equation}{0}
		\renewcommand\theequation{C.\arabic{equation}}
		
		\setcounter{theorem}{0}
		\renewcommand\thetheorem{C.\arabic{theorem}}
		
		\setcounter{proposition}{0}
		\renewcommand\theproposition{C.\arabic{proposition}}
		
		\setcounter{corollary}{0}
		\renewcommand\thecorollary{C.\arabic{corollary}}
		
		\setcounter{assumption}{0}
		\renewcommand\theassumption{C.\arabic{assumption}}
		
		In this appendix, we summarize Appendix I.2 of \citet{Possebom2022}, which provides a detailed explanation of how the dataset used in our empirical application was constructed. We explain the specific crime types included in our sample, the classification algorithms used to define which defendants were punished, and the fuzzy matching algorithm used to define which defendants recidivate. At the end, we explain how we constructed the extra covariates used to assess the validity of our identifying assumptions.
		
		The final dataset was created from three initial datasets.
		\begin{enumerate}
			\item CPOPG (``Consulta de Processos de Primeiro Grau''): It contains information about all criminal cases in the Justice Court System in the State of São Paulo (TJ-SP) between 2010 and 2019.
			
			\item CJPG (``Consulta de Julgados de Primeiro Grau''): It contains information about the last decision made by a trial judge in all criminal cases in TJ-SP between 2010 and 2019.
			
			\item CPOSG (``Consulta de Processos de Segundo Grau''): It contains information about all appealing criminal cases in TJ-SP between 2010 and 2019.
		\end{enumerate}
		
		Starting from the CPOPG dataset, we implement the following steps.
		\begin{enumerate}
			\item We only keep cases that are currently in the Appeals Court, closed, or whose status is empty. Those cases are already associated with a trial judge's sentence.
			
			\item We only keep cases whose crime types are associated with sentences that must be less than four years of incarceration.
			
			\item We only keep cases that aim to analyze whether a defendant is guilty or not.
			
			\item We only keep cases that were randomly assigned to trial judges.
			
			\item We only keep cases whose starting date is after January 1\textsuperscript{st}, 2010.
		\end{enumerate}
		
		After these steps, our dataset contains 98,552 cases. We then merged it with the CJPG dataset using cases' \texttt{id} codes. Since some cases do not have \texttt{id} codes, our dataset now contains 98,422 cases.
		
		After this step, we randomly select 35 cases per year (2010-2019) for manual classification. We manually classify them into five categories: ``defendant died during the trial'', ``defendant is guilty'', ``defendant accepted a non-prosecution agreement'' (``transação penal'' in Portuguese), ``case was dismissed'' (``processo suspenso'' in Portuguese) and ``defendant was acquitted''. Since some sentences are missing or incomplete, we are able to manually classify only 325 sentences.
		
		Now, we use those 325 manually classified cases to train a classification algorithm. To do so, we divide them into a training sample (216 cases) and a validation sample (109 sentences).
		
		First, we design an algorithm to identify which defendants died during the trial. To do so, we check whether the sentence contains any reference to the first paragraph of Article 107 from the Brazilian Criminal Code. This deterministic algorithm perfectly classifies cases into the category ``defendant died during the trial''.
		
		Second, we design an algorithm to identify which cases were dismissed. To do so, we check whether the sentence contains any reference to Article 89 in Law n. 9099/95. This deterministic algorithm correctly classifies 98\% of the cases into the category ``case was dismissed''.
		
		Third, we design an algorithm to identify which defendants accepted a non-prosecution agreement. To do so, we check whether the sentence contains any expression connected to a non-prosecution agreement. This deterministic algorithm correctly classified almost all the cases into the category ``defendant accepted a non-prosecution agreement'', making only three mistakes.
		
		Finally, we design an algorithm to classify the remaining cases into two categories: ``defendant is guilty'' and ``defendant was acquitted''. To do so, we define a bag of words that were found to be strong signals of acquittal and guilt when manually classifying the cases in our samples. We then count how many times each one of those expressions appears in each sentence, and we normalize those counts to be between 0 and 1.
		
		Using the normalized counts, we train an L1-Regularized Logistic Regression using our training sample. We then validate this algorithm using our validation sample and find that it correctly classifies 98.8\% of the cases. Given this high success rate, we use the L1-Regularized Logistic Regression algorithm to define the treatment variable in our full sample.
		
		Having designed the above algorithm, we use it to define the trial judge's treatment variable $T$ in the full sample. First, we find which defendants died during their trials and drop them from our sample. We then use the second and third algorithms to define which cases were dismissed and which cases are associated with a non-prosecution agreement. Moreover, we use the trained L1-regularized Logistic Regression algorithm to classify the remaining cases into the categories ``defendant is guilty'' and ``defendant was acquitted''. Finally, we combine the categories ``defendant was acquitted'' and ``case was dismissed'' into the untreated group (``not punished'', $T = 0$) and the categories ``defendant accepted a non-prosecution agreement'' and ``defendant is guilty'' into the treated group (``punished'', $T = 1$). At the end, our dataset contains 96,225 cases.
		
		Now, we merge our current dataset with the CPOSG dataset using each case's \texttt{id} code. When merging these datasets, we create an indicator variable that denotes which cases went to the Appeals Court, i.e., which cases were matched. We then randomly select 50 cases per year for manual classification (2010-2019) and divide them into three categories: ``cases that went to the Appeals Court, but were immediately returned due to bureaucratic errors'', ``cases whose trial judge's sentences were affirmed'' and ``cases whose trial judge's sentences were reversed''.
		
		Now, we use those 500 manually classified cases to train a classification algorithm. To do so, we divide them into a training sample (300 cases) and a validation sample (200 sentences).
		
		First, we design an algorithm to identify which cases went to the Appeals Court but were immediately returned. To do so, we simply check whether the Appeals Court's \texttt{decision} is empty.
		
		Finally, we design an algorithm to classify the non-empty cases into two categories: ``cases whose trial judge's sentences were affirmed'' and ``cases whose trial judge's sentences were reversed''. To do so, we define a bag of words that were found to be strong signals of sentence reversal when manually classifying the cases in our sample. We then count how many times each one of those expressions appears in each sentence, and we normalize those counts to be between 0 and 1.
		
		Using the normalized counts, we train an L1-Regularized Logistic Regression using our training dataset. We then validate this algorithm using our validation sample and find that it correctly classifies 96.2\% of the cases. Given this high success rate, we use the L1-Regularized Logistic Regression to define the treatment variable in our full sample.
		
		Having designed the above algorithms, we use them to define the final treatment variable $D$ in the full sample. First, we set $D = T$ if a case did not go to the Appeals Court or if a case went to the Appeals Court, but was immediately returned. Second, we use the trained L1-Regularized Logistic Regression algorithm to classify the remaining cases into the categories ``reversed trial judge's sentence'' and ``affirmed trial judge's sentence''. We, then, set $D = T$ if the trial judge's sentence was affirmed and $D = 1 - T$ if the trial judge's sentence was reversed. Moreover, we also drop the cases whose dates (starting date, trial judge's sentence date and Appeal Court's decision date) are not appropriately ordered. At the end, our dataset contains 95,119 cases.
		
		Now, our goal is to find the defendants' names in each case. To do so, we use the variable \texttt{parties} from the CPOPG dataset and search for names listed as defendants. Finally, we delete names that are not a person's name --- such as district attorney offices, public defender offices and ``unknown author''. Our sample now contains 103,423 case-defendant pairs.
		
		Furthermore, we repeat the steps in the last paragraph to find defendants' names in a dataset that contains all cases from the CPOPG dataset, including cases that are still open and cases with severe crimes. This dataset contains 1,027,120 case-defendants pairs.
		
		Now, we use these two datasets to define our outcome variable ($Y = $ ``time to recidivism''). A defendant $i$ in a case $j$ in the smaller dataset recidivates if and only if defendant $i$'s full name appears in a case $\bar{j}$ in the larger dataset. To match defendants' names across cases, we use the Jaro–Winkler similarity metric and we define a match if the similarity between full names in two different cases is greater than or equal to 0.95. If we find a match, we define the outcome variable as the time difference between the second case's start date and the first case's final date.
		
		Furthermore, we delete the case-defendant pairs whose cases started in 2018 and 2019. Consequently, our dataset contains 43,468 case-defendants pairs.
		
		{
			Lastly, we use the CPOPG dataset to create four covariates that are used to assess the validity of our identifying assumptions in Appendix \ref{Sdescriptive}. Following \citet{Possebom2022}, we use the defendant's name to find whether the defendant has a typically male name according to the Brazilian 2010 Census (\emph{R package} \texttt{genderBR}). Following \citet{Laneuville2024}, we use each case's events to find which defendants were caught red-handed when committing a crime (\emph{in flagrante delicto}), and we use each case's list of parties to find which defendants used public defenders. Moreover, we also use information on the type of crime to create an indicator variable of whether the defendant was accused of theft.
			
		}
		
		\newpage
		\section{Additional Empirical Results} \label{AppExtraEmpirical}
		
		\setcounter{table}{0}
		\renewcommand\thetable{D.\arabic{table}}
		
		\setcounter{figure}{0}
		\renewcommand\thefigure{D.\arabic{figure}}
		
		\setcounter{equation}{0}
		\renewcommand\theequation{D.\arabic{equation}}
		
		\setcounter{theorem}{0}
		\renewcommand\thetheorem{D.\arabic{theorem}}
		
		\setcounter{assumption}{0}
		\renewcommand\theassumption{D.\arabic{assumption}}
		
		This appendix contains additional empirical results. Appendix \ref{Sdescriptive} assesses the plausibility of three of our identifying assumptions. Appendix \ref{Sfirststage} presents the first stage results. Moreover, Appendix \ref{AppQMTEgivenV} presents the $QMTE$ as a function of the quantiles for a small set of values of punishment resistance while Appendix \ref{AppCIs} presents the confidence sets around our main estimates. Furthermore, Appendix \ref{AppRobustness} presents our results for a subsample of cases that start in 2014. Appendix \ref{AppDescriptiveCourtJudge} presents descriptive statistics about court districts and judges. Lastly, Appendix \ref{AppDescriptiveOutcome} presents descriptive statistics for the outcome variable, repeated offenders, and defendants' names.
		
		\subsection{Assessing the plausibility of our assumptions}\label{Sdescriptive}
		{
			The credibility of any causal inference procedure depends on the plausibility of its underlying identification assumptions. In this subsection, we use some descriptive statistics to assess the plausibility of three of our identifying assumptions: Random Assignment (Assumption \ref{AsIndependence}), Monotonicity Condition (Equation \eqref{EqTreatment}), and Random Censoring (Assumption \ref{AsCensoring}). We focus on these three assumptions because the others are not controversial in this paper.\footnote{The full set of our identifying assumptions, including the parametric restrictions in our estimation procedure, can be jointly tested in a similar manner as in \cite{arai2022testingQE}. Since, for any particular $y$, the $DMTR$ has to be between $0$ and $1$, the local instrumental variable estimand, $\dfrac{\partial \mathbb{P}\left[\left. Y\leq y, D = d \right\vert P\left(Z, C \right) = p, C = y + \delta \right]}{\partial p}$, has to be between $0$ and $1$ for any $y,d,p,\delta$. This restriction is testable in the data using a type of KS-statistic similar to the one developed by the aforementioned authors. Formally deriving this test is beyond the scope of this article.} For instance, Assumptions \ref{AsRank} is plausible as we use the judge's leave-one-out rate of punishment (or ``leniency rate'') as an instrument for the trial judge's decision, and we have 525 judges and a good amount of variation in this variable as further discussed around Table \ref{TabFirstStage} and Figure \ref{FigPS}.
			
			\subsubsection{Random Assignment (Assumption \ref{AsIndependence})}
			
			Table \ref{TabTestAssignment} focuses on assessing the plausibility of Assumption \ref{AsIndependence}. Although it is not possible to directly test that the instrument $\left(Z\right)$ is independent of the potential uncensored time-to-recidivism outcomes $\left(Y^{*}\left(0\right), Y^{*}\left(1\right) \right)$ and the latent heterogeneity $\left(V\right)$ given the censoring variable $\left(C\right)$, it is possible to test whether $Z$ is correlated with four excluded covariates: an indicator for whether the defendant has a male name, an indicator for whether the defendant used a public defender instead of hiring a private lawyer, and indicator for whether the defendant was accused of theft, and an indicator for whether the defendant was caught red-handed when committing a crime (``in flagrante delicto'').\footnote{Appendix \ref{AppData} explains how we measure these covariates.} If we find that the instrument is uncorrelated with these four covariates after conditioning on the censoring variable and court-district fixed effects, this could serve as indirect support for Assumption \ref{AsIndependence}.
			
			\begin{table}[!htb]
				\centering
				\caption{{Testing the Random Assignment Assumption}} \label{TabTestAssignment}
				\begin{lrbox}{\tablebox}
					\begin{tabular}{lccccc}
						\hline \hline
						& Male Name & Public Defender & Theft & Red-Handed & C \\ \hline
						Final Ruling $(D)$ & 0.049*** & 0.161*** & 0.007 & 0.132*** & 0.018*** \\
						& (0.009) & (0.023) & (0.015) & (0.011) & (0.004) \\
						Punishment Rate $(Z)$ & 0.001 & 0.003 & -0.003 & -0.002 & -0.001 \\
						& (0.002) & (0.008) & (0.005) & (0.003) & (0.001) \\ \hline
					\end{tabular}
				\end{lrbox}
				\usebox{\tablebox}\\
				\justifying
				\hspace{-.6cm}\scriptsize{Note: The first row reports the estimated coefficients of regressing the treatment variable ($D = $``punished according to the final ruling in the case'') on the binary covariates indicated in the columns, the censoring variable ($C$ is measured in years for readability) and court-district fixed effects. The second row reports the estimated coefficients of regressing the instrumental variable ($Z = $``trial judge's punishment rate'') on the binary covariates indicated in the columns, the censoring variable ($C$ is measured in years for readability) and court-district fixed effects. The standard errors are reported in parentheses and are clustered at the court district level. The covariates indicate whether the defendant has a male name, whether the defendant used a public defender instead of hiring a private lawyer, whether the defendant was accused of theft, and whether the defendant was caught red-handed when committing a crime (``in flagrante delicto'').}
			\end{table}
			
			Table \ref{TabTestAssignment} reports the estimated coefficients of regressing the treatment variable (first row) and the instrumental variable (second row) on the four excluded covariates indicated in the columns, the censoring variable and court-district fixed effects. The standard errors are reported in parentheses and are clustered at the court district level. We find that our four excluded covariates are correlated with the final ruling in each case (treatment variable), but are not correlated with the punishment rate (instrumental variable). As discussed above, this finding indirectly supports Assumption \ref{AsIndependence}.
			
			\subsubsection{Monotonicity Condition (Equation \eqref{EqTreatment})}
			
			Table \ref{TabTestMonotonicity} focuses on assessing the plausibility of the monotonicity condition (Equation \eqref{EqTreatment}). Although it is not possible to directly test that any change in the instrument changes all agents in the same direction with respect to treatment take-up, it is possible to test whether, on average, $Z$ moves agents with different covariate values towards treatment take-up. To implement this test, we adapt the strategy proposed by \citet{Bhuller2019} to our context and regress treatment take-up (final punishment) on the instrument (punishment rate), the censoring variable, and court-district fixed effect for different samples based on values of excluded covariates. We use nine different samples for this test: all defendants, only defendants whose first names are characteristically male, only defendants whose first names are not characteristically male, only defendants who used public defenders, only defendants who hired private lawyers, only defendants accused of theft, only defendants who were not accused of theft, only defendants who were caught red-handed when committing a crime (``in flagrante delicto''), and only defendants who were not caught red-handed. If the estimated punishment rate's coefficient has the same sign in the regressions for all subsamples, this could serve as indirect support for the monotonicity condition (Equation \eqref{EqTreatment}).
			{\linespread{1.5}
				\begin{table}[!htbp]
					\centering
					\caption{{Testing the Monotonicity Condition}} \label{TabTestMonotonicity}
					\begin{lrbox}{\tablebox}
						\begin{tabular}{lccc}
							\hline \hline
							& Punishment Rate $(Z)$ & Censoring Variable $(C)$ & Sample Size \\
							& (1) & (2) & (3) \\ \hline
							\multirow{2}{*}{(1) All} & 0.772*** &  0.013*** & \multirow{2}{*}{43,468} \\
							& (0.046) & (0.004) & \\
							\multirow{2}{*}{(2) Male} & 0.761*** & 0.012***& \multirow{2}{*}{37,304}   \\
							& (0.051) & (0.004) & \\
							\multirow{2}{*}{(3) Female} & 0.799*** & 0.020***& \multirow{2}{*}{6,164}   \\
							& (0.080) & (0.006) & \\
							\multirow{2}{*}{(4) Public Defender} & 0.718*** & 0.021**& \multirow{2}{*}{5,101}   \\
							& (0.152) & (0.010) & \\
							\multirow{2}{*}{(5) Private Lawyer} & 0.766*** & 0.022***& \multirow{2}{*}{38,367}   \\
							& (0.054) & (0.004) & \\
							\multirow{2}{*}{(6) Theft} & 0.765*** & 0.031***& \multirow{2}{*}{24,200}   \\
							& (0.071) & (0.005) & \\
							\multirow{2}{*}{(7) Other Crimes} & 0.750*** & -0.006 & \multirow{2}{*}{19,268}  \\
							& (0.082) & (0.005) & \\
							\multirow{2}{*}{(8) Red-Handed} & 0.766*** & 0.030***& \multirow{2}{*}{15,147}   \\
							& (0.076) & (0.005) & \\
							\multirow{2}{*}{(9) Not Red-Handed} & 0.789*** & 0.005 & \multirow{2}{*}{28,321}  \\
							& (0.065) & (0.004) & \\ \hline
						\end{tabular}
					\end{lrbox}
					\usebox{\tablebox}\\
					\justifying
					\hspace{-.6cm}\scriptsize{Note: This table reports the estimated coefficients of regressing the treatment variable ($D = $``punished according to the final ruling in the case'') on the the instrumental variable ($Z = $``trial judge's punishment rate'' in Column (1)), the censoring variable ($C$ is measured in years for readability in Column (2)) and court-district fixed effects for different samples. The standard errors are reported in parentheses and are clustered at the court district level. The first row uses our entire sample, the second row uses only defendants whose first names are characteristically male, the third row uses only defendants whose first names are not characteristically male, the fourth row uses only defendants who used public defenders, the fifth row uses only defendants who hired private lawyers, the sixth row uses only defendants accused of theft, the seventh row uses only defendants who were not accused of theft, the eight row uses only defendants who were caught red-handed when committing a crime (``in flagrante delicto''), and the nineth row uses only defendants who were not caught red-handed. Column(3) reports the number of observations in each sample.}
				\end{table}
			}
			
			Table \ref{TabTestMonotonicity} reports the estimated coefficients of regressing the treatment variable ($D = $``punished according to the final ruling in the case'') on the instrumental variable ($Z = $``trial judge's punishment rate'' in Column (1)), the censoring variable ($C$ is measured in years for readability in Column (2)) and court-district fixed effects for the sample indicated in the rows. The standard errors are reported in parentheses and are clustered at the court district level. Column(3) reports the number of observations in each sample. We find that the estimated punishment rate's coefficient has the same sign and similar magnitude in the regressions for all subsamples, indirectly supporting the monotonicity condition (Equation \eqref{EqTreatment}).
			
		}
		
		\subsubsection{Random Censoring (Assumption \ref{AsCensoring})}
		
		{
			
			This appendix offers three ways to indirectly assess the plausibility of the Random Censoring Condition (Assumption \ref{AsCensoring}). The first uses the distribution of the realized uncensored time-to-recidivism outcome ($Y^{*}$) given the censoring variable $C$. The second uses the distribution of an excluded covariate given the the censoring variable $C$. The third uses the fact that, under Assumption \ref{AsCensoring}, the function $\gamma_{d}$ in Equation \eqref{Eq_gamma} identifies the same object for a continuum of values of the censoring variable.
			
		}
		
\begin{figure}[!htb]
			\begin{center}
				\begin{subfigure}[t]{0.47\textwidth}
					\centering
					\includegraphics[width = \textwidth]{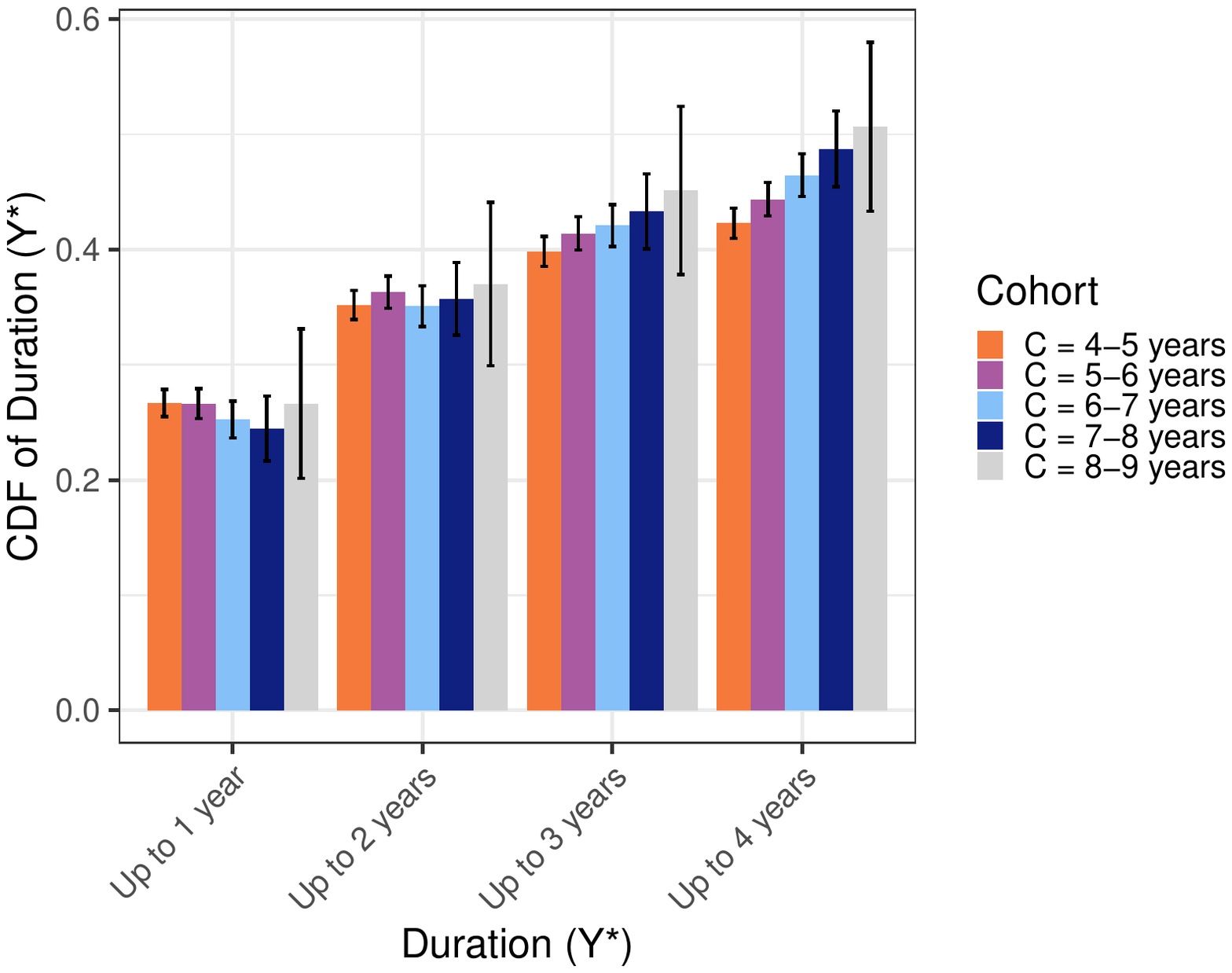}
					\caption{CDF of the Uncensored Outcome given the Defendant's Cohort: $\mathbb{P}\left[\left. Y^{*} \leq y \right\vert C\right]$}
					\label{FigCDF}
				\end{subfigure}
				\hfill
				\begin{subfigure}[t]{0.47\textwidth}
					\begin{center}
						\includegraphics[width = \textwidth]{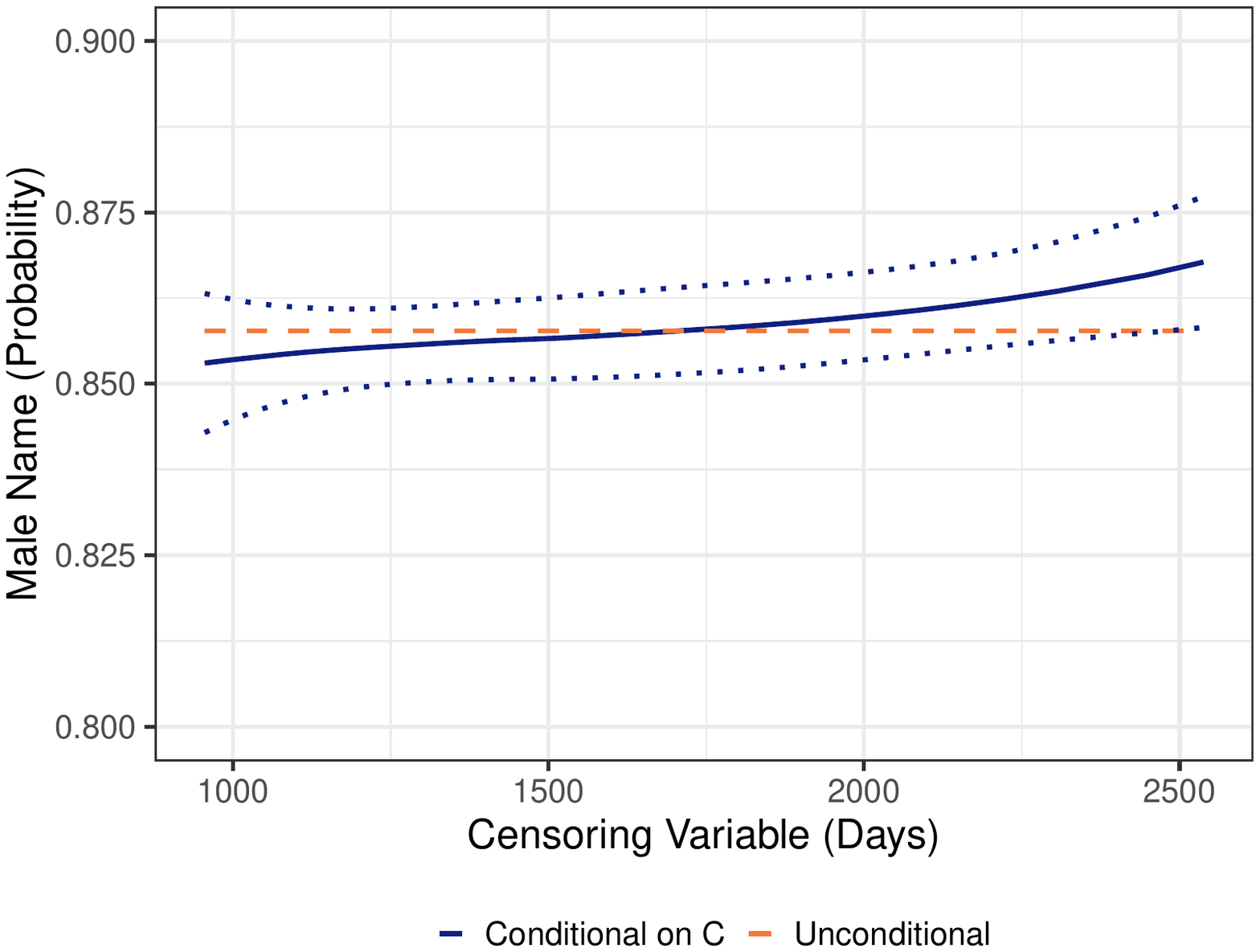}
						\caption{Probability of having a typically male name given the defendant's censoring variable: $\mathbb{P}\left[\left. \text{Male Name} \right\vert C\right]$}
						\label{FigMale}
					\end{center}
				\end{subfigure}
				\caption[]{Assessing the Plausibility of Assumption \ref{AsCensoring}}
				\label{FigDescriptive}
			\end{center}
			\justifying
			\vspace{-.5cm}\scriptsize{Notes: Figure \ref{FigCDF} shows the cumulative distribution function of the uncensored potential outcome ($Y^{*}$) given cohorts based on the censoring variable. Each color denotes a different cohort: orange denotes defendants who are observed for at least four years and at most five years during our sampling period, purple denotes defendants who are observed for at least five years and at most six years, light blue denotes defendants who are observed for at least six years and at most seven years, dark blue denotes defendants who are observed for at least seven years and at most eight years, and gray denotes defendants who are observed for at least eight years and at most nine years. These conditional CDFs are evaluated at four values of the uncensored potential outcome (one, two, three, or four years), and these evaluation points are denoted in the x-axis. The y-axis denotes the value of the CDF, while black lines denote point-wise 99\%-confidence intervals around the values of the CDF.
				
				\hspace{-.6cm}Figure \ref{FigMale} shows, as a solid dark blue line, the probability that the defendant has a typically male name as a function of their censoring variable. This nonparametric function was estimated using a local linear regression with an Epanechnikov kernel based on \citet{Calonico2019}. The bandwidth was optimally selected according to the IMSE criterion. The dotted dark blue lines are robust bias-corrected 99\%-confidence intervals. The dashed orange line is the unconditional probability of having a typically male name.}
		\end{figure}

		Figure \ref{FigDescriptive} focuses on assessing the plausibility of Assumption \ref{AsCensoring}. Although it is not possible to directly test that the potential uncensored time-to-recidivism outcomes $Y^{*}\left(0\right)$ and $Y^{*}\left(1\right)$ are independent of the censoring variable $C$ (time-to-follow-up) given the value of the latent heterogeneity (or punishment resistance) $V$, it is possible to test whether the realized uncensored time-to-recidivism outcome ($Y^{*}$) is (approximately) independent of $C$. If we find no strong evidence against the independence assumption between $Y^*$ and $C$, this could serve as indirect support for (or lack of support against) Assumption \ref{AsCensoring}. We emphazie that this statistic is only an indirect and suggestive test of random censoring.
		
		Figure \ref{FigCDF} shows the cumulative distribution function (CDF) of the uncensored time-to-recidivism outcome given cohorts defined based on the time-to-follow-up variable. If $Y^{*}$ is independent of $C$, then this CDF should not vary across cohorts. Taking into account the sampling uncertainty, Figure \ref{FigCDF} shows that the CDFs are not very different across censoring cohorts, indirectly suggesting that the censoring variable may be independent of the potential outcomes as required by Assumption \ref{AsCensoring}.\footnote{This figure also suggests that if there is dependence, the potential outcomes would be negatively regression dependent on the censoring variable, as discussed in Appendix \ref{Scid}. Moreover, as a robustness check against violations of the Random Censoring assumption, we re-do our analysis focusing solely on the period starting in 2014. These results are presented in Appendix \ref{AppRobustness} and are similar to the ones using the entire sample period, indicating that our results are robust to possible violations of the Random Censoring Assumption.}
		
		Another practical test we can entertain is to analyze the relationship between the censoring variable and an excluded covariate: an indicator for whether the defendant has a male name. Under random censoring, we would expect to see that the information about $C$ should not affect the distribution of the excluded covariate. Figure \ref{FigMale} entertains this exercise and shows the probability of having a typically male name given the defendant's censoring variable (dark blue line). Regardless of the censoring variable, this probability is relatively close to the unconditional share of male names (orange line). This, again, can serve as indirect evidence of the plausibility of our random censoring assumption.
		
		{
			
			The last indirect test of random censoring uses the fact that the function $\gamma_{d}$ in Equation \eqref{Eq_gamma} identifies the same object for a continuum of values of the censoring variable.\footnote{This test was suggested by an anonymous referee.} To see this, note that the function $\gamma_{d}$ is the left-hand side of Equations \eqref{EqPYD1A6derivative} and \eqref{EqPYD0A6derivative} after we explicitly condition on a full set of court district dummies:
			\begin{equation*}
				\gamma_{1}(y,v,c,x) = \dfrac{\partial \mathbb{P}\left[\left. Y\leq y, D = 1 \right\vert P\left(Z,C\right) = v, C = c, X = x\right]}{\partial v} = \mathbb{P}\left[\left. Y^*(1) \leq y  \right\vert V = v, X = x\right]
			\end{equation*}
			and
			\begin{equation*}
				\gamma_{0}(y,v,c,x) = \dfrac{\partial \mathbb{P}\left[\left. Y\leq y, D = 0 \right\vert P\left(Z,C\right) = v, C = c, X = x\right]}{\partial z} = - \mathbb{P}\left[\left.  Y^*(0) \leq y  \right\vert V = v, X = x\right]
			\end{equation*}
			for any $y \in \mathcal{Y}$ such that $y < \gamma_{C}$, $c \in \mathcal{C}$ such that $c > y$ and $x \in \mathcal{X}$. When we add these equations, we have that
			\begin{align*}
				DMTE\left(y,v,x\right) & \coloneqq \mathbb{P}\left[\left. Y^*(1) \leq y  \right\vert V = v, X = x\right] - \mathbb{P}\left[\left.  Y^*(0) \leq y  \right\vert V = v, X = x\right] \\
				& = \gamma_{1}(y,v,c,x) + \gamma_{0}(y,v,c,x),
			\end{align*}
			which can be estimated using Equation \eqref{EqEstimated-Derivatives} as
			\begin{align*}
				\widehat{DMTE}_{c}\left(y,v,x\right) & \coloneqq \widehat{\gamma}_{1}(y,v,c,x) + \widehat{\gamma}_{0}(y,v,c,x)
			\end{align*}
			for any $c \in \mathcal{C}$ such that $c > y$. We can then aggregate this object across court districts using the procedure described in Section \ref{AppCovariateAggregation} and find that
			\begin{align}
				\widehat{DMTE}_{c}^{ave}\left(y,v\right) & = \sum_{x\in \mathcal{X}}\widehat{w}_x \cdot \widehat{DMTE}_{c}\left(y,v,x\right) \nonumber \\
				& = \sum_{x\in \mathcal{X}}\widehat{w}_x \cdot \left\lbrace \widehat{\gamma}_{1}(y,v,c,x) + \widehat{\gamma}_{0}(y,v,c,x) \right\rbrace \label{EqDMTEc}.
			\end{align}
			Since this equation holds for any $c \in \mathcal{C}$ such that $c > y$, we can estimate $\widehat{DMTE}_{c}^{ave}$ for different values of $c$ and compare these estimates as an indirect test of the random censoring condition. If these estimates are close to each other, there is suggestive evidence for (or lack of evidence against) the plausibility of Assumption \ref{AsCensoring}.
			
			Figure \ref{FigDMTE_censoring} implements this test informally.\footnote{Formalizing this test is beyond the scope of this manuscript. However, if one would do so, they may use the following test statistic: $$\hat{\phi} \coloneqq \max_{y \in \mathcal{Y}} \left\lbrace \max_{v \in \mathcal{P}} \left\lbrace \max_{\left(c, c^{\prime}\right) \in \mathcal{C}^{2} \text{ such that } c > y \text{ and } c^{\prime} > y}\left\lbrace \widehat{DMTE}_{c}^{ave}\left(y,v\right) - \widehat{DMTE}_{c^{\prime}}^{ave}\left(y,v\right) \right\rbrace \right\rbrace \right\rbrace.$$ To conduct inference around this test statistic, one may use the weighted bootstrapped procedure explained in Algorithm \ref{algo:bootstrap}. We highlight that deriving the large-sample asymptotic properties of this test is an interesting path for future research. Since these theoretical results do not yet exist, we do not report the estimated value of this test statistic for our sample.} Solid lines are the point estimates for the average $\widehat{DMTE}_{c}^{ave}\left(y,\cdot\right)$ functions (Equation \eqref{EqDMTEc}) indicated in the caption of each subfigure, where $y =$ 1, 2, 3, 4, 5 and 6 years. Each color represents one value of the censoring variable, where $c =$ 2, 4, 6 and 8 years. Since $c$ must be strictly larger than $y$, the first two subfigures have four solid lines, the next two subfigures have three solid lines and the last two subfigures have only two solid lines.
			
			These estimates indirectly offer suggestive evidence that Assumption \ref{AsCensoring} is plausible in our empirical setting. The estimates in Figures \ref{DMTE1_censoring} and \ref{DMTE2_censoring} are so close to each other that we cannot distinguish them at this scale. The estimates in Figures \ref{DMTE3_censoring} and \ref{DMTE5_censoring} exhibit larger differences but remain close to each other. The largest differences are found in Figures \ref{DMTE4_censoring} and \ref{DMTE6_censoring}. Yet, these differences are still small compared with the width of the confidence sets reported in Appendix \ref{AppCIs}.
			
			After analyzing the three test types outlined in this appendix, we conclude that Assumption \ref{AsCensoring} is plausible in our setting.
			
			\begin{landscape}
				\begin{figure}[!htbp]
					\begin{center}
						\begin{subfigure}[t]{0.32\textwidth}
							\centering
							\includegraphics[width = \textwidth]{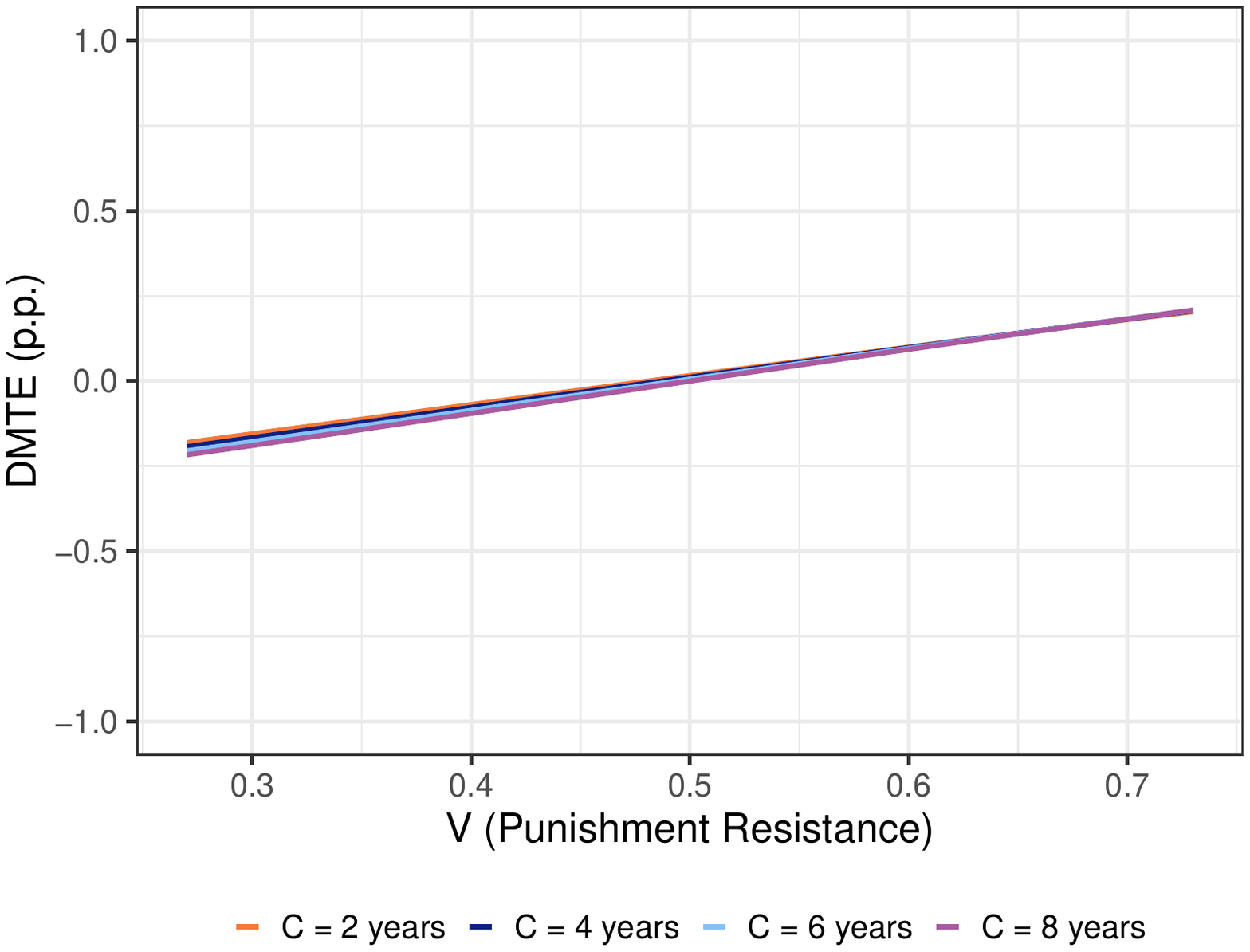}
							\caption{$\widehat{DMTE}_{c}^{ave}\left(1,\cdot\right)$}
							\label{DMTE1_censoring}
						\end{subfigure}
						\hfill
						\begin{subfigure}[t]{0.32\textwidth}
							\begin{center}
								\includegraphics[width = \textwidth]{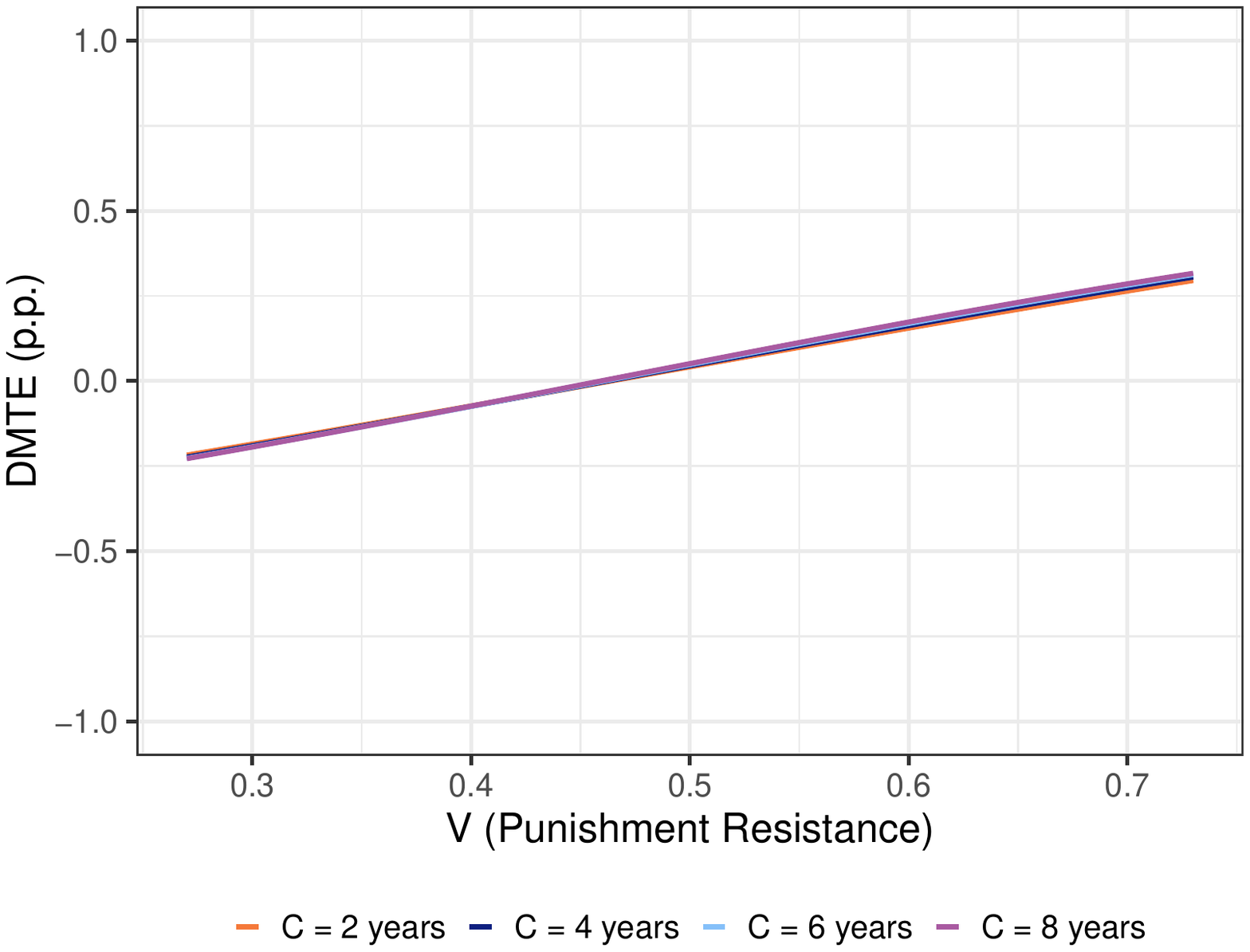}
								\caption{$\widehat{DMTE}_{c}^{ave}\left(2,\cdot\right)$}
								\label{DMTE2_censoring}
							\end{center}
						\end{subfigure}
						\hfill
						\begin{subfigure}[t]{0.32\textwidth}
							\centering
							\includegraphics[width = \textwidth]{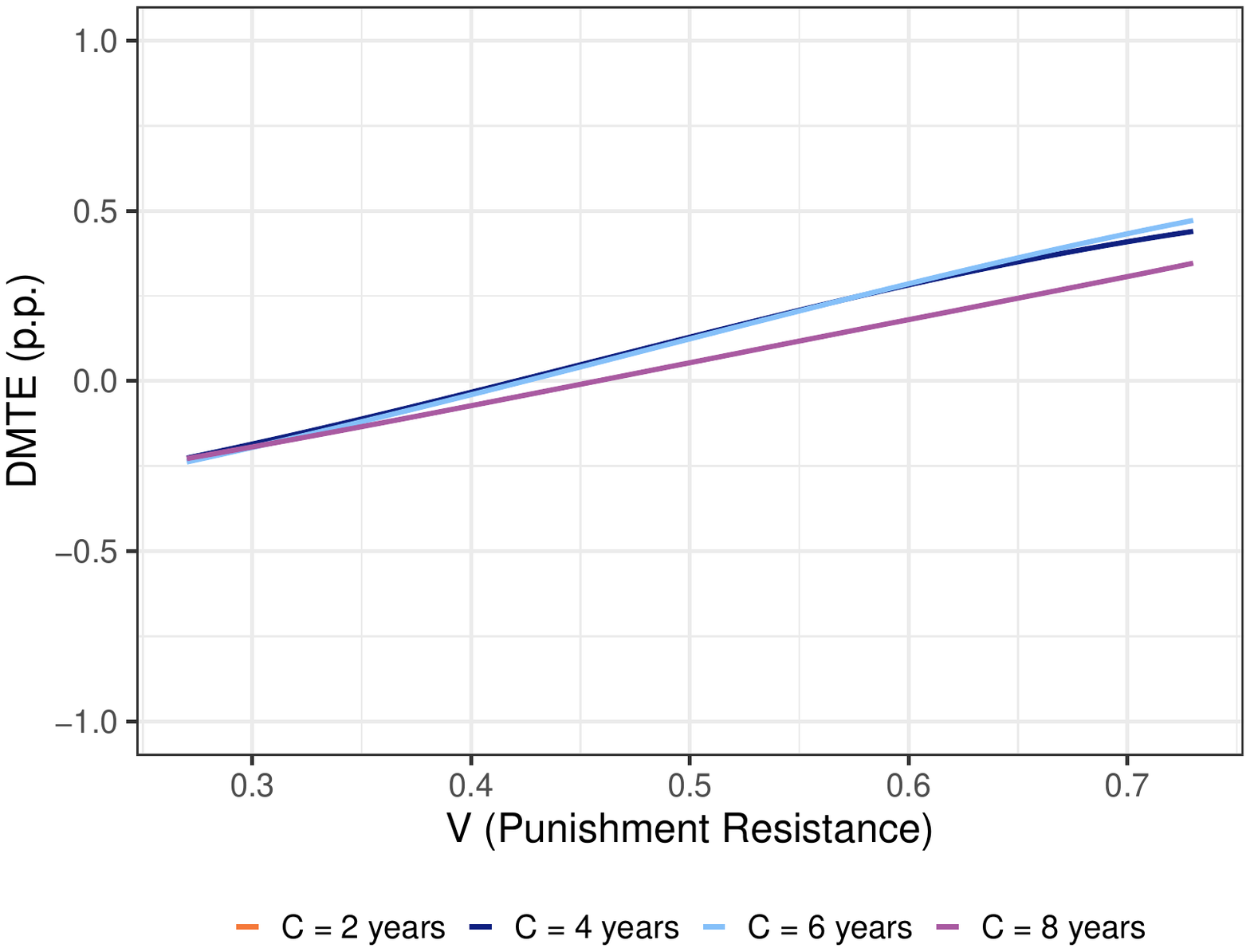}
							\caption{$\widehat{DMTE}_{c}^{ave}\left(3,\cdot\right)$}
							\label{DMTE3_censoring}
						\end{subfigure}
						
						\begin{subfigure}[t]{0.32\textwidth}
							\begin{center}
								\includegraphics[width = \textwidth]{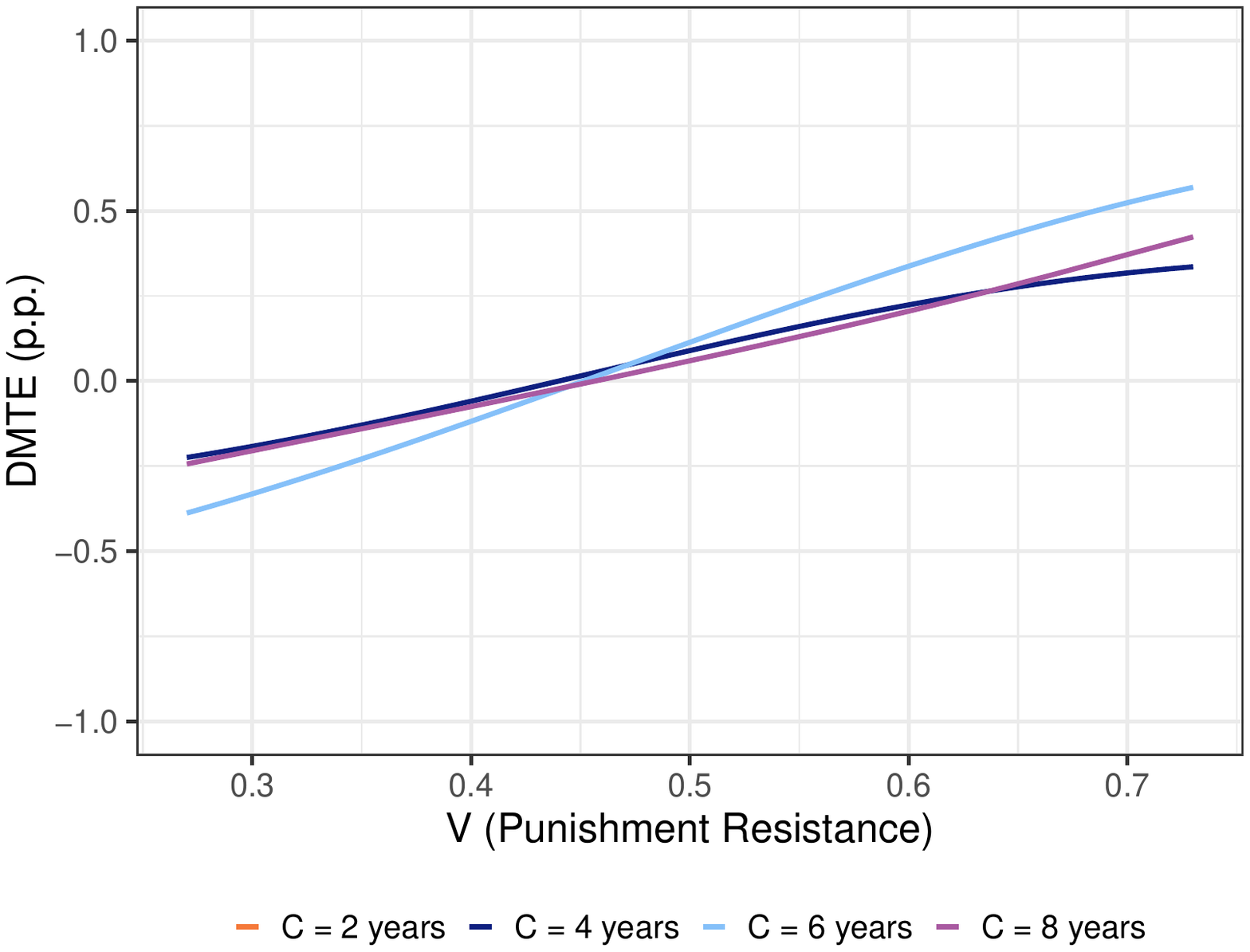}
								\caption{$\widehat{DMTE}_{c}^{ave}\left(4,\cdot\right)$}
								\label{DMTE4_censoring}
							\end{center}
						\end{subfigure}
						\hfill
						\begin{subfigure}[t]{0.32\textwidth}
							\begin{center}
								\includegraphics[width = \textwidth]{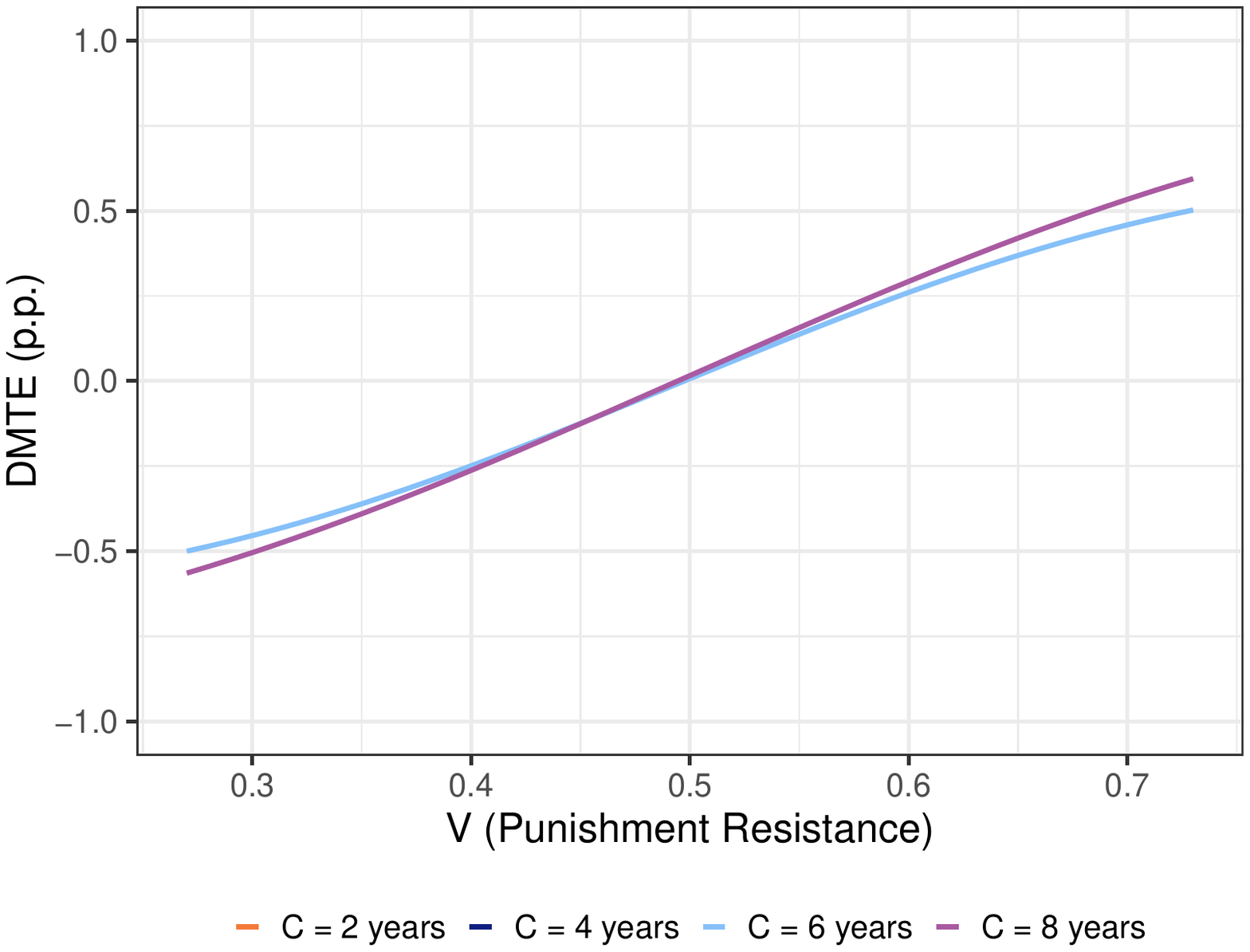}
								\caption{$\widehat{DMTE}_{c}^{ave}\left(5,\cdot\right)$}
								\label{DMTE5_censoring}
							\end{center}
						\end{subfigure}
						\hfill
						\begin{subfigure}[t]{0.32\textwidth}
							\begin{center}
								\includegraphics[width = \textwidth]{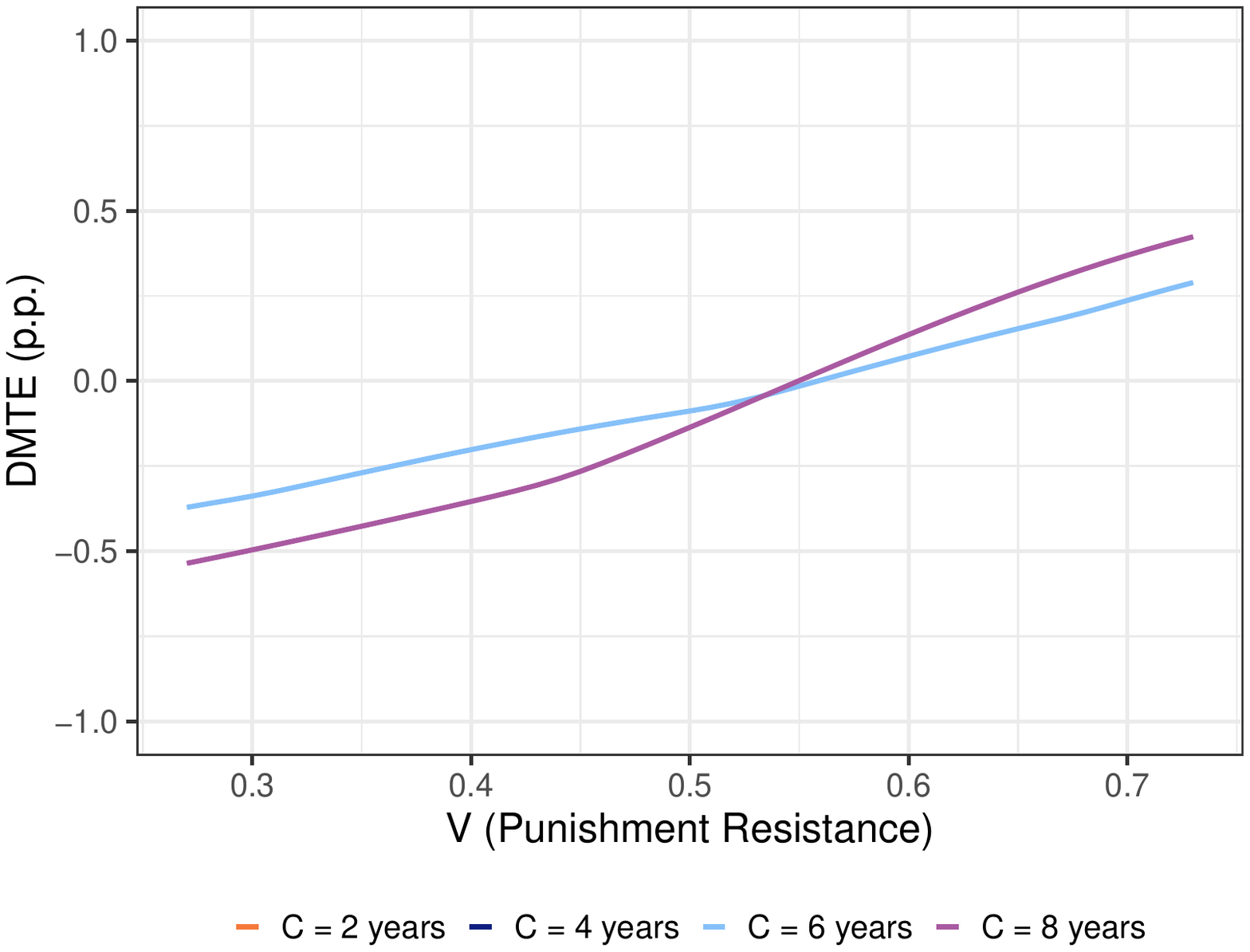}
								\caption{$\widehat{DMTE}_{c}^{ave}\left(6,\cdot\right)$}
								\label{DMTE6_censoring}
							\end{center}
						\end{subfigure}
						\caption{Using $\widehat{DMTE}_{c}^{ave}$ to indirectly test Assumption \ref{AsCensoring}.}
						\label{FigDMTE_censoring}
					\end{center}
					\justifying
					\vspace{-.5cm}\scriptsize{Notes: Solid lines are the point estimates for the average $\widehat{DMTE}_{c}^{ave}\left(y,\cdot\right)$ functions (Equation \eqref{EqDMTEc}) indicated in the caption of each subfigure, where $y =$ 1, 2, 3, 4, 5 and 6 years. Each color represents one value of the censoring variable, where $c =$ 2, 4, 6 and 8 years. Since $c$ must be strictly larger than $y$, the first two subfigures have four solid lines, the next two subfigures have three solid lines and the last two subfigures have only two solid lines.
					}
				\end{figure}
			\end{landscape}

		}
		
		\clearpage
		
		\subsection{First stage results}\label{Sfirststage}
		
		In this appendix, we present the results of the first stage regression in our empirical analysis. In our model, the treatment variable $D$ (``final ruling'') is a function of the instrument $Z$ (``trial judge's punishment rate''), the censoring variable $C$, and court district fixed effects. Following Subsection \ref{SestSemiPara}, we use a polynomial series to approximate the propensity score and report the estimated coefficients of a quadratic model in Table \ref{TabFirstStage}. Note that our instrument is strong according to the F-statistic of the first-stage regression. This result suggests that Assumption \ref{AsRank} is plausible.
		
		\begin{table}[!htb]
			\centering
			\caption{{First Stage Results}} \label{TabFirstStage}
			\begin{lrbox}{\tablebox}
				\begin{tabular}{cccc}
					\hline \hline
					& $Z$ & $Z^{2}$ & $C$ \\ \hline
					Coefficient & 0.663*** & 0.096 & 0.012*** \\
					Clusterized S.E. & (0.235) & (0.208) & (0.004) \\
					F-statistic & \multicolumn{2}{c}{817} & \\ \hline
				\end{tabular}
			\end{lrbox}
			\usebox{\tablebox}\\
			\justifying
			\hspace{-.6cm}\scriptsize{Note: The left-hand side variable is our treatment variable, i.e., $D = $``punished according to the final ruling in the case''. The standard errors are clusterized at the court district level. The third line reports the F-Statistic of a hypothesis test whose null is that the coefficients associated with $Z$ and $Z^{2}$ are equal to zero. The first stage regression controls for court district fixed effects. To improve readability, we multiply the coefficient of the censoring variable (and its standard error) by 365. This transformation is equivalent to measuring the censoring variable in years instead of days.}
		\end{table}
		
\begin{figure}[!htbp]
			\begin{center}
				\includegraphics[width = 0.5\textwidth]{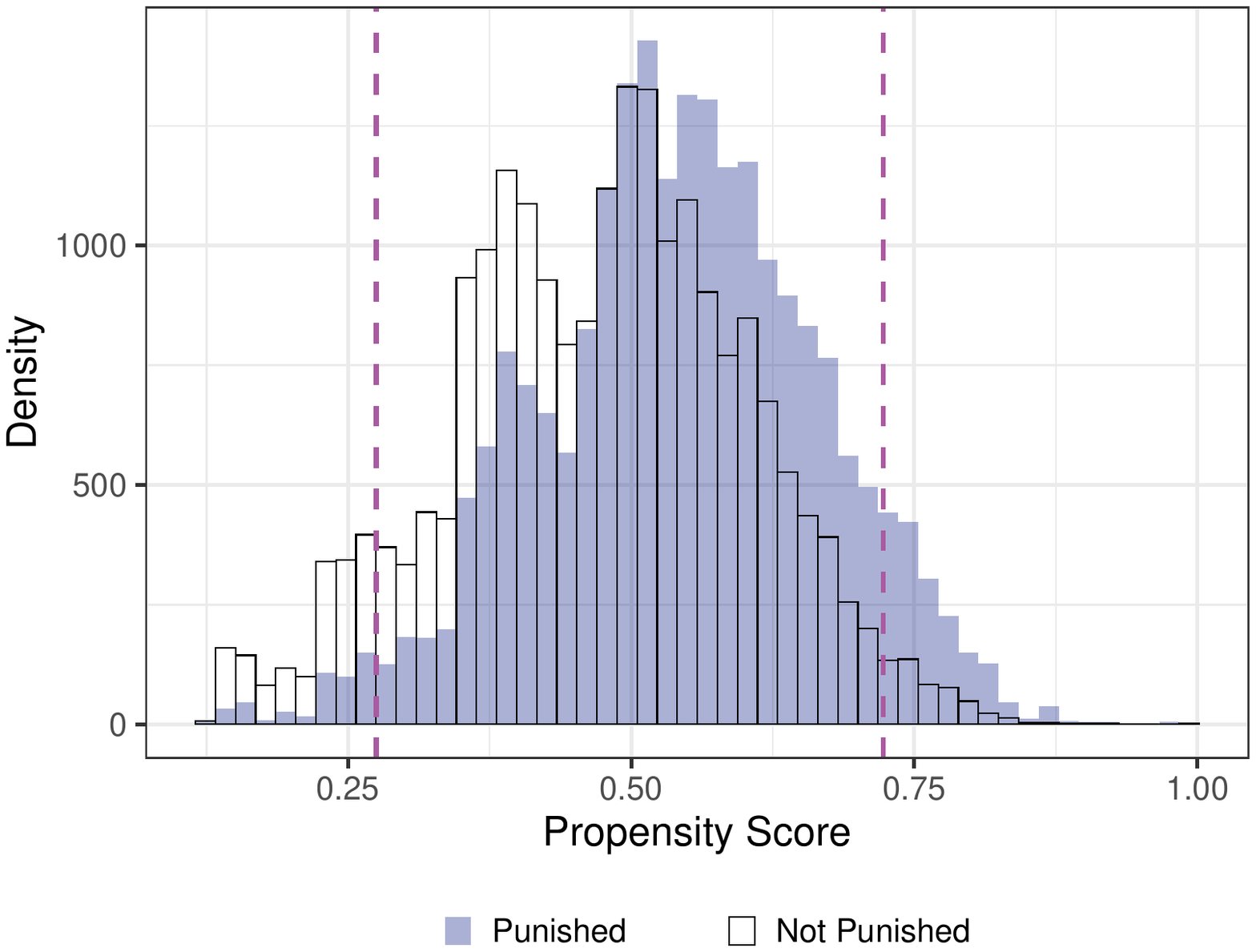}
				\caption{Distribution of the estimated propensity score among treatment groups}
				\label{FigPS}
			\end{center}
			\justifying
			\vspace{-.5cm}\scriptsize{Notes: {The blue histogram shows the distribution of the estimated propensity score among defendants who were punished (treated group). The white histogram shows the distribution of the estimated propensity score among defendants who were not punished (comparison group). The vertical lines denote the unconditional 5\textsuperscript{th} and 95\textsuperscript{th} percentiles of the estimated propensity score.}}
		\end{figure}

		We also report the distribution of the estimated propensity score in Figure \ref{FigPS}. The blue histogram shows the distribution of the estimated propensity score given that defendant was punished (treated group) while the white histogram shows the distribution of the estimated propensity score given that defendant was not punished (comparison group). We find that most defendants have a probability of being punished around 50\%. However, some defendants are more unlikely to be punished (estimated propensity score around 30\%) and others are more likely to be punished (estimated propensity score around 70\%). These widely spread propensity score distributions are positive for identification and estimation because they allow us to discuss DMTE and QMTE functions evaluated at many different points of the latent heterogeneity variable.
		
		The vertical lines in Figure \ref{FigPS} denote the unconditional 5\textsuperscript{th} and 95\textsuperscript{th} percentiles of the estimated propensity score. When discussing our results about the DMTE and QMTE functions, we only report the estimates for latent heterogeneity values between these two percentiles. We do so to avoid extrapolation bias and to ensure the plausibility of Assumption \ref{AsPositive}.
		
		\newpage
		\subsection{$QMTE$ for a given value of punishment resistance}\label{AppQMTEgivenV}
		\begin{figure}[!htb]
			\begin{center}
				\includegraphics[width = 0.7\textwidth]{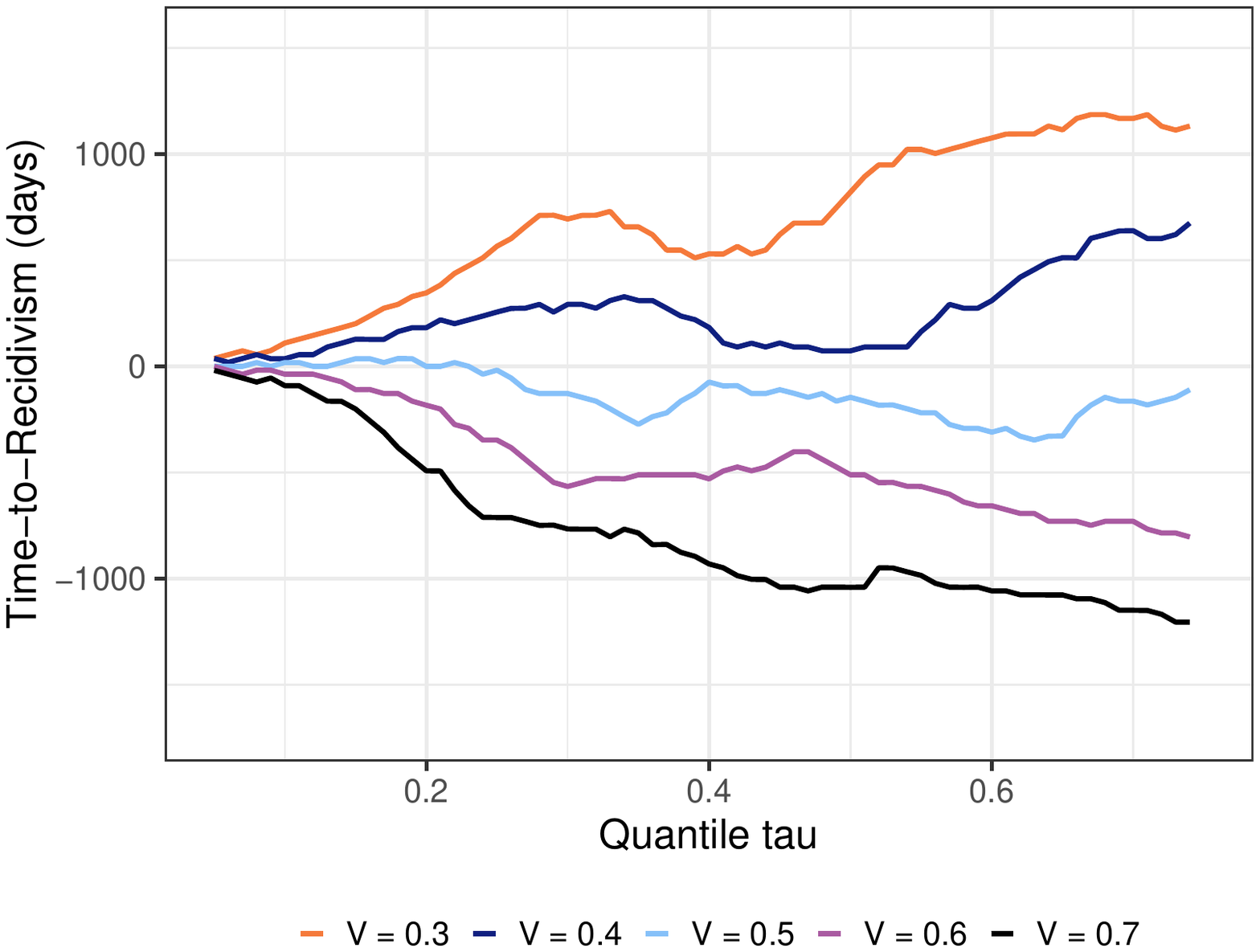}
				\caption{$QMTE\left(\cdot, v\right)$ for $v \in \left\lbrace .3, .4, \ldots, .7 \right\rbrace$}
				\label{FigQMTEfunctionTau}
			\end{center}
			\justifying
			\vspace{-.5cm}\scriptsize{Notes: Solid lines are the point estimates for the average $QMTE\left(\cdot, v\right)$ functions indicated in the legend. These results are based on Corollary \ref{cor:semi}.}
		\end{figure}

		\newpage
		\subsection{Confidence Intervals for DMTE, QMTE and RMTE functions}\label{AppCIs}
		
		\begin{figure}[!htbp]
			\begin{center}
				\begin{subfigure}[t]{0.47\textwidth}
					\centering
					\includegraphics[width = \textwidth]{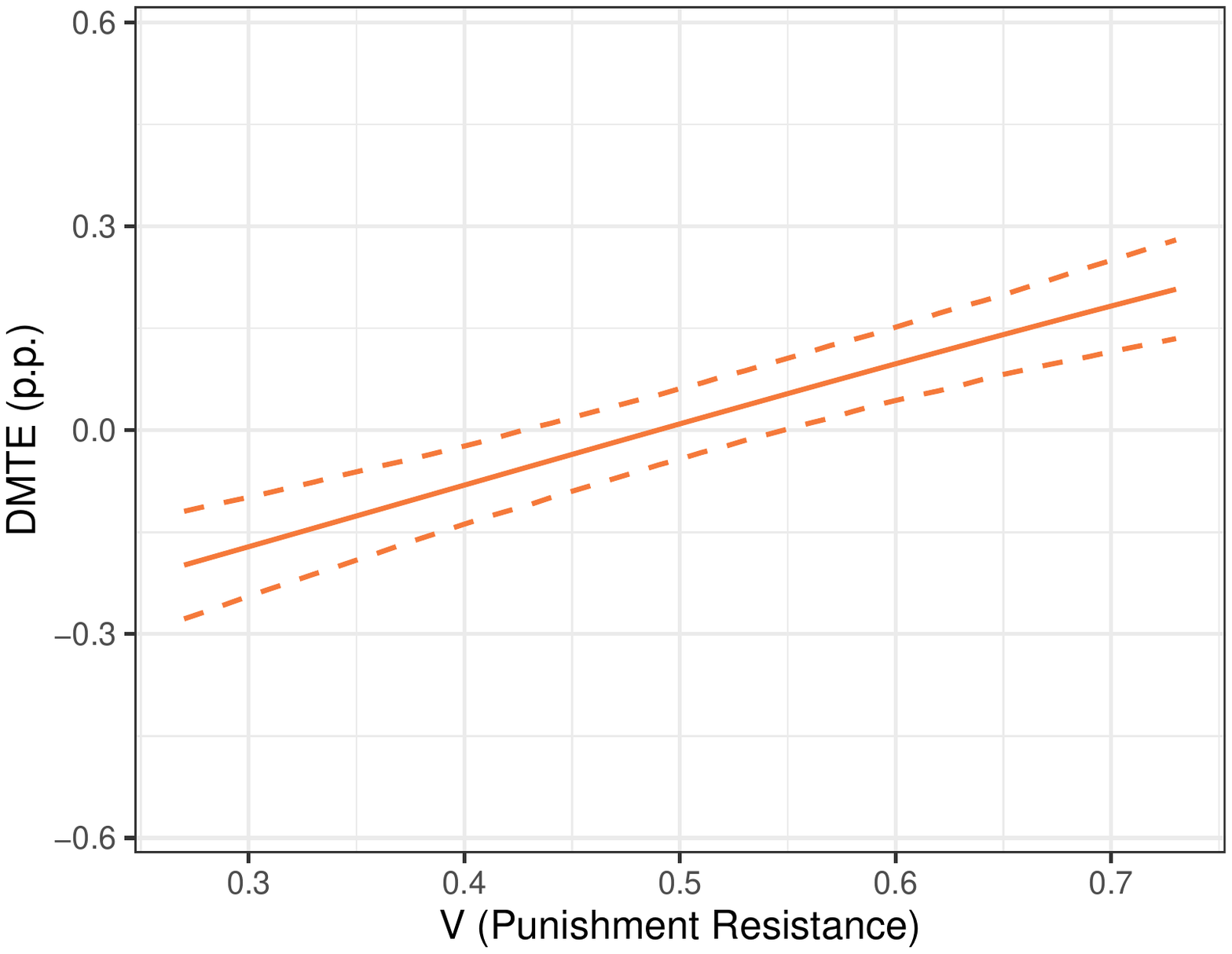}
					\caption{$DMTE\left(1,\cdot\right)$}
					\label{DMTE1-CI}
				\end{subfigure}
				\hfill
				\begin{subfigure}[t]{0.47\textwidth}
					\begin{center}
						\includegraphics[width = \textwidth]{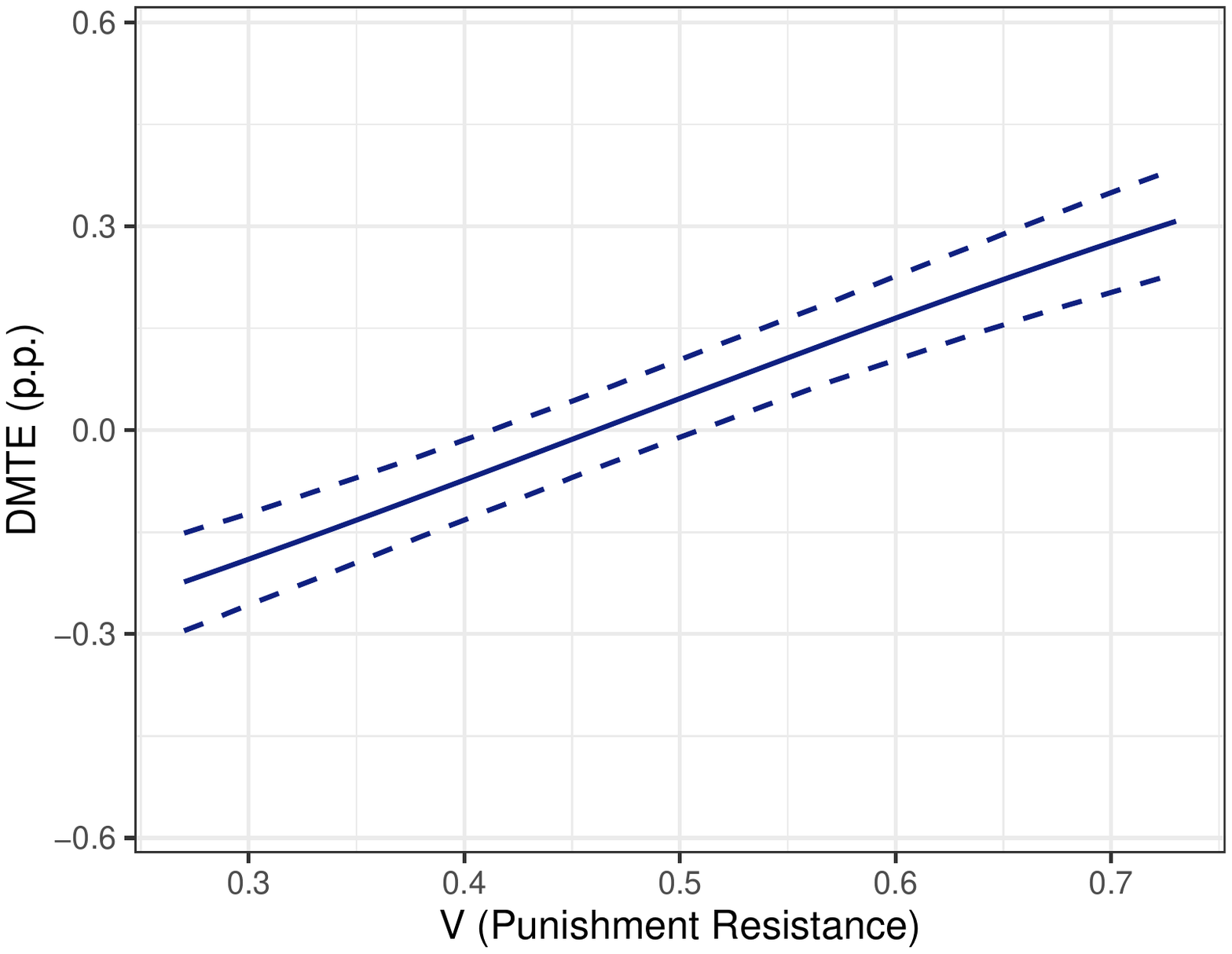}
						\caption{$DMTE\left(2,\cdot\right)$}
						\label{DMTE2-CI}
					\end{center}
				\end{subfigure}
				\begin{subfigure}[t]{0.47\textwidth}
					\centering
					\includegraphics[width = \textwidth]{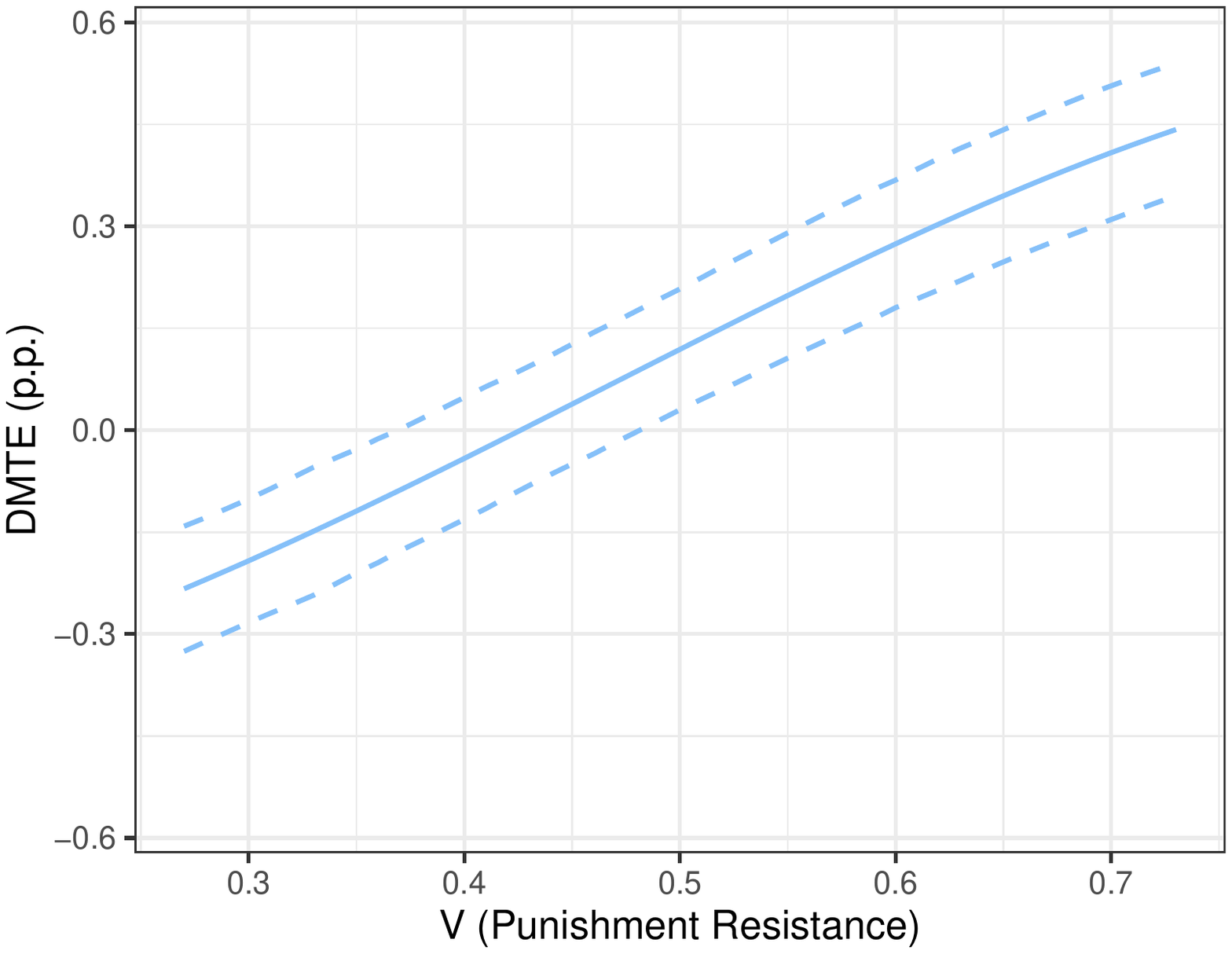}
					\caption{$DMTE\left(3,\cdot\right)$}
					\label{DMTE3-CI}
				\end{subfigure}
				\hfill
				\begin{subfigure}[t]{0.47\textwidth}
					\begin{center}
						\includegraphics[width = \textwidth]{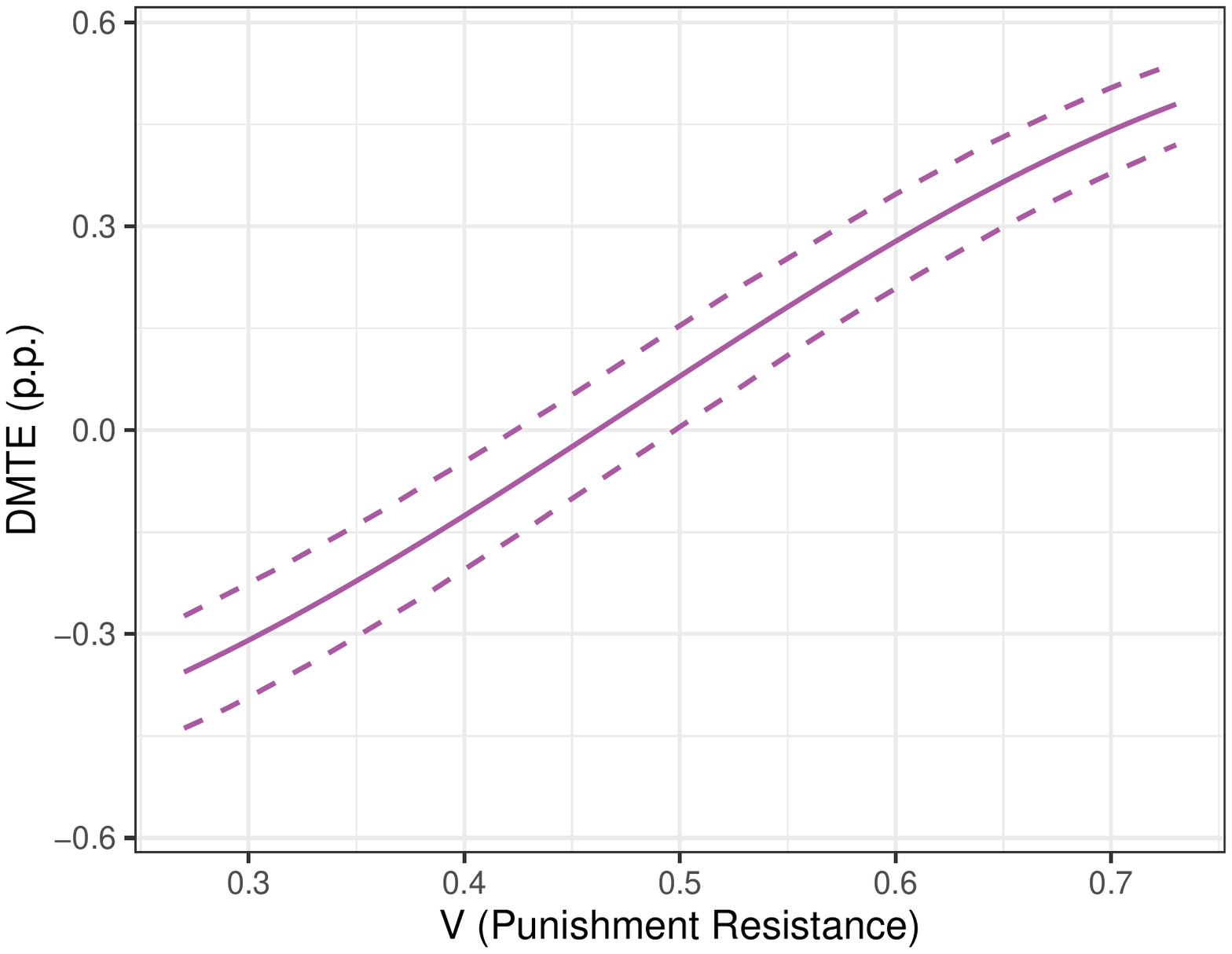}
						\caption{$DMTE\left(4,\cdot\right)$}
						\label{DMTE4-CI}
					\end{center}
				\end{subfigure}
				\caption{90\%-Confidence Intervals for $DMTE\left(y,\cdot\right)$ for $y \in \left\lbrace 1, 2, 3, 4 \right\rbrace$}
				\label{FigDMTE-CI-1-4}
			\end{center}
			\justifying
			\vspace{-.5cm}\scriptsize{Notes: Solid lines are the point estimates for the average $DMTE\left(y,\cdot\right)$ functions indicated in the caption of each subfigure. These results are based on Corollary \ref{cor:semi}. Moreover, point-wise 90\%-confidence intervals are reported using dashed lines. These confidence intervals were computed using the weighted bootstrap clusterized at the court district level (Algorithm \ref{algo:bootstrap}) using 399 repetitions.
			}
		\end{figure}

		\begin{figure}[!htbp]
			\begin{center}
				\begin{subfigure}[t]{0.47\textwidth}
					\centering
					\includegraphics[width = \textwidth]{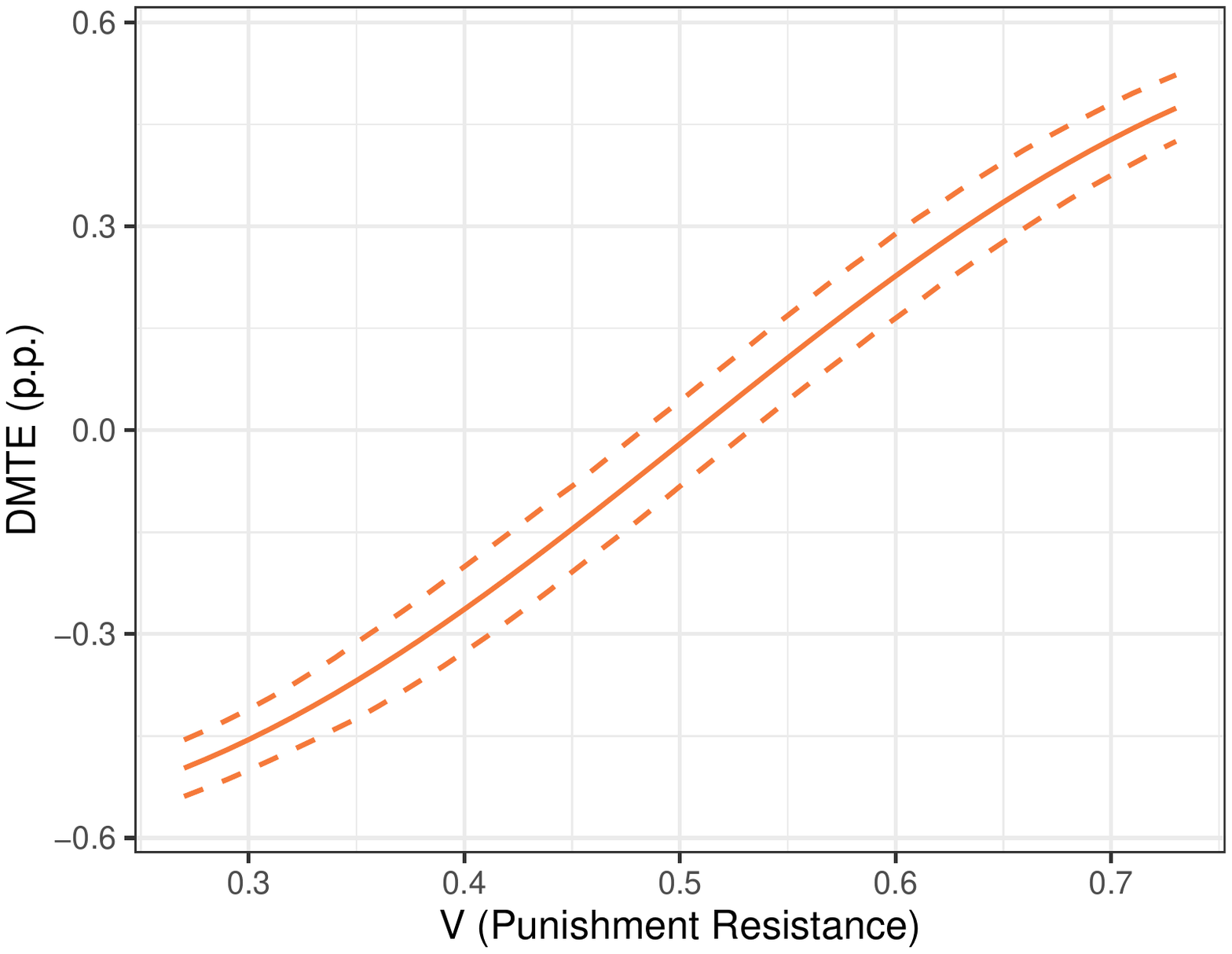}
					\caption{$DMTE\left(5,\cdot\right)$}
					\label{DMTE5-CI}
				\end{subfigure}
				\hfill
				\begin{subfigure}[t]{0.47\textwidth}
					\begin{center}
						\includegraphics[width = \textwidth]{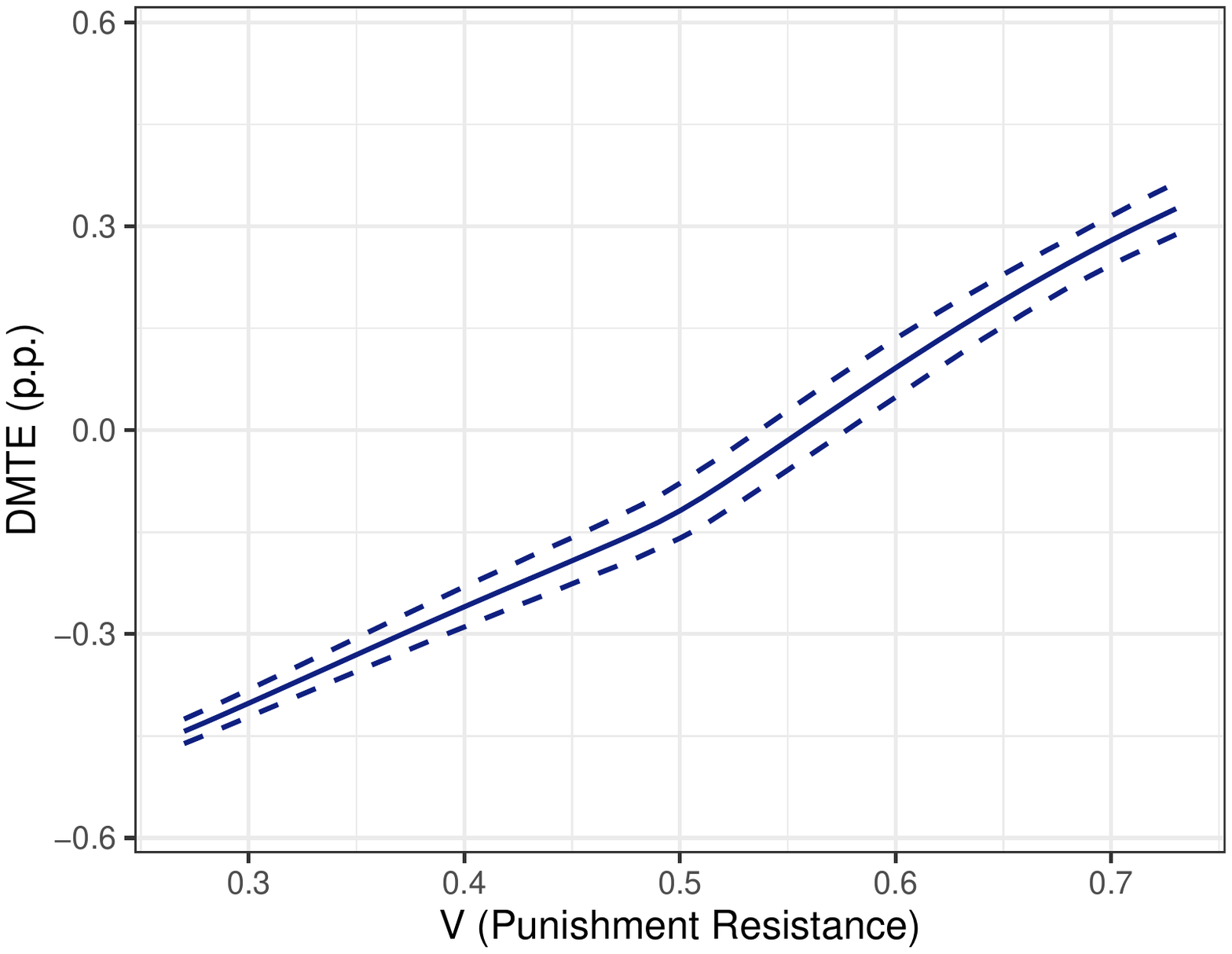}
						\caption{$DMTE\left(6,\cdot\right)$}
						\label{DMTE6-CI}
					\end{center}
				\end{subfigure}
				\begin{subfigure}[t]{0.47\textwidth}
					\centering
					\includegraphics[width = \textwidth]{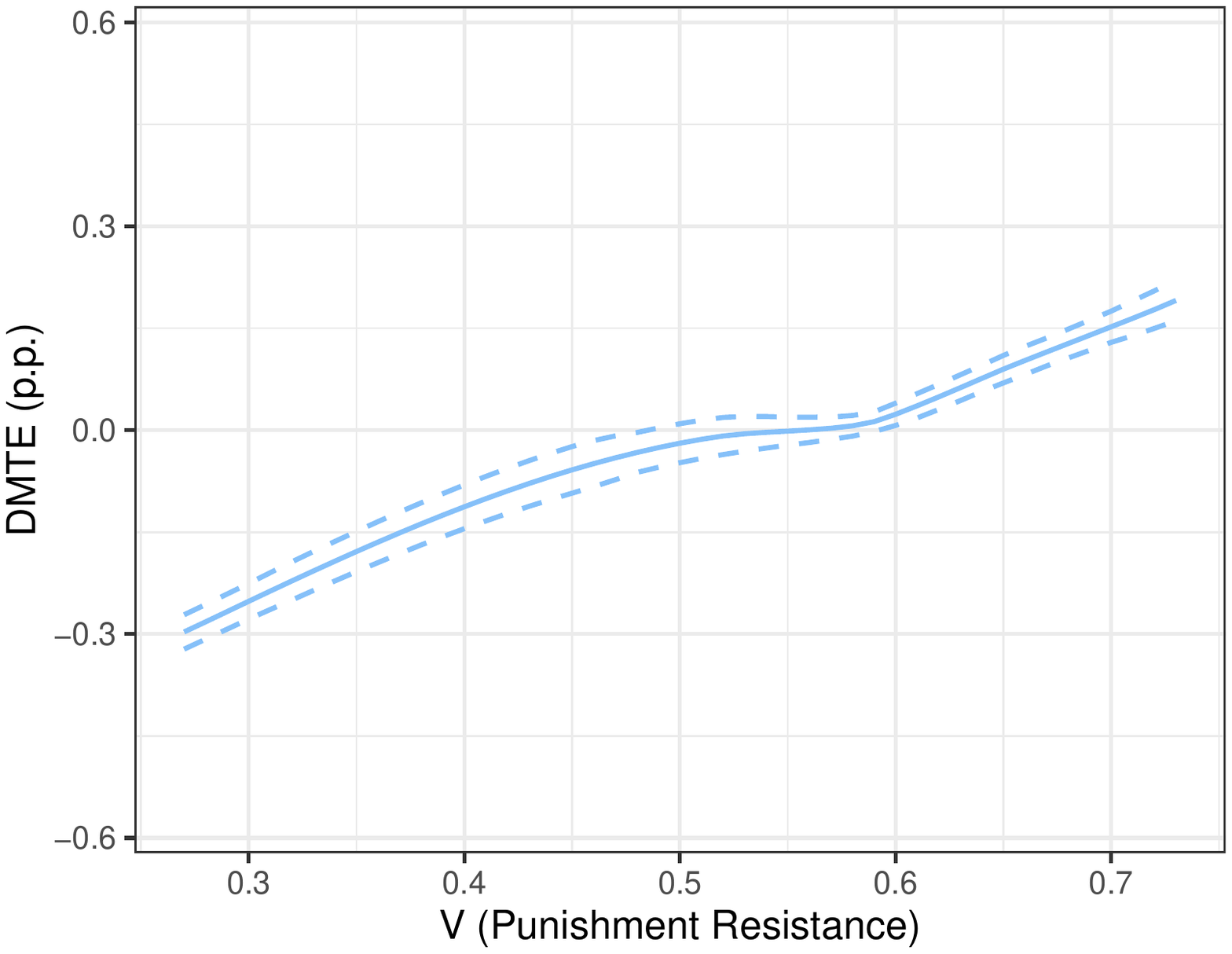}
					\caption{$DMTE\left(7,\cdot\right)$}
					\label{DMTE7-CI}
				\end{subfigure}
				\hfill
				\begin{subfigure}[t]{0.47\textwidth}
					\begin{center}
						\includegraphics[width = \textwidth]{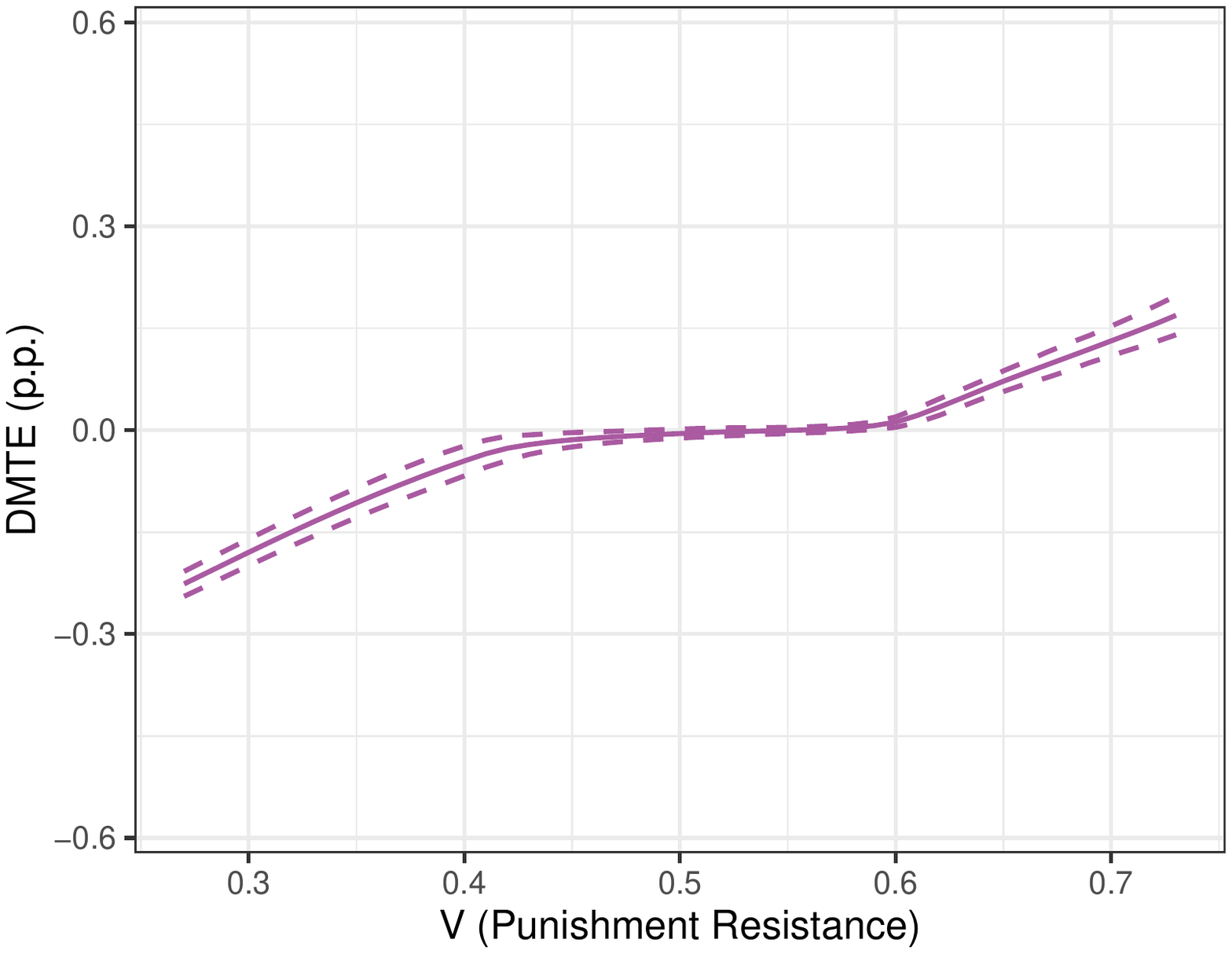}
						\caption{$DMTE\left(8,\cdot\right)$}
						\label{DMTE8-CI}
					\end{center}
				\end{subfigure}
				\caption{90\%-Confidence Intervals for $DMTE\left(y,\cdot\right)$ for $y \in \left\lbrace 5, 6, 7, 8 \right\rbrace$}
				\label{FigDMTE-CI-5-8}
			\end{center}
			\justifying
			\vspace{-.5cm}\scriptsize{Notes: Solid lines are the point estimates for the average $DMTE\left(y,\cdot\right)$ functions indicated in the caption of each subfigure. These results are based on Corollary \ref{cor:semi}. Moreover, point-wise 90\%-confidence intervals are reported using dashed lines. These confidence intervals were computed using the weighted bootstrap clusterized at the court district level (Algorithm \ref{algo:bootstrap}) using 399 repetitions.
			}
		\end{figure}

		\begin{figure}[!htbp]
			\begin{center}
				\begin{subfigure}[t]{0.47\textwidth}
					\centering
					\includegraphics[width = \textwidth]{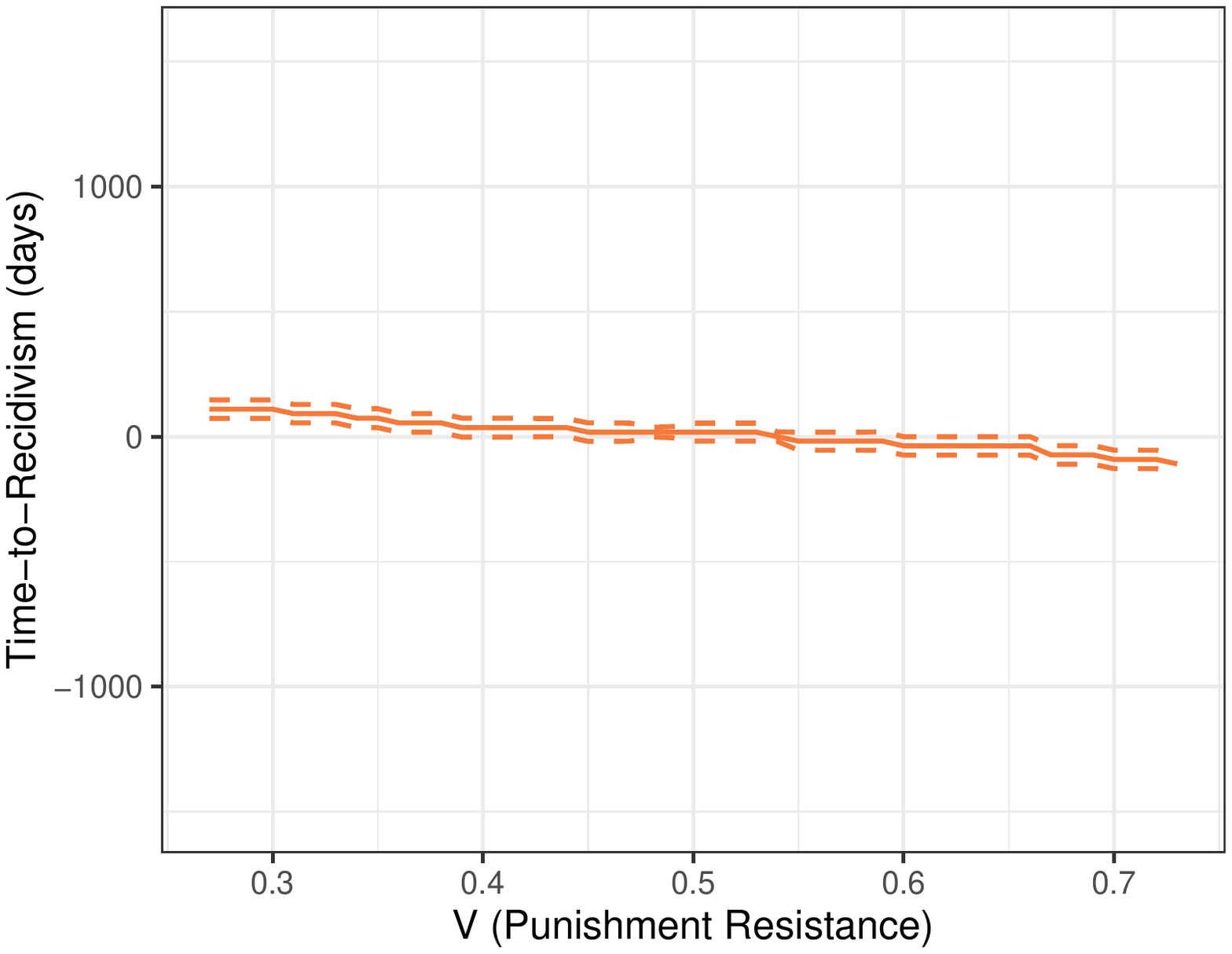}
					\caption{$QMTE\left(.10,\cdot\right)$}
					\label{QMTE10-CI}
				\end{subfigure}
				\hfill
				\begin{subfigure}[t]{0.47\textwidth}
					\begin{center}
						\includegraphics[width = \textwidth]{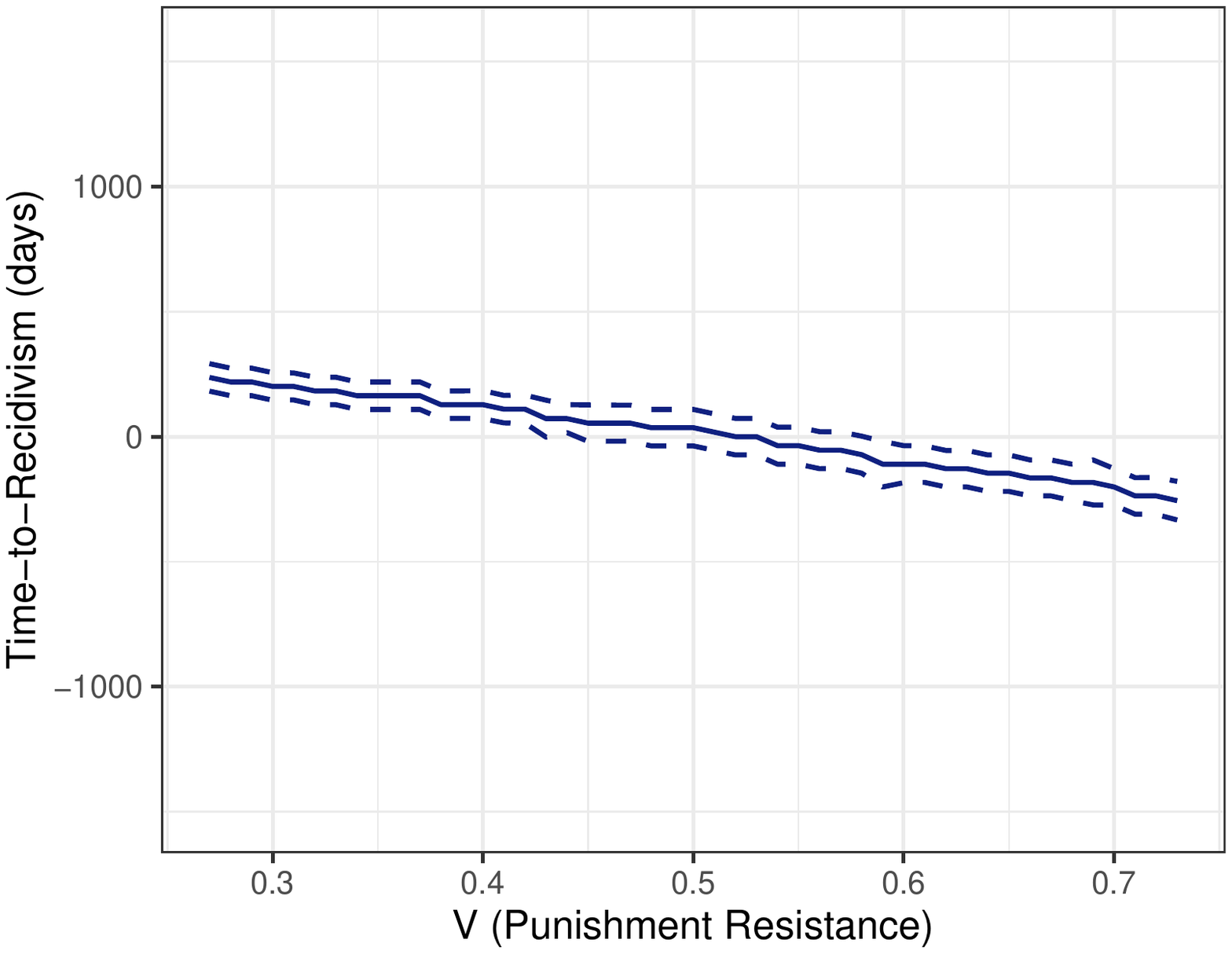}
						\caption{$QMTE\left(.15,\cdot\right)$}
						\label{QMTE15-CI}
					\end{center}
				\end{subfigure}
				\begin{subfigure}[t]{0.47\textwidth}
					\centering
					\includegraphics[width = \textwidth]{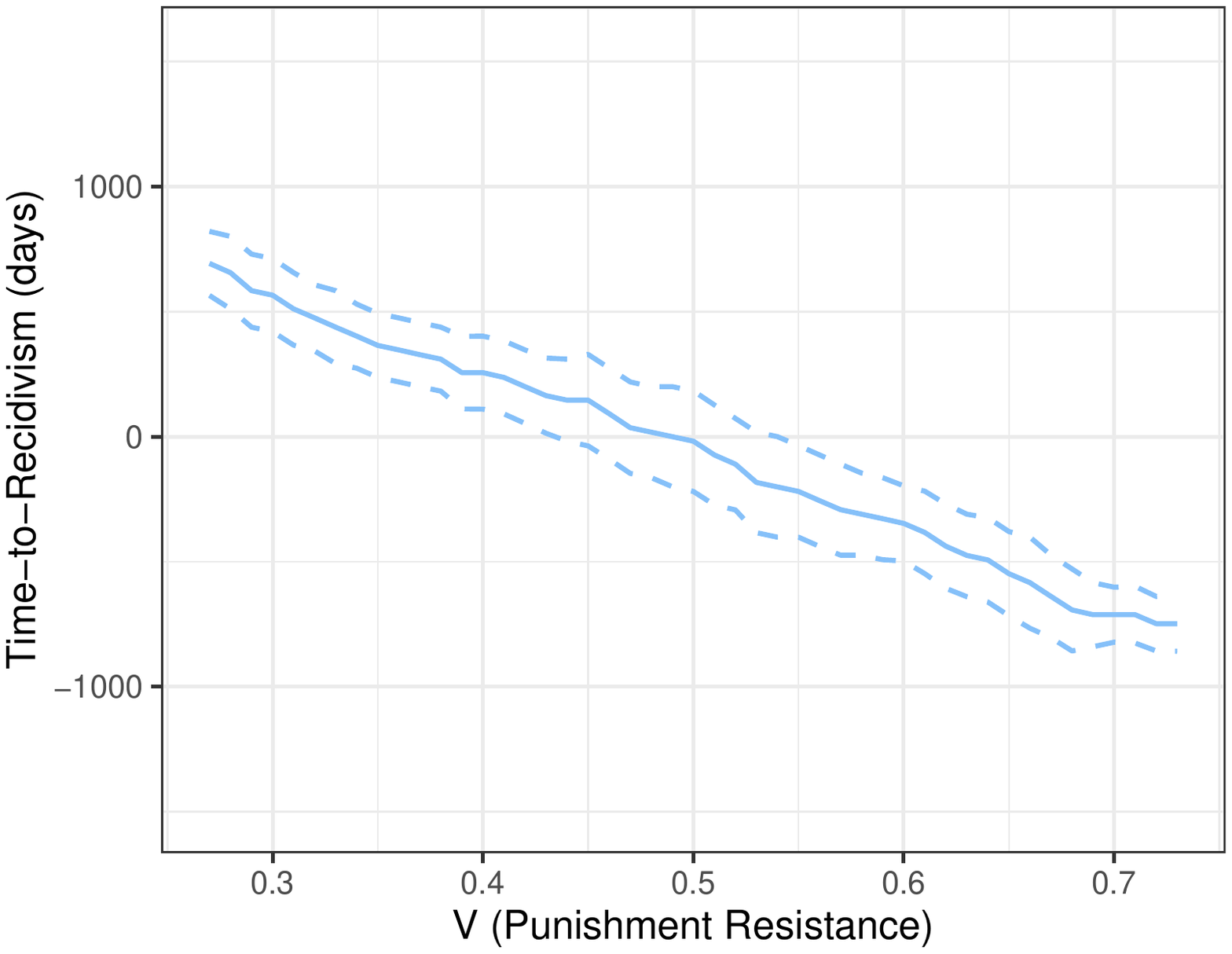}
					\caption{$QMTE\left(.25,\cdot\right)$}
					\label{QMTE25-CI}
				\end{subfigure}
				\hfill
				\begin{subfigure}[t]{0.47\textwidth}
					\begin{center}
						\includegraphics[width = \textwidth]{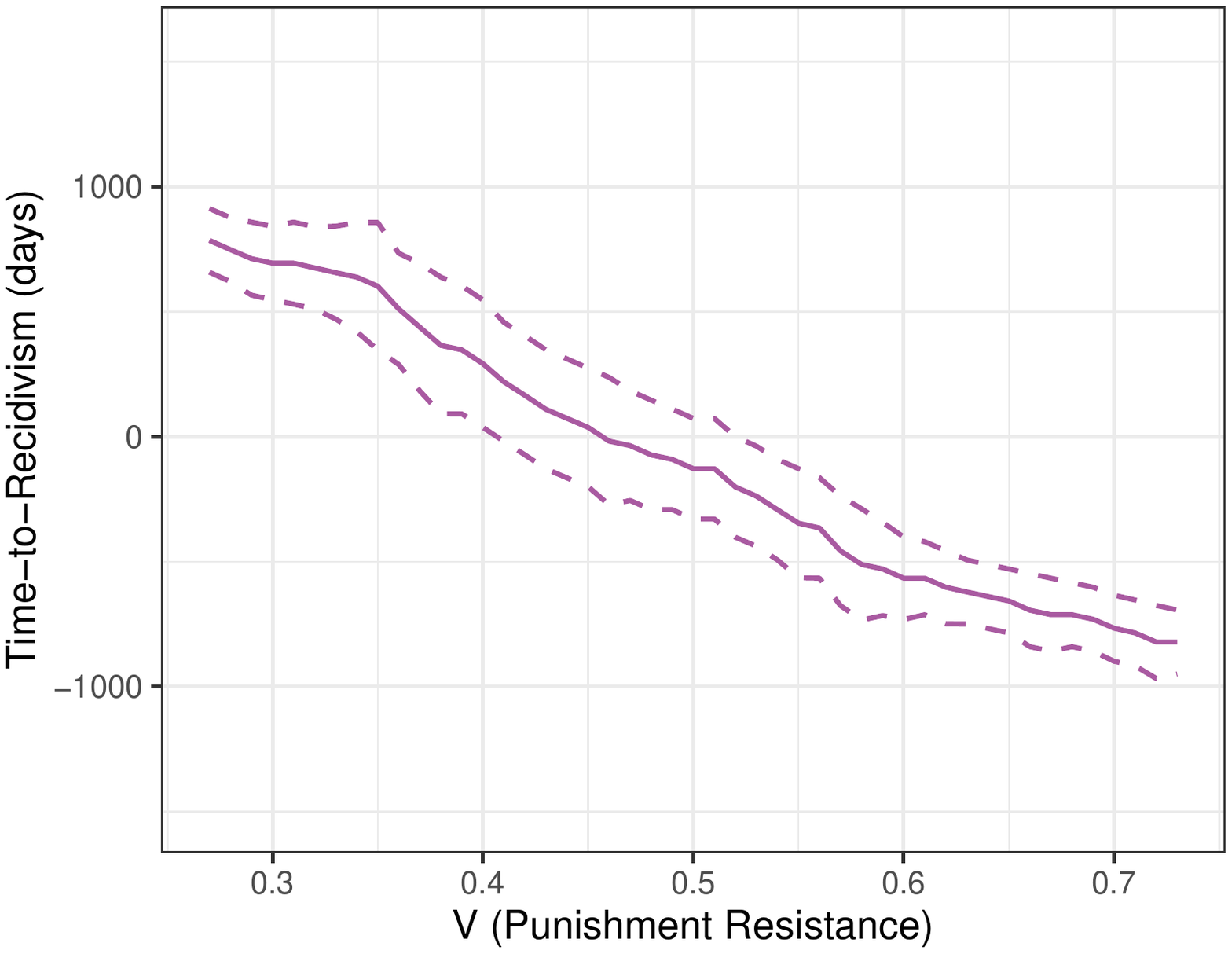}
						\caption{$QMTE\left(.30,\cdot\right)$}
						\label{QMTE30-CI}
					\end{center}
				\end{subfigure}
				\caption{90\%-Confidence Intervals for $QMTE\left(\tau,\cdot\right)$ for $\tau \in \left\lbrace .10, .15, .25, .30 \right\rbrace$}
				\label{FigQMTE-CI-10-30}
			\end{center}
			\justifying
			\vspace{-.5cm}\scriptsize{Notes: Solid lines are the point estimates for the average $QMTE\left(\tau,\cdot\right)$ functions indicated in the caption of each subfigure. These results are based on Corollary \ref{cor:semi}. Moreover, point-wise 90\%-confidence intervals are reported using dashed lines. These confidence intervals were computed using the weighted bootstrap clusterized at the court district level (Algorithm \ref{algo:bootstrap}) using 399 repetitions.
			}
		\end{figure}

		\begin{figure}[!htb]
			\begin{center}
				\begin{subfigure}[t]{0.47\textwidth}
					\centering
					\includegraphics[width = \textwidth]{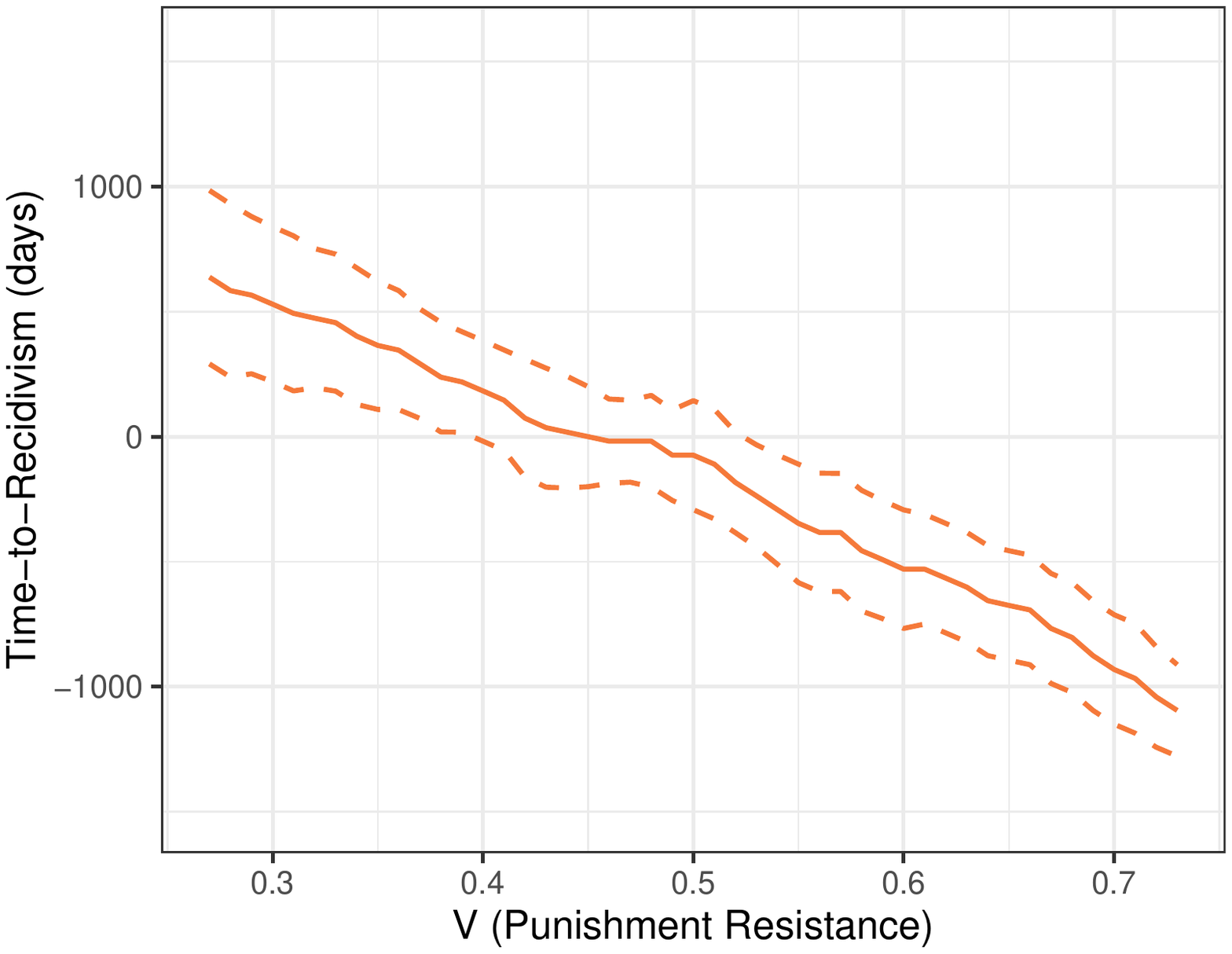}
					\caption{$QMTE\left(.40,\cdot\right)$}
					\label{QMTE40-CI}
				\end{subfigure}
				\hfill
				\begin{subfigure}[t]{0.47\textwidth}
					\begin{center}
						\includegraphics[width = \textwidth]{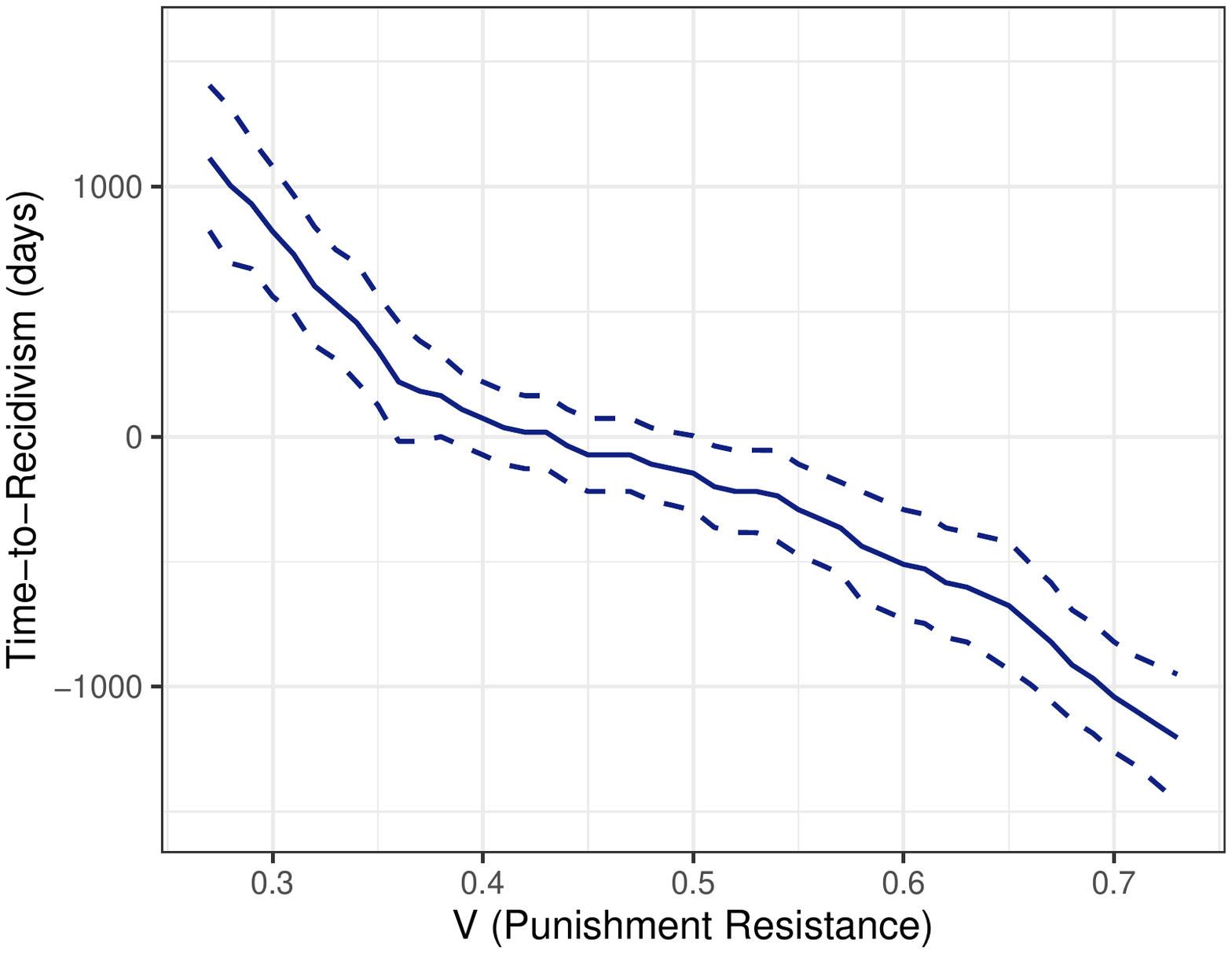}
						\caption{$QMTE\left(50,\cdot\right)$}
						\label{QMTE50-CI}
					\end{center}
				\end{subfigure}
				\begin{subfigure}[t]{0.47\textwidth}
					\centering
					\includegraphics[width = \textwidth]{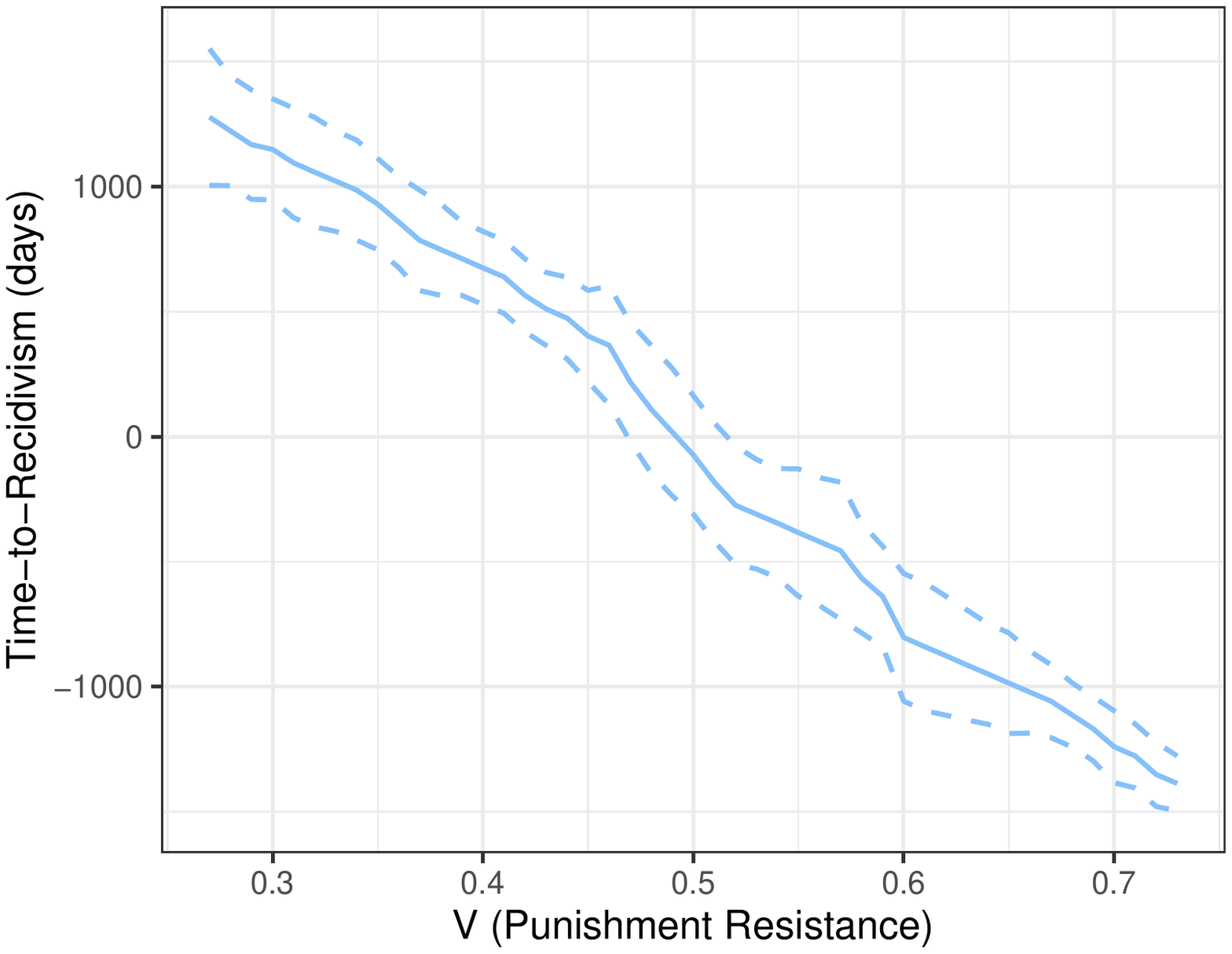}
					\caption{$QMTE\left(.75,\cdot\right)$}
					\label{QMTE75-CI}
				\end{subfigure}
				\hfill
				\begin{subfigure}[t]{0.47\textwidth}
					\begin{center}
						\includegraphics[width = \textwidth]{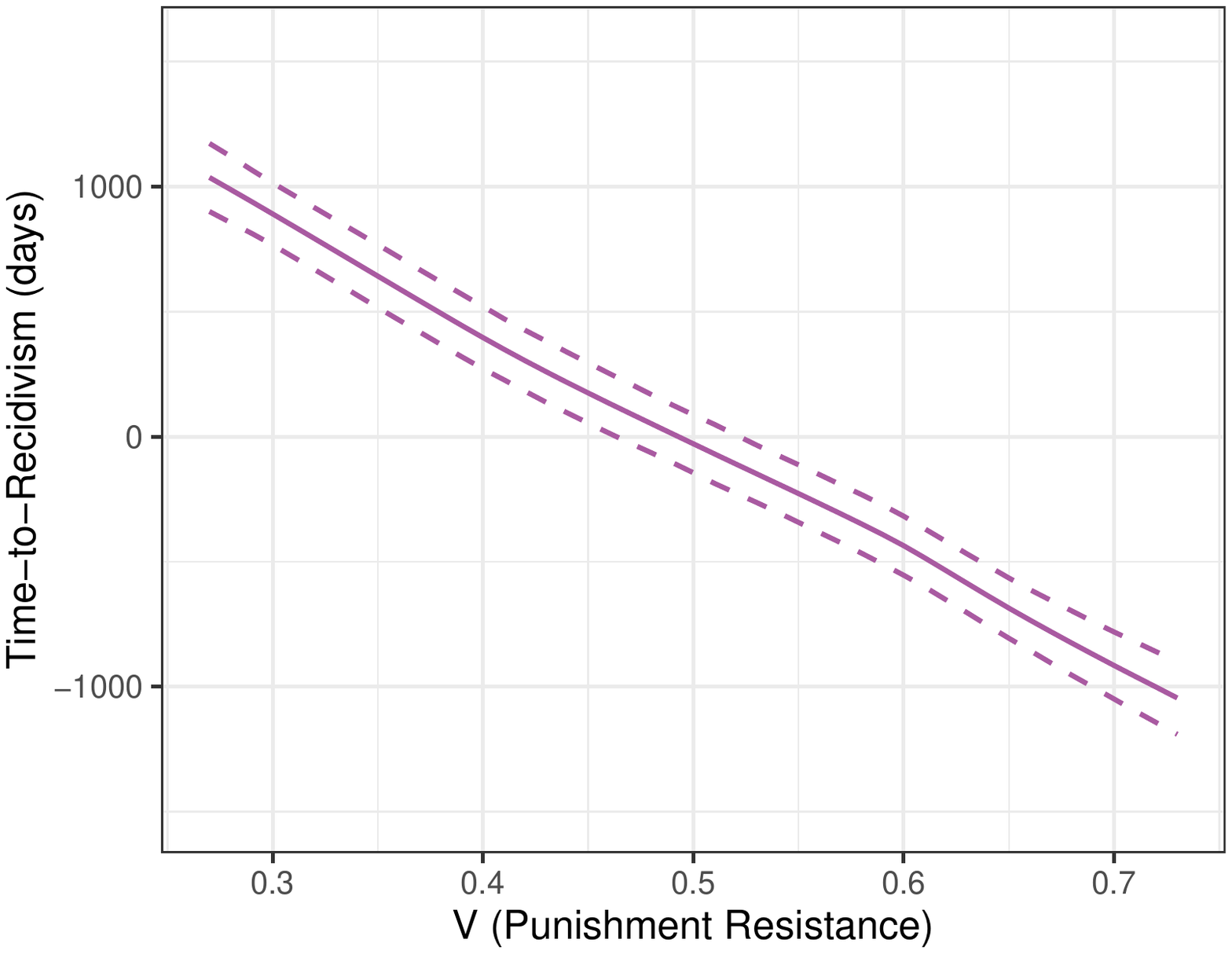}
						\caption{$RMTE\left(\cdot\right)$}
						\label{RMTE-CI}
					\end{center}
				\end{subfigure}
				\caption{90\%-Confidence Intervals for $QMTE\left(\tau,\cdot\right)$ for $\tau \in \left\lbrace .40, .50, .75 \right\rbrace$ and $RMTE\left(\cdot\right)$}
				\label{FigQMTE-RMTE-CI}
			\end{center}
			\justifying
			\vspace{-.5cm}\scriptsize{Notes: Solid lines are the point estimates for the average $QMTE\left(\tau,\cdot\right)$ and $RMTE\left(\cdot\right)$ functions indicated in the caption of each subfigure. These results are based on Corollary \ref{cor:semi}. Moreover, point-wise 90\%-confidence intervals are reported using dashed lines. These confidence intervals were computed using the weighted bootstrap clusterized at the court district level (Algorithm \ref{algo:bootstrap}) using 399 repetitions.
			}
		\end{figure}
		
		\newpage
		\subsection{Robustness Check: Cases after 2014}\label{AppRobustness}
		
		{
			
			In this exercise, we verify the robustness of our empirical results to violations of Assumption \ref{AsCensoring} (Random Censoring). Recall that this assumption, when combined with Assumption \ref{AsContinuous}, implies that $C$ is unconditionally independent of the uncensored potential outcomes, i.e., $C \independent \left(Y^{*}\left(0\right), Y^{*}\left(1\right)\right)$. In our empirical application, this restriction imposes that the case's sentence date is independent of the defendant's decision to commit another crime in the future, i.e., potential recidivism is stationary. Consequently, our main concern in this appendix is that the distribution of potential recidivism statuses varies over time.
			
			If this is the case, then the $DMTE$ and $QMTE$ functions may vary when we change the sampling period. There are at least two approaches to verify this possibility. The first way is fully statistical: we may split the sample in the middle of our sampling period (i.e., in the year 2015) or in a year that roughly has half of the observations before it (i.e., in the year 2013). The second is to look for changes in the ``production function of Justice'' and split the sample according to any relevant changes in its ``functional form'' or its inputs. In this appendix, we follow the second approach.
			
			The ``production function of Justice'' includes legislation, jurisprudence, and the work of police officers, defense attorneys, public defenders, district attorneys, court clerks and judges. Below, we describe our investigation of changes in any of those inputs.
			
			To the best of our knowledge, the main legal change in our sampling period (2010--2019) related to the types of crime included in our sample occurred in 2018. This year, the Brazilian Superior Justice Court (the main criminal justice court in Brazil) decided that sentences involving domestic violence crimes must include incarceration.\footnote{This decision is the ``Súmula 588'' of the ``Superior Tribunal de Justiça''.} Nevertheless, this decision does not violate our assumptions because our sample includes cases starting between 2010--2017 and we use the years of 2018 and 2019 only to measure recidivism.\footnote{If a case started in 2017, it is not impacted by the Superior Justice Court's decision because, according to the Brazilian Constitution, criminal law changes cannot negatively impact the defendants of cases that started before the change.}
			
			Next, we investigate changes in judges' work during our sample period. To the best of our knowledge, the main change in the judicial career in the State of São Paulo occurred in 2024 when Law n. 1.414 created many new positions at the top of the judicial career. In a cascading effect, this law led to the hiring of 126 junior judges in 2024 and 117 in 2025. Nevertheless, this change does not violate our assumptions because it happens outside of our sampling period.
			
			Additionally, we examine police officer hiring during our sampling period. If the number of police officers increases significantly, the distribution of potential recidivism may change because more crimes will be detected by the police. After analyzing many news articles on the number of police officers over time in the State of São Paulo, we did not find any relevant changes. Similarly, we did not find any changes in the hiring of district attorneys over time.
			
			Lastly, we examine public defender hiring during our sample period. The number of public defenders may affect the distribution of potential recidivism because greater access to high-quality publicly provided defense may improve defendants' perceptions of fairness and incentivize them to behave properly after the trial. If this is the case, the potential recidivism distribution will differ between scenarios with a small or a large number of public defenders. This setting is particularly relevant in our empirical context because the number of public defenders varies substantially in the state of São Paulo during our sample period: 500 in 2011, 501 in 2012, 610 in 2013, 719 in 2014, 718 in 2015, 717 in 2016, 721 in 2017, 749 in 2018 and 748 in 2019.

			Having in mind the discrete jump in the number of public defenders in 2014, we re-estimate our target parameter using a sample that starts in 2014. Naturally, we cannot re-estimate all the $DMTE$ and $QMTE$ functions that we estimated in Sections \ref{SresultsDMTE} and \ref{SresultsQMTE} because we have a shorter sample period. For this reason, we re-estimate the $DMTE$ function for the first four years and the $QMTE$ function for the first decile, the first quartile and the median.
			
			Figure \ref{FigResultsLater} presents these results. They are very similar to the results in Sections \ref{SresultsDMTE} and \ref{SresultsQMTE}, indicating the robustness of our results against violations of Assumption \ref{AsCensoring} (Random Censoring).
			
			\begin{figure}[htbp]
				\begin{center}
					\begin{subfigure}[t]{0.47\textwidth}
						\centering
						\includegraphics[width = \textwidth]{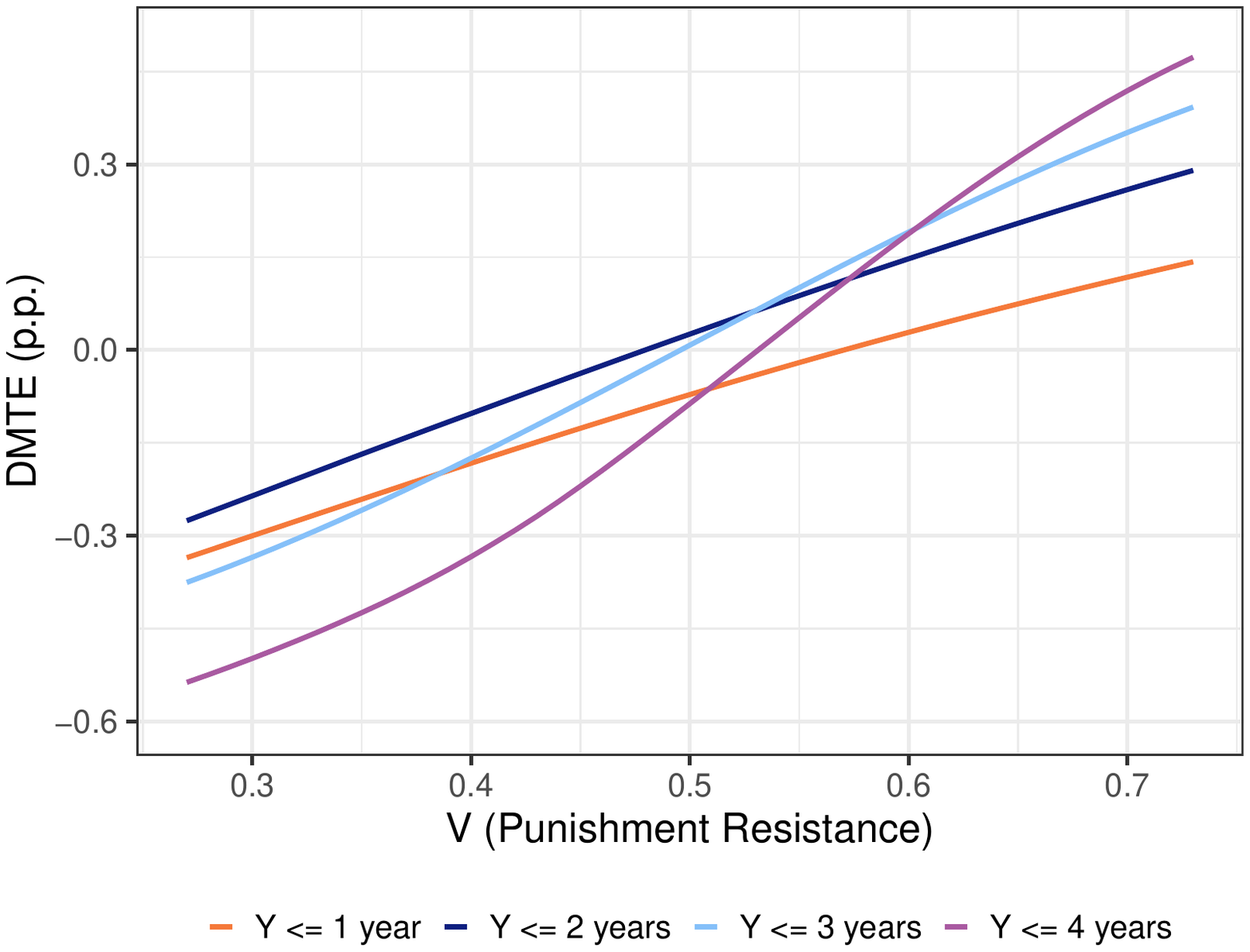}
						\caption{$DMTE\left(y,\cdot\right)$ for $y \in \left\lbrace 1, 2, 3, 4 \right\rbrace$}
						\label{FigDMTElater}
					\end{subfigure}
					\hfill
					\begin{subfigure}[t]{0.47\textwidth}
						\begin{center}
							\includegraphics[width = \textwidth]{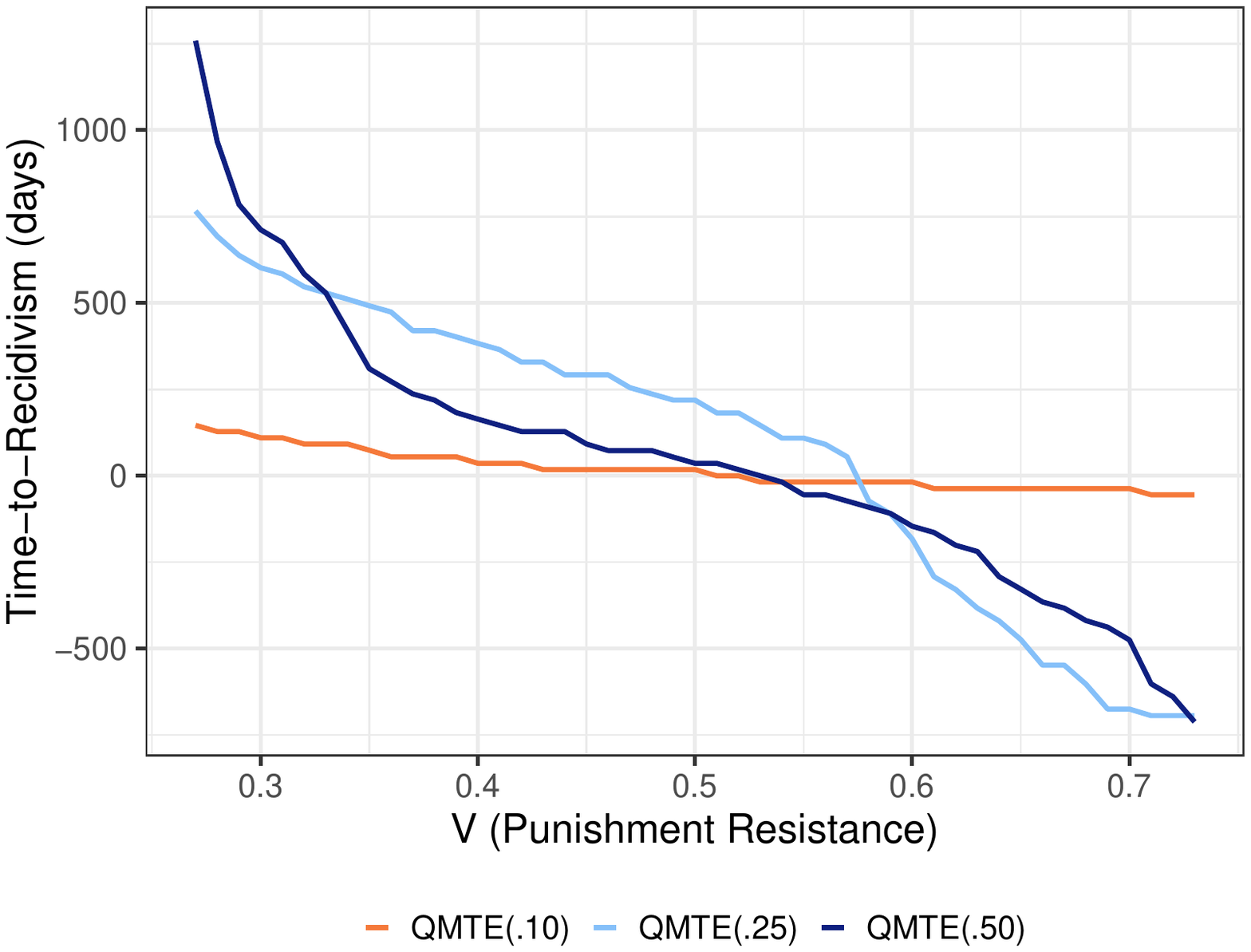}
							\caption{$QMTE\left(\tau,\cdot\right)$ for $\tau \in \left\lbrace .10, .25, .50 \right\rbrace$}
							\label{FigQMTElater}
						\end{center}
					\end{subfigure}
					\caption[]{$DMTE$ and $QMTE$ results using cases that started after 2014}
					\label{FigResultsLater}
				\end{center}
				\justifying
				\vspace{-.5cm}\scriptsize{Notes: In Figure \ref{FigDMTElater}, solid lines are the point estimates for the average $DMTE\left(y,\cdot\right)$ functions indicated in the legend. In Figure \ref{FigQMTElater}, solid lines are the point estimates for the average $QMTE\left(\tau,\cdot\right)$ functions indicated in the legend. All results are based on Corollary \ref{cor:semi}.}
			\end{figure}

			\subsection{Descriptive Statistics: Court Districts and Judges}\label{AppDescriptiveCourtJudge}
			
			This appendix shows descriptive statistics about the distribution of judges per court district and cases per judge. Figure \ref{FigSample} shows the distribution of the number of judges for every court district in São Paulo (Figure \ref{FigNCourt}) and the distribution of the number of cases for every judge in São Paulo (Figure \ref{FigJudges}).\footnote{Appendix \ref{AppDescriptiveYear} shows the distribution of the number of judges per court district in each year of our sample period.} The average number of cases per judge is 120 in our full sample and 159 after we focus on judges who analyzed more than 20 cases. Furthermore, the average number of judges per court district is 2.2 in the full sample and 3.0 in our restricted sample.
			
			\begin{figure}[!htb]
				\begin{center}
					\begin{subfigure}[t]{0.47\textwidth}
						\centering
						\includegraphics[width = .9\textwidth]{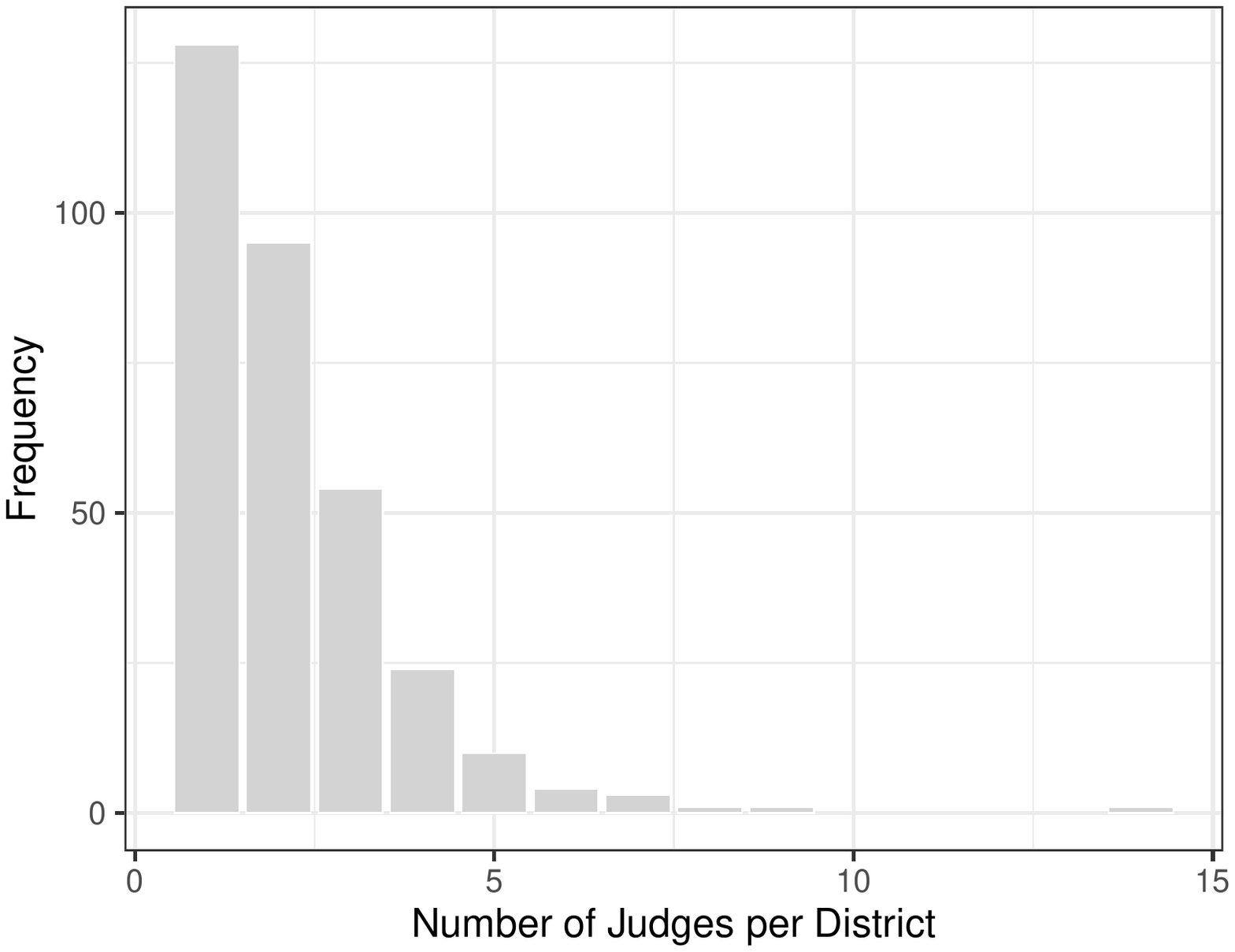}
						\caption{Number of Judges per Court District}
						\label{FigNCourt}
					\end{subfigure}
					\hfill
					\begin{subfigure}[t]{0.47\textwidth}
						\centering
						\includegraphics[width = .9\textwidth]{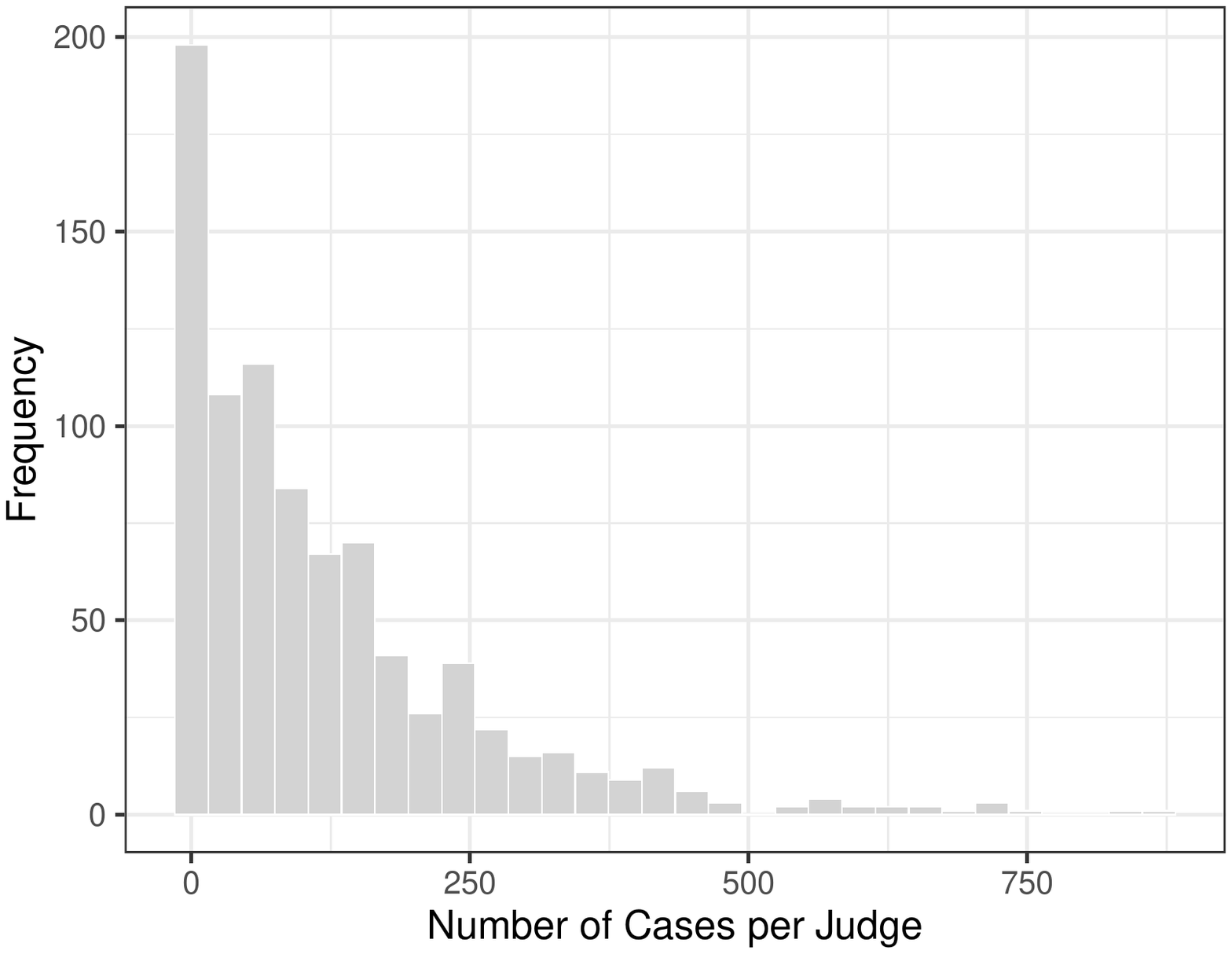}
						\caption{Number of Cases per Trial Judge}
						\label{FigJudges}
					\end{subfigure}
					\caption[]{Descriptive Statistics for the Number of Judges and the Number of Cases}
					\label{FigSample}
				\end{center}
				\justifying
				\scriptsize{Notes: Figure \ref{FigNCourt} plots the histogram of the number of judges per court district in our full sample, while Figure \ref{FigJudges} plots the histogram of the number of cases per trial judge in our full sample.}
			\end{figure}

			\subsubsection{Descriptive Statistics per Year}\label{AppDescriptiveYear}
			
			In this appendix, we report the distribution of the number of judges per court district in each year of our sample period. First, we discuss why it is sufficient to look at this distribution for the entire sample period as we do in Figure \ref{FigNCourt}. Second, we explain why it is interesting to look at these distributions for each year as we do in this appendix. Lastly, we interpret these annual distributions.
			
			First, it is important to understand why analyzing the distribution of judges per court district for the entire sample is sufficient. In our estimation procedure (Section \ref{SestSemiPara}), we condition on a full set of court district dummies, implying that the relevant exogenous variation comes from judges of different punishment rates within court districts. Consequently, the distribution of the number of judges per court district across our entire sample period is key to ensure that we have enough variation to identify our target parameters.
			
			However, our estimation procedure also conditions on the censoring variable. Since this variable is a one-to-one map to the sentencing dates, it is interesting to understand the distribution of judges per court district after conditioning on calendar time. The simplest way to understand this conditional distribution is to subsample each year in our sample and compute a histogram of the number of judges per court district for each year.
			
			Figures \ref{FigJudgesPerYear1} and \ref{FigJudgesPerYear2} present these histograms. They show that these distributions are relatively stable over time.
			
			\begin{figure}[!htbp]
				\begin{center}
					\begin{subfigure}[t]{0.47\textwidth}
						\centering
						\includegraphics[width = \textwidth]{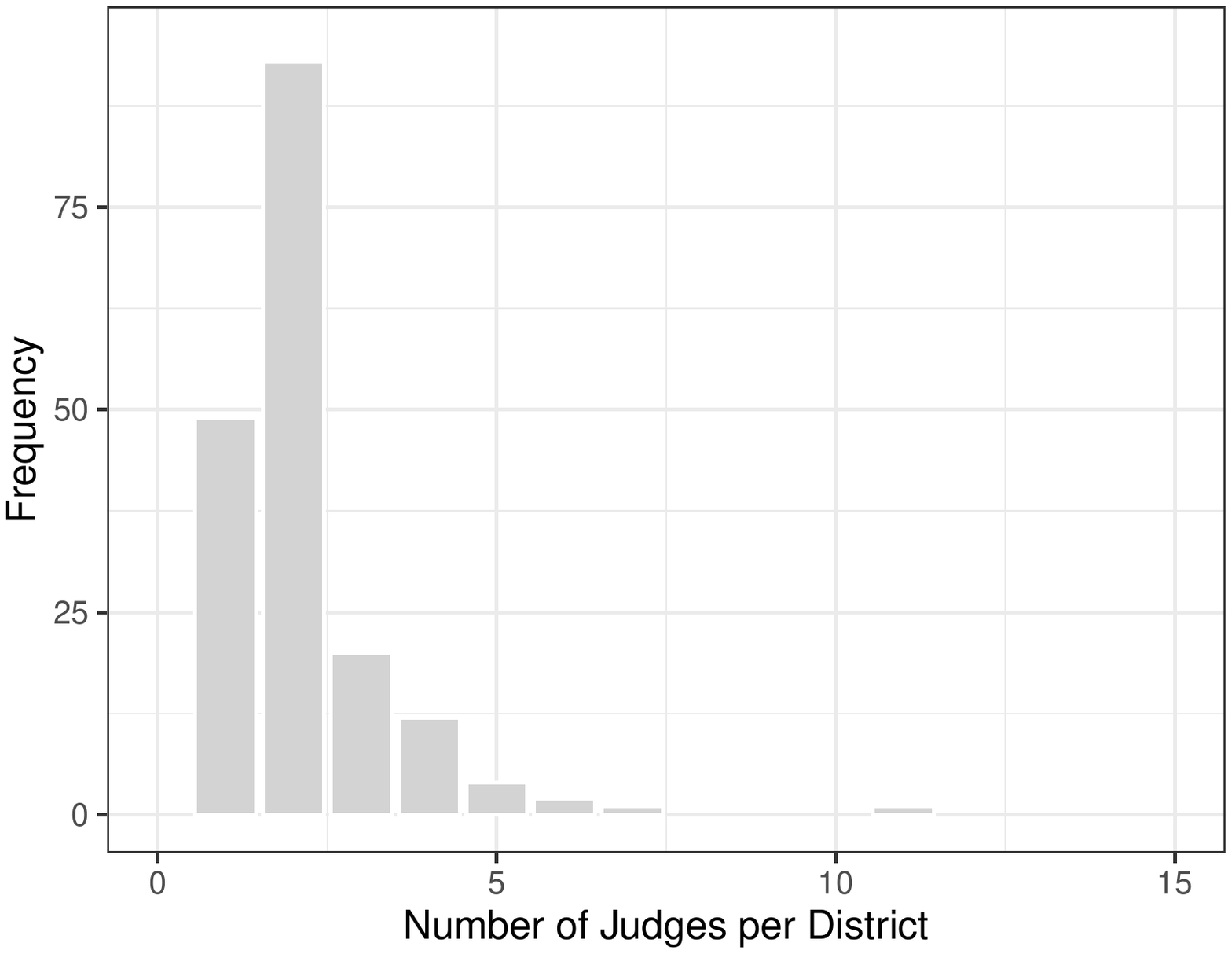}
						\caption{Year = 2010}
						\label{FigJudgesIn2010}
					\end{subfigure}
					\hfill
					\begin{subfigure}[t]{0.47\textwidth}
						\begin{center}
							\includegraphics[width = \textwidth]{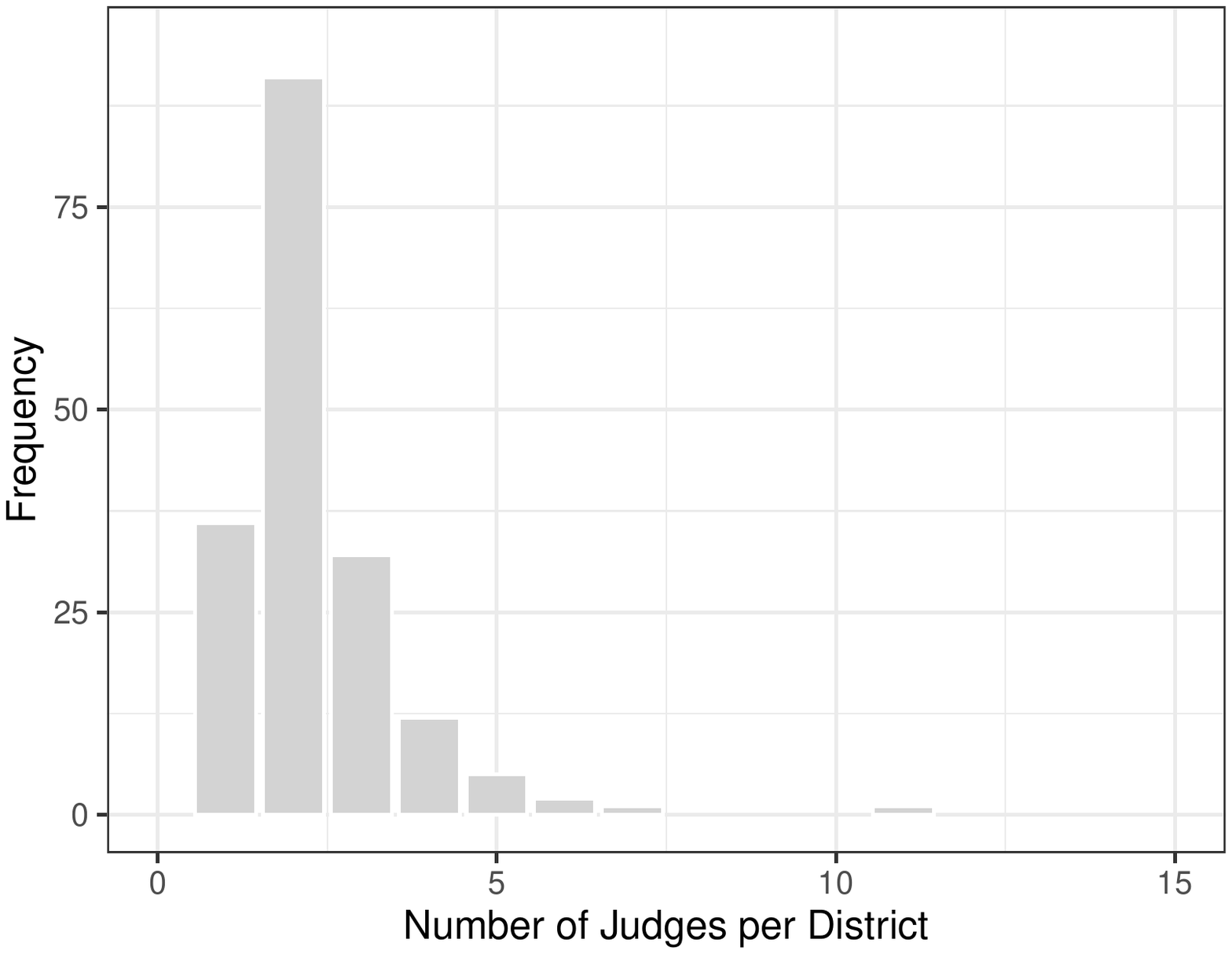}
							\caption{Year = 2011}
							\label{FigJudgesIn2011}
						\end{center}
					\end{subfigure}
					\begin{subfigure}[t]{0.47\textwidth}
						\centering
						\includegraphics[width = \textwidth]{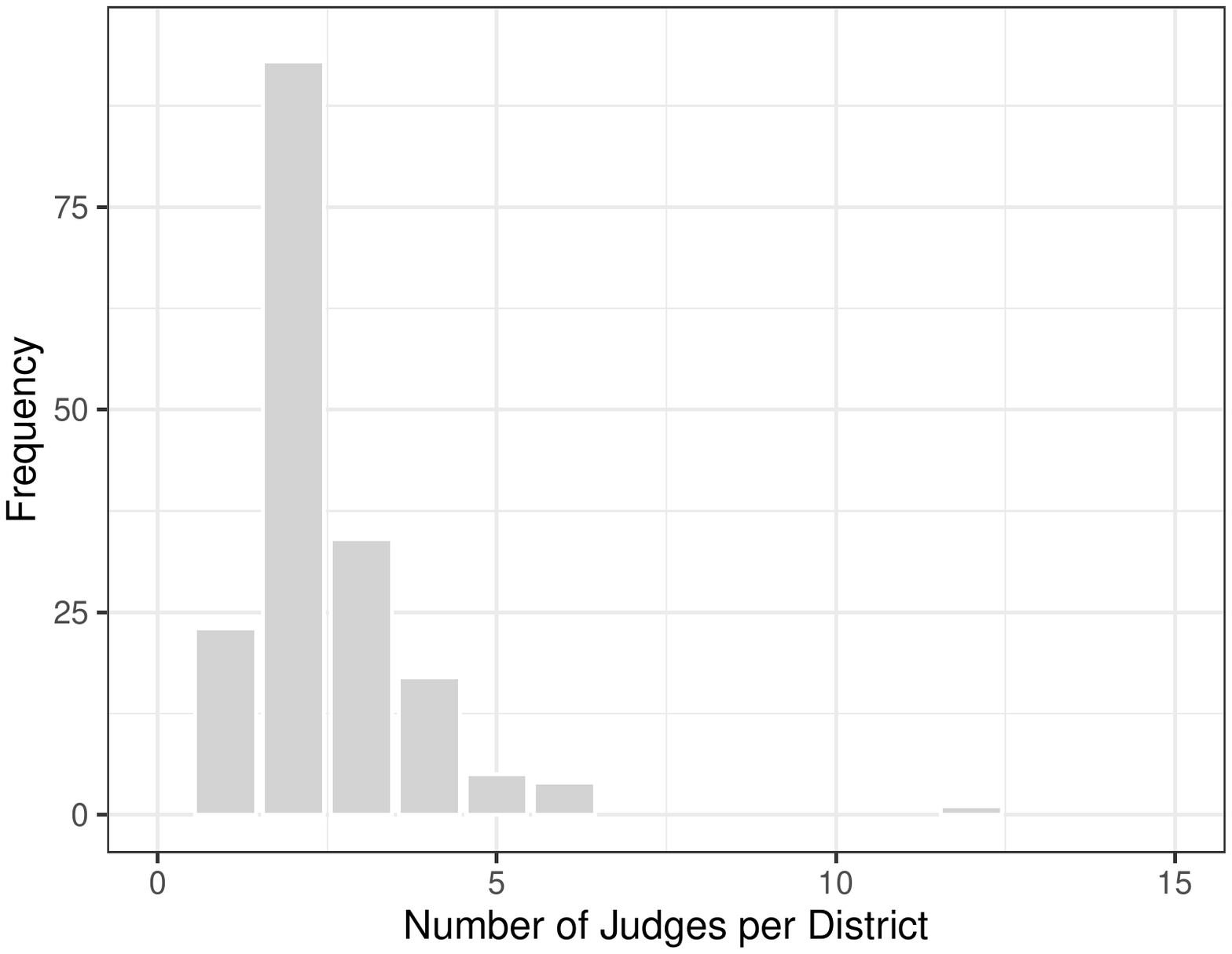}
						\caption{Year = 2012}
						\label{FigJudgesIn2012}
					\end{subfigure}
					\hfill
					\begin{subfigure}[t]{0.47\textwidth}
						\begin{center}
							\includegraphics[width = \textwidth]{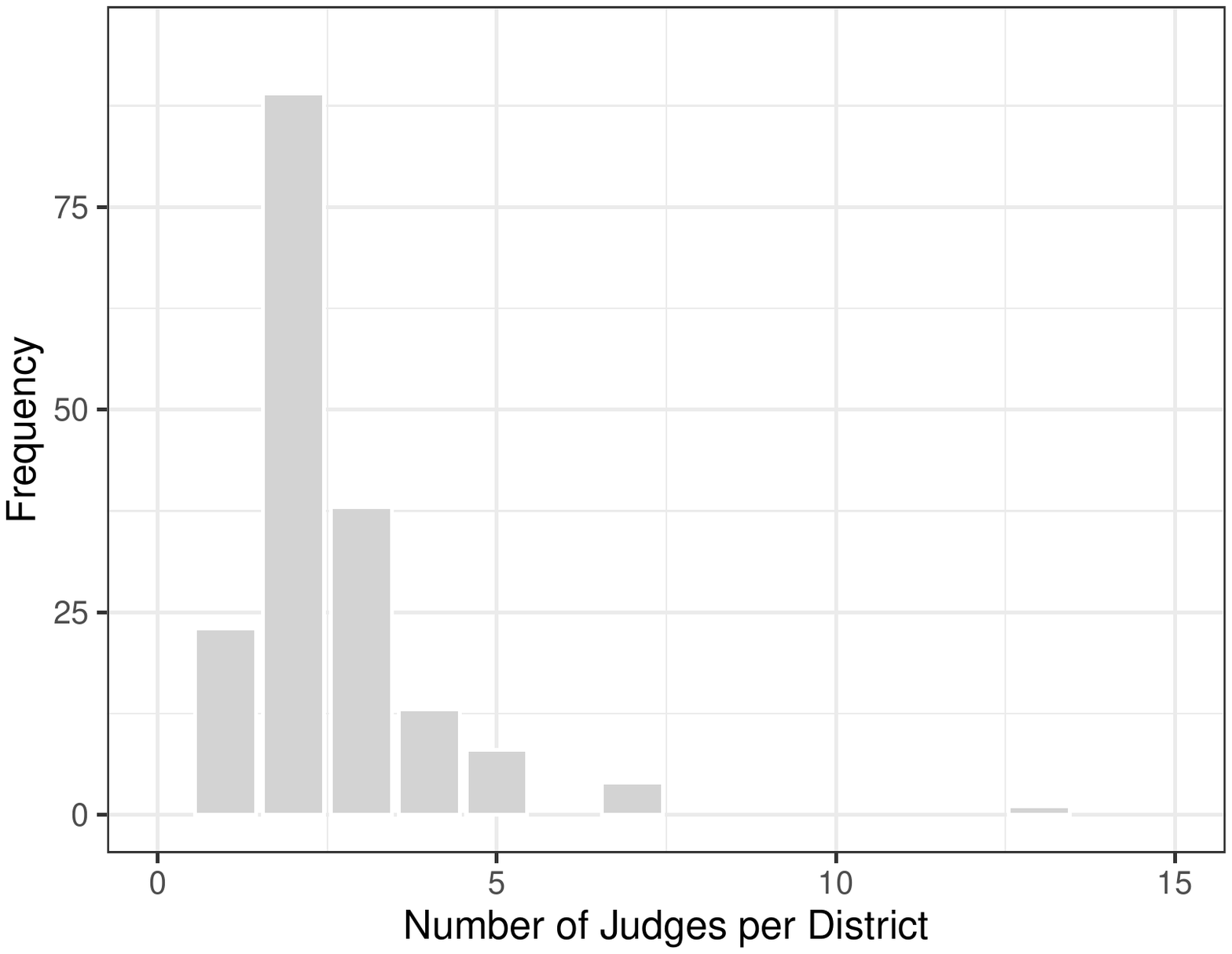}
							\caption{Year = 2013}
							\label{FigJudgesIn2013}
						\end{center}
					\end{subfigure}
					\caption{Histograms of the Number of Judges per Court District in each Year (2010-2013)}
					\label{FigJudgesPerYear1}
				\end{center}
				\justifying
				\vspace{-.5cm}\scriptsize{Notes: Each figure plots the histogram of the number of judges per court district in each year of our sample period. The years are indicated in the title of each figure.
				}
			\end{figure}

			\begin{figure}[!htbp]
				\begin{center}
					\begin{subfigure}[t]{0.47\textwidth}
						\centering
						\includegraphics[width = \textwidth]{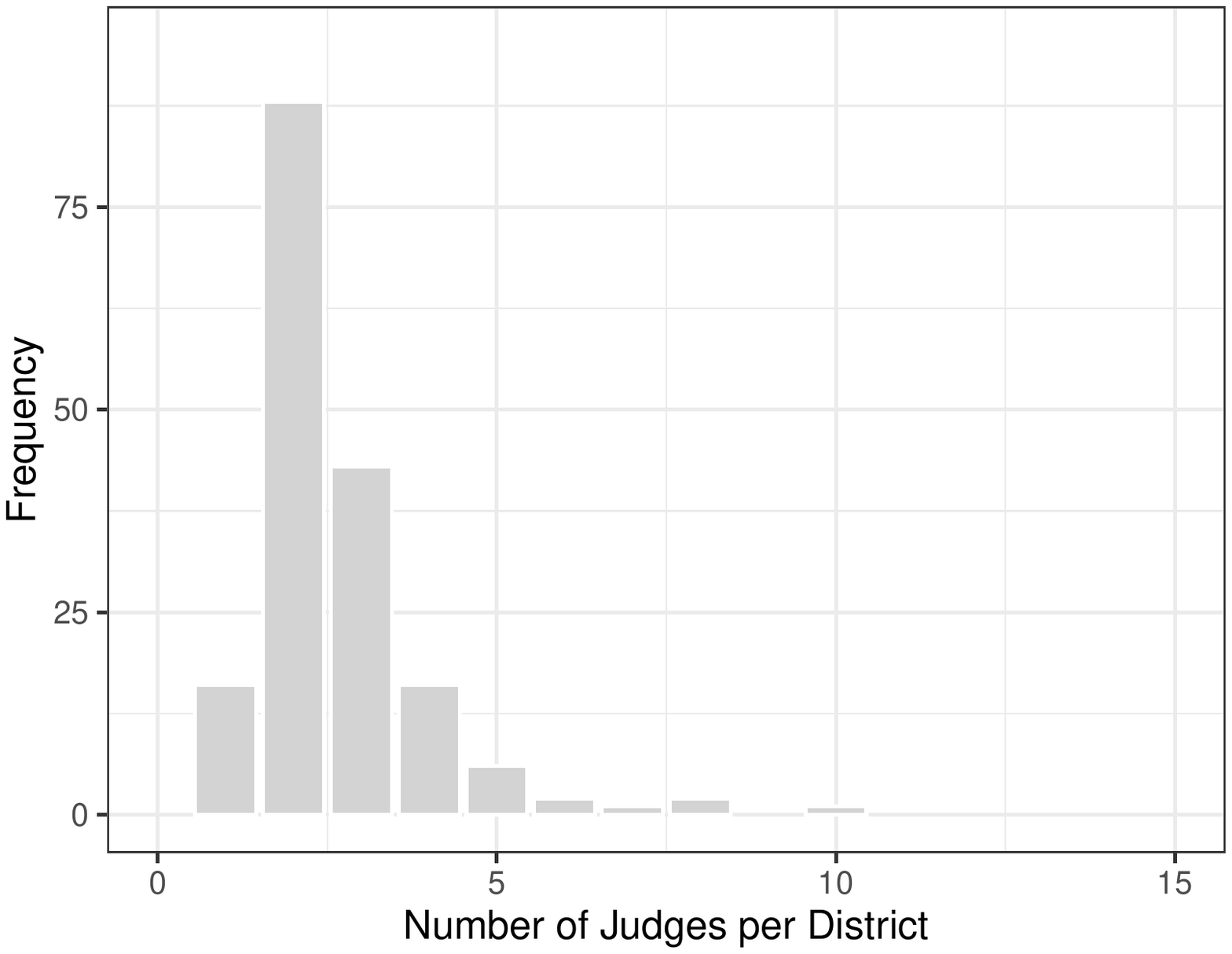}
						\caption{Year = 2014}
						\label{FigJudgesIn2014}
					\end{subfigure}
					\hfill
					\begin{subfigure}[t]{0.47\textwidth}
						\begin{center}
							\includegraphics[width = \textwidth]{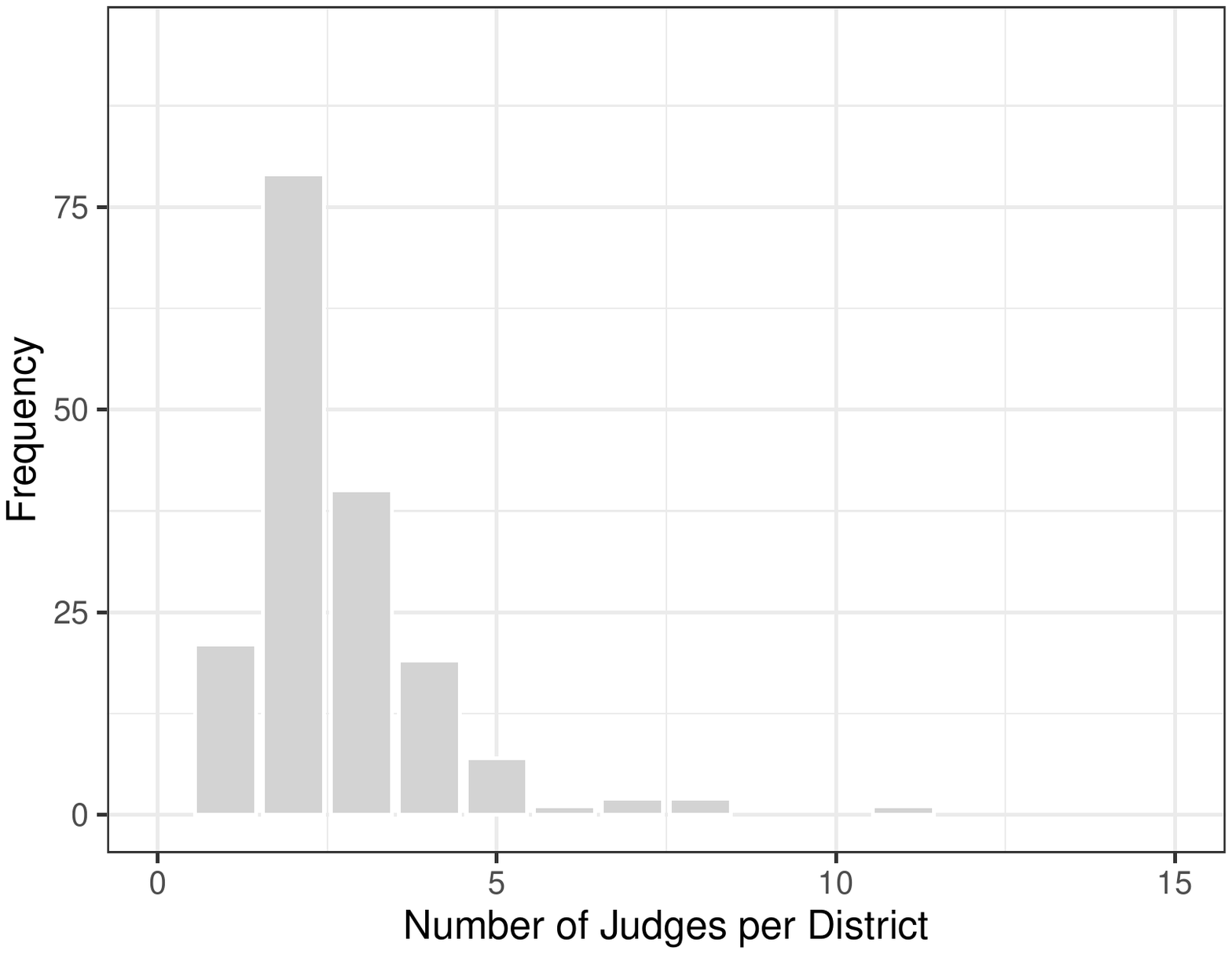}
							\caption{Year = 2015}
							\label{FigJudgesIn2015}
						\end{center}
					\end{subfigure}
					\begin{subfigure}[t]{0.47\textwidth}
						\centering
						\includegraphics[width = \textwidth]{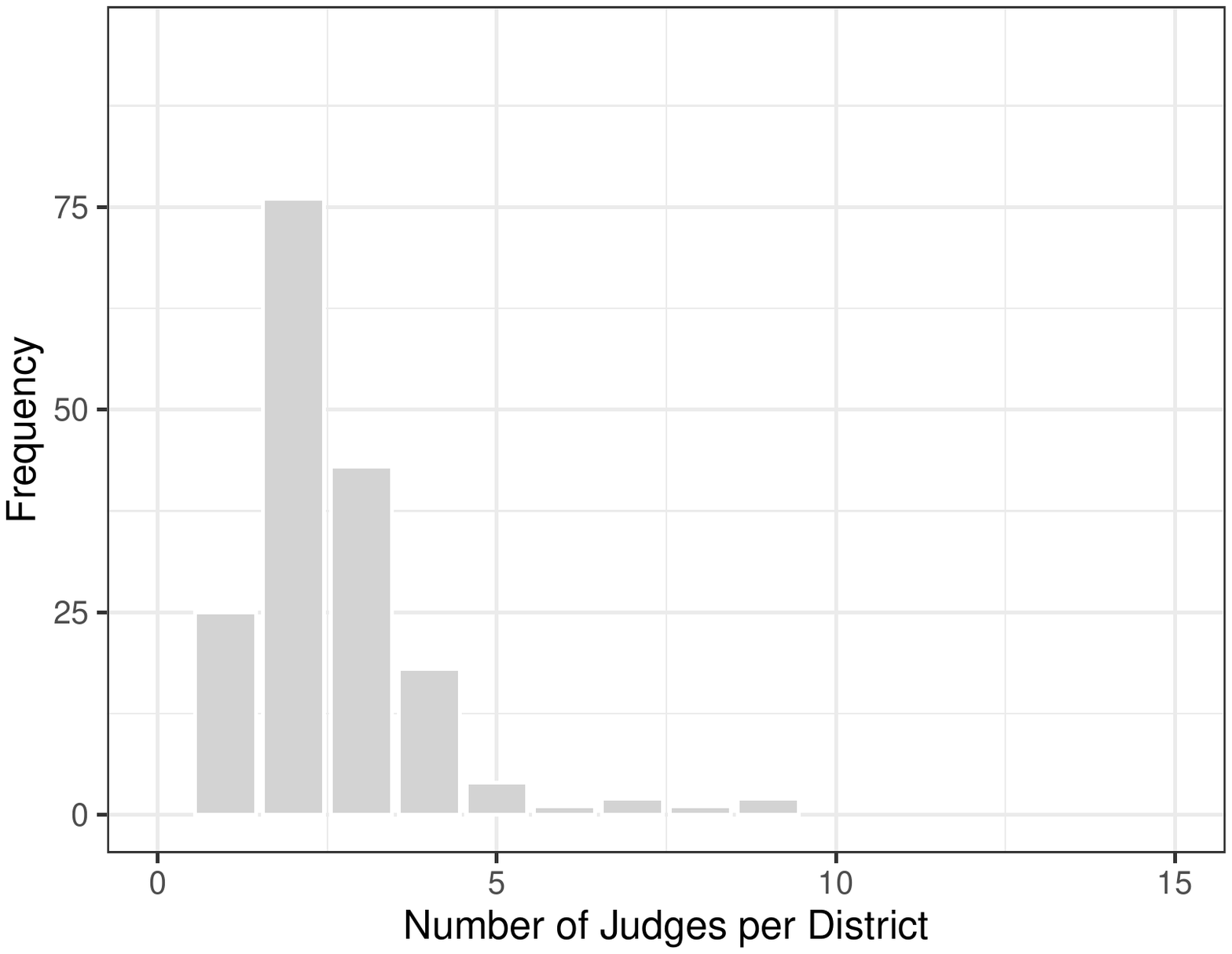}
						\caption{Year = 2016}
						\label{FigJudgesIn2016}
					\end{subfigure}
					\hfill
					\begin{subfigure}[t]{0.47\textwidth}
						\begin{center}
							\includegraphics[width = \textwidth]{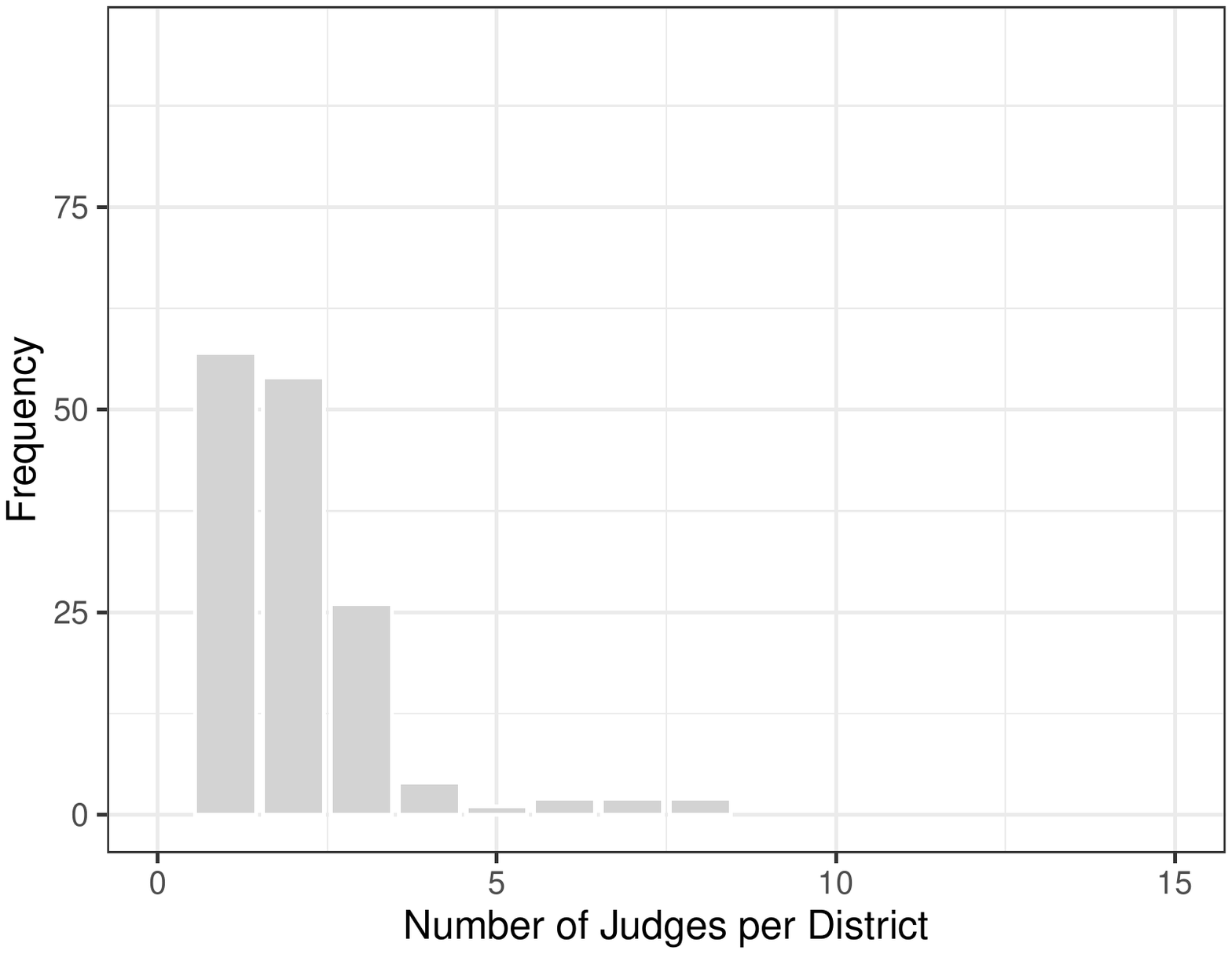}
							\caption{Year = 2017}
							\label{FigJudgesIn2017}
						\end{center}
					\end{subfigure}
					\caption{Histograms of the Number of Judges per Court District in each Year (2014-2017)}
					\label{FigJudgesPerYear2}
				\end{center}
				\justifying
				\vspace{-.5cm}\scriptsize{Notes: Each figure plots the histogram of the number of judges per court district in each year of our sample period. The years are indicated in the title of each figure.
				}
			\end{figure}
			
		}
		
		{
			
			\subsection{Additional Descriptive Statistics }\label{AppAdditionalDescriptive}
			
			\subsubsection{Distribution of the Outcome Variable}\label{AppDescriptiveOutcome}
			
			Table \ref{TabDescriptive} presents some summary statistics of our final sample. It shows the outcome's mean, 1\textsuperscript{st} decile, 1\textsuperscript{st} quartile and median for all defendants, for the defendants who were punished (treated group), and for the defendants who were not punished (comparison group). It also shows the sample size of each one of these three groups. The comparison between the treated and comparison groups suggests that being punished slightly harms defendants. However, this naive comparison ignores endogenous selection-into-treatment, right-censoring, and heterogeneous treatment effects. As such, one should be careful with such a comparison, as these descriptive statistics may not have a causal interpretation.
			
			\begin{table}[!htb]
				\centering
				\caption{{Descriptive Statistics --- Outcome Variable}} \label{TabDescriptive}
				\begin{lrbox}{\tablebox}
					\begin{tabular}{cccc}
						\hline \hline
						& Unconditional & Treated Group & Comparison Group \\ \hline
						Mean & 1,081 & 1,047 & 1,116 \\
						1\textsuperscript{st} Decile & 77 & 69 & 86 \\
						1\textsuperscript{st} Quartile & 364 & 321 & 430 \\
						Median & 1082 & 1047 & 1127 \\
						Number of Observations & 43,468 & 22,060 & 21,408 \\ \hline
					\end{tabular}
				\end{lrbox}
				\usebox{\tablebox}\\
				
				\justifying
				\scriptsize{\hspace{-.6cm}Note: The treated group receives a punishment, i.e., its defendants were fined or sentenced to community services because they were either convicted or signed a non-prosecution agreement. The comparison group did not receive a punishment, i.e., its defendants were acquitted or its cases were dismissed. The outcome variable measures the number of days between the case's final ruling's date and the first recidivism event if the defendant recidivates or the number of days between the case's final ruling's date and the end of the sampling period if the defendant did not recidivate. An observation is a case-defendant pair.}
			\end{table}
			
			\subsubsection{Analyzing Repeated Offenders}\label{AppRepeatedOffenders}
			
			When analyzing recidivism, there are, at least, two ways to construct the dataset with respect to repeated offenders. First, one may treat each case-defendant pair as a separate observation. As a consequence, repeated offenders appear in the dataset more than once and receive greater weight in any estimation procedure. Second, one may remove repeated offenders and include only the first case of each defendant. As a consequence, all defendants receive the same weight and the overall probability of recidivism decreases.
			
			In our analysis, we follow the first approach and treat each case-defendant pair as a separate observation, even if the defendant is a repeated offender. We do so for two reasons. First, pragmatically, it is impossible to know whether a defendant is a first-time or repeated offender because we have no information about whether the defendant committed a crime before 2010. Second, the Criminal Justice System faces repeated offenders frequently, and these defendants are part of the population of interest for any policymaker aiming to reduce crime. For this reason, policy-relevant parameters could place greater weight on repeated offenders, as they represent a larger share of Justice-involved individuals.
			
			To understand how much weight is placed on repeated offenders in our sample, we may compare the number of repeated offenders against the total number of defendants. In Table \ref{TabRepeatedOffenders}, the first row reports the number of defendants who show up once, twice, three times, ..., fourteen times, while the second row shows the same information as a percentage of the total number of defendants. We find that 89.1\% of defendants show up only once in our dataset.\footnote{Part of these defendants do not recidivate, while part of these defendants recidivate during the last two years of our sample period. Consequently, this percentage is not the share of defendants who do not recidivate during our sample period.} Moreover, 9.0\% and 1.2\% of defendants show up twice or three times in our dataset. The remaining 0.7\% of defendants appear at least 4 times in our dataset.
			
			\begin{table}[!htb]
				\centering
				\caption{{Times Each Defendant Appears in our Sample}} \label{TabRepeatedOffenders}
				\begin{lrbox}{\tablebox}
					\begin{tabular}{lcccccccccccccc}
						\hline \hline
						& 1 & 2 & 3 & 4 & 5 & 6 & 7 & 8 & 9 & 10 & 11 & 12 & 13 & 14 \\ \hline
						Obs. & 33,960 & 3,412 & 473 & 155 & 53 & 16 & 12 & 7 & 6 & 3 & 3 & 0 & 1 & 1 \\
						Share (\%) & 89.1 & 9.0 & 1.2 & 0.4 & 0.1 & 0.0 & 0.0 & 0.0 & 0.0 & 0.0 & 0.0 & 0.0 & 0.0 & 0.0 \\ \hline
					\end{tabular}
				\end{lrbox}
				\usebox{\tablebox}\\
				\justifying
				\hspace{-.6cm}\scriptsize{Note: The first row reports the number of defendants who show up once, twice, three times, ..., fourteen times in our dataset. The second row shows the same information as a percentage of the total number of defendants.}
			\end{table}
			
			Another way to think about this weighting choice is to focus on the number of case-defendant pairs rather than the number of defendants. This approach reveals the actual weight given to each defendant in our dataset. In Table \ref{TabRepeatedCaseDefendant}, the first row reports the number of case-defendant pairs whose defendant shows up once, twice, three times, ..., fourteen times in our dataset, while the second row shows the same information as a percentage of the total number of case-defendant pairs. We find that 78.1\% of case-defendant pairs have defendants who show up only once in our dataset. Moreover, 15.7\%, 3.3\% and 1.4\% of case-defendant pairs have defendants who show up twice, three or four times in our dataset. The remaining 1.5\% of case-defendant pairs have defendants who appear at least 5 times in our dataset.
			
			\begin{table}[!htb]
				\centering
				\caption{{Case-Defendant Pairs with Repeated Offenders}} \label{TabRepeatedCaseDefendant}
				\begin{lrbox}{\tablebox}
					\begin{tabular}{lcccccccccccccc}
						\hline \hline
						& 1 & 2 & 3 & 4 & 5 & 6 & 7 & 8 & 9 & 10 & 11 & 12 & 13 & 14 \\ \hline
						Obs. & 33,960 & 6,824 & 1,419 & 620 & 265 & 96 & 84 & 56 & 54 & 30 & 33 & 0 & 13 & 14 \\
						Share (\%) & 78.1 & 15.7 & 3.3 & 1.4 & 0.6 & 0.2 & 0.2 & 0.1 & 0.1 & 0.1 & 0.1 & 0.0 & 0.0 & 0.0 \\ \hline
					\end{tabular}
				\end{lrbox}
				\usebox{\tablebox}\\
				\justifying
				\hspace{-.6cm}\scriptsize{Note: The first row reports the number of case-defendant pairs whose defendant shows up once, twice, three times, ..., fourteen times in our dataset. The second row shows the same information as a percentage of the total number of case-defendant pairs.}
			\end{table}
			
			\subsubsection{Defendants: Words per Name}\label{AppNames}
			
			This appendix presents information about the number of words per name for all defendants in our sample. Since we measure recidivism through a fuzzy matching of defendants' names, this type of discussion is important for understanding the likelihood of false positives in our outcome variable.
			
			Table \ref{TabNames} shows the distribution of the number of words per name for all the defendants in our sample. To construct this table, we first record every name that appears in our sample and find that we have 38,102 names in our dataset. Then, the first row counts the number of names with one, two, ..., six, and seven or more words, while the second row shows the same information as a percentage of the total number of names that appear in our sample.
			
			\begin{table}[!htb]
				\centering
				\caption{{Number of Words per Name for the Defendants}} \label{TabNames}
				\begin{lrbox}{\tablebox}
					\begin{tabular}{lccccccc}
						\hline \hline
						& 1 & 2 & 3 & 4 & 5 & 6 & 7 or More \\ \hline
						Obs. & 23 & 2,526 & 12,941 & 16,458 & 5,425 & 567 & 162 \\
						Share (\%) & 0.1 & 6.6 & 34.0 & 43.2 & 14.2 & 1.5 & 0.43 \\ \hline
					\end{tabular}
				\end{lrbox}
				\usebox{\tablebox}\\
				\justifying
				\hspace{-.6cm}\scriptsize{Note: The first row reports the number of names with one, two, ..., six, and seven or more words in our dataset. The second row shows the same information as a percentage of the total number of names that appear in our sample.}
			\end{table}
			
			We have 23 names with only one word. When we look at these names, we find that these are incomplete names because we either observe only the first name (e.g., ``edinaldo'' or ``alessandra'') or one surname (e.g., ``moreira''). The small number of incomplete cases shows that it is extremely uncommon for São Paulo's Court System to make severe mistakes when recording defendants' names.
			
			Two-word names represent 6.6\% of our sample of names. This type of name (e.g., ``sebastiao pimenta'' and ``zacarias ferreira'') faces the highest risk of creating false positive matches when measuring recidivism through fuzzy matching of names. For this reason, their small share suggests that false-positive recidivism events are unlikely to be a relevant problem in our data.
			
			Three-word and four-word names represent 34.0\% and 43.2\% of our sample of names. These names include examples like ``fernando augusto tostes'' and ``norton henrique de souza'' since prepositions like ``de'', ``da'' and ``do'' count as one word. Having such a large share of three-word and four-word names also suggests that false-positive recidivism events are unlikely to be a relevant problem in our data.
			
			Lastly, names with five, six, seven or more words represent more than 15\% of our sample. These names include examples like ``johnny willian teixeira de castro e silva'' and ``maria de fatima da silva pinheiro de oliveira''. The fact that these types of names account for a larger share of our sample than two-word names also suggests that false-positive recidivism events are unlikely to be a relevant problem in our data.
			
		}
		
		\newpage
		\section{Relevance of MTE for Duration Outcomes} \label{AppTM}
		
		\setcounter{table}{0}
		\renewcommand\thetable{E.\arabic{table}}
		
		\setcounter{figure}{0}
		\renewcommand\thefigure{E.\arabic{figure}}
		
		\setcounter{equation}{0}
		\renewcommand\theequation{E.\arabic{equation}}
		
		\setcounter{theorem}{0}
		\renewcommand\thetheorem{E.\arabic{theorem}}
		
		\setcounter{assumption}{0}
		\renewcommand\theassumption{E.\arabic{assumption}}

		In this appendix, we justify focusing on the marginal treatment effect (MTE) for duration outcomes using two arguments.
		
		In Appendix \ref{AppModelJustification}, we develop a theoretical model in which a policymaker selects a treatment assignment rule that minimizes the cost of recidivism for the target population of defendants.\footnote{The decision model implemented in this appendix focuses on time-to-recidivism because this outcome variable easily illustrates the censoring issues we want to address with our proposed method. However, this decision model could be extended to include total crime counts or severity-weighted offenses as outcomes of interest for a policymaker. As a consequence, this appendix's analysis is only one small step in better understanding the relationship between recidivism and social welfare.}
		
		In Appendix \ref{AppIllustration}, we provide a simple example where the treatment benefits most agents in our population. In this example, our proposed focus on quantile treatment effects for duration outcomes correctly highlights that this treatment benefits society. However, focusing on short-time horizons, as usually done in the crime economics literature, leads to the opposite conclusion.

		\subsection{Theoretical justification of relevance of MTE for duration outcomes}\label{AppModelJustification}
		
		Following \citet{Kitagawa2018}, the policymaker has to choose a treatment rule that determines whether individuals with variables $W=\{Z,V, C\}$ in our target population will be assigned to the treatment group or the comparison group. The policymaker chooses non-randomized treatment rules described by decision sets $G \subset \mathcal{W}$, where $\mathcal{W}$ is the support of $W$. These decision sets determine the group of individuals $\{W \in G\}$ to whom treatment is assigned.  We denote the collection of candidate treatment rules by $\mathcal{G} = \{G \subset \mathcal{W} \}$.
		
		The policymaker's goal in our context is to select a treatment assignment rule that minimizes the cost of recidivism for the target population of defendants. Assuming that the policymaker discounts cost inter-temporally, she chooses the treatment rule that maximizes $Y^{*}$ for each individual in the target population.
		
		Specifically, we impose that the policymaker chooses the decision set $G \in \mathcal{G}$ that minimizes
		\begin{equation*}
			K\left(G\right) \coloneqq \mathbb{E}\left[\ln \left\lbrace b^{\left[Y^{*}\left(1\right) \cdot \mathbf{1}\left\lbrace W \in G \right\rbrace + Y^{*}\left(0\right) \cdot \mathbf{1}\left\lbrace W \notin G \right\rbrace \right]} \cdot k \right\rbrace \right]
		\end{equation*}
		where $k \in \mathbb{R}_{+}$ is the fixed cost of recidivism and $b \in \left(0, 1 \right)$ is the policymaker's discount rate.\footnote{This simple model imposes that all criminal occurrences have the same cost $k$. A more complete model would consider the full stream of criminal behavior, including total crime counts and the severity of each crime. Since we do not analyze these more complex variables and time-to-recidivism is only a partial measure of social welfare, our model is simply one small step in fully understanding the relationship between recidivism and social welfare.} Rearranging the last equation, we find that
		\begin{align*}
			K\left(G\right) & = \ln \left\lbrace b \right\rbrace \cdot \mathbb{E}\left[ Y^{*}\left(1\right) \cdot \mathbf{1}\left\lbrace W \in G \right\rbrace + Y^{*}\left(0\right) \cdot \mathbf{1}\left\lbrace W \notin G \right\rbrace \right]  + \ln\left\lbrace k \right\rbrace \\
			& = \ln \left\lbrace b \right\rbrace \cdot \mathbb{E}\left[ \left(Y^{*}\left(1\right) - Y^{*}\left(0\right)\right) \cdot \mathbf{1}\left\lbrace W \in G \right\rbrace \right] + \ln \left\lbrace b \right\rbrace \cdot \mathbb{E}\left[Y^{*}\left(0\right)\right]  + \ln\left\lbrace k \right\rbrace
		\end{align*}
		Consequently, the policymaker's problem is equivalent to
		\begin{equation*}
			\max_{G \in \mathcal{G}} \mathbb{E}\left[ \left(Y^{*}\left(1\right) - Y^{*}\left(0\right)\right) \cdot \mathbf{1}\left\lbrace W \in G \right\rbrace \right].
		\end{equation*}
		
		Moreover, note that
		\begin{align*}
			& \mathbb{E}\left[ \left(Y^{*}\left(1\right) - Y^{*}\left(0\right)\right) \cdot \mathbf{1}\left\lbrace W \in G \right\rbrace \right] \\
			& \hspace{20pt} = \mathbb{E}\left[ \mathbb{E} \left[\left. \left(Y^{*}\left(1\right) - Y^{*}\left(0\right)\right) \cdot \mathbf{1}\left\lbrace W \in G \right\rbrace \right\vert V, Z, C \right] \right] \\
			& \hspace{40pt} \text{by the Law of Iterated Expectations} \\
			& \hspace{20pt} = \mathbb{E}\left[ \mathbb{E} \left[\left. \left(Y^{*}\left(1\right) - Y^{*}\left(0\right)\right) \right\vert V, Z, C \right]  \cdot \mathbf{1}\left\lbrace W \in G \right\rbrace \right] \\
			& \hspace{20pt} = \mathbb{E}\left[ \mathbb{E} \left[\left. \left(Y^{*}\left(1\right) - Y^{*}\left(0\right)\right) \right\vert V, C \right]  \cdot \mathbf{1}\left\lbrace W \in G \right\rbrace \right] \\
			& \hspace{40pt} \text{by Assumption \ref{AsIndependence}} \\
			& \hspace{20pt} = \mathbb{E}\left[ \mathbb{E} \left[\left. \left(Y^{*}\left(1\right) - Y^{*}\left(0\right)\right) \right\vert V \right]  \cdot \mathbf{1}\left\lbrace W \in G \right\rbrace \right] \\
			& \hspace{40pt} \text{by Assumption \ref{AsCensoring}} \\
			& \hspace{20pt} = \mathbb{E}\left[ MTE\left(V\right)  \cdot \mathbf{1}\left\lbrace W \in G \right\rbrace \right].
		\end{align*}
		
		Therefore, the policymaker's problem is equivalent to
		\begin{equation*}
			\max_{G \in \mathcal{G}} \mathbb{E}\left[ MTE\left(V\right)  \cdot \mathbf{1}\left\lbrace W \in G \right\rbrace \right],
		\end{equation*}
		implying that focusing on the MTE of duration outcomes is relevant when the policymaker wishes to minimize the cost of recidivism over time. { Importantly, we discuss how to identify the MTE of duration outcomes in Appendix \ref{AppAverage}.}
		
		{
			\subsection{Illustrating the relevance of duration outcomes}\label{AppIllustration}
			
			When analyzing the impact of judicial decisions on recidivism, many authors \citep{Agan2021,Bhuller2019,Giles2021,Huttunen2020,Klaassen2021,Possebom2022} focus on a short time horizon, using a small set of outcome variables that indicate whether the defendant recidivated within a pre-specified number of years. In this paper, we advocate moving beyond this short time horizon and focusing on quantile or average treatment effects of duration outcomes.
			
			In this appendix, we illustrate why focusing on duration outcomes may provide more information than the standard approach in the empirical literature in crime economics. To do so, we abstract from the MTE heterogeneity (variable $V$) and focus exclusively on the heterogeneity arising from the distribution of the potential outcomes $\left(Y^{*}\left(0\right), Y^{*}\left(1\right) \right)$.
			
			We illustrate the relevance of quantile and average treatment effects of duration outcomes by analyzing a simple example with discrete random variables. In this example, focusing on short-term outcomes or long-term quantile treatment effects lead to different conclusions about our policy of interest.
			
			We denote potential time-to-recidivism by $Y^{*}\left(0\right)$ and $Y^{*}\left(1\right)$ and measure it in years. Table \ref{TabIllustation} shows the joint probability mass function of $\left(Y^{*}\left(0\right), Y^{*}\left(1\right) \right)$ and their marginal distributions.
			
			\begin{table}[!htb]
				\centering
				\caption{{Joint Probability of $\left(Y^{*}\left(0\right), Y^{*}\left(1\right) \right)$ and their Marginal Distributions}} \label{TabIllustation}
				\begin{lrbox}{\tablebox}
					\begin{tabular}{cccccccc}
						&                                                                                                         & \multicolumn{5}{c}{$Y^{*}\left(0\right) =$}       &                                                       \\
						& \multicolumn{1}{c|}{$\mathbb{P}\left[Y^{*}\left(0\right) = \cdot, Y^{*}\left(1\right) = \cdot \right]$} & 1   & 2   & 3   & 10  & \multicolumn{1}{c|}{20} & $\mathbb{P}\left[Y^{*}\left(1\right) = \cdot \right]$ \\ \cline{2-8} 
						\multirow{5}{*}{$Y^{*}\left(1\right) = $} & \multicolumn{1}{c|}{1}                                                                                  & .10 & 0   & .10 & 0   & \multicolumn{1}{c|}{0}  & .20                                                   \\
						& \multicolumn{1}{c|}{2}                                                                                  & 0   & .10 & .10 & 0   & \multicolumn{1}{c|}{0}  & .20                                                   \\
						& \multicolumn{1}{c|}{3}                                                                                  & 0   & 0   & 0   & 0   & \multicolumn{1}{c|}{0}  & 0                                                     \\
						& \multicolumn{1}{c|}{10}                                                                                  & 0   & 0   & 0   & .10 & \multicolumn{1}{c|}{0}  & .10                                                   \\
						& \multicolumn{1}{c|}{20}                                                                                  & .05 & .05 & 0   & .40 & \multicolumn{1}{c|}{0}  & .50                                                   \\ \cline{2-8} 
						& \multicolumn{1}{c|}{$\mathbb{P}\left[Y^{*}\left(0\right) = \cdot \right]$}                              & .15 & .15 & .20 & .50 & \multicolumn{1}{c|}{0}  & 1                                                    
					\end{tabular}
				\end{lrbox}
				\usebox{\tablebox}\\
				\settowidth{\tableboxwidth}{\usebox{\tablebox}} \parbox{\tableboxwidth}{\scriptsize{Note: The last column reports the marginal distribution of $Y^{*}\left(1\right)$. The last row reports the marginal distribution of $Y^{*}\left(0\right)$. The cells in the center of the table report the joint distribution of $\left(Y^{*}\left(0\right), Y^{*}\left(1\right) \right)$.}
				}
			\end{table}
			
			Note that, in this example, our judicial decision benefits most defendants. For instance, this treatment strictly increases time-to-recidivism for 50\% of the defendants $\left(Y^{*}\left(0\right) < Y^{*}\left(1\right)\right)$. Moreover, only 20\% of the defendants are harmed by this treatment $\left(Y^{*}\left(0\right) > Y^{*}\left(1\right)\right)$.
			
			However, a short-time horizon analysis would conclude that this treatment is harmful. For example, this treatment increases the probability of recidivism within one year by 5 p.p. and the probability of recidivism within two years by 10 p.p, i.e., $\mathbb{P}\left[Y^{*}\left(1\right) \leq 1 \right] - \mathbb{P}\left[Y^{*}\left(0\right) \leq 1 \right] = 0.05$ and $\mathbb{P}\left[Y^{*}\left(1\right) \leq 2 \right] - \mathbb{P}\left[Y^{*}\left(0\right) \leq 2 \right] = 0.1$.
			
			Differently from the standard empirical analysis, we advocate for focusing on quantile and average treatment effects of duration outcomes. For example, the Quantile Treatment Effect on the Median is equal to seven years because the median of $Y^{*}\left(1\right)$ equal ten years and the median of $Y^{*}\left(0\right)$ equals three years. Moreover, the average treatment effect equals 5.55 years in this example.
			
			Therefore, our proposed analysis would correctly highlight that this treatment benefits at least some agents in our society.
			
		}

		\newpage
		\section{Identification without Restrictions on Censoring } \label{AppAltAs}
		
		\setcounter{table}{0}
		\renewcommand\thetable{F.\arabic{table}}
		
		\setcounter{figure}{0}
		\renewcommand\thefigure{F.\arabic{figure}}
		
		\setcounter{equation}{0}
		\renewcommand\theequation{F.\arabic{equation}}
		
		\setcounter{theorem}{0}
		\renewcommand\thetheorem{F.\arabic{theorem}}
		
		\setcounter{proposition}{0}
		\renewcommand\theproposition{F.\arabic{proposition}}
		
		\setcounter{assumption}{0}
		\renewcommand\theassumption{F.\arabic{assumption}}
		
		{
			
			In this appendix, we focus on which parameters can be point-identified when  we do not impose any restriction on the relationship between the censoring variable and the potential outcomes. To compensate for not imposing Assumption \ref{AsCensoring} nor Assumptions \ref{AsIVCCensoring}-\ref{AsContinuousRelax}, we need to allow the $DMTR$ function to depend on the censoring variable.
			
			Specifically, our causal parameter is given by:
			\begin{equation*}
				DMTR_{d}\left(y,v,c\right) \coloneqq \mathbb{P}\left[\left.Y^*(d) \leq y \right\vert V = v, C = c\right]
			\end{equation*}
			for any $d \in \left\lbrace 0, 1 \right\rbrace$, $y < \gamma_{C}$, $v \in \left[0,1\right]$ and $c \in \mathcal{C}$. Note that our causal parameter is interpretable as a conditional distributional marginal treatment response. In particular, the censoring variable $C$ acts similarly to a covariate in the standard MTE analysis \citep{Carneiro2011}.
			
			In our empirical application, conditioning on the censoring variable is equivalent to conditioning on the defendant cohort or time-fixed effects. Considering that most studies about judicial decisions \citep{Agan2021,Bhuller2019,Huttunen2020,Klaassen2021} condition on district-by-time fixed effects, they identify the conditional $DMTR$ function for a pre-specified value of $y$. In this appendix, we discuss how to extend their analysis to consider conditional quantile marginal treatment effects and marginal treatment effects (Remark \ref{RemarkQTE}). { We also discuss the advantages of formally addressing censoring by imposing Assumption \ref{AsCensoring} (Remark \ref{RemarkFormal}). Lastly, we discuss the implicit assumptions behind this standard approach in the empirical literature (Remark \ref{RemarkSeparability}).}
			
			To point-identify the conditional $DMTR$ function, we eliminate Assumption \ref{AsCensoring} and impose Assumptions \ref{AsIndependence}-\ref{AsPositive} only.
			\begin{proposition}\label{AppLemmaID}
				If Assumptions \ref{AsIndependence}-\ref{AsPositive} hold, then
				\begin{equation*}
					DMTR_{d}\left(y, p, y + \delta \right) = \left(2 d - 1 \right) \cdot \dfrac{\partial \mathbb{P}\left[\left. Y\leq y, D = d \right\vert P\left(Z, C \right) = p, C = y + \delta \right]}{\partial v}
				\end{equation*}
				for any $d \in \left\lbrace 0, 1 \right\rbrace$, $y < \gamma_{C}$, $v \in \mathcal{P}$ and $\delta \in \mathbb{R}_{++}$ such that $y + \delta \in \mathcal{C}$.
			\end{proposition}
			
			\begin{remark}\label{RemarkQTE}
				A direct consequence of Proposition \ref{AppLemmaID} is the identification of the quantile marginal treatment response function $QMTR_{d}\left(\tau, p, y + \delta \right)$ conditional on the censoring variable for any $\tau \in \left(0, \overline{\tau}_{d}\left(p, y + \delta\right)\right)$, where $\overline{\tau}_{d}\left(p, y + \delta\right) \coloneqq DMTR_{d}\left(\gamma_{C}, p, y + \delta \right)$. Additionally, if we impose Assumptions \ref{AsFinite} and \ref{AsSupport}, then we straight-forwardly identify the MTE function conditional on the censoring variable.
			\end{remark}
			
			\begin{remark}\label{RemarkFormal}
				The comparison between Propositions \ref{PropDMTR} and \ref{AppLemmaID} illustrates the identifying power of Assumption \ref{AsCensoring}. It allows us to combine multiple values of the censoring variable to identify a single point of the $DMTR$ function through the integral of $\dfrac{\partial \mathbb{P}\left[\left. Y\leq y, D = d \right\vert P\left(Z, C \right) = p, C = y + \delta \right]}{\partial v}$ over different values of $\delta$. In our empirical application, it means that we can combine multiple defendant cohorts to identify a single evaluation point in the $DMTR$ function, increasing statistical power by being transparent about censoring assumptions.
				
				{ In contrast, the standard empirical practice frequently defines a set of time periods, creates binary outcome variables for each time horizon, and only keeps observations that are observed for a period longer than their longest time horizon of interest. Due to this desire to keep the sample consistent across outcome variables, the existing empirical practice effectively reduces power when analyzing shorter horizons.}
				
			\end{remark}
			
			\begin{remark}\label{RemarkSeparability}
				{  Proposition \ref{AppLemmaID} controls for time effects non-linearly without relying on implicit separability assumptions. In contrast, existing empirical strategies focusing on time to event-date make use of linear models with time-fixed effects. Hence, they rely on implicit separability assumptions. In case of failure of such assumptions, the estimands from these time-fixed effects models are unlikely to be a relevant causal parameter.
					
					Differently from this standard empirical practice, our approach recovers causal effects of interest by bypassing separability assumptions and relying on Random Censoring. We do so by connecting time effects and treatment effects in an explicit and transparent way under the assumption that there are no endogenous time effects once we control for the individual latent punishment resistance. 
				}
			\end{remark}
			
			\begin{proof}
				For brevity, we show the proof of Proposition \ref{AppLemmaID} when $d = 1$.
				
				Fix $y  < \gamma_{C}$, $v \in \mathcal{P}$ and $\delta \in \mathbb{R}_{++}$ such that $y + \delta \in \mathcal{C}$. Note that
				\begin{align*}
					& \mathbb{P}\left[\left. Y\leq y, D = 1 \right\vert P\left(Z,C\right) = v, C = y+\delta\right] \\
					& \hspace{20pt} = \mathbb{E}\left[\left. \mathbf{1}\left\lbrace Y \leq y \right\rbrace\mathbf{1}\left\lbrace P\left(Z, C\right) \geq V \right\rbrace \right\vert P\left(Z,C\right) = v, C = y+\delta\right] \\
					& \hspace{40pt} \text{by \eqref{EqTreatment}} \\
					& \hspace{20pt} = \mathbb{E}\left[\left. \mathbf{1}\left\lbrace Y^*(1) \leq y \right\rbrace\mathbf{1}\left\lbrace p \geq V \right\rbrace \right\vert P\left(Z,y + \delta\right) = p, C = y+\delta\right] \\
					& \hspace{40pt} \text{because $Y_{1}^{*}$ is not censored when $C > y$} \\
					& \hspace{20pt} = \int_0^1\mathbb{E}\left[\left. \mathbf{1}\left\lbrace Y^*(1) \leq y \right\rbrace\mathbf{1}\left\lbrace p \geq v \right\rbrace \right\vert P\left(Z,y + \delta\right) = p, C = y+\delta, V=v\right]dv \\
					& \hspace{40pt} \text{by the Law of Iterated Expectations and Assumption \ref{AsContinuous}} \\
					& \hspace{20pt} = \int_0^1 \mathbf{1}\left\lbrace p \geq v \right\rbrace \mathbb{E}\left[\left. \mathbf{1}\left\lbrace Y^*(1) \leq y \right\rbrace \right\vert P\left(Z,y + \delta\right) = p, C = y+\delta, V=v\right]dv \\
					& \hspace{20pt} = \int_0^{p} \mathbb{E}\left[\left. \mathbf{1}\left\lbrace Y^*(1) \leq y \right\rbrace \right\vert P\left(Z,y + \delta\right) = p, C = y+\delta, V=v\right]dv \\
					& \hspace{20pt} = \int_0^{ p} \mathbb{P}\left[\left. Y^*(1) \leq y  \right\vert C = y+\delta, V=v\right]dv \\
					& \hspace{40pt} \text{by Assumption \ref{AsIndependence}}.
				\end{align*}
				
				Consequently, the Leibniz Integral Rule implies that
				\begin{align*}
					\dfrac{\partial \mathbb{P}\left[\left. Y\leq y, D = 1 \right\vert P\left(Z, C \right) = p, C = y + \delta \right]}{\partial v} & = \mathbb{P}\left[\left. Y^*(1) \leq y  \right\vert C = y+\delta, V = p \right] \\
					& = DMTR_{1}\left(y, p, y + \delta \right).
				\end{align*}
				
				We can prove the same result for $d = 0$ analogously.
			\end{proof}
		}

		\newpage
		\section{Partial Identification Strategies}\label{AppPartialID}
		
		\setcounter{table}{0}
		\renewcommand\thetable{G.\arabic{table}}
		
		\setcounter{figure}{0}
		\renewcommand\thefigure{G.\arabic{figure}}
		
		\setcounter{equation}{0}
		\renewcommand\theequation{G.\arabic{equation}}
		
		\setcounter{theorem}{0}
		\renewcommand\thetheorem{G.\arabic{theorem}}
		
		\setcounter{assumption}{0}
		\renewcommand\theassumption{G.\arabic{assumption}}
		
		\setcounter{proposition}{0}
		\renewcommand\theproposition{G.\arabic{proposition}}
		
		\setcounter{corollary}{0}
		\renewcommand\thecorollary{G.\arabic{corollary}}
		
		In some empirical applications, Assumption \ref{AsCensoring} may be too strong, while in others, it may be plausible. Since this is very context-specific, it is worth coming up with alternative identification strategies that accommodate dependent censoring mechanisms. In this appendix, we discuss two alternative assumptions that restrict the dependence between the censoring variable and the latent heterogeneity, but not to the point of imposing censoring independence. Section \ref{Scid} imposes that potential outcomes are negatively regression-dependent on the censoring variable, while Section \ref{AppPartialContinuous} intuitively imposes that the censoring problem is not too severe.
		
		To illustrate possible violations of Assumption \ref{AsCensoring}, we briefly discuss our empirical application. As discussed in Appendix \ref{AppRobustness}, potential recidivism may not be stationary if legislation or inputs to the production of Justice change over time. These inputs may include the number of police officers, judges, or public defenders. Moreover, the potential outcomes may fail to be stationary if crime rates increase or decrease substantially during the sampling period, possibly changing the incentives to recidivate.\footnote{In São Paulo, the homicide rate (the only criminal measure that is observable for a long period of time) increased substantially during the 1990s and decreased substantially during the 2000s. However, it is relatively stable during the 2010s, our sampling period. If anything, the murder rate decreased slightly during the 2010s, justifying Assumption \ref{AsIVCCensoring}.} Overall, we do not believe these factors changed substantially during our sampling period, supporting the validity of Assumption \ref{AsCensoring} in our empirical application.
		
		\subsection{Partial identification under regression dependence}\label{Scid}
		
		In this subsection, we impose that potential outcomes are negatively regression-dependent on the censoring variable. This alternative assumption restricts the relationship between the latent heterogeneity, the censoring variable, and the potential outcomes.\footnote{Related assumptions have been used by \citet{Chesher2005}, \citet{Jun2011}, and \citet{Kedagni2014} in different contexts. For more information on the definition of regression dependence and other concepts of statistical dependence, see \citet{Lehmann1966}.}
		
		\begin{assumption}[Regression Dependence]\label{AsIVCCensoring}
			Conditional on $V$, the potential outcomes are negatively regression dependent on the censoring variable, i.e., $\mathbb{P}\left[\left.  Y^*(d) \leq y  \right\vert C = \tilde{c}, V = v \right]  \geq \mathbb{P}\left[\left.  Y^*(d) \leq y  \right\vert C =c, V = v \right]$ for any $d \in \left\lbrace 0, 1 \right\rbrace$, any $v \in (0,1)$ and any $\left(c,\tilde{c}\right) \in \mathcal{C}^{2}$ such that $c \leq \tilde{c} $.
		\end{assumption}
		
		In our empirical application, Assumption \ref{AsIVCCensoring} imposes that the potential outcomes of more recent cases first-order stochastically dominate the potential outcomes of older cases. Intuitively, this restriction implies that defendants are committing fewer crimes over time, and is plausible given that the state of São Paulo became safer during our sampling period.\footnote{In a different empirical context, positive regression dependence may be more plausible than negative regression dependence. Similar bounds can be derived based on this alternative assumption.}
		
		To derive bounds around the $DMTR$ functions, define the following auxiliary quantities:
		\begin{eqnarray*}
			LB_d(y,v,\delta) &\coloneqq& \mathbb{P}(y+\delta \leq C) \cdot \left(2 d - 1\right) \cdot \gamma_{d}(y,v,y+\delta)\\
			UB_d(y,v,\delta) &\coloneqq &\mathbb{P}(C \leq y) + \mathbb{P}(y + \delta \leq C) \\
			&& +\linebreak \mathbb{P}(y \leq C \leq y+\delta) \cdot \left(2 d - 1\right) \cdot \gamma_{d}(y,v,y+\delta),
		\end{eqnarray*}
		where $\delta \in \mathbb{R}_{++}$. The next proposition describes the bounds around the $DMTR$ functions when potential outcomes are negatively regression-dependent on the censoring variable.
		
		\begin{proposition}\label{PropDMTRunderCID}
			Suppose that Assumptions \ref{AsIndependence}-\ref{AsPositive} and \ref{AsIVCCensoring} hold. Then,
			\begin{align*}
				& DMTR_{d}\left(y, v\right) \in \left[\max_{\delta \in \mathcal{D}} LB_d(y,v,\delta), \min_{\delta \in \mathcal{D}} UB_d(y,v,\delta) \right]
			\end{align*}
			for any $d \in \left\lbrace 0, 1 \right\rbrace$, $y < \gamma_{C}$ and $v \in \mathcal{P}$, where $\mathcal{D} \coloneqq \left\lbrace \delta \in \mathbb{R}_{++} \colon y + \delta \in \mathcal{C} \right\rbrace$.
		\end{proposition}
		
		\begin{proof}
			See Appendix \ref{AppProofPropDMTRunderCID}.
		\end{proof}
		
		First, one can see that the bounds in Proposition \ref{PropDMTRunderCID} do not ``collapse'' to point identification when Assumption \ref{AsCensoring} holds. That is because the regression-dependence in Assumption \ref{AsIVCCensoring} is compatible with dependent censoring but does not directly restrict the amount of dependence between the potential outcomes and the censoring random variable. In order words, the nature of Assumption \ref{AsIVCCensoring} is different from Assumption \ref{AsCensoring} and does not constitute ``continuous relaxations'' of the independence assumption.
		
		Second, Assumption \ref{AsIVCCensoring} allows us to exploit the information in $ \gamma_{d}(y,v,y+\delta)$ even under dependent censoring because of the (stochastic) monotonicity in $C$. Since this monotonicity property holds for different values of $C$ less than the $\tilde{c}$, we take the supremum and infimum over $\delta$, so the bounds are tighter.
		
		Third, the identification region will functionally depend on the propensity score (and thus the instrument), as reflected in the presence of $\gamma_{d}(y,v,y+\delta)$ in the bounds. This will determine the shape of it.
		
		Furthermore, the bounds' length depends on the proportion of censored observations close to the value of the particular $y$, reflecting that regions with heavier censoring data tend to have wider bounds.
		
		From the partial identification of the $DMTR_{d}\left(y, v\right)$ functions, it also follows the partial identification of a range of $QMTE\left(\tau,v\right)$ across $\tau$ and the $RMTE(v)$, just like before. If one further imposes Assumptions \ref{AsFinite} and \ref{AsSupport}, partial identification results for the MTE function will also follow. For these functions, though, it is important to ensure that the lower and upper bounds in Proposition \ref{PropDMTRunderCID} are monotone in $y$, which can be enforced using a similar approach as in \citet{Manski2021}. More specifically, for a grid of weakly increasing $y$'s, if $ \max_{\delta \in \mathcal{D}}LB_d(y_{k+1},c,\delta) < \max_{\delta \in \mathcal{D}}LB_d(y_{k},c,\delta)$, we can simply redefine $\max_{\delta \in \mathcal{D}}LB_d(y_{k+1},c,\delta) = \max_{\delta \in \mathcal{D}}LB_d(y_{k},c,\delta)$; the analogous is true for the upper bound. Alternatively, one can use the rearrangement procedure as in \citet{Chernozhukov2009b}.  We state these results as corollaries for convenience.
		
		\begin{corollary}\label{CorQMTEpartial}
			Suppose that Assumptions \ref{AsIndependence}-\ref{AsPositive}, \ref{AsRC4} and \ref{AsIVCCensoring} hold. Then, \begin{enumerate}
				\item[(a)] $QMTE\left(\tau,v\right)$ is partially identified for any $v \in \mathcal{P}$ and $\tau \in \left(0, \overline{\tau}\left(v\right)\right)$.
				\item[(b)] the $RMTE(v)$ function is partially-identified for any $v \in \mathcal{P}$.
			\end{enumerate}
		\end{corollary}
		
		\begin{corollary}\label{CorMTEpartial}
			If Assumptions Assumptions \ref{AsIndependence}-\ref{AsPositive}, \ref{AsIVCCensoring}, \ref{AsFinite} and \ref{AsSupport} hold, then $MTE\left(v\right)$ is  partially identified for any $v \in \mathcal{P}$.
		\end{corollary}
		
		Finally, the bounds in Proposition \ref{PropDMTRunderCID} can be estimated using methods similar to the methods described in Subsection \ref{SestSemiPara}. The main difference between the estimators of the bounds and the point-estimators (Subsection \ref{SestSemiPara}) is that, when estimating the bounds, we take either the maximum or the minimum over values of $c$ in Step \ref{StepMean} instead of taking the mean. Consequently, these estimators will converge in probability to the bounds in Proposition \ref{PropDMTRunderCID}.
		
		\subsubsection{Proof of Proposition \ref{PropDMTRunderCID}}\label{AppProofPropDMTRunderCID}
		
		Fix $d \in \left\lbrace 0, 1 \right\rbrace$, $y  < \gamma_{C}$, $v \in \mathcal{P}$ and $\delta \in \mathbb{R}_{++}$ such that $y + \delta \in \mathcal{C}$.
		
		Note that Equations \eqref{EqPYD1} and \eqref{EqPYD0} imply that
		\begin{equation}\label{EqPYD1derivativeCID}
			\dfrac{\partial \mathbb{P}\left[\left. Y\leq y, D = 1 \right\vert P\left(Z,C\right) = v, C = y+\delta\right]}{\partial v} = \mathbb{P}\left[\left. Y^*(1) \leq y  \right\vert C = y+\delta, V = v\right]
		\end{equation}
		and
		\begin{equation}\label{EqPYD0derivativeCID}
			\dfrac{\partial \mathbb{P}\left[\left. Y\leq y, D = 0 \right\vert P\left(Z,C\right) = v, C = y+\delta\right]}{\partial v} = - \mathbb{P}\left[\left.  Y^*(0) \leq y  \right\vert C = y+\delta, V = v\right]
		\end{equation}
		according to the Leibniz Integral Rule.
		
		Combining the last two equations, we have that
		\begin{equation}\label{EqPYCVcid}
			\mathbb{P}\left[\left.  Y^*(d) \leq y  \right\vert C = y+\delta, V = v\right] = \left(2d - 1\right) \cdot \dfrac{\partial \mathbb{P}\left[\left. Y\leq y, D = d \right\vert P\left(Z,C\right) = v, C = y+\delta\right]}{\partial v}.
		\end{equation}
		Moreover, observe that:
		\begin{align*}
			& \mathbb{P}\left[\left.  Y^*(d) \leq y  \right\vert V=v\right] \\
			& \hspace{20pt} = \int \mathbb{P}\left[\left.  Y^*(d) \leq y  \right\vert C = \tilde{c}, V = v \right] f_{C\vert V}\left(\tilde{c} \vert v \right) d\tilde{c} \\
			& \hspace{40pt} \text{by the Law of Iterated Expectations} \\
			& \hspace{20pt} = \int \mathbb{P}\left[\left.  Y^*(d) \leq y  \right\vert C = \tilde{c}, V = p \right] f_{C}\left(\tilde{c} \right) d\tilde{c} \\
			& \hspace{40pt} \text{because $V \independent C$ by Assumption \ref{AsContinuous}} \\
			& \hspace{20pt} = \int_{0}^{y} \mathbb{P}\left[\left.  Y^*(d) \leq y  \right\vert C = \tilde{c}, V = p \right] f_{C}\left(\tilde{c} \right) d\tilde{c} \\
			& \hspace{40pt} + \int_{y}^{y+\delta} \mathbb{P}\left[\left.  Y^*(d) \leq y  \right\vert C = \tilde{c}, V = p \right] f_{C}\left(\tilde{c} \right) d\tilde{c}
			\\
			& \hspace{40pt} + \int_{y+\delta}^{+ \infty} \mathbb{P}\left[\left.  Y^*(d) \leq y  \right\vert C = \tilde{c}, V = p \right] f_{C}\left(\tilde{c} \right) d\tilde{c},
		\end{align*}
		implying, by Assumption \ref{AsIVCCensoring}, that
		\begin{align}\label{Aux1}
			\mathbb{P}\left[\left.  Y^*(d) \leq y  \right\vert V= p\right]
			& \leq   \mathbb{P}(C \leq y) + \mathbb{P}(y+\delta \leq C) \nonumber \\
			& \hspace{20pt} + \mathbb{P}(y \leq C \leq y+\delta)\mathbb{P}\left[\left.  Y^*(d) \leq y  \right\vert C = y+\delta, V = p \right]
		\end{align}
		and
		\begin{align}\label{Aux2}
			\mathbb{P}\left[\left.  Y^*(d) \leq y  \right\vert V= p\right]
			& \geq  \mathbb{P}(y+\delta \leq C)\mathbb{P}\left[\left.  Y^*(d) \leq y  \right\vert C = y+\delta, V = p \right]
		\end{align}
		Thus, combining Equations \eqref{Aux1} and \eqref{Aux2} with \eqref{EqPYCVcid}, we have that
		\begin{align*}
			& DMTR_{d}\left(y, v\right) \\
			& \hspace{20pt} \in \left[\mathbb{P}(y+\delta \leq C)\cdot \left(2 d - 1\right) \cdot \dfrac{\partial \mathbb{P}\left[\left. Y\leq y, D = d \right\vert P\left(Z,C\right) = v, C = y + \delta \right]}{\partial v}, \right. \\
			& \hspace{40pt} \left. \begin{array}{l}
				\mathbb{P}(C \leq y) + \mathbb{P}(y+\delta \leq C) \\ \hspace{20pt} + \mathbb{P}(y \leq C \leq y+\delta)\cdot \left(2 d - 1\right) \cdot \dfrac{\partial \mathbb{P}\left[\left. Y\leq y, D = d \right\vert P\left(Z,C\right) = v, C = y + \delta \right]}{\partial v}
			\end{array} \right].
		\end{align*}
		
		Since the bounds above hold for any $\delta \in \mathbb{R}_{++}$ such that $y + \delta \in \mathcal{C}$, we have that
		\begin{align*}
			& DMTR_{d}\left(y, v\right) \\
			& \hspace{20pt} \in \left[\max_{\delta \in \mathcal{D}} \left\lbrace \mathbb{P}(y+\delta \leq C) \cdot \left(2 d - 1\right) \cdot \dfrac{\partial \mathbb{P}\left[\left. Y\leq y, D = d \right\vert P\left(Z,C\right) = v, C = y + \delta \right]}{\partial v} \right\rbrace, \right. \\
			& \hspace{40pt} \left. \min_{\delta \in \mathcal{D}} \left\lbrace \begin{array}{l}
				\mathbb{P}(C \leq y) + \mathbb{P}(y + \delta \leq C) + \mathbb{P}(y \leq C \leq y+\delta) \\
				\hspace{20pt} \cdot \left(2 d - 1\right) \cdot \dfrac{\partial \mathbb{P}\left[\left. Y\leq y, D = d \right\vert P\left(Z,C\right) = v, C = y + \delta \right]}{\partial v}
			\end{array} \right\rbrace \right],
		\end{align*}
		where $\mathcal{D} \coloneqq \left\lbrace \delta \in \mathbb{R}_{++} \colon y + \delta \in \mathcal{C} \right\rbrace$.

		\subsection{Partial identification under a continuous violation of random censoring}\label{AppPartialContinuous}
		
		In this subsection, we impose that the conditional distribution of the potential outcomes given the censoring variable and the latent heterogeneity variable is close to the conditional distribution of the potential outcomes given only the latent heterogeneity variable.\footnote{A similar assumption is used by \citet{Kline2013} in a sample selection context.} Differently from Assumption \ref{AsIVCCensoring}, the following assumption constitutes a ``continuous relaxation'' of censoring independence (Assumption \ref{AsCensoring}).
		
		\begin{assumption}[``Continuous Relaxation'']\label{AsContinuousRelax}
			The conditional distribution of the potential outcomes given the censoring variable and the latent heterogeneity variable is similar to the conditional distribution of the potential outcomes given only the latent heterogeneity variable i.e., there exists $\overline{B} \in \mathbb{R}_{++}$ such that $$\left\vert \mathbb{P}\left[\left. Y^{*}\left(d\right) \leq y \right\vert C = y + \delta, V = v\right] - \mathbb{P}\left[\left. Y^{*}\left(d\right) \leq y \right\vert V = v\right]\right\vert \leq \overline{B}$$ for any $y \in \mathcal{Y}$, $v \in \left[0,1\right]$, $\delta \in \mathcal{D} \coloneqq \left\lbrace \delta \in \mathbb{R}_{++} \colon y + \delta \in \mathcal{C} \right\rbrace$ and $d \in \left\lbrace 0, 1\right\rbrace$.
		\end{assumption}
		
		Using this assumption, we can derive bounds around the $DMTR$ functions.
		
		\begin{proposition}\label{PropDMTRcontinuousRelax}
			Suppose that Assumptions \ref{AsIndependence}-\ref{AsPositive} and \ref{AsContinuousRelax} hold. Then,
			\begin{align*}
				& DMTR_{d}\left(y, v\right) \in \left[LB_{d}\left(y, v, \overline{B}\right), UB_{d}\left(y, v, \overline{B}\right) \right]
			\end{align*}
			where $$LB_{d}\left(y, v, \overline{B}\right) \coloneqq - \overline{B} + \max_{\delta \in \mathcal{D}} \left\lbrace \left(2 \dot d - 1\right) \cdot \gamma_{d}(y,v,y+\delta) \right\rbrace$$ and $$UB_{d}\left(y, v, \overline{B}\right) \coloneqq \overline{B} + \min_{\delta \in \mathcal{D}} \left\lbrace \left(2 \dot d - 1\right) \cdot \gamma_{d}(y,v,y+\delta) \right\rbrace$$ for any $d \in \left\lbrace 0, 1 \right\rbrace$, $y < \gamma_{C}$ and $v \in \mathcal{P}$.
		\end{proposition}
		
		\begin{proof}
			See Appendix \ref{AppProofContinuousRelax}.
		\end{proof}
		
		First, the bounds in Proposition \ref{PropDMTRcontinuousRelax} can be estimated using methods similar to the methods described in Subsection \ref{SestSemiPara}. The main difference between the estimators of the bounds and the point-estimators (Subsection \ref{SestSemiPara}) is that, when estimating the bounds, we take either the maximum or the minimum over values of $c$ in Step \ref{StepMean} instead of taking the mean. Consequently, these estimators will converge in probability to the bounds in Proposition \ref{PropDMTRcontinuousRelax}.
		
		Second, differently from Proposition \ref{PropDMTRunderCID}, the bounds in Proposition \ref{PropDMTRcontinuousRelax} ``collapse'' to point identification when Assumption \ref{AsCensoring} holds. In this case, we have that $\overline{B} = 0$ and $\max_{\delta \in \mathcal{D}} \gamma_{d}(y,v,y+\delta) = \min_{\delta \in \mathcal{D}} \gamma_{d}(y,v,y+\delta)$.
		
		Third, we take the supremum and infimum over $\delta$ to tighten the bounds because our continuous relaxation of censoring independence holds for every value of $C$ greater than $y$.
		
		Finally, the identification region will functionally depend on the propensity score (and thus the instrument), as reflected in the presence of $\gamma_{d}(y,v,y+\delta)$ in the bounds. This will determine the shape of it.
		
		Furthermore, the bounds' length depends on the choice of $\overline{B}$. We recommend choosing $\overline{B}$ based on the bounds around the $DMTE$ function and defined in Corollary \ref{CorDMTEcontinuousRelax}.
		\begin{corollary}\label{CorDMTEcontinuousRelax}
			Suppose that Assumptions \ref{AsIndependence}-\ref{AsPositive} and \ref{AsContinuousRelax} hold. Then,
			\begin{align*}
				& DMTE\left(y, v\right) \in \left[\underline{\Delta}\left(y, v, \overline{B}\right), \overline{\Delta}\left(y, v, \overline{B}\right) \right]
			\end{align*}
			where $$\underline{\Delta}\left(y, v, \overline{B}\right) \coloneqq LB_{1}\left(y, v, \overline{B}\right) - UB_{0}\left(y, v, \overline{B}\right)$$ and $$\overline{\Delta}\left(y, v, \overline{B}\right) \coloneqq UB_{1}\left(y, v, \overline{B}\right) - LB_{0}\left(y, v, \overline{B}\right)$$ for any $y < \gamma_{C}$ and $v \in \mathcal{P}$.
		\end{corollary}
		Corollary \ref{CorDMTEcontinuousRelax} may be used to choose $\overline{B}$ according to a breakdown analysis \citep{Kline2013,Masten2018}. For example, the researcher may be particularly interested in $DMTE\left(\overline{y}, \cdot \right)$ for some value of $\overline{y} \in \mathcal{Y}$. If, based on Proposition \ref{PropDMTR}, $DMTE\left(\overline{y}, v \right) \neq 0$ for some $v \in \mathcal{P}$, then the researcher can choose the smallest value of $\overline{B}\left(\overline{y}\right)$ such that the bounds in Corollary \ref{CorDMTEcontinuousRelax} contain the zero function, i.e., $0 \in \left[\underline{\Delta}\left(\overline{y}, v, \overline{B}\left(\overline{y}\right)\right), \overline{\Delta}\left(\overline{y}, v, \overline{B}\left(\overline{y}\right)\right) \right]$ for every $v \in \mathcal{P}$. This value of $\overline{B}\left(\overline{y}\right)$ is known as the breakdown point.\footnote{Conducting inference about the breakdown point is beyond the scope of this paper.}
		
\begin{figure}[!htb]
			\begin{center}
				\includegraphics[width = 0.7\textwidth]{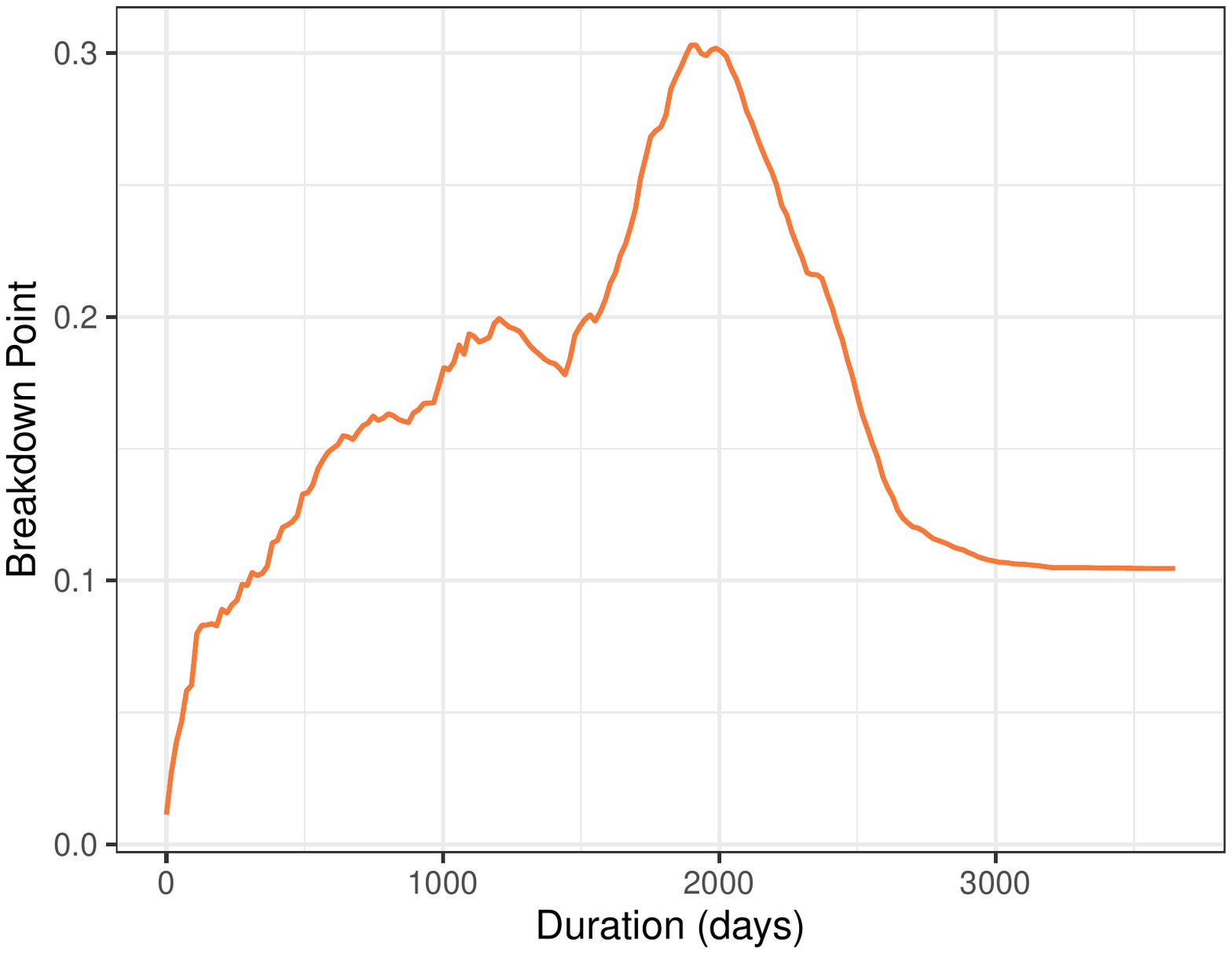}
				\caption{Breakdown Point $\overline{B}\left(y\right)$ for $y < \gamma_{C}$}
				\label{FigBreakdownPoint}
			\end{center}
			\justifying
			\vspace{-.5cm}\scriptsize{Notes: For $y < 8 \text{ years}$, the breakdown point $\overline{B}\left(y\right)$ is the smallest value of $\overline{B}\left(y\right)$ such that the bounds in Corollary \ref{CorDMTEcontinuousRelax} contain the zero function, i.e., $0 \in \left[\underline{\Delta}\left(y, v, \overline{B}\left(y\right)\right), \overline{\Delta}\left(y, v, \overline{B}\left(y\right)\right) \right]$ for every $v \in \mathcal{P}$.}
		\end{figure}

		Figure \ref{FigBreakdownPoint} shows the Breakdown Points $\overline{B}\left(y\right)$ for $y < 8 \text{ years}$. Note that the breakdown points for short-run recidivism (small values of $y$) are smaller than the breakdown point for mid-run recidivism (values of $y$ around 2000 days or 5 years approximately). This finding suggests that our results for mid-run recidivism are more robust to violations of the random censoring assumption than our results for short-run recidivism.
		
		More importantly, Figure \ref{FigBreakdownPoint} suggests that the results in Section \ref{SresultsDMTE} are robust to violations of the random censoring assumption. Considering that the largest possible value of the breakdown point is one, we believe that a breakdown point larger than 0.1 (i.e., $\overline{B}\left(y\right) \geq 0.1$) suggests that the estimated $DMTE\left(y,\cdot\right)$ is robust against continuous relaxations of Assumption \ref{AsCensoring}. We highlight that this condition holds for most time horizons in Figure \ref{FigBreakdownPoint}.
		
\begin{figure}[!htbp]
			\begin{center}
				\begin{subfigure}[t]{0.47\textwidth}
					\centering
					\includegraphics[width = \textwidth]{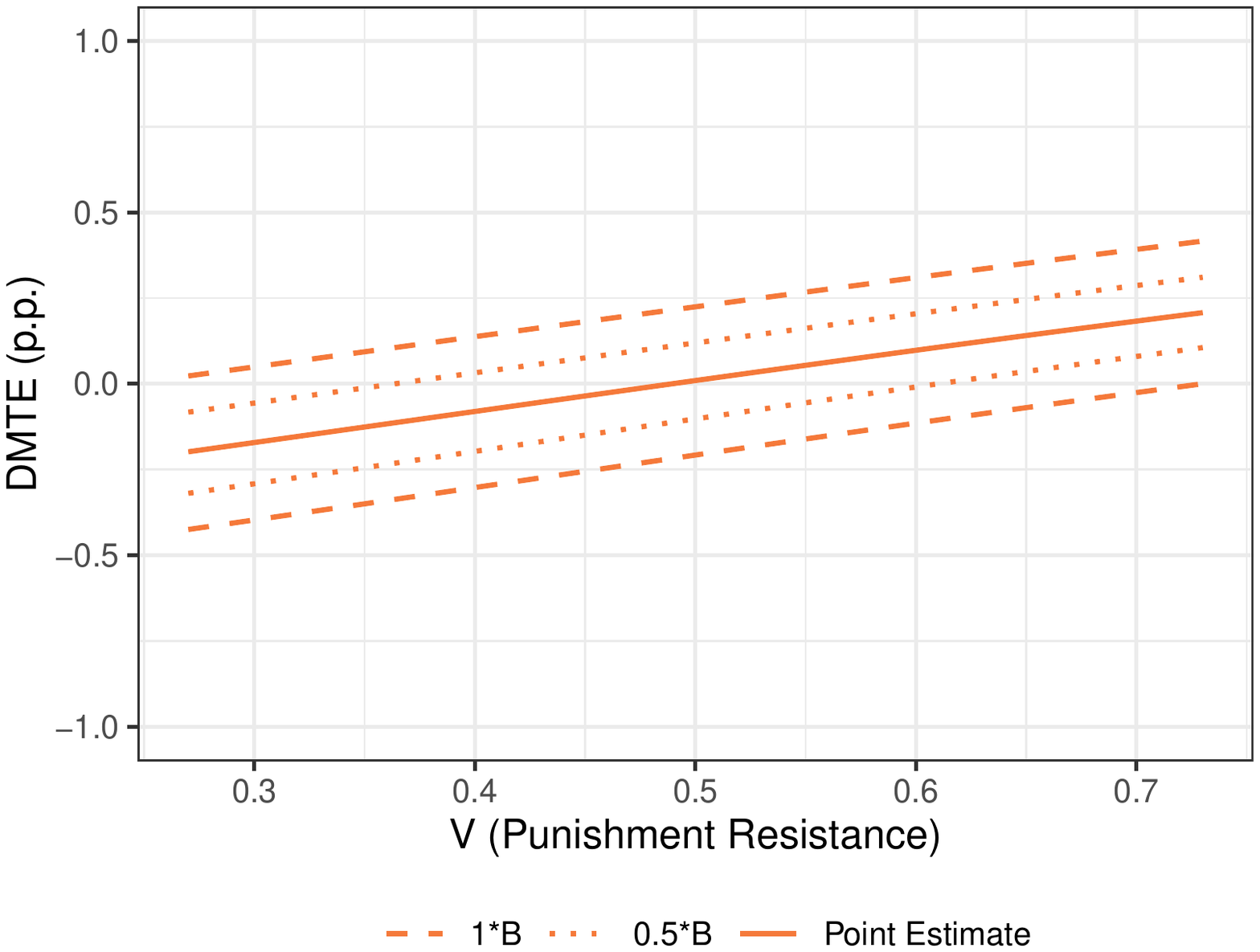}
					\caption{$DMTE\left(1,\cdot\right)$}
					\label{DMTE1-bounds}
				\end{subfigure}
				\hfill
				\begin{subfigure}[t]{0.47\textwidth}
					\begin{center}
						\includegraphics[width = \textwidth]{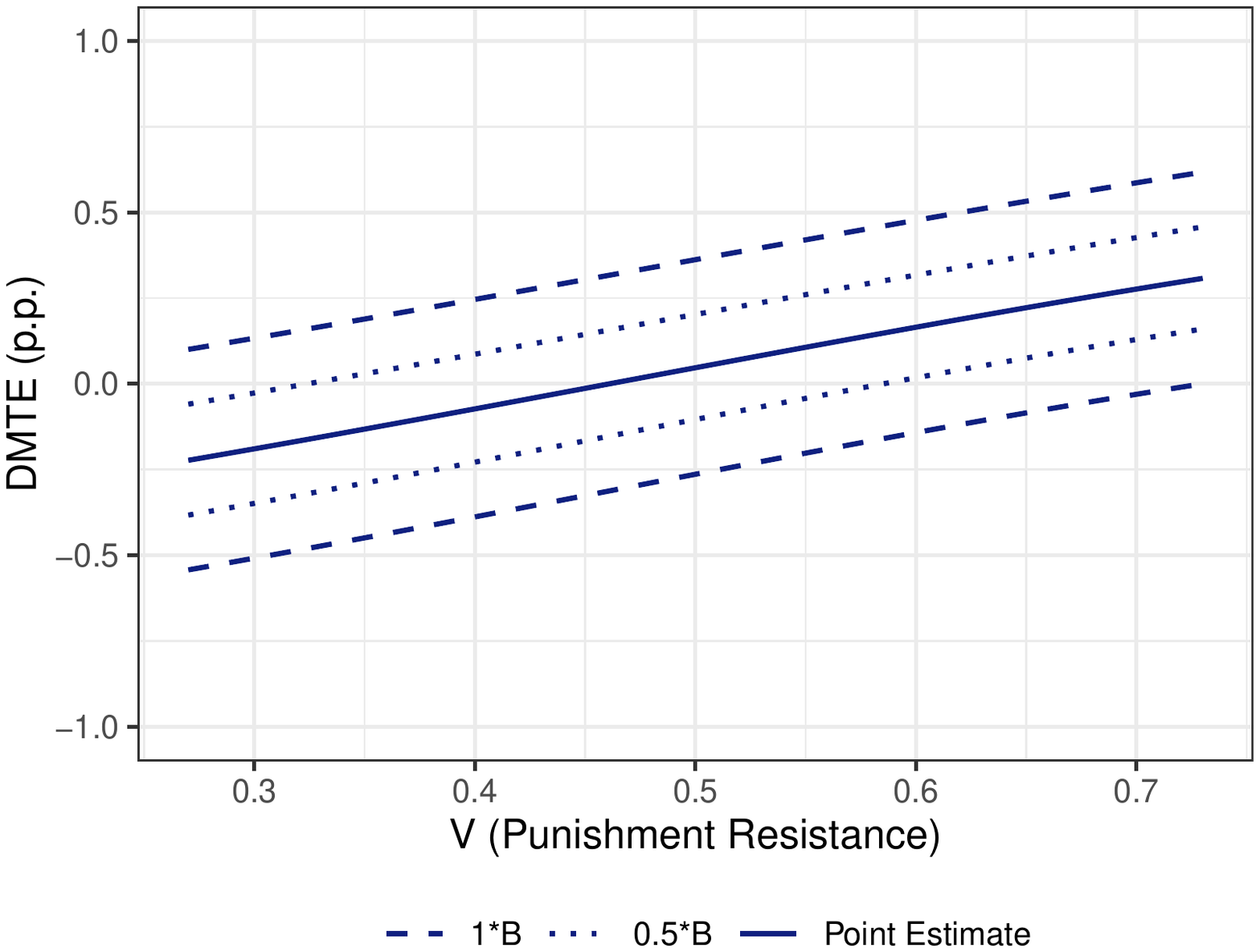}
						\caption{$DMTE\left(2,\cdot\right)$}
						\label{DMTE2-bounds}
					\end{center}
				\end{subfigure}
				\begin{subfigure}[t]{0.47\textwidth}
					\centering
					\includegraphics[width = \textwidth]{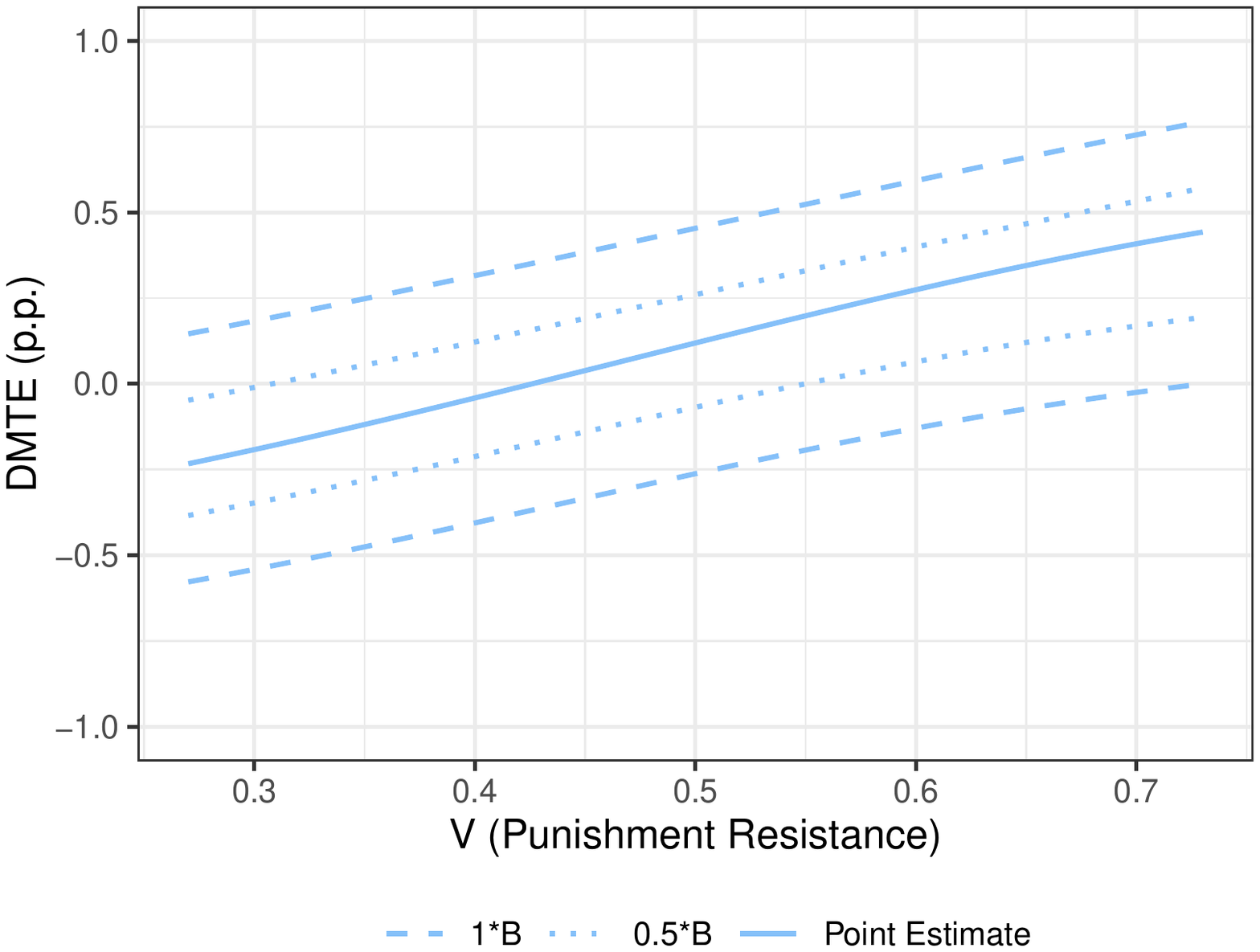}
					\caption{$DMTE\left(3,\cdot\right)$}
					\label{DMTE3-bounds}
				\end{subfigure}
				\hfill
				\begin{subfigure}[t]{0.47\textwidth}
					\begin{center}
						\includegraphics[width = \textwidth]{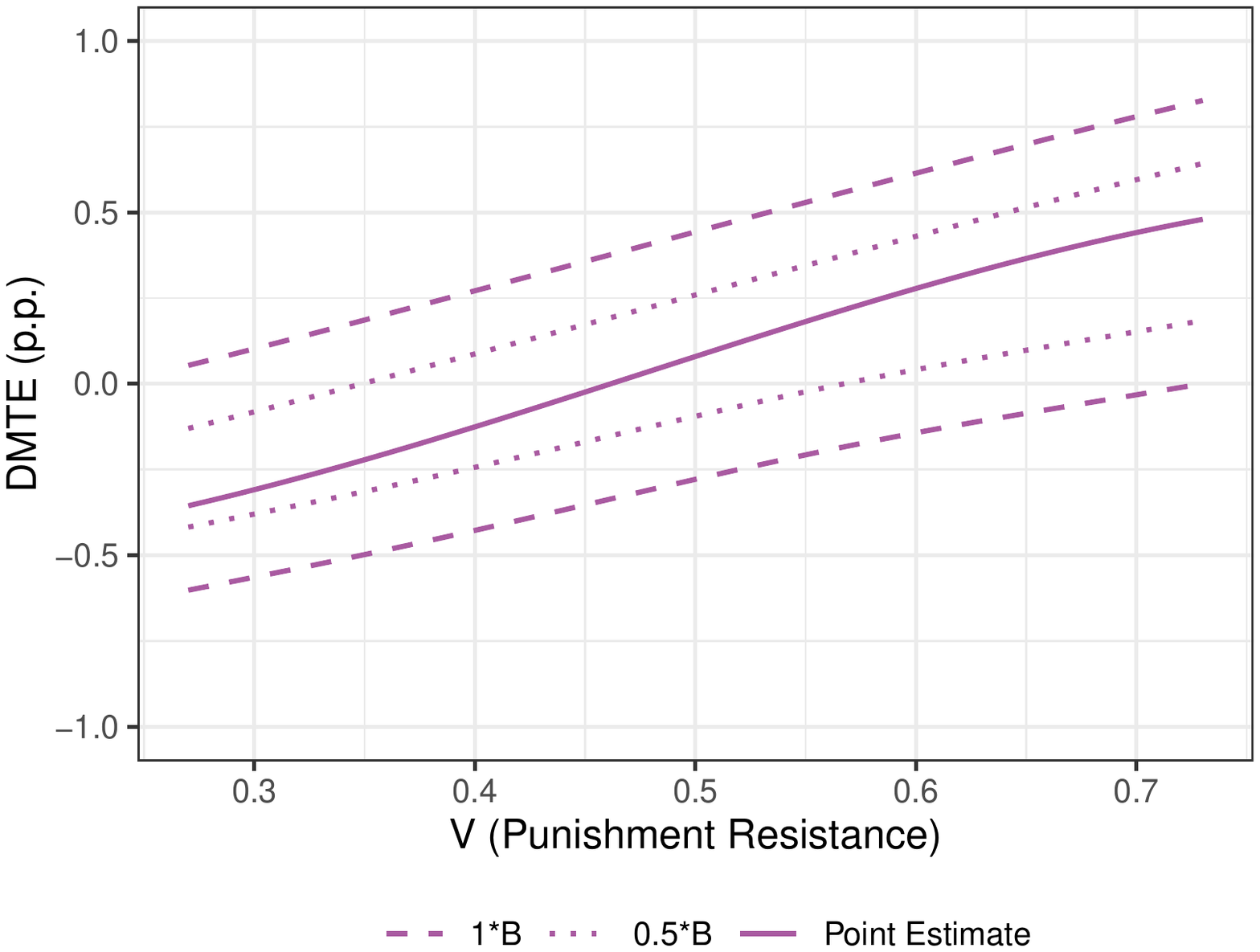}
						\caption{$DMTE\left(4,\cdot\right)$}
						\label{DMTE4-bounds}
					\end{center}
				\end{subfigure}
				\caption{Bounds around $DMTE\left(y,\cdot\right)$ for $y \in \left\lbrace 1, 2, 3, 4 \right\rbrace$}
				\label{FigDMTE-bounds-1-4}
			\end{center}
			\justifying
			\vspace{-.5cm}\scriptsize{Notes: Solid lines are the point estimates for the average $DMTE\left(y,\cdot\right)$ functions indicated in the caption of each subfigure. These results are based on Corollary \ref{cor:semi}. Dashed lines are the bounds around the average $DMTE\left(y,\cdot\right)$ functions that use the estimated breakdown point in their construction, i.e., $\underline{\Delta}\left(y, \cdot, \overline{B}\left(y\right)\right)$ and $\overline{\Delta}\left(y, \cdot, \overline{B}\left(y\right)\right)$. Dotted lines are the bounds around the average $DMTE\left(y,\cdot\right)$ functions that use half of the estimated breakdown point in their construction, i.e., $\underline{\Delta}\left(y, \cdot, \sfrac{\overline{B}\left(y\right)}{2}\right)$ and $\overline{\Delta}\left(y, \cdot, \sfrac{\overline{B}\left(y\right)}{2}\right)$. All bounds are based on  Corollary \ref{CorDMTEcontinuousRelax}.
			}
		\end{figure}

\begin{figure}[!htbp]
			\begin{center}
				\begin{subfigure}[t]{0.47\textwidth}
					\centering
					\includegraphics[width = \textwidth]{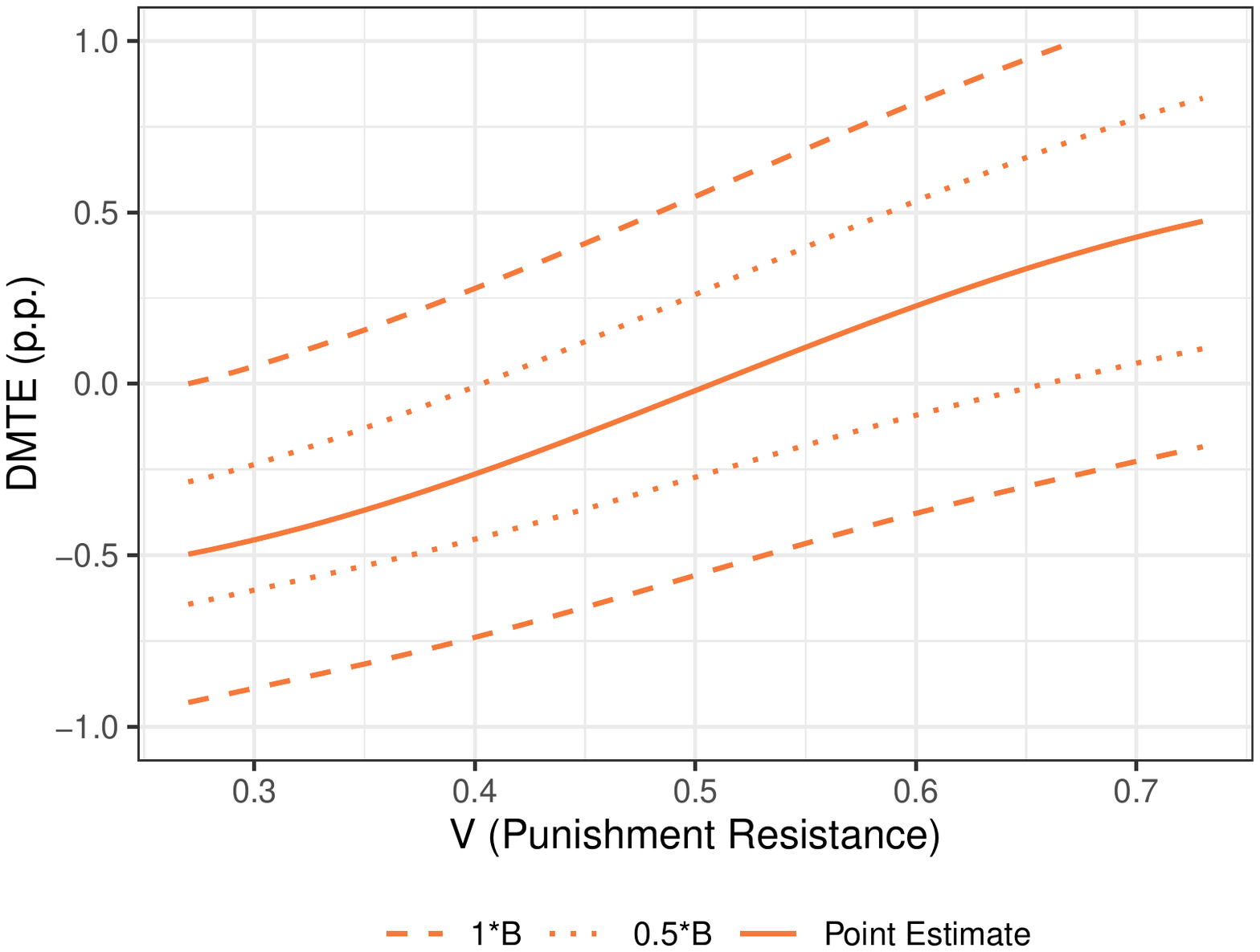}
					\caption{$DMTE\left(5,\cdot\right)$}
					\label{DMTE5-bounds}
				\end{subfigure}
				\hfill
				\begin{subfigure}[t]{0.47\textwidth}
					\begin{center}
						\includegraphics[width = \textwidth]{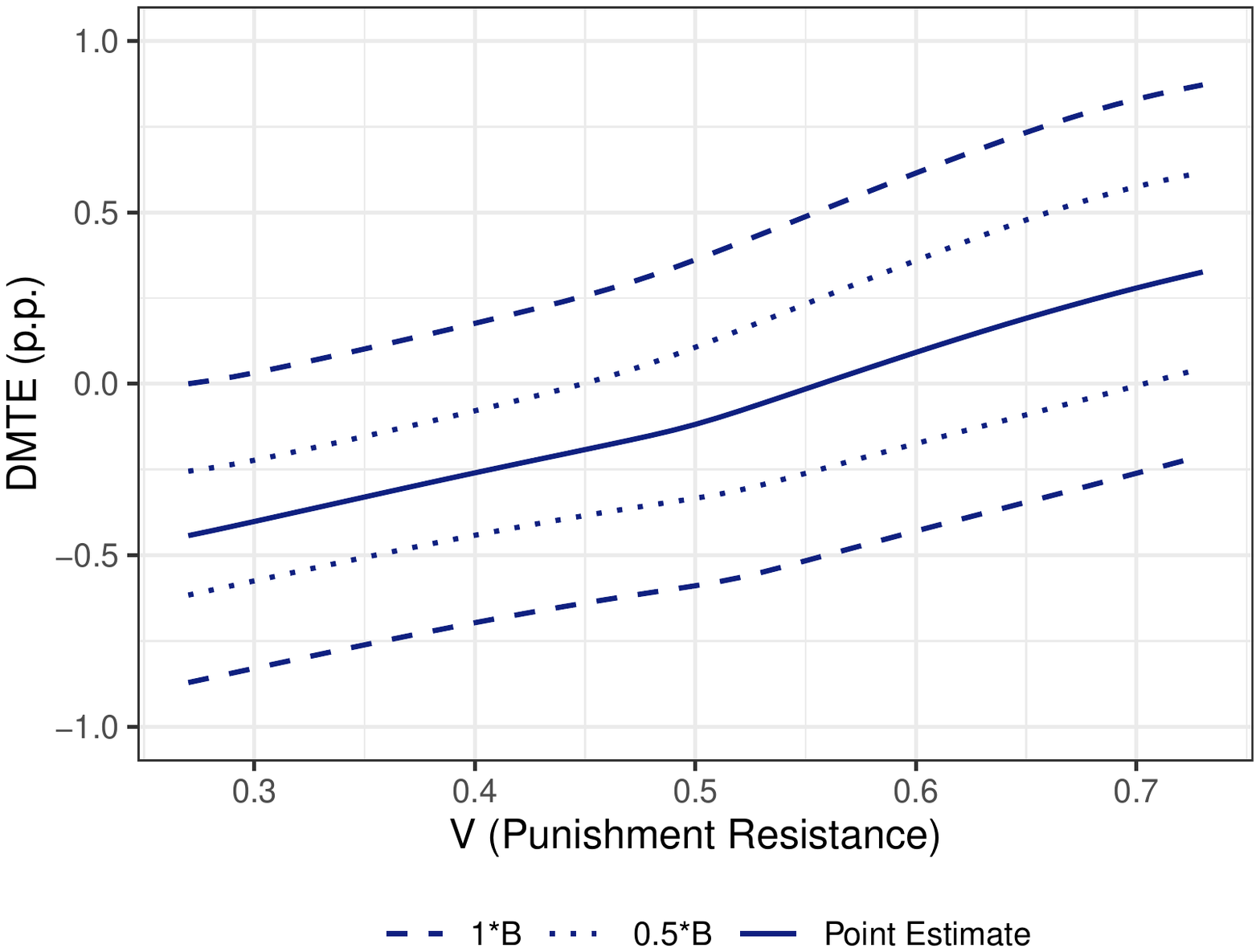}
						\caption{$DMTE\left(6,\cdot\right)$}
						\label{DMTE6-bounds}
					\end{center}
				\end{subfigure}
				\begin{subfigure}[t]{0.47\textwidth}
					\centering
					\includegraphics[width = \textwidth]{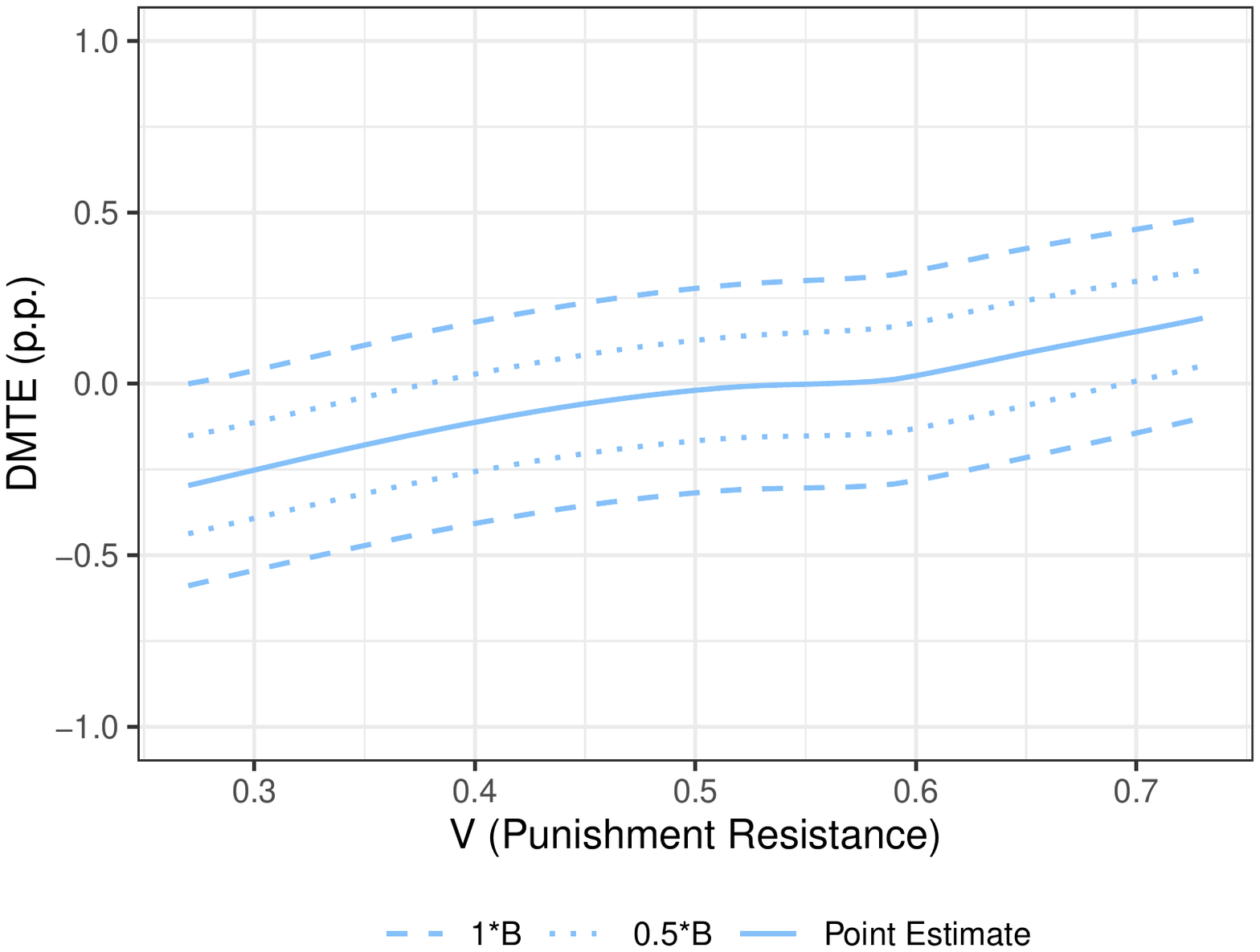}
					\caption{$DMTE\left(7,\cdot\right)$}
					\label{DMTE7-bounds}
				\end{subfigure}
				\hfill
				\begin{subfigure}[t]{0.47\textwidth}
					\begin{center}
						\includegraphics[width = \textwidth]{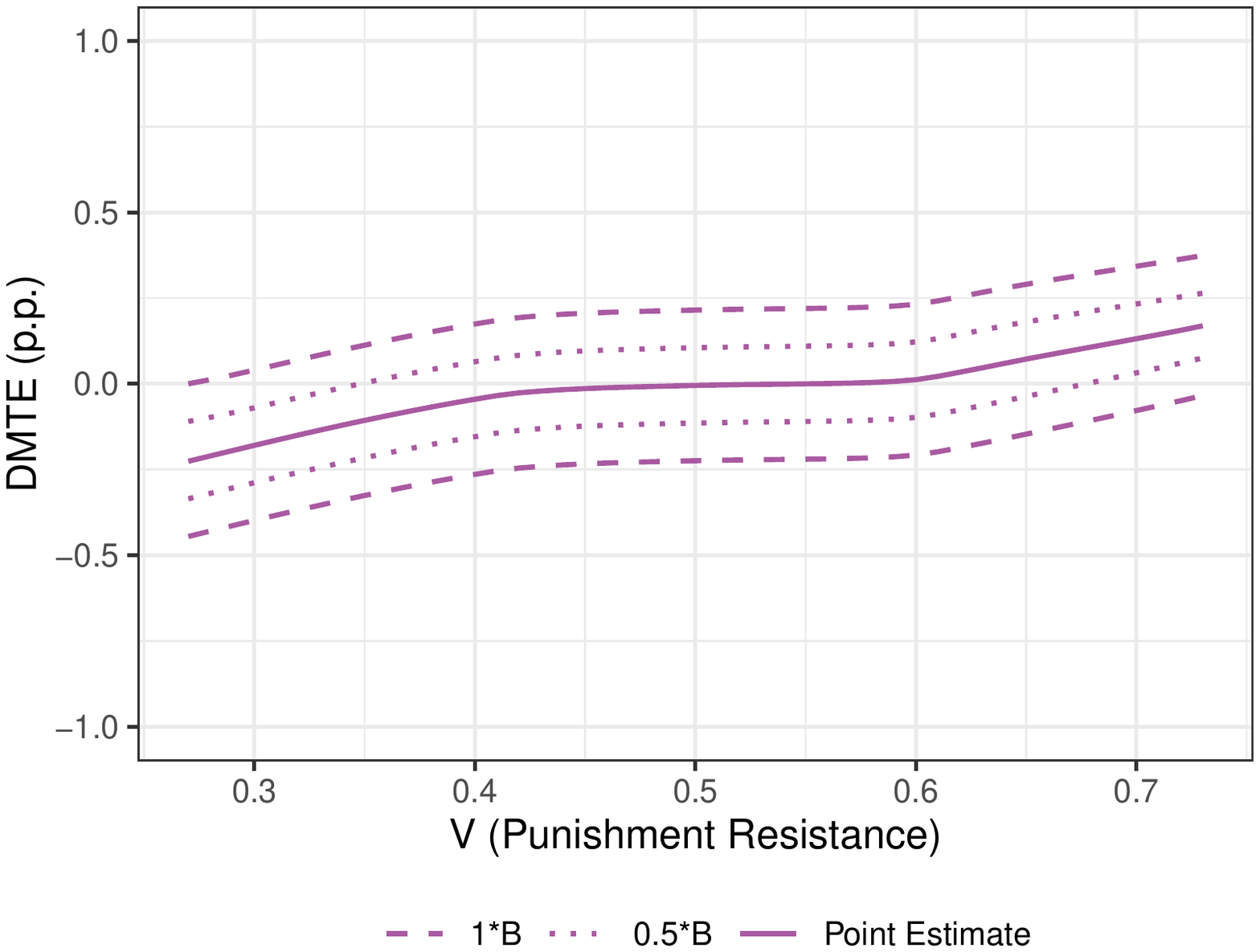}
						\caption{$DMTE\left(8,\cdot\right)$}
						\label{DMTE8-bounds}
					\end{center}
				\end{subfigure}
				\caption{Bounds around $DMTE\left(y,\cdot\right)$ for $y \in \left\lbrace 5, 6, 7, 8 \right\rbrace$}
				\label{FigDMTE-bounds-5-8}
			\end{center}
			\justifying
			\vspace{-.5cm}\scriptsize{Notes: Solid lines are the point estimates for the average $DMTE\left(y,\cdot\right)$ functions indicated in the caption of each subfigure. These results are based on Corollary \ref{cor:semi}. Dashed lines are the bounds around the average $DMTE\left(y,\cdot\right)$ functions that use the estimated breakdown point in their construction, i.e., $\underline{\Delta}\left(y, \cdot, \overline{B}\left(y\right)\right)$ and $\overline{\Delta}\left(y, \cdot, \overline{B}\left(y\right)\right)$. Dotted lines are the bounds around the average $DMTE\left(y,\cdot\right)$ functions that use half of the estimated breakdown point in their construction, i.e., $\underline{\Delta}\left(y, \cdot, \sfrac{\overline{B}\left(y\right)}{2}\right)$ and $\overline{\Delta}\left(y, \cdot, \sfrac{\overline{B}\left(y\right)}{2}\right)$. All bounds are based on  Corollary \ref{CorDMTEcontinuousRelax}.
			}
		\end{figure}

		To directly inspect the bounds in Corollary \ref{CorDMTEcontinuousRelax}, we can plot them for a few values of $\overline{B}$. Figures \ref{FigDMTE-bounds-1-4} and \ref{FigDMTE-bounds-5-8} plot $\underline{\Delta}\left(y, \cdot, \overline{B}\right)$ and $\overline{\Delta}\left(y, \cdot, \overline{B}\right)$ for $\overline{B} \in \left\lbrace \sfrac{\overline{B}\left(y\right)}{2} , \overline{B}\left(y\right)\right\rbrace$. Solid lines are the point estimates for the average $DMTE\left(y,\cdot\right)$ functions indicated in the caption of each subfigure (Corollary \ref{cor:semi}). Dashed lines are the bounds around the average $DMTE\left(y,\cdot\right)$ functions that use the estimated breakdown point in their construction, i.e., $\underline{\Delta}\left(y, \cdot, \overline{B}\left(y\right)\right)$ and $\overline{\Delta}\left(y, \cdot, \overline{B}\left(y\right)\right)$. Dotted lines are the bounds that use half of the estimated breakdown point in their construction, i.e., $\underline{\Delta}\left(y, \cdot, \sfrac{\overline{B}\left(y\right)}{2}\right)$ and $\overline{\Delta}\left(y, \cdot, \sfrac{\overline{B}\left(y\right)}{2}\right)$.
		
		When focusing on $\underline{\Delta}\left(y, \cdot, \sfrac{\overline{B}\left(y\right)}{2}\right)$ and $\overline{\Delta}\left(y, \cdot, \sfrac{\overline{B}\left(y\right)}{2}\right)$ (dotted lines), we find that the results in Section \ref{SresultsDMTE} are robust to violations of the random censoring assumption. For example, it is not possible to fit a flat line between these bounds. This finding suggests that there is heterogeneity with respect to the punishment resistance even when we allow for continuous relaxations of Assumption \ref{AsCensoring}.
		
		Beyond discussing bounds around the $DMTE$ functions and their breakdown points, the partial identification of the $DMTR_{d}\left(y, v\right)$ functions (Proposition \ref{PropDMTRcontinuousRelax}) also implies the partial identification of a range of $QMTE\left(\tau,v\right)$ across $\tau$ and the $RMTE(v)$, just like before. If one further imposes Assumptions \ref{AsFinite} and \ref{AsSupport}, partial identification results for the MTE function will also follow. For these functions, though, it is important to ensure that the lower and upper bounds in Proposition \ref{PropDMTRcontinuousRelax} are monotone in $y$, which can be enforced using a similar approach as in \citet{Manski2021}. Alternatively, one can use the rearrangement procedure as in \citet{Chernozhukov2009b}.  We state these results as corollaries for convenience.
		
		\begin{corollary}\label{CorQMTEcontinuousRelax}
			Suppose that Assumptions \ref{AsIndependence}-\ref{AsPositive}, \ref{AsRC4} and \ref{AsContinuousRelax} hold. Then, \begin{enumerate}
				\item[(a)] $QMTE\left(\tau,v\right)$ is partially identified for any $v \in \mathcal{P}$ and $\tau \in \left(0, \overline{\tau}\left(v\right)\right)$.
				\item[(b)] the $RMTE(v)$ function is partially-identified for any $v \in \mathcal{P}$.
			\end{enumerate}
		\end{corollary}
		
		\begin{corollary}\label{CorMTEcontinuousRelax}
			If Assumptions Assumptions \ref{AsIndependence}-\ref{AsPositive}, \ref{AsContinuousRelax}, \ref{AsFinite} and \ref{AsSupport} hold, then $MTE\left(v\right)$ is  partially identified for any $v \in \mathcal{P}$.
		\end{corollary}
		
		\subsubsection{Proof of Proposition \ref{PropDMTRcontinuousRelax}}\label{AppProofContinuousRelax}
		
		Fix $d \in \left\lbrace 0, 1 \right\rbrace$, $y  < \gamma_{C}$, $v \in \mathcal{P}$ and $\delta \in \mathbb{R}_{++}$ such that $y + \delta \in \mathcal{C}$.
		
		To derive the upper bound, observe that
		\begin{align*}
			DMTR_{d}\left(y,v\right) & = \mathbb{P}\left[\left.  Y^*(d) \leq y  \right\vert V=v\right] \\
			& \hspace{20pt} \text{by definition} \\
			& \leq \overline{B} + \mathbb{P}\left[\left.  Y^*(d) \leq y  \right\vert C = y + \delta,  V=v\right] \\
			& \hspace{20pt} \text{according to Assumption \ref{AsContinuousRelax}} \\
			& = \overline{B} + \left(2d - 1\right) \cdot \dfrac{\partial \mathbb{P}\left[\left. Y\leq y, D = d \right\vert P\left(Z,C\right) = v, C = y+\delta\right]}{\partial v} \\
			& \hspace{20pt} \text{according to \eqref{EqPYCVcid}} \\
			& = \overline{B} + \left(2d - 1\right) \cdot \gamma_{d}(y,v,y+\delta) \\
			& \hspace{20pt} \text{by definition}.
		\end{align*}
		
		Since the bounds above hold for any $\delta \in \mathbb{R}_{++}$ such that $y + \delta \in \mathcal{C}$, we have that $$DMTR_{d}\left(y,v\right) \leq \overline{B} + \min_{\delta \in \mathcal{D}} \left\lbrace \left(2 \dot d - 1\right) \cdot \gamma_{d}(y,v,y+\delta) \right\rbrace.$$
		
		We can derive the lower bound analogously.
		
		\newpage
		\section{Exploring Average Marginal Treatment Effects}\label{AppAverage}
		
		\setcounter{table}{0}
		\renewcommand\thetable{H.\arabic{table}}
		
		\setcounter{figure}{0}
		\renewcommand\thefigure{H.\arabic{figure}}
		
		\setcounter{equation}{0}
		\renewcommand\theequation{H.\arabic{equation}}
		
		\setcounter{theorem}{0}
		\renewcommand\thetheorem{H.\arabic{theorem}}
		
		\setcounter{assumption}{0}
		\renewcommand\theassumption{H.\arabic{assumption}}
		
		\setcounter{proposition}{0}
		\renewcommand\theproposition{H.\arabic{proposition}}
		
		\setcounter{corollary}{0}
		\renewcommand\thecorollary{H.\arabic{corollary}}
		
		In this appendix, we explore two types of average marginal treatment effects. Section \ref{AppMTEDefinition} defines truncated and untruncated versions of the marginal treatment effect. Section \ref{AppMTEAssumption} states the assumptions that are necessary to identify the untruncated version of the marginal treatment effect function. Section \ref{AppMTEIdentification} describes how to identify both types of marginal treatment effect functions, while Section \ref{AppMTEEstimation} describes how to estimate them. Lastly, Section \ref{AppMTEResults} analyzes our empirical application through the lenses of these types of average marginal treatment effect functions.
		
		\subsection{Definition}\label{AppMTEDefinition}
		
		We define the average marginal treatment response function as
		\begin{eqnarray}
			AMTR_{d}\left( v\right) &\coloneqq& \mathbb{E}\left[\left.Y^*(d) \right\vert V = v\right],\label{EqAMTR}
		\end{eqnarray}
		where $d \in \{0, 1 \}$ and $v \in \left[0, 1\right]$. These counterfactual parameters provide the average time-to-recidivism under treatment $d$ among defendants with punishment resistance $v$.
		
		Based on these counterfactual objects, it is straightforward to define the Average Marginal Treatment Effect function:
		\begin{eqnarray}
			MTE\left(v\right) &\coloneqq& AMTR_{1}\left( v\right) - AMTR_{0}\left( v\right)= \int_{0}^{1} QMTE\left(\tau, v\right) \, d\tau . \label{EqMTE}
		\end{eqnarray}
		We also note that one can express $MTE(v)$ as a function of the $DMTE\left(y,v\right)$,\footnote{This follows from the fact that, for any non-negative random variable $I$, $\expe{I} = \int_{\mathbb{R}_+} (1 - \mathbb{P}[I \leq u])du$.}
		$$MTE(v)= -\int_{\mathbb{R}_+} DMTE\left(y,v\right)dy.$$
		
		Positive values of the MTE functions indicate that punishment by fines and community services increases the defendant's time-to-recidivism compared to no punishment (so treatment is working as intended).
		
		Since we are dealing with a duration outcome subject to right-censoring, it is important to recognize that recovering $MTE(v)$ may be challenging, as it requires identifying the potential outcomes' entire (conditional) counterfactual distribution. To somehow sidestep this limitation, it is common in the survival analysis literature to focus on a restricted version of the mean.\footnote{See, e.g., \citet{Karrison1987}, \citet{Zucker1998}, \citet{Chen2001}, \citet{Zhang2012}, among many others.} Following this rationale, we introduce a restricted version of the $MTE(v)$ function below, though we recognize that it is potentially less interesting than the $MTE(v)$. 
		
		Let $\gamma_{C}$ denote the upper-bound of the support of the censoring variable $C$, that is, $\gamma_{C} \coloneqq \inf\left\lbrace c \in \overline{\mathbb{R}} \colon \mathbb{P}\left[C \leq c\right] = 1 \right\rbrace$.  Define the restricted AMTR function as
		\begin{eqnarray}
			RAMTR_{d}\left( v\right) \coloneqq \mathbb{E}\left[\left. \min\{Y^*(d), \gamma_{C}\} \right\vert V = v\right],\label{EqTAMTR}
		\end{eqnarray}
		and the restricted average marginal treatment effect as 
		\begin{eqnarray}
			RMTE\left(v\right) &\coloneqq& RAMTR_{1}\left( v\right) - RAMTR_{0}\left( v\right).\label{EqRMTE}
		\end{eqnarray}
		
		The RMTE is also connected with the DMTE function: $$RMTE(v)= -\int_{0}^{\gamma_C} DMTE\left(y,v\right)dy,$$
		which follows from the fact that, for a generic non-negative outcome $W$, $\mathbb{E}\left[\left. \min\{W, \gamma_{C}\} \right\vert V = v\right] = \int_{0}^{\gamma_C} \left( 1 - \mathbb{P}(W \leq y | V=v\right) dy$. Of course, if $\gamma_C = \infty$ or if the support of $Y^*(d)$ is contained in the support of $C$ (for the given $v)$, then $RMTE\left(v\right) = MTE\left(v\right)$.

		\subsection{Assumptions}\label{AppMTEAssumption}
		
		Although Assumptions \ref{AsIndependence}-\ref{AsCensoring} are sufficient to identify the RMTE parameter, we require two extra assumptions to identify the MTE function. These additional support restrictions (Assumptions \ref{AsFinite} and \ref{AsSupport}) guarantee the identification of the entire DMTE and QMTE functions, implying that the MTE is also identified.\footnote{The last support assumption may be restrictive in most applications, but we include it for completeness.}
		
		\begin{assumption}[Finite Moments]\label{AsFinite}
			Conditional on $C$, the potential outcome variables have finite first moments, i.e., $\mathbb{E}\left[\left. \left\vert Y\left(d\right) \right\vert \right\vert V = v, C =c \right] < \infty$ for any $d \in \left\lbrace 0, 1 \right\rbrace$, any $v \in \left[0, 1\right]$ and any $c \in \mathcal{C}$.
		\end{assumption}
		
		Assumption \ref{AsFinite} is a regularity condition that allows us to apply standard integration theorems and ensures that average treatment effects are well-defined.
		
		\begin{assumption}[Support Restriction]\label{AsSupport}
			The support of the uncensored potential outcomes is smaller than the support of the censoring variable, i.e., $\gamma_{C} = + \infty$ or $\gamma_{d} < \gamma_{C}$ for any $d \in \left\lbrace 0, 1 \right\rbrace$, where $\gamma_{C} \coloneqq \inf\left\lbrace c \in \overline{\mathbb{R}} \colon \mathbb{P}\left[C \leq c\right] = 1 \right\rbrace$ and $\gamma_{d} \coloneqq \inf\left\lbrace y \in \overline{\mathbb{R}} \colon \mathbb{P}\left[Y^{*}\left(d\right) \leq y\right] = 1 \right\rbrace$ for any $d \in \left\lbrace 0, 1 \right\rbrace$.
		\end{assumption}
		
		Assumption \ref{AsSupport} restricts the support of the potential outcomes of interest to be smaller than the censoring variable's support. In our empirical application, this assumption imposes that all defendants recidivate within ten years, which is the longest observation period in our sample. Formally, this restriction imposes that $\gamma_{d} < \gamma_{C} = 10 \text{ years}$ for any $d \in \left\lbrace 0, 1 \right\rbrace$. This rule out the possibility of defendants not recidivating until they die, and it is therefore not very plausible in our specific contest. We still present results using this assumption as they may be appropriate in empirical contexts different from ours.

		\subsection{Identification}\label{AppMTEIdentification}
		
		In this section, we state two identification results. The first one identifies the restricted marginal treatment effect function using only Assumptions \ref{AsIndependence}-\ref{AsCensoring} and a regularity condition. Its proof is a direct consequence of the Proposition \ref{PropDMTR} and the relationship between distribution functions and expected values of non-negative variables. The second result identifies the average marginal treatment effect function after imposing additional support restrictions. 
		
		\begin{corollary}\label{CorRMTE}
			Suppose that Assumptions \ref{AsIndependence}-\ref{AsCensoring} and Assumption \ref{AsRC4} listed in the Appendix \ref{Ssemiparaconsistency} hold. Then, the $RMTE(v)$ function (Equation \eqref{EqRMTE}) is point-identified for any $v \in \mathcal{P}$.
		\end{corollary}
		
		Notice that, under Assumptions \ref{AsIndependence}-\ref{AsCensoring}, we cannot point-identify the MTE function (Equation \eqref{EqMTE}). The rationale for this ``negative'' result is that we may never observe realizations of $Y^*$ beyond $\gamma_C$ when the support of $C$ is smaller than the support of $Y^*$. 
		In those cases, we cannot identify the right-tail of the distributional marginal treatment response, i.e., we cannot identify $DMTR_{d}\left(y, v\right)$ for $y\geq \gamma_C$. Of course, when the support of $C$ is contained in the support of $Y^*(d)$, this situation does not arise, and we can identify the MTE function as long as it is well-defined. This is precisely what Assumptions \ref{AsFinite} and \ref{AsSupport} impose. We summarize this result in the next corollary.
		
		\begin{corollary}\label{CorMTE}
			Suppose that Assumptions \ref{AsIndependence}-\ref{AsSupport} and Assumption \ref{AsRC4} listed in the Appendix \ref{Ssemiparaconsistency} hold. Then, $MTE\left(v\right)$ is point-identified for any $v \in \mathcal{P}$. 
		\end{corollary}

		\subsection{Estimation and Inference}\label{AppMTEEstimation}

		In this section, we provide algorithms on how to semiparametrically estimate the MTE and RMTE functions based on the identification results described in Corollary \ref{CorMTE}. We also provide practical methods to conduct point-wise inference around these two target parameters.
		
		Using the $DMTE$ estimator described in Algorithm \ref{algo:semi}, we can estimate the $RMTE(v,x)$ function according to 
		\begin{eqnarray}
			\widehat{RMTE}(v,x) &=& -\int_0^{\gamma_C}\widehat{DMTE}(y,v,x)dy \label{RMTE_semi}.
		\end{eqnarray}
		Importantly, this estimator is asymptotically normal as formalized by Theorem \ref{ThmAsymptoticRMTE}
		
		\begin{theorem}\label{ThmAsymptoticRMTE}
			Suppose that Assumptions \ref{AsIndependence}-\ref{AsCensoring} and Assumptions \ref{AsSP1}-\ref{AsRC4} listed in Appendix \ref{Ssemiparaconsistency} hold. Then, as $n \rightarrow \infty$, for each fixed  $v \in \mathcal{P}$, and $x \in \mathcal{X}$, $$\sqrt{n}\left( \widehat{RMTE}(v,x) - {RMTE}(v,x) \right) {\overset{d}{\rightarrow}} N(0,V_{v,x}^{rmte}),$$
			with $\widehat{RMTE}(v,x)$ as defined in Equation \eqref{RMTE_semi} and $V_{v,x}^{rmte}$ as defined in Appendix \ref{Ssemiparaconsistency}.
		\end{theorem}
		
		Although Theorem \ref{ThmAsymptoticRMTE} indicates that one can potentially conduct inference using plug-in estimates of the variance, this procedure would involve estimating additional nuisance functions and could be cumbersome in practice. To avoid this issue, we propose using a weighted bootstrap procedure as in \citet{Ma2005} and \citet{Chen2009}. This bootstrap procedure is very straightforward to implement, as described in the next algorithm.
		
		\begin{algorithm}[Weighted-Bootstrap Implementation]\phantom{a}
			\begin{enumerate}
				\item Estimate DMTE and RMTE according to Algorithm \ref{algo:semi} and Equation \eqref{RMTE_semi}.
				\item Generate $\{\omega_{i}, i=1,\dots,n\}$ as a sequence of independent and identically distributed non-negative random variables with mean one, variance one, and finite third moment (e.g., $\omega_i \sim Exp\left(1\right)$).
				\item Compute the propensity score coefficients associated with Equation \eqref{EqPartiallinear} by minimizing the weighted least squares function, i.e, \begin{equation}
					\hat{\theta}^{fs, *} =\underset{\theta^{fs} \in \Theta^{fs} }{\arg \min }~ n^{-1}\sum_{i=1}^n \omega_i \left( D_i -  {\alpha}_{0} - X_{i}'{\alpha}_{X} -  C_{i}{\alpha}_{C} - \psi^L(Z_i)' {\alpha}_Z,\right)^2
				\end{equation}%
				where $\hat{\theta}^{fs,*} = (\widehat{\alpha}_0^*, \widehat{\alpha}^{*,\prime}_X, \widehat{\alpha}^*_C, \widehat{\alpha}^{*}_Z))'$.
				Denote its trimmed fitted propensity score values by $\widehat{P}_i^*$ as defined in Equation \eqref{EqLogit_trimmed}, but with $\hat{\theta}^{fs, *}$ in place of $\hat{\theta}^{fs}$.
				
				\item Consider the same grid of values for the duration outcome $Y$ as defined in Step 2 of Algorithm \ref{algo:semi}.
				
				\item For each $k \in \left\lbrace 0, \ldots, K \right\rbrace$ and each $d \in \left\lbrace 0, 1 \right\rbrace$, estimate the conditional distribution function of $Y \cdot \mathbf{1}\left\lbrace D = d \right\rbrace$ given $P\left(Z,C\right)$, $C$, and $X$ using the distribution regression model (Equation \eqref{EqDistReg}) with estimated coefficients 
				\begin{equation}
					\hat{\theta}^*\left( y,d\right) =\underset{\theta \in \Theta }{\arg \max }%
					\text{ }\frac{1}{n}\sum_{i=1}^{n} \omega_i~\ln
					\ell_\theta(\mathbf{1}\{Y_i \leq y, D_i=d \}, X_i, C_i, \widehat{P}_i^*; y,d) .
				\end{equation}
				\item Follow Steps 4-9 of Algorithm \ref{algo:semi} and Equation \eqref{RMTE_semi} using $\hat{\theta}^*\left( y,d\right)$ instead of $\hat{\theta}\left( y,d\right)$. Denote by $\widehat{DMTE}^*(y_k,v,x)$ and $\widehat{RMTE}^*(v,x)$ the distributional and restricted marginal treatment effects estimates.
				
				\item Repeat Steps 2-6 $B$ times, e.g., $B=399$, and collect $\left\{  \left(\widehat{DMTE}^*(y_k,v,x)\right)  _{b},b=1\dots,B\right\}$. Do the same for the $\widehat{RMTE}^*(v,x)$.
				
				\item Obtain the $\left(  1-\alpha\right)  $ quantile of $\left\{ \left| \left(\widehat{DMTE}^*(y_k,v,x) - \widehat{DMTE}(y_k,v,x)\right)_{b}\right|,b=1\dots,B\right\}$, $c^{dmte,\ast}(y_k,v,x;{\alpha})$. Compute the analogous critical values based on $\widehat{RMTE}^*(v,x)$.
				
				\item Construct the $1-\alpha$ (pointwise) confidence interval for ${DMTE}(y_k,v,x)$ as $\widehat{C}^{dmte}(y_k,v,x) = [\widehat{DMTE}(y_k,v,x) \pm c^{dmte,\ast}(y_k,v,x;{\alpha})]$. Define $\widehat{C}^{rmte}(v,x;\alpha)$ analogously.
			\end{enumerate}
		\end{algorithm}
		
		The next theorem establishes that the above bootstrap procedure has asymptotically correct coverage.
		
		\begin{theorem}\label{ThmBootRMTE}
			Under the assumptions of Theorem \ref{ThmAsymptoticRMTE}, for any $0<\alpha<1$, and for each $v\in \mathcal{P}$, $x \in \mathcal{X}$, $y < \gamma_C$, and $\tau \in \left( 0,\overline{\tau}(v,x) \right)$, for $n \rightarrow \infty$, $\mathbb{P} \left(RMTE(v,x) \in \widehat{C}^{rmte}(v,x;\alpha) \right) \rightarrow 1 - \alpha$.
		\end{theorem}
		
		Note that the functionals in Theorems \ref{ThmAsymptoticRMTE} and \ref{ThmBootRMTE} provide a covariate-specific treatment effect. In our application's context, we can get court-district-specific RMTE estimates of the effect of fines and community service sentences on time-to-recidivism. However, it may be desirable to further aggregate the MTE functionals as a way to summarize these effects.
		
		As explained at the end of Section \ref{SestSemiPara}, we decided to aggregate the court-district-specific DMTE functionals across court districts using the proportion of cases per court district as weights. Analogously, let $RMTE^{avg}(v) = -\int_0^{\gamma_C}DMTE^{avg}(y,v) dy $ be the average restricted marginal treatment effect. This functional can be straightforwardly estimated using functionals of $$ \widehat{DMTR}_d^{avg}(y,v) = \sum_{x\in \mathcal{X}}\widehat{w}_x~\widehat{DMTR}_d (y,v,x),$$ with  $\widehat{DMTR}_d (y,v,x)$ as in Equation \eqref{EqDMTR_d_estimator}, just like in Equations \eqref{DMTE_semi} and \eqref{RMTE_semi}. Their large-sample properties follow from the delta method and are summarized in the following corollary.
		
		\begin{corollary}\label{CorRMTEaverage}
			Suppose that Assumptions \ref{AsIndependence}-\ref{AsCensoring} and Assumptions \ref{AsSP1}-\ref{AsRC4} listed in Appendix \ref{Ssemiparaconsistency} hold. Then,  as $n \rightarrow \infty$, for each fixed  $v \in \mathcal{P}$, $$\sqrt{n}\left( \widehat{RMTE}^{avg}(v) - {RMTE}^{avg}(v) \right) {\overset{d}{\rightarrow}} N(0,V_{v}^{rmte,avg}).$$
		\end{corollary}
		
		It is also straightforward to construct a weighted-bootstrap confidence interval for these functionals by using $\widehat{w}^*_{x} = n^{-1}\sum_{i=1}^n \omega_i~\mathbf{1}\{X_i=x\}$ as weights for the MTE functionals. We omit a detailed description to avoid repetition.

		\subsection{Empirical Results}\label{AppMTEResults}
		
		\subsubsection{Assessing the plausibility of our assumptions}
		
		The interpretation of the RMTE function relies strongly on Assumption \ref{AsSupport}. If such an assumption is not plausible, we must interpret our results accordingly and avoid jumping from restricted MTE functionals to overall MTE ones. 
		
\begin{figure}[!htbp]
			\centering
			\includegraphics[width = .5 \textwidth]{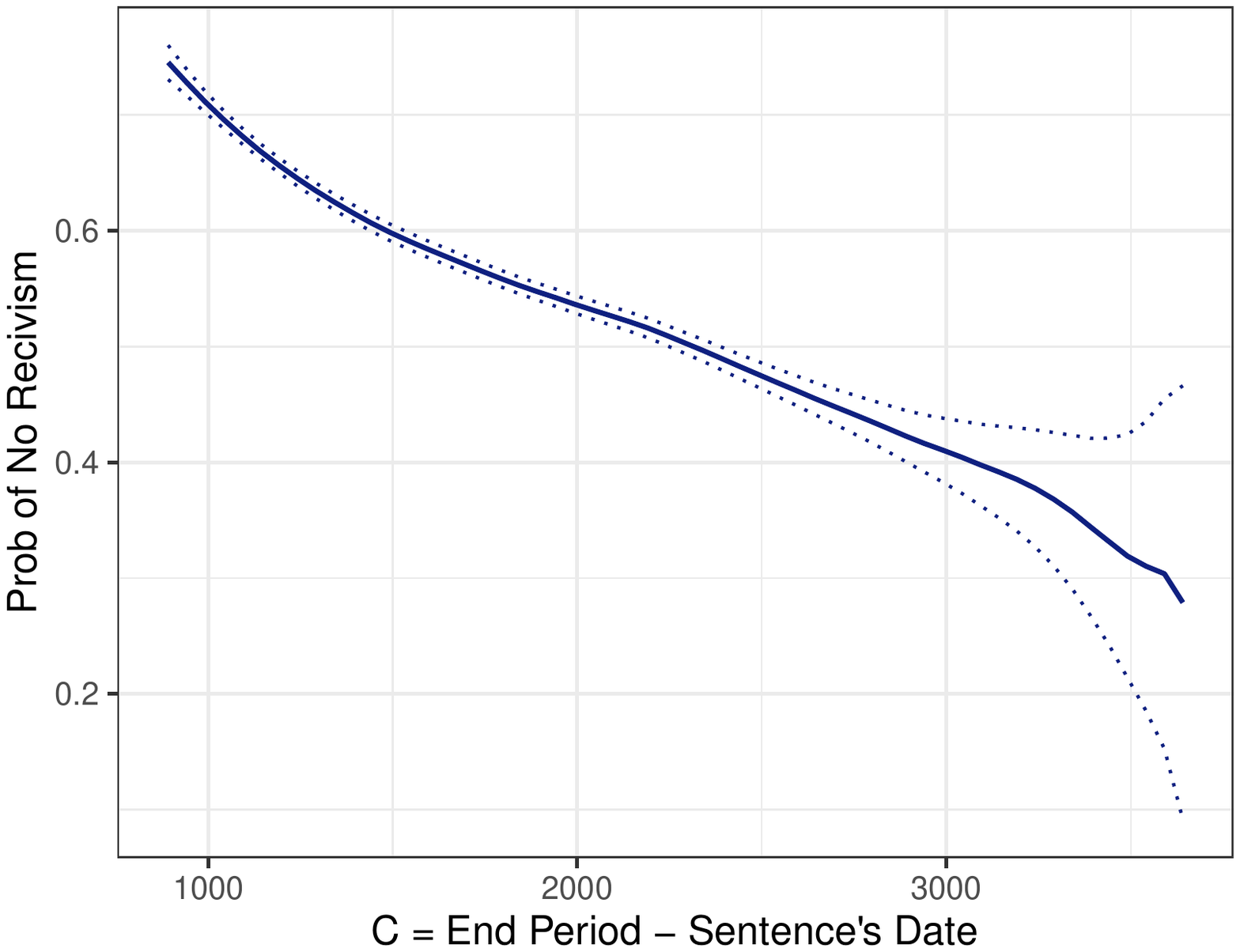}
			\caption{Probability of No Recidivism during the Sampling Period: $\mathbb{P}\left[\left. Y^{*} > C \right\vert C\right]$}
			\label{FigNoRecidivism}
			\justifying
			\hspace{-.6cm}\scriptsize{Notes: Figure \ref{FigNoRecidivism} shows the probability that a defendant does not recidivate during our sampling period given the value of her censoring variable. This nonparametric function was estimated using a local linear regression with an Epanechnikov kernel based on \citet{Calonico2019}. The bandwidth was optimally selected according to the IMSE criterion. The dotted lines are robust bias-corrected 95\% confidence intervals.}
		\end{figure}

		In this subsection, we use descriptive statistics to assess the plausibility of our support restriction.\footnote{The credibility of our other identifying assumptions is discussed in Appendix \ref{Sdescriptive}.} To do so, we analyze how the conditional share of no recidivism varies with the censoring variable. More precisely, Figure \ref{FigNoRecidivism} shows the probability that a defendant does not recidivate during our sampling period, given the value of her censoring variable. Conditioning on the defendants who stay the longest in our sample (large values of $C$), we still find a 30\% probability that they do not recidivate during the observation period. This result suggests that our support restriction in Assumption \ref{AsSupport} is not valid in this context, implying that we should interpret our restricted marginal treatment effect estimates accordingly. In any event, this result does not invalidate our DMTE and QMTE procedures since they rely on Assumptions \ref{AsIndependence}-\ref{AsCensoring} only. It just stresses that censoring is an important complication we must address.
		
		\subsubsection{Estimated RMTE function}\label{SresultsRMTE}
		
		To estimate the RMTE functions in our empirical application, we flexibly account for court district fixed effects. More precisely, we estimate 193 district-specific functions for each of our treatment effect parameters (Theorem \ref{ThmAsymptoticRMTE}). Although very flexible, this strategy makes it challenging to concisely report a summary result. The way we proceeded was to average these district-specific functions over court districts using the proportion of cases per court district as weights, as in Corollary \ref{CorRMTEaverage}. In this section, we report the average RMTE function and compare our proposed methods against standard methods in the literature.
		
\begin{figure}[!htbp]
			\begin{center}
				\includegraphics[width = .5\textwidth]{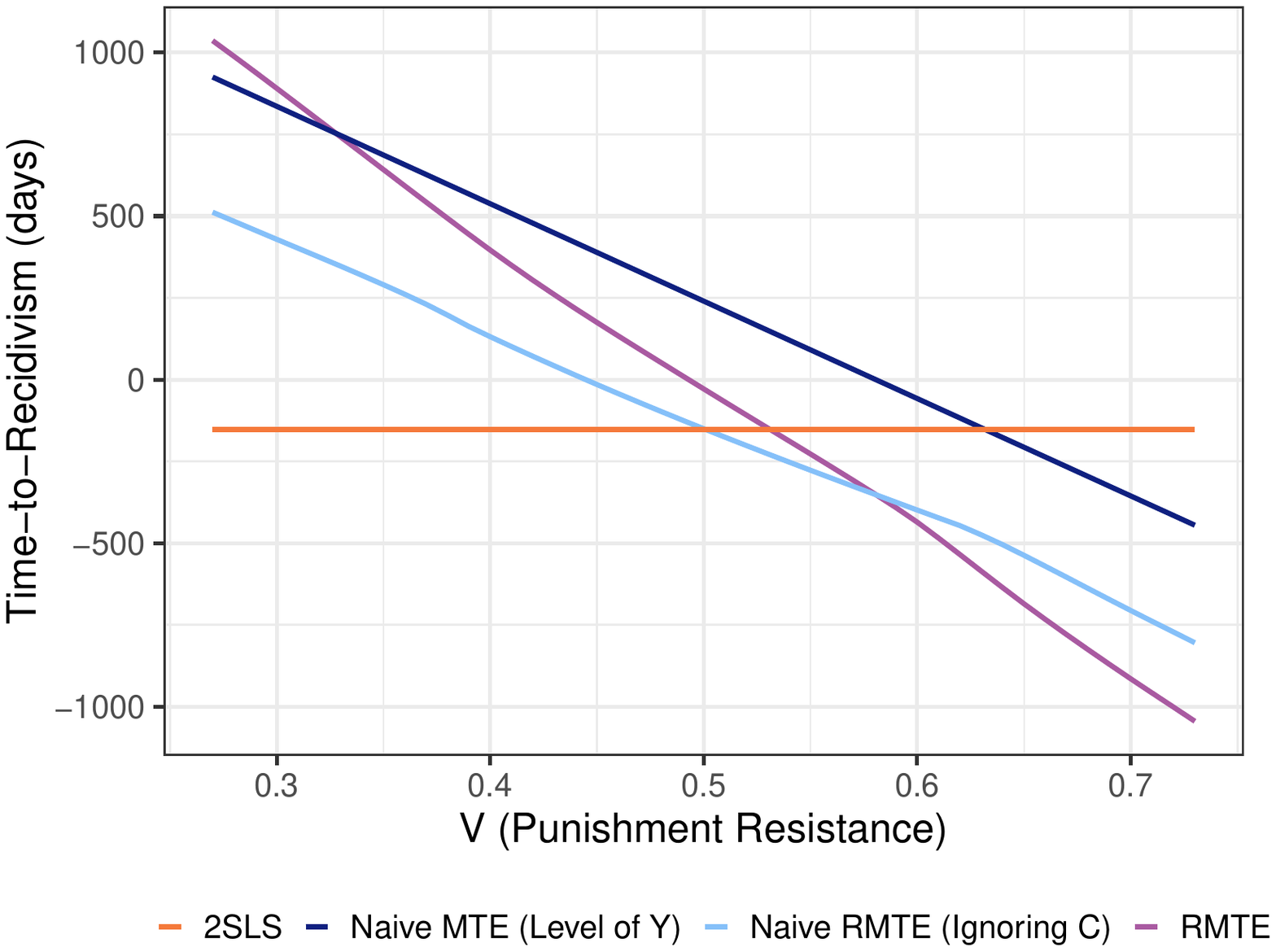}
				\caption{RMTE against Standard Methods}
				\label{RMTEcomparison}
			\end{center}
			\justifying
			\vspace{-.5cm}\scriptsize{Notes: The purple line shows the average RMTE function (Corollary \ref{cor:semi}). The light blue line denotes a naive version of our estimators that ignores censoring. The orange line is the treatment coefficient of a 2SLS regression. The dark blue line is the estimated average MTE function based on a parametric estimator \citep[Appendix B.2]{Cornelissen2016} that directly uses the level of the censored outcome variable.
				
			}
		\end{figure}

		The main advantage of the $RMTE\left(\cdot\right)$ in comparison to the $DMTE$ and $QMTE$ functions is its ability to summarize all results in one single function. Figure \ref{RMTEcomparison} plots the point estimates of this function in purple. Before discussing this result, we must understand how restricted is the RMTE function (Equations \eqref{EqTAMTR} and \eqref{EqRMTE}) compared to the overall MTE function. If the support of $C$ is ``too small'' compared to the support of the time-to-event outcome, RMTE may be further away from the MTE function, affecting its interpretability.
		
\begin{figure}[!htb]
			\begin{center}
				\includegraphics[width = 0.5\textwidth]{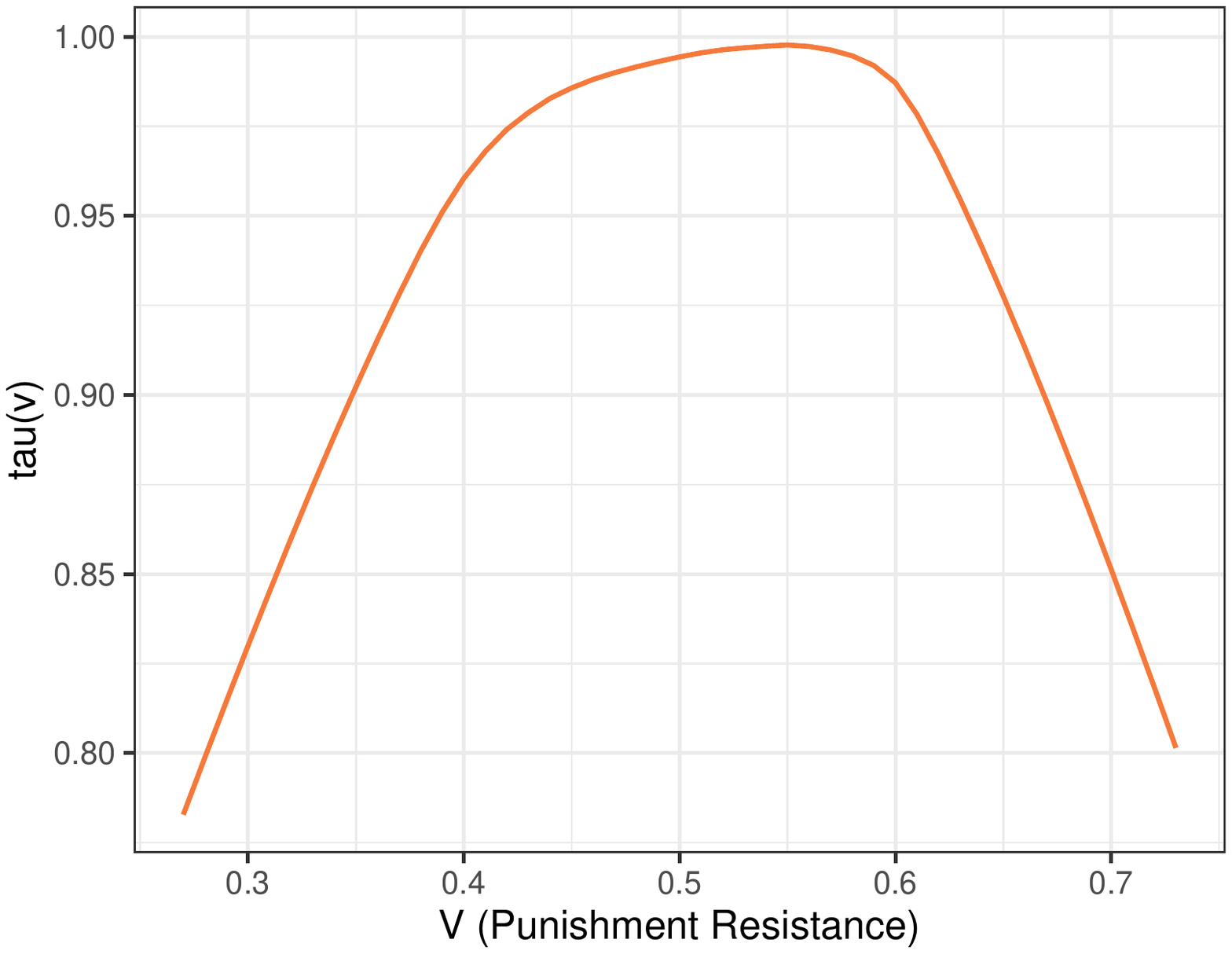}
				\caption{Maximum Identifiable Quantile: $\overline{\tau}\left(v\right)$}
				\label{FigTau}
			\end{center}
			\justifying
			\vspace{-.5cm}\scriptsize{Notes: The orange line plots the estimated maximum identifiable quantile, $\overline{\tau}\left(v\right)$, for each value of the unobserved resistance to treatment. The definition of $\overline{\tau}\left(v\right)$ can be found in Corollary \ref{CorQMTE}.}
		\end{figure}

		Figure \ref{FigTau} plots the estimated maximum identifiable quantile: $\overline{\tau}\left(v\right) \coloneqq \min\left\lbrace \overline{\tau}_{0}\left(v\right), \overline{\tau}_{1}\left(v\right) \right\rbrace$ where $\overline{\tau}_{d}\left(v\right) \coloneqq DMTR_{d}\left(\gamma_{C}, v\right)$ for any $d \in \left\lbrace 0, 1 \right\rbrace$ and $\gamma_{C} \coloneqq \inf\left\lbrace c \in \overline{\mathbb{R}} \colon \mathbb{P}\left[C \leq c\right] = 1 \right\rbrace$. If the maximum identifiable quantile, $\overline{\tau}\left(\cdot\right)$ is ``far away'' from 1, then the support of $C$ is ``small'' compared to the support of time-to-event outcome.
		
		Figure \ref{FigTau} shows that the RMTE function is almost an unrestricted mean for $v \in \left(.5,.6\right)$. However, the censoring problem is binding for small and large values of the unobserved resistance to treatment. Consequently, the RMTE is further away from the MTE function for extreme values of punishment resistance.
		
		Now, analyzing the point estimates of the RMTE function (purple line in Figure \ref{RMTEcomparison}), we find that the estimated restricted average marginal treatment effects decrease with the unobserved resistance to treatment. This result is statistically significant at the 10\% significance level according to Figure \ref{RMTE-CI}). This result is similar to the QMTE results in Section \ref{SresultsQMTE}.
		
		Furthermore, Figure \ref{RMTEcomparison} compares our proposed methods against other available methods in the literature. Differently from our approach, these estimates ignore that the outcome variable is right-censored and provide different conclusions when compared against our proposed estimator. The light blue line denotes a ``naive'' version of our estimator that follows the same steps as described in Section \ref{SestSemiPara} but does not condition on the censoring variable. The orange line is the treatment coefficient of a two-stage least squares (2SLS) regression that uses the censored outcome variable as the left-hand side variable, controls for court district fixed effects, and uses the judge's punishment rate as the instrument for the defendant being punished. The dark blue line is the estimated average MTE function based on a parametric estimator \citep[Appendix B.2]{Cornelissen2016} that imposes a linear MTE curve and ignores censoring concerns by directly using the level of the censored outcome variable.
		
		Figure \ref{RMTEcomparison} highlights three interesting results. First, the standard MTE method's estimates (dark blue line) tend to be larger than the estimated $RMTE$ function that accounts for censoring (purple line). Second, the naive estimator (light blue line) finds a $RMTE$ function that is less steep. Importantly, both blue lines do not lie entirely within the 90\%-confidence intervals of the estimated $RMTE$ function that accounts for censoring (Figure \ref{RMTE-CI}). Third, the 2SLS estimate (orange line) finds a negative effect of punishment on time-to-recidivism, suggesting that punishing defendants with fines and community service has led to faster recidivism. This 2SLS result does not capture that punishing defendants with alternative sentences may increase time-to-recidivism for some defendant types, as suggested by our preferred $RMTE$ estimate (purple line). Importantly, the 2SLS estimate does not lie entirely within the 90\%-confidence intervals of the estimated $RMTE$ function (Figure \ref{RMTE-CI}).

		\newpage

		\section{What can we identify if we treat the censoring variable as a common covariate?}\label{AppModelJustification2}
		
		\setcounter{table}{0}
		\renewcommand\thetable{I.\arabic{table}}
		
		\setcounter{figure}{0}
		\renewcommand\thefigure{I.\arabic{figure}}
		
		\setcounter{equation}{0}
		\renewcommand\theequation{I.\arabic{equation}}
		
		\setcounter{theorem}{0}
		\renewcommand\thetheorem{I.\arabic{theorem}}
		
		\setcounter{assumption}{0}
		\renewcommand\theassumption{I.\arabic{assumption}}
		
		\setcounter{proposition}{0}
		\renewcommand\theproposition{I.\arabic{proposition}}
		
		\setcounter{corollary}{0}
		\renewcommand\thecorollary{I.\arabic{corollary}}
		
		In this appendix, we discuss which object is identified when we treat the censoring variable as a common covariate. More specifically, we analyze the local instrumental variable estimand that directly conditions on the censoring variable and uses the censored outcome variable as its left-hand-side variable. We show that this estimand has no clear causal interpretation. Moreover, we show that focusing on the distribution of the censored outcome variable through this local instrumental variable estimand achieves a causally interpretable parameter, but it ignores the integration step present in Proposition \ref{PropDMTR}. These two results should serve as a cautionary tale for researchers to take special care when considering the censoring variable.
		
		Under our assumptions, one could consider the censoring variable as a covariate $C$ affecting an outcome $Y$. In this case, we define $Y(d,C) \coloneqq \min\{Y^*(d),C\}$ and impose the following model:
		\begin{align*}
			Y & \coloneqq Y(1,C) \cdot D+Y(0,C) \cdot (1-D) \\
			D & \coloneqq \mathbf{1}\left\lbrace P\left(Z, C\right) \geq V \right\rbrace.
		\end{align*}
		
		Following steps similar to the ones used to prove Equations \eqref{EqPYD1A6derivative} and \eqref{EqPYD0A6derivative}, we can show the following result for the local instrumental variable estimand that treats the censoring variable as a common covariate:
		\begin{equation*}
			\frac{\partial \mathbb{E}\left[\left. Y  \right\vert P(Z,C) = v, C = c\right]}{\partial v} =  \mathbb{E}\left[\left.Y(1,c)-Y(0,c)  \right\vert V = v\right].
		\end{equation*}
		To interpret this estimand, we substitute $Y(d,c)=\min\{Y^*(d),c\}$ into the right-hand side of this equation to find:
		\begin{align*}
			& \frac{\partial \mathbb{E}\left[\left. Y  \right\vert P(Z,C) = v, C = c\right]}{\partial v} \\
			& \hspace{20pt} = \mathbb{E}\left[\min\{Y^*(1),c\} - \min\{Y^*(0),c\} \right\vert V = v] \\
			& \hspace{20pt} = \mathbb{E}\left[\min\{Y^*(1),c\} \right\vert V = v]  - \mathbb{E}\left[\min\{Y^*(0),c\} \right\vert V = v] \\
			& \hspace{60pt} \text{by linearity of the expectation operator} \\
			& \hspace{20pt} = c \cdot \mathbb{P}\left[Y^*(1)>c|V=v \right] +  \mathbb{E}\left[Y^*(1) \right\vert V = v, Y^*(1)\leq c ] \cdot \mathbb{P}\left[Y^*(1)\leq c|V=v\right] \\
			& \hspace{40pt} - c \cdot \mathbb{P}\left[Y^*(0)>c|V=v\right] -  \mathbb{E}\left[Y^*(0) \right\vert V = v, Y^*(0)>c ] \cdot \mathbb{P}\left[Y^*(0)\leq c|V=v\right] \\
			& \hspace{60pt} \text{by Law of Iterated Expectations} \\
			& \hspace{20pt} = \mathbb{E}\left[Y^*(1) \right\vert V = v, Y^*(1)\leq c ] \cdot \mathbb{P}\left[Y^*(1)\leq c|V=v\right] \\
			& \hspace{40pt} - \mathbb{E}\left[Y^*(0) \right\vert V = v, Y^*(0)>c ] \cdot \mathbb{P}\left[Y^*(0)\leq c|V=v\right] \\
			& \hspace{40pt} + c \cdot \left\lbrace \mathbb{P}\left[Y^*(1)>c|V=v \right] - \mathbb{P}\left[Y^*(0)>c|V=v\right] \right\rbrace.
		\end{align*}
		Hence, we can conclude that the local instrumental variable estimand that treats the censoring variable as a common covariate --- $\frac{\partial \mathbb{E}\left[\left. Y  \right\vert P(Z,C) = v, C = c\right]}{\partial v}$ --- does not have a causal interpretation in general. 
		
		Since our strategy focuses on the distribution marginal treatment effect instead of the marginal treatment effect, we also analyze what would happen if one focused on the distribution of the censored outcome variable through a local instrumental variable estimand that handles the censoring variable as a common covariate. In this case, we follow the same steps used to prove Equations \eqref{EqPYD1A6derivative} and \eqref{EqPYD0A6derivative} to find that, for any $c > y$,
		\begin{align*}
			& \frac{\partial \mathbb{P}\left[\left. Y \leq y  \right\vert P(Z,C) = v, C = c\right]}{\partial v} \\
			& \hspace{20pt} = \mathbb{P}[\min\{Y^*(1),c\}\leq y\vert V = v]-\mathbb{P}[\min\{Y^*(0),c\}\leq y\vert V = v], \\
			& \hspace{20pt} = \mathbb{P}[Y^*(1) \leq y \vert V = v] - \mathbb{P}[Y^*(0) \leq y\vert V = v],
		\end{align*}
		which is a causally interpretable parameter.
		
		However, this last approach ignores that this identification result holds for any $c > y$. As a consequence, it does not integrate over values of the censoring variable satisfying $C > y$ as we do in Proposition \ref{PropDMTR}. This integration step is key in right-censoring problems because it fully leverages the problem's duration structure and the random censoring restriction (Assumption \ref{AsCensoring}). We also emphasize that this integration step is fundamental in \citeauthor{Frandsen2015}'s (\citeyear{Frandsen2015}) approach to identifying the local average treatment effect with duration outcomes. 
		
		Acknowledging that this identification result holds for any $c > y$ has important consequences for estimation and testing. Our estimation method takes this integration step into account in Equation \eqref{EqDMTR_d_estimator}, possibly increasing the efficiency of our estimation algorithm. For testing, this argument appears in Equation \eqref{EqTesting}. Since the left-hand side of this equation does not depend on $C$ while its right-hand side depends on the censoring variable, it is possible to jointly test our identification assumptions by comparing the right-hand side across different values of $C$. We implement this test in Figure \ref{FigDMTE_censoring}.
		
		\newpage
		
		\section{Controlling for Case Processing Time as an Extra Covariate}\label{AppCaseProcessingTime}
		
		\setcounter{table}{0}
		\renewcommand\thetable{J.\arabic{table}}
		
		\setcounter{figure}{0}
		\renewcommand\thefigure{J.\arabic{figure}}
		
		\setcounter{equation}{0}
		\renewcommand\theequation{J.\arabic{equation}}
		
		\setcounter{theorem}{0}
		\renewcommand\thetheorem{J.\arabic{theorem}}
		
		\setcounter{assumption}{0}
		\renewcommand\theassumption{J.\arabic{assumption}}
		
		\setcounter{proposition}{0}
		\renewcommand\theproposition{J.\arabic{proposition}}
		
		\setcounter{corollary}{0}
		\renewcommand\thecorollary{J.\arabic{corollary}}
		
		{
			
			When we think of our judicial setting as a natural experiment where the outcome is a duration variable, we have to decide whether we should start measuring time when the experiment begins (i.e., case enters the court system) or when the individual receives the treatment (i.e., when the defendant is sentenced). In most applications, this question is irrelevant because these two options are identical. However, in our application, these two ways to measure time differ because judges spend different amounts of time analyzing each case. Consequently, case processing time varies across case-defendant pairs.
			
			We believe the most appropriate approach is to start measuring time when the defendant is sentenced. Since potential outcomes are defined by sentence decisions, the natural way to measure time-to-recidivism --- $Y\left(1\right)$ and $Y\left(0\right)$ --- is using the sentencing date as the starting date. This is our choice throughout the entire text. Doing otherwise (measuring time-to-recidivism using case entry date as the starting point) would allow potential outcomes to differ even before treatment was known, which is inconsistent with the treatment's definition.
			
			However, we recognize that longer case-processing times may impact defendants differently. To account for this possible heterogeneity, we redo our entire analysis controlling for case processing times as an extra covariate. In this appendix, Section \ref{AppCaseProcessingTimeDescriptive} presents descriptive statistics related to case-processing times, Section \ref{AppCaseProcessingTimeEstimation} explains the changes made to the estimation procedure to account for case-processing time as an extra covariate, Section \ref{AppCaseProcessingTimeMainResults} discusses the estimates of our main target parameters controlling for case-processing time as an extra covariate, and Section \ref{AppCaseProcessingTimeIdentification} presents our main exogeneity test controlling for case-processing time as an extra covariate. Importantly, all results in the main text are similar to the results in this appendix.
			
			\subsection{Descriptive Statistics about Case-processing Times}\label{AppCaseProcessingTimeDescriptive}
			
			This section defines case-processing time for each case-defendant pair and presents descriptive statistics about this variable.
			
			There are four key calendar dates in our empirical setting. The first one, $t_{1,i}$, is the day a case enters the judicial court system and a judge is assigned to preside over it. The second one, $t_{2,i}$, is the day the presiding judge makes a decision on the case and sentences the defendant. The third one, $t_{3,i}$, is the day the defendant recidivates if recidivism occurs at all. The fourth one, $t_{4}$, is the day we finish collecting our data. We have that the first three dates differ for each case-defendant pair $i$, while the fourth one is common to all case-defendant pairs. Our censoring variable is defined as $C_{i} \coloneqq t_{4} - t_{2,i}$ while our possibly censored outcome is defined as $Y^{*}_{i} \coloneqq t_{3,i} - t_{2,i}$. Consequently, our observed outcome is defined as $Y_{i} \coloneqq \min\left\lbrace t_{3,i} - t_{2,i}, t_{4} - t_{2,i} \right\rbrace$. Based on these calendar dates, we can also define case-processing time: $T_{i} \coloneqq t_{2,i} - t_{1,i}$.
			
			Case-processing times vary by case-defendant pairs and by judge. First, we analyze the distribution of case-processing times across case-defendant pairs. Table \ref{TabDescriptiveProcessing} shows descriptive statistics for $T_{i}$. It shows this variable's mean, 1\textsuperscript{st} decile, 1\textsuperscript{st} quartile and median for all defendants, for the defendants who were punished (treated group), and for the defendants who were not punished (comparison group). It also shows the sample size of each of these three groups. The comparison between the treated and comparison groups shows that judges are faster when they decide to punish a defendant than when they decide not to punish. This finding is due to the fact that non-prosecution agreements are faster than other types of decisions.
			
			\begin{table}[!htb]
				\centering
				\caption{{Descriptive Statistics --- Case-processing Time}} \label{TabDescriptiveProcessing}
				\begin{lrbox}{\tablebox}
					\begin{tabular}{cccc}
						\hline \hline
						& Unconditional & Treated Group & Comparison Group \\ \hline
						Mean & 900 & 718 & 1088 \\
						1\textsuperscript{st} Decile & 209 & 173 & 302 \\
						1\textsuperscript{st} Quartile & 409 & 312 & 617 \\
						Median & 812 & 590 & 1063 \\
						Number of Observations & 43,468 & 22,060 & 21,408 \\ \hline
					\end{tabular}
				\end{lrbox}
				\usebox{\tablebox}\\
				
				\justifying
				\scriptsize{\hspace{-.6cm}Note: The treated group receives a punishment, i.e., its defendants were fined or sentenced to community services because they were either convicted or signed a non-prosecution agreement. The comparison group did not receive a punishment, i.e., its defendants were acquitted or its cases were dismissed. The case-processing time variable measures the number of days between the case's entry date and the case's final ruling's date. An observation is a case-defendant pair.}
			\end{table}
			
			Lastly, we analyze the average case-processing time for each judge. Figure \ref{FigProcessing} plots the histogram of the average case-processing time for each judge in our full sample. It shows that some judges analyze cases much faster than other judges even though our sample only includes misdemeanor cases. 
			
			\begin{figure}[!ht]
				\begin{center}
					\includegraphics[width = .55\textwidth, keepaspectratio]{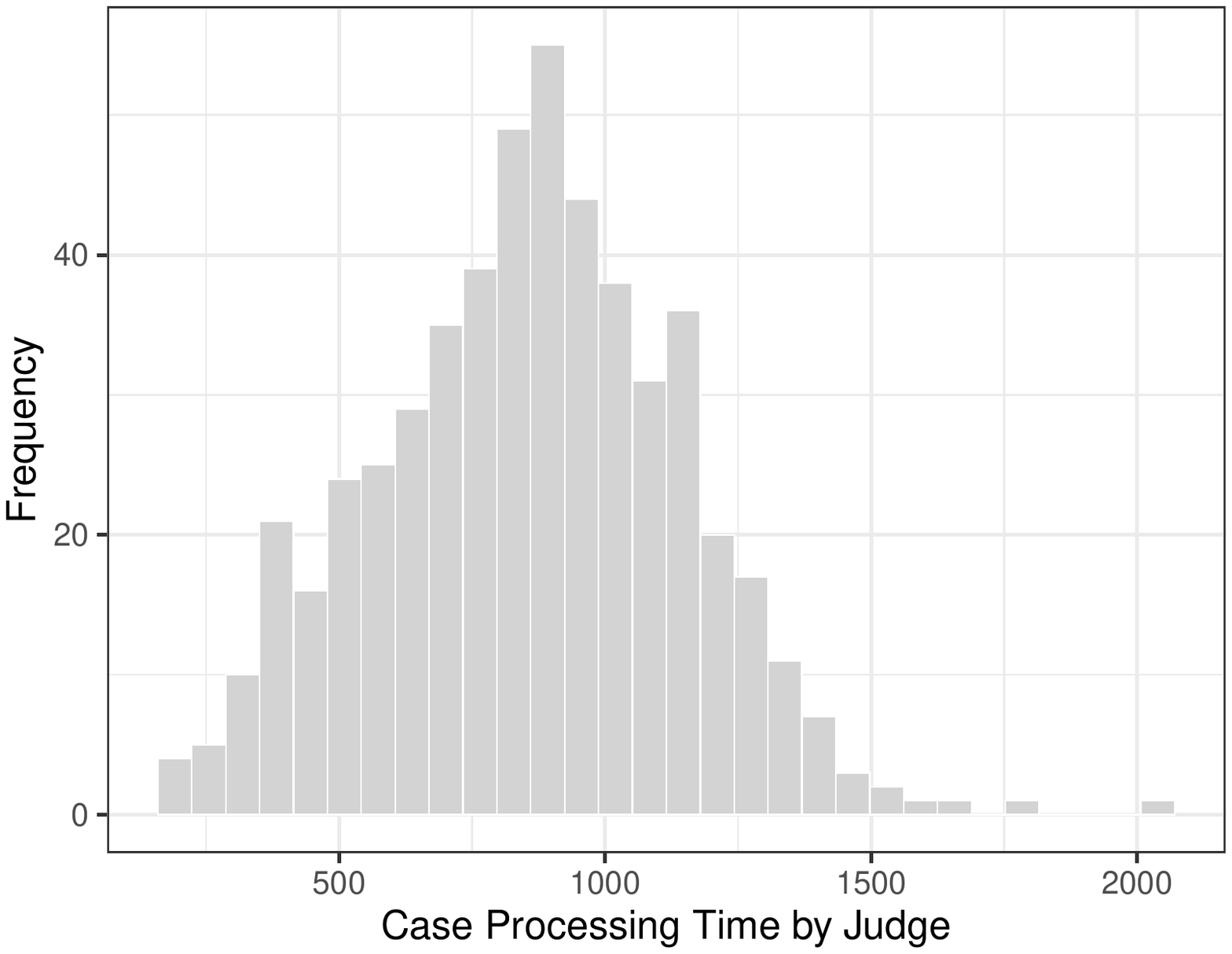}
					\caption{PDF of the Uncensored Outcome given the Defendant's Cohort}
					\label{FigProcessing}
				\end{center}
				\justifying
				\scriptsize{Notes: This figure plots the histogram of the average case-processing time for each judge in our full sample.}
			\end{figure}
			
			\subsection{Estimation Procedure when Controlling for Case-processing Time}\label{AppCaseProcessingTimeEstimation}
			
			This appendix's goal is to estimate the target parameters described in Section \ref{Squestion} while controlling for case-processing times. To include this extra covariate, we have to make small changes in the estimation procedure described in Section \ref{SestSemiPara}. The current section explains these changes.
			
			First, we have to estimate the conditional distribution function of $Y \cdot \mathbf{1}\left\lbrace D = d \right\rbrace$ given $P, C, X, T$ for $d \in \{0,1\}$ in Equation \eqref{EqDistReg} while controlling for case-processing time. To do so, we impose the following model:
			\begin{align}
				& {\Gamma}({P},C,X;y,d,T) \nonumber \\
				& \hspace{20pt} \coloneqq \mathbb{E}\left[ \mathbf{1}\left\lbrace Y \leq y, D = d \right\rbrace\vert P, C,X,T\right] \nonumber \\
				&\hspace{20pt} = \Lambda \left(	\beta_0 \left(y,d\right) +X^{\prime }\beta_{X}(y,d) + C \beta_C(y,d) + P \beta_P (y,d) + T \beta_T (y,d) \right) 
				\text{ a.s.}  \label{EqDistRegProcessing}
			\end{align}
			where $\theta_0(\cdot, \cdot) = (\beta_0 \left(\cdot,\cdot\right), \beta_X \left(\cdot,\cdot\right)', \beta_C \left(\cdot,\cdot\right), \beta_P \left(\cdot,\cdot\right), \beta_T \left(\cdot,\cdot\right))' \mapsto \Theta
			\subseteq \mathbb{R}^{4+k_X}$ is a vector of nonparametric functions, $k_X$ is the dimension of $X$, and $\Lambda$ is a known link function. For concreteness,  we focus on a logistic link function, $\Lambda(\cdot) = \exp(\cdot)\big{/}(1 + \exp(\cdot))$.
			
			Second, note that the derivative of $\Gamma$ (Equation \eqref{EqDistRegProcessing}) with respect to $P$ depends on case-processing time: 
			\begin{eqnarray}
				\gamma_{d}(y,v,c,x,t) &=& \beta_{P}(y,d) \cdot {\Gamma}(v,c,x;y,d,t) \cdot (1 - {\Gamma}(v,c,x;y,d,t)),\label{Eq_gammaProcessing}
			\end{eqnarray}
			because $\Lambda$ is the logistic link function. Denote the estimated fitted values of $\gamma_{d}(y,v,c,x,t)$ by $\widehat{\gamma}_{d}(y,v,c,x,t)$.
			
			Third, we need to estimate the distribution marginal response function for each court-district. To do so, let $n_{d,x,y} = \sum_{i=1}^n \mathbf{1}\{D_i=d, X_i=x, C_{i} > y \}$ denote the sample size with treatment status $d$, covariate value $x$, and censoring variable above $y$. Our proposed estimator for $DMTR_d (y,v,x)$ is given by 
			\begin{eqnarray}
				\widehat{DMTR}_d (y,v,x) = (2d - 1) \dfrac{\sum_{i=1}^n \mathbf{1}\{D_i=d, X_i=x, C_{i} > y \}~ \widehat{\gamma}_{d}(y,v,C_i,x,T_{i})}{n_{d,x,y}}. \label{EqDMTR_d_estimatorProcessing}
			\end{eqnarray}
			Consequently, $\widehat{DMTR}_d (y,v,x)$ estimates the average distributional marginal treatment response for each duration spell $y$, punishment resistance $v$ and court-district $x$ over the distribution of censoring values $C_{i}$ above $y$ and case-processing times $T_{i}$.\footnote{Alternatively, we could take the summation in Equation \eqref{EqDMTR_d_estimatorProcessing} after fixing the case-processing time in a specific value, e.g., its mean.}
			
			Based on Equation \eqref{EqDMTR_d_estimatorProcessing}, we can then estimate $DMTE(y,v,x) \coloneqq DMTR_1(y,v,x) - DMTR_0(y,v,x)$ using \begin{eqnarray}
				\widehat{DMTE}(y,v,x) \coloneqq \widehat{DMTR}_1(y,v,x) - \widehat{DMTR}_0(y,v,x). \label{DMTE_semiProcessing}
			\end{eqnarray}
			Analogously, one can estimate $QMTE(\tau,v,x)$ functionals using 
			\begin{eqnarray}
				\widehat{QMTE}(\tau,v,x) &=& \widehat{QMTR}_1(\tau,v,x) - \widehat{QMTR}_0(\tau,v,x), \label{QMTE_semiProcessing}
			\end{eqnarray}
			where $\widehat{QMTR}_d(\tau,v,x) = \inf\{y\in \mathbb{R}_+ \colon \widehat{DMTR}_d(y,v,x)\geq \tau \}$.
			
			Fourth, we need to aggregate these parameters over court-districts. To do so, we use the proportion of cases per court district as weights.
			Let $\widehat{w}_{x} = n^{-1}\sum_{i=1}^n \mathbf{1}\{X_i=x\}$ be the plug-in estimator of the proportion of cases assigned to a court district $x$. For each $d \in \{0,1\}$, $y \in \mathcal{Y}$ and $v\in \mathcal{P}$, let $$ \widehat{DMTR}_d^{avg}(y,v) = \sum_{x\in \mathcal{X}}\widehat{w}_x~\widehat{DMTR}_d (y,v,x),$$ with  $\widehat{DMTR}_d (y,v,x)$ as in Equation \eqref{EqDMTR_d_estimatorProcessing}. After this step, we compute
			\begin{equation}
				\label{EqDMTEaverageProcessing}
				\widehat{DMTE}^{avg}(y,v) = \widehat{DMTR}_1^{avg}(y,v) - \widehat{DMTR}_0^{avg}(y,v)
			\end{equation}
			and
			\begin{equation}
				\label{EqDQTEaverageProcessing}
				\widehat{QMTE}^{avg}(\tau,v) = \widehat{QMTR}_1^{avg}(\tau,v) - \widehat{QMTR}_0^{avg}(\tau,v),
			\end{equation}
			where $ \widehat{QMTR}_{d}^{avg}\left(\tau, v\right) \coloneqq \inf\{y\in \mathbb{R}_+ \colon \widehat{DMTR}_d^{avg}(y,v)\geq \tau \}.$
			
			\subsection{Main Results when Controlling for Case-processing Time}\label{AppCaseProcessingTimeMainResults}
			
			In this section, we present our estimated results after controlling for case processing time. To estimate the average DMTE and QMTE functions, we use the procedure detailed in Appendix \ref{AppCaseProcessingTimeEstimation}. Importantly, the results in this appendix are very similar to those in Section \ref{Sresults}, implying that our results are robust to controlling for case-processing time as an additional covariate.
			
			Figure \ref{FigDMTEprocessing} shows the estimated average $DMTE\left(y,\cdot\right)$ functions for $y \in \left\lbrace 1, 2, \ldots, 8 \right\rbrace$ after we control for case-processing time. This figure uses the same y-axis as Figures \ref{DMTE1-4} and \ref{DMTE5-8} to facilitate the comparison of our main results with the estimates that control for case-processing time. When we compare these two figures, we find that the magnitude and shape of our estimated functions are not affected by case-processing time as an additional covariate. Moreover, when we compare Figure \ref{FigDMTEprocessing} against the confidence bands reported in Figures \ref{FigDMTE-CI-1-4} and \ref{FigDMTE-CI-5-8}, we find that the functions reported in Figure \ref{FigDMTEprocessing} are inside the point-wise confidence bands of our main results.
			
			\begin{figure}[!htbp]
				\begin{center}
					\begin{subfigure}[t]{0.47\textwidth}
						\centering
						\includegraphics[width = \textwidth]{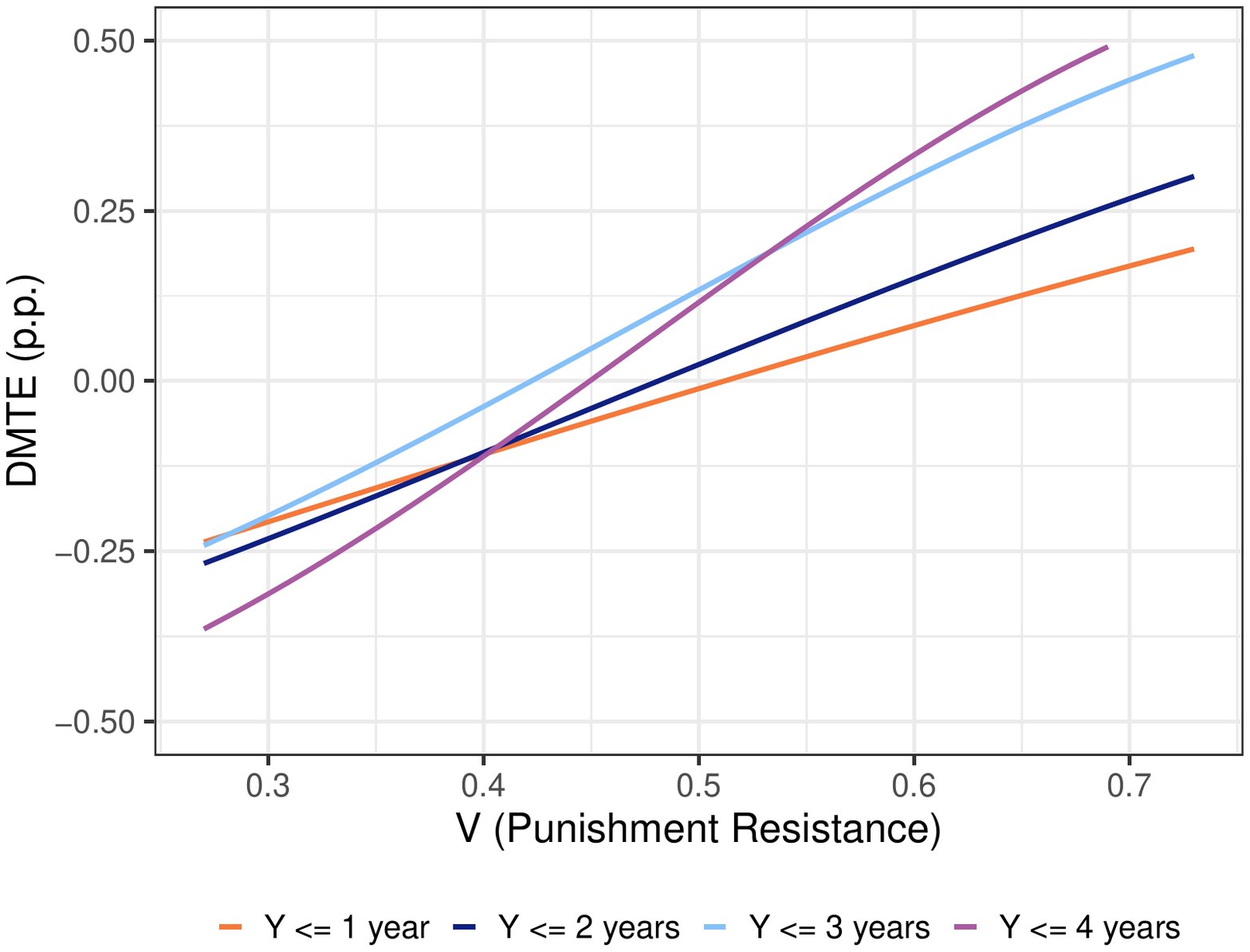}
						\caption{$DMTE\left(y,\cdot\right)$ for $y \in \left\lbrace 1, 2, 3, 4 \right\rbrace$}
						\label{DMTE1-4processing}
					\end{subfigure}
					\hfill
					\begin{subfigure}[t]{0.47\textwidth}
						\begin{center}
							\includegraphics[width = \textwidth]{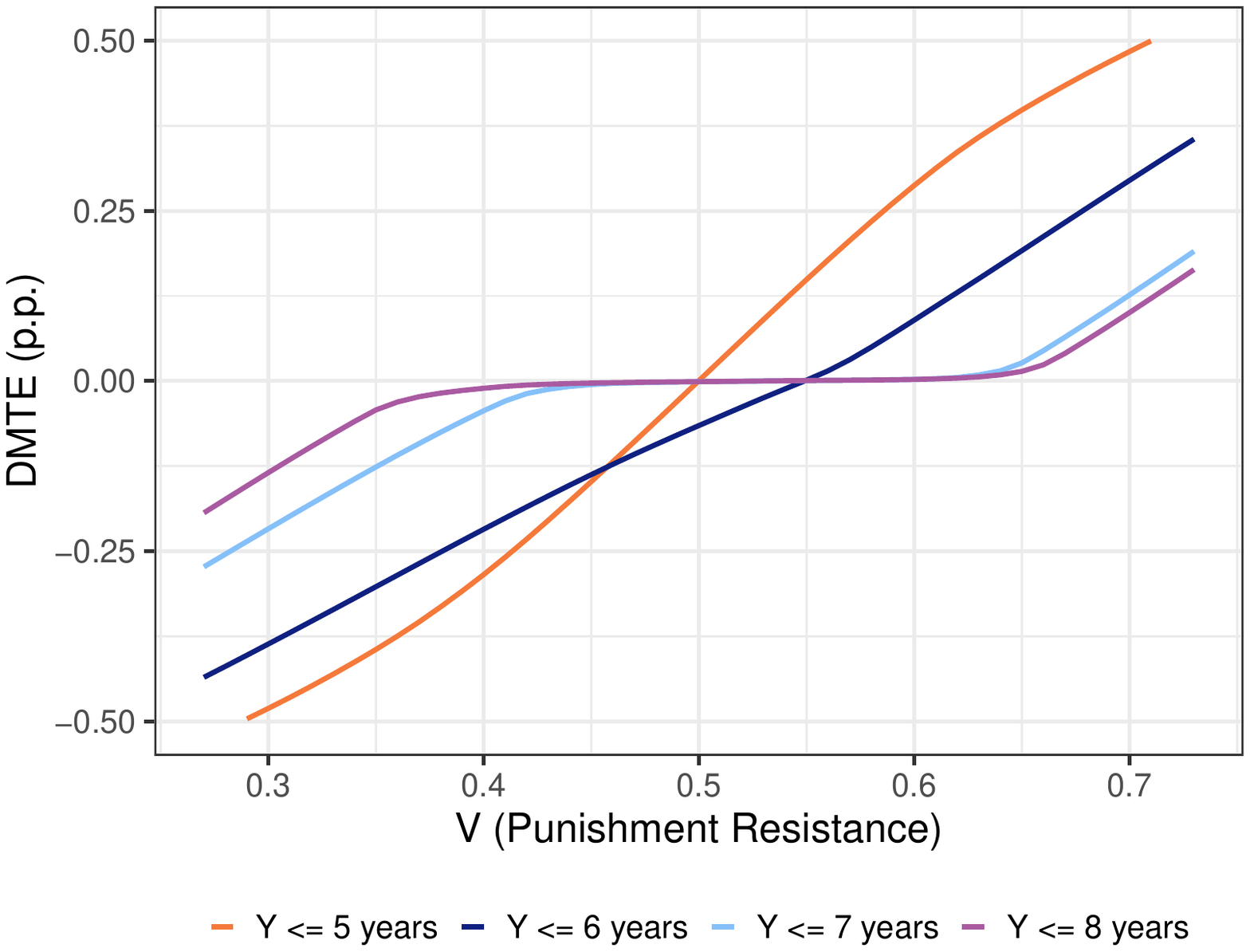}
							\caption{$DMTE\left(y,\cdot\right)$ for $y \in \left\lbrace 5, 6, 7, 8 \right\rbrace$}
							\label{DMTE5-8processing}
						\end{center}
					\end{subfigure}
					\caption{$DMTE\left(y,\cdot\right)$ for $y \in \left\lbrace 1, 2, \ldots, 8 \right\rbrace$}
					\label{FigDMTEprocessing}
				\end{center}
				\justifying
				\scriptsize{Notes: Solid lines are the point estimates for the average $DMTE\left(y,\cdot\right)$ functions indicated in the legend of each subfigure. These results are based on the estimation procedure explained in Appendix \ref{AppCaseProcessingTimeEstimation}.
					
				}
			\end{figure}
			
			Figure \ref{FigQMTE-RMTEProcessing} shows the estimated average QMTE$\left(\tau,\cdot\right)$ functions for $\tau \in \left\lbrace .10, .15, .25, .30, .40, .50, .75 \right\rbrace$ after we control for case-processing time. This figure uses the same y-axis as Figures \ref{QMTE10-30} and \ref{QMTE40-75-RMTE} to facilitate the comparison of our main results with the estimates that control for case-processing time. When we compare these two figures, we find that the magnitude and shape of our estimated functions are not affected by case-processing time as an additional covariate. Moreover, when we compare Figure \ref{FigQMTE-RMTEProcessing} against the confidence bands reported in Figures \ref{FigQMTE-CI-10-30} and \ref{FigQMTE-RMTE-CI}, we find that the functions reported in Figure \ref{FigQMTE-RMTEProcessing} are inside the point-wise confidence bands of our main results.
			
			\begin{figure}[!htb]
				\begin{center}
					\begin{subfigure}[t]{0.47\textwidth}
						\centering
						\includegraphics[width = \textwidth]{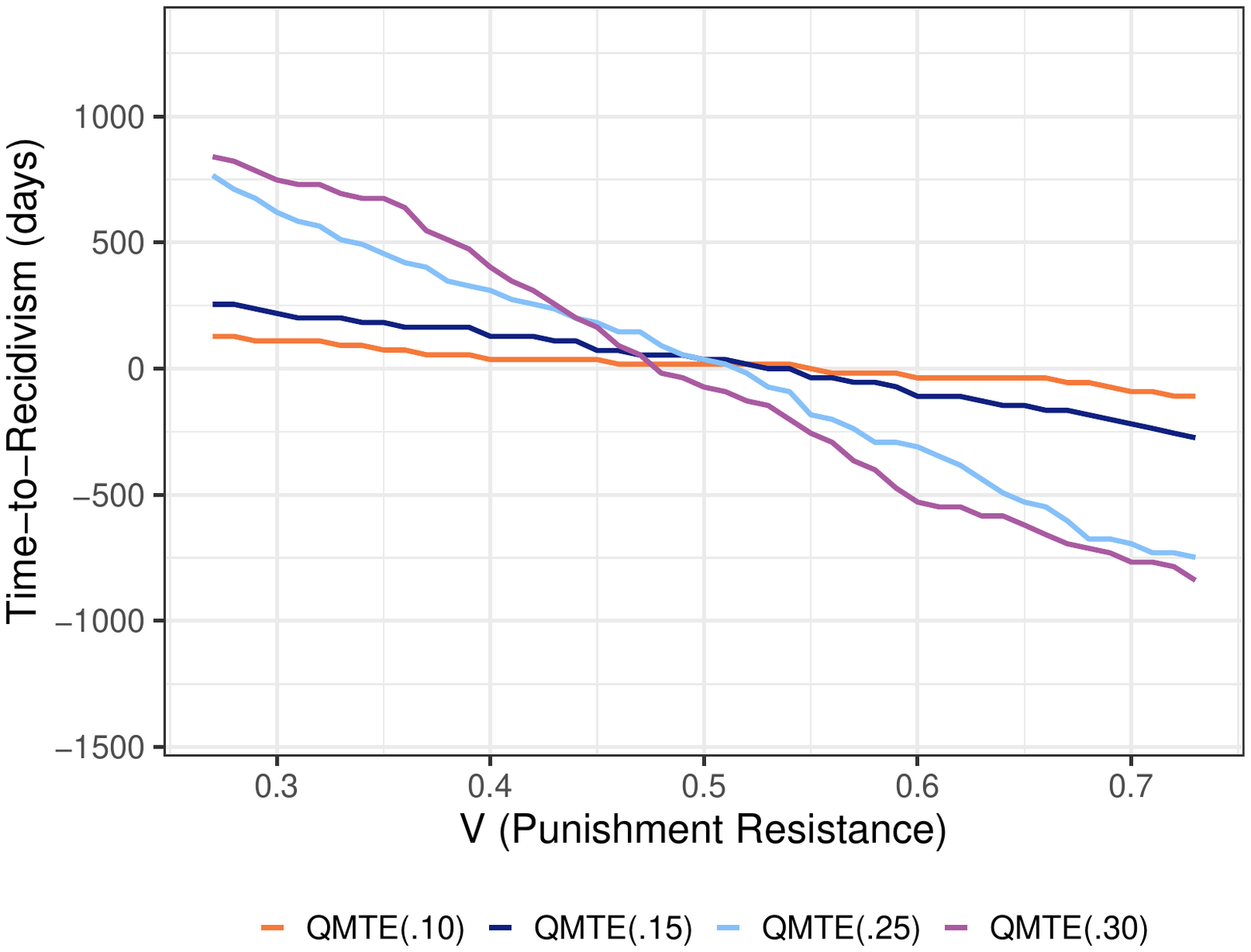}
						\caption{$QMTE\left(\tau,\cdot\right)$ for $\tau \in \left\lbrace .10, .15, .25, .30 \right\rbrace$}
						\label{QMTE10-30Processing}
					\end{subfigure}
					\hfill
					\begin{subfigure}[t]{0.47\textwidth}
						\begin{center}
							\includegraphics[width = \textwidth]{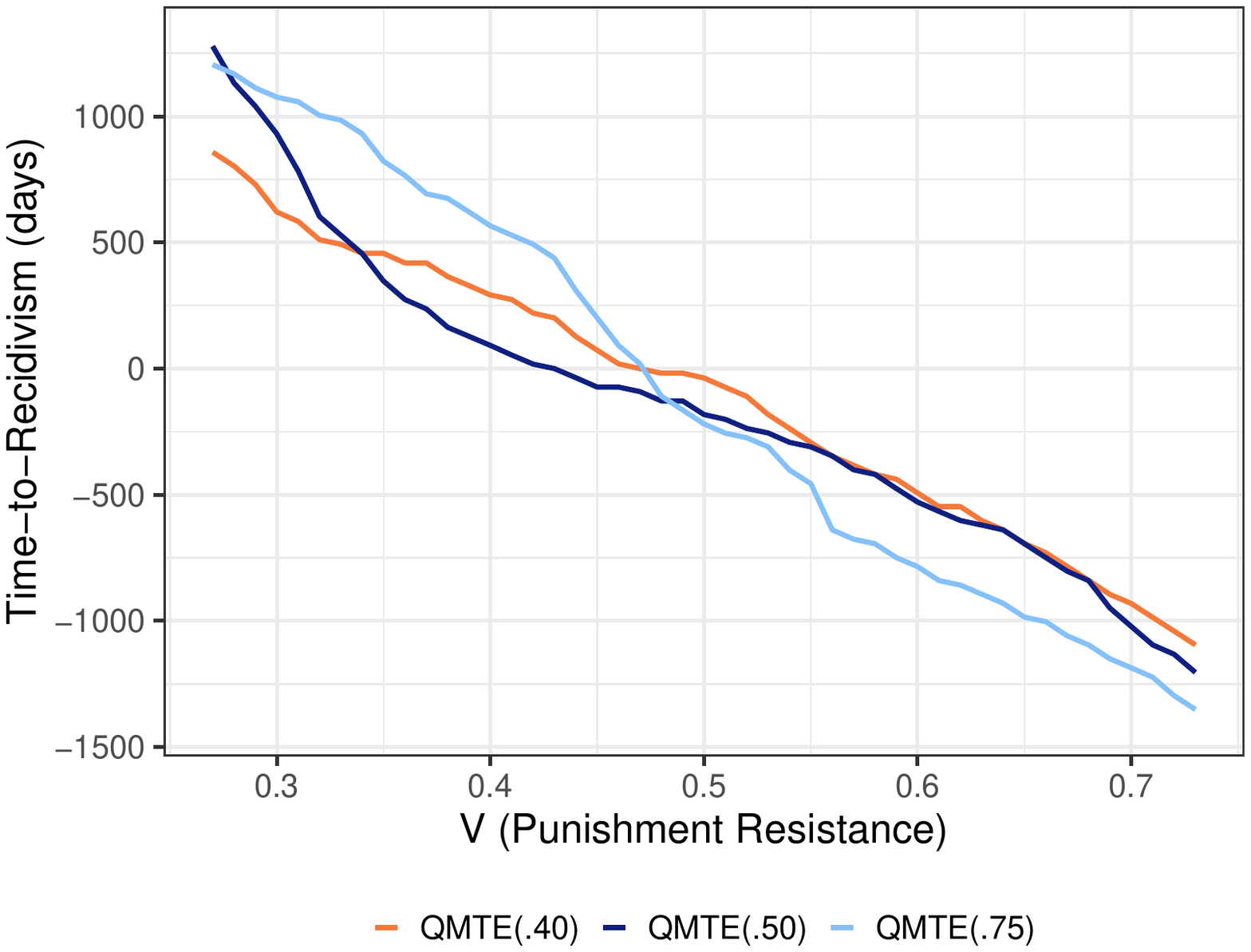}
							\caption{$QMTE\left(\tau,\cdot\right)$ for $\tau \in \left\lbrace .40, .50, .75, \right\rbrace$}
							\label{QMTE40-75-RMTEProcessing}
						\end{center}
					\end{subfigure}
					\caption{$QMTE\left(\tau,\cdot\right)$ for $\tau \in \left\lbrace .10, .15, .25, .30, .40, .50, .75 \right\rbrace$}
					\label{FigQMTE-RMTEProcessing}
				\end{center}
				\justifying
				\scriptsize{Notes: Solid lines are the point estimates for the average $QMTE\left(\tau,\cdot\right)$ functions indicated in the legend of each subfigure. These results are based on the estimation procedure explained in Appendix \ref{AppCaseProcessingTimeEstimation}.
					
				}
			\end{figure}
			
			Consequently, the overall conclusion of this appendix is that our main results are robust to the inclusion of case-processing time as an additional covariate.
			
			\subsection{Assessing Main Identification Assumptions when Controlling for Case-processing Time}\label{AppCaseProcessingTimeIdentification}
			
			In this appendix, we discuss the plausibility of our two main identification assumptions --- Random Assignment (Assumption \ref{AsIndependence}) and Random Censoring (Assumption \ref{AsCensoring}) --- when we control for case-processing time as an extra covariate. 
			
			To analyze the validity of Assumption \ref{AsIndependence}, we reproduce Table \ref{TabTestAssignment} after conditioning on case processing time as an additional covariate. Table \ref{TabTestAssignmentProcessing} reports the estimated coefficients of regressing the treatment variable (first row) and the instrumental variable (second row) on the four excluded covariates indicated in the columns, the censoring variable, case-processing time and court-district fixed effects. The standard errors are reported in parentheses and are clustered at the court district level. We find that our four excluded covariates are correlated with the final ruling in each case (treatment variable), but only one is correlated with the punishment rate (instrumental variable) after controlling for case-processing time. Even in this case, this excluded covariate (``red-handed'') has a much smaller coefficient (in magnitude) in the second regression. These findings indirectly support Assumption \ref{AsIndependence}.
			
			\begin{table}[!htb]
				\centering
				\caption{{Testing the Random Assignment Assumption Given Case-processing Time}} \label{TabTestAssignmentProcessing}
				\begin{lrbox}{\tablebox}
					\begin{tabular}{ccccccc}
						\hline \hline
						& Male & Public Def. & Theft & Red-Handed & C & T \\ \hline
						Final & 0.035*** & 0.075*** & 0.106*** & 0.044*** & -0.010** & -0.105*** \\
						Ruling $(D)$ & (0.008) & (0.019) & (0.013) & (0.009) & (0.004) & (0.003) \\ \hline
						Punishment & 0.000 & 0.001 & 0.001 & -0.006** & -0.002 & -0.005*** \\
						Rate $(Z)$ & (0.002) & (0.008) & (0.005) & (0.003) & (0.001) & (0.001) \\ \hline
					\end{tabular}
				\end{lrbox}
				\usebox{\tablebox}\\
				\justifying
				\hspace{-.6cm}\scriptsize{Note: The first row reports the estimated coefficients of regressing the treatment variable ($D = $``punished according to the final ruling in the case'') on the binary covariates indicated in the columns, the censoring variable ($C$ is measured in years for readability), the case-processing time ($T$ is measured in years for readability) and court-district fixed effects. The second row reports the estimated coefficients of regressing the instrumental variable ($Z = $``trial judge's punishment rate'') on the binary covariates indicated in the columns, the censoring variable ($C$ is measured in years for readability), the case-processing time ($T$ is measured in years for readability) and court-district fixed effects. The standard errors are reported in parentheses and are clustered at the court district level. The covariates indicate whether the defendant has a male name, whether the defendant used a public defender instead of hiring a private lawyer, whether the defendant was accused of theft, and whether the defendant was caught red-handed when committing a crime (``in flagrante delicto'').}
			\end{table}
			
			Lastly, we discuss the plausibility of Random Censoring (Assumption \ref{AsCensoring}) when we control for case-processing time as an extra covariate. When we control for case-processing time, Assumption \ref{AsCensoring} implies that potential recidivism is stationary, given the time the presiding judge spent analyzing the case. If we assume that case-processing time proxies for a case's complexity, adding this covariate increases our confidence in the Random Censoring condition.
			
			Despite this argument, we do not believe this approach is better than the main text's approach for two reasons. First, conditioning on court districts is sufficient to ensure the validity of Assumption \ref{AsIndependence} (Random Assignment). Second, case-processing time might be affected by each judge's punishment rate, as shown by the last column in Figure \ref{TabTestAssignmentProcessing}. This dependence structure may cause identification issues when conditioning on case-processing time as suggested by the significant coefficient of the ``red-handed'' covariate (Second Row of Table \ref{TabTestAssignmentProcessing}).
			
		}
		
		\newpage
		
		{
			\section{Identifying Effects on True Criminal Behavior}\label{AppLeeBounds}
			
			\setcounter{table}{0}
			\renewcommand\thetable{K.\arabic{table}}
			
			\setcounter{figure}{0}
			\renewcommand\thefigure{K.\arabic{figure}}
			
			\setcounter{equation}{0}
			\renewcommand\theequation{K.\arabic{equation}}
			
			\setcounter{theorem}{0}
			\renewcommand\thetheorem{K.\arabic{theorem}}
			
			\setcounter{assumption}{0}
			\renewcommand\theassumption{K.\arabic{assumption}}
			
			\setcounter{proposition}{0}
			\renewcommand\theproposition{K.\arabic{proposition}}
			
			\setcounter{corollary}{0}
			\renewcommand\thecorollary{K.\arabic{corollary}}
			
			In the main text, we are interested in the impact of judicial punishment, $D$, on time-to-recidivism, $Y^{*}$, where recidivism is measured as being prosecuted for a new crime. In this appendix, we discuss how to identify the effect of judicial punishment, $D$, on time to committing a new crime. In this setting, this new crime may or may not be observed by the police and, consequently, may or may not result in a judicial case. Such an analysis is relevant because, from society's perspective, any crime is costly, regardless of whether it results in prosecution.
			
			To answer this new question, we need to extend our data-generating process to encompass the occurrence of new crimes and the actions of the police and the district attorney. First, define $\tilde{Y}\left(d\right)$ as the potential time to commit a new crime. This variable captures, for each treatment status, the length of time between the initial case's sentencing date and the occurrence of a new crime committed by the defendant. The occurrence of this crime is known to the defendant, but it may not be known to the judicial system. Second, define $S\left(d\right)$ as the potential indicator that equals one when the defendant is captured by the police and the district attorney decides to charge the defendant. Lastly, note that the definition of time-to-recidivism in the main text implies that $Y^{*}\left(d\right) = S\left(d\right) \cdot \tilde{Y}\left(d\right)$.\footnote{Implicitly, we assume that the police and the district attorney act immediately and that defendants are caught by the police and prosecuted for a crime on the same day. Including investigation time is well beyond the scope of this manuscript.}
			
			We will identify the distributional marginal treatment effect of judicial punishment on time-to-committing a new crime for the defendants who commit a new crime regardless of treatment status:
			\begin{align*}
				& DMTE_{OO}\left(y,v\right) \\
				& \hspace{20pt} \coloneqq \mathbb{P}\left[\left. \tilde{Y}\left(1\right) \leq y \right\vert V = v, S\left(0\right) = 1, S\left(1\right) = 1 \right] - \mathbb{P}\left[\left. \tilde{Y}\left(0\right) \leq y \right\vert V = v, S\left(0\right) = 1, S\left(1\right) = 1 \right]
			\end{align*}
			for $y \leq \gamma_{C}$, where $\gamma_{C}$ is defined in Proposition \ref{PropDMTR}. To do so, we need to follow three steps.
			
			First, we identify the distribution of $\left. \left( \tilde{Y}\left(d\right), S\left(d\right) = 1\right) \right\vert V$. To do so, note that $Y = \min \left\lbrace Y^{*}, C \right\rbrace \neq C$ if and only if $S = 1$, and that $Y^{*}\left(d\right) = \tilde{Y}\left(d\right)$ if and only if $S = 1$. This result implies that the methods in the main text identify the distribution of $\left. \left( Y^{*}\left(d\right), S\left(d\right) \right) \right\vert V$ up to $\gamma_{C}$ and, in particular, the distribution of $\left. \left( \tilde{Y}\left(d\right), S\left(d\right) = 1 \right) \right\vert V$ up to $\gamma_{C}$.
			
			Second, we identify the distribution of $\left. S\left(d\right) \right\vert V$. To do so, we can use the methods proposed by \cite{Heckman2006} directly.
			
			To identify the function $DMTE_{OO}\left(y,v\right)$, we can combine the distributions of $\left. \left( \tilde{Y}\left(d\right), S\left(d\right) = 1\right) \right\vert V$ and $\left. S\left(d\right) \right\vert V$ through the methods proposed by \citet[Proposition 1]{Bartalotti2021}.\footnote{Alternatively, \citet[p. 576]{Bartalotti2021} show that we can identify the unconditional distributional marginal treatment effect of judicial punishment on time-to-committing a new crime --- i.e, $\mathbb{P}\left[\left. \tilde{Y}\left(1\right) \leq y \right\vert V = v \right] - \mathbb{P}\left[\left. \tilde{Y}\left(0\right) \leq y \right\vert V = v \right]$ --- if we are willing to impose that the police's and the district attorney's behavior are independent from potential time-to-commit a new crime, i.e., $\left. \left(S\left(0\right), S\left(1\right) \right) \independent \left(\tilde{Y}\left(0\right), \tilde{Y}\left(1\right), V\right) \right\vert C.$} To do so, we need to strengthen our Random Assignment Assumption to $$\left. Z \independent \left(\tilde{Y}\left(0\right), \tilde{Y}\left(1\right), S\left(0\right), S\left(1\right), V\right) \right\vert C.$$ Under this stronger random assignment restriction and Assumptions \ref{AsRank}-\ref{AsCensoring}, the methods proposed by \cite{Bartalotti2021} partially identify the function $DMTE_{OO}\left(y,v\right)$ up to $y = \gamma_{C}$. We can tighten these bounds \citep[Proposition 2]{Bartalotti2021} if we are willing to impose that the police observe crimes committed by convicted defendants more easily than they observe crimes committed by acquitted defendants, i.e., $S\left(1\right) \geq S\left(0\right)$. Lastly, we can point-identify the function $DMTE_{OO}\left(y,v\right)$ \citep[p. 576]{Bartalotti2021} if we are willing to impose that the police observe crimes committed by convicted defendants as easily as they observe crimes committed by acquitted defendants, i.e., $S\left(1\right) = S\left(0\right)$. 
			
		}

\end{document}